\RequirePackage{fix-cm}
\documentclass[twocolumn,epjc3]{svjour3}
\RequirePackage{graphicx}
\usepackage{epsfig}
\usepackage{dcolumn}
\usepackage{array}
\usepackage{afterpage}
\usepackage{bm}
\usepackage{fancyhdr}
\usepackage{longtable}
\usepackage{multirow}
\usepackage{float}
\usepackage{rotating}
\usepackage{hyperref}
\usepackage{afterpage}  
\usepackage{color}
\usepackage{tabularx}
\usepackage{adjustbox}

\journalname{Eur. Phys. J. C}
\setlongtables
\usepackage{placeins}
\usepackage[dvipsnames]{xcolor}
\definecolor{myblue}{RGB}{0, 0, 255}

\newcommand\g{$\gamma$}
\newcommand\be{$\beta$}

\newcommand\T{\rule{0pt}{2.6ex}}       % Top strut
\newcommand\B{\rule[-1.2ex]{0pt}{0pt}} % Bottom strut

\begin{document}
%\setcounter{page}{1}
%%Proposal by Vivian (07/12/2021): new title
% Improving Fission Product Nuclear Structure and Decay Data Relevant to Applications: Part I: Reactor Decay Heat %(and anti-neutrino spectra??) % calculations

\title{
%\qquad \\ \qquad \\ \qquad \\  \qquad \\  \qquad \\ \qquad \\ 
Improving Fission-product Decay Data for Reactor Applications: Part I - Decay Heat \\ %(Part 2: Reactor Anti-neutrinos) % calculations
%Potentially Suitable Fission-product Radionuclides for Study by Means of Total Absorption Gamma-ray Spectroscopy (TAGS/TAS), \g~Singles, and \g–\g~Coincidence as a Combined Aid to Improve Nuclear Structure and Decay Data 
%Alan, 5 Dec 2022 - I have added Karny and Rasco, plus Karny's Warsaw address.  I have added both of them, although they were never members of the TAGS-related IAEA cosultants' meetings (but there again neither was Giot). 
}

\author{ A.L. Nichols\thanksref{1,2}  \and P. Dimitriou\thanksref{e1,4} \and A. Algora\thanksref{3,5} \and M. Fallot\thanksref{6} \and L. Giot\thanksref{6} \and \\ F.G. Kondev\thanksref{7} \and T. Yoshida\thanksref{8} \and M. Karny\thanksref{9} \and G. Mukherjee\thanksref{10} \and B.C. Rasco\thanksref{11} \and K.P. Rykaczewski\thanksref{11} \and A.A. Sonzogni\thanksref{12} \and J.L. Tain\thanksref{3}}

\thankstext{e1}{e-mail: P.Dimitriou@iaea.org}

\institute{Department of Physics, University of Surrey,
Guildford, GU2 7XH, Surrey, UK \label{1} \and
Manipal Academy of Higher Education, Manipal,
Karnataka 576104, India\label{2} \and Nuclear Data Section, International Atomic Energy Agency, A-1400 Vienna, Austria\label{4} \and Instituto de Fisica Corpuscular (IFIC), CSIC-Universidad de Valencia, 46071 Valencia, Spain\label{3} \and Institute of Nuclear Research (ATOMKI), Debrecen, Hungary\label{5} \and Laboratoire Subatech, University of Nantes, CNRS/IN2P3, Institut Mines Telecom Atlantique, 44307 Nantes, France\label{6} \and Physics Division, Argonne National Laboratory, Lemont, Illinois 60439, USA\label{7} \and Institute of Innovative Research, Tokyo Institute of Technology, Tokyo 152-8550, Japan\label{8} \and Physics Faculty, University of Warsaw, PL-02-093, Warsaw, Poland\label{9} \and Variable Energy Cyclotron Centre, Kolkata 700064, India\label{10} \and Physics Division, Oak Ridge National Laboratory, Oak Ridge, Tennessee 37831, USA\label{11} \and 
 Nuclear Science and Technology Department, Building 817, Brookhaven National Laboratory, Upton, NY 19973-5000, USA\label{12}}

\date{Received: date / Revised version: date}
\maketitle

\begin{abstract}
%%V: needs re-writing. 
Effort has been expended to assess the relative merits of undertaking further decay-data measurements of the main fission-product contributors to the decay heat of neutron-irradiated fissile fuel and related actinides by means of Total Absorption Gamma-ray Spectroscopy (TAGS/TAS) and Discrete Gamma-ray Spectroscopy (DGS). This review has been carried out following similar  
%in the latter half of 2017 and early 2018, following on from similar 
work performed under the auspices of OECD/WPEC-Subgroup 25 (2005-2007) and the International Atomic Energy Agency (2010, 2014), and various highly relevant TAGS measurements completed as a consequence of such assessments. We present our recommendations for new decay-data evaluations, along with possible requirements for total absorption and discrete high-resolution gamma-ray spectroscopy studies that cover approximately 120 fission products and various isomeric states.
\end{abstract}

\section{Introduction}\label{intro}
% Changes by Vivian based on suggestions from Tadashi Y. (07/12/2021):
% I was wondering if we should not define slightly better decay heat here, it is implicit anyway (AA, 07/06/22), something like what I included in blue.

Decay heat is generally defined as the energy released from the direct and indirect fission reaction products, neglecting the contribution of antineutrinos that 
%are not released} 
% modified by [TY:20220919]
do not deposit any energy in the reactor core.
%ALN to Vivian, 10 Oct 2022: do antineutrinos deposit any energy at all anywhere, or do they pass through it all without any form of deposition?
The main contributors to reactor decay heat are the fission products that accumulate during reactor operation, followed by the actinides generated via neutron capture along with their heavy-element decay products 
%arise from immediate fission (based on fission-product yields and their decay data), delayed-neutron fission, spontaneous fission, heavy elements and actinides, 
and various activation products \cite{Tobias1980,Nichols2002}. Such decay heat is extremely significant at short and intermediate cooling times after a planned or accidental shutdown of a nuclear reactor, and is most immediately dependent on the resulting inventory of fission products, actinides and their decay products as a function of the time after shutdown.
%, and (b) fission induced within a few minutes by delayed neutrons from specific short-lived fission products following reactor shutdown. 
Accurate estimates of the resulting decay heat are required in safety assessments of all types of reactor and fuel-handling plant, the storage of spent fuel, the transport of fuel-storage flasks, and the intermediate-term management of any resulting radioactive waste. Most of the above phenomena are also associated with the related emission of antineutrinos that provide an important means of monitoring reactor operations in a benign manner, as well as defining the constraining fundamental physics quantities that determine their oscillation properties.   

Both the actinide and fission-product inventories of neutron-irradiated fuel are calculated for known operational conditions and all subsequent cooling periods. These inventory data are used in conjunction with radionuclidic half-lives, and all relevant heavy-particle, light-particle and electromagnetic radiation characteristics to determine the
total energy release rates for heavy particles, light particles and electromagnetic radiations. 
% By Vivian (07/12/2021): Tadashi proposes we define heavy particles; see addition below
Heavy particles are primarily identified with alpha particles, light particles are defined as \be$^{-}$, \be$^{+}$, and internal-conversion and Auger electrons, and electromagnetic radiation consists of \g, X-rays, annihilation radiation and internal bremsstrahlung. Along with neutron-induced cross sections and fission yields, the radioactive decay data constitute key input to the summation calculations used to determine the release of decay heat during reactor operation and as a function of time after the termination of fuel irradiation within the reactor core \cite{Tobias1980,Nichols2002,Tobias1989}. These calculations require the inclusion of mean $\alpha$, \be~and \g~energies derived normally from the decay schemes of a significant number of radionuclides dominated by fission products at early and intermediate cooling times, and the results need to be compared with experimental decay-heat benchmarks. However, the determination of \be$^{-}$ emission probabilities has long been problematic in decay-scheme studies. Although \g-ray emission probabilities and internal conversion coefficients can be used to derive \be$^{-}$ feeding to daughter nuclear levels, experiments based on Ge detectors suffer from low efficiency for the detection and quantification of high-energy \g~rays above $\sim$ 1.5 MeV that undermines such an approach. 
%all forms of singles-based Ge detector possess low intrinsic efficiency for the detection of high-energy \g~rays above $\sim$ 1.5 MeV that undermines such an approach. 
%Furthermore, 
The determination of direct \be$^{-}$ decay to the ground state of the daughter nucleus can also pose serious problems. Hardy $et~al.$~proposed that $\sim$ 20$\%$ of the true \g-ray intensity above 1.7 MeV for a fictional radionuclide (Pandemonium) may remain undetected \cite{Hardy1977}, impacting significantly on the use of \g-ray singles data to calculate \be$^{-}$ transitions by means of gamma population-depopulation balances of the proposed nuclear levels. This problem also exists in experiments based on coincident data even in the case of relatively high-efficiency Ge detector arrays \cite{Hu2000,Algora2003}. 
%Vivian 12 Dec 2022: adding extra comment from CR as formulated by Alan
Furthermore, defining the nature of each individual \be$^{-}$ transition can result in differing energy distributions  to create distinct total \be$^{-}$ energy spectra. Such unknown beta shape factors may impact on the average \be$^{-}$ energy of many of the radionuclides under consideration.
% Vivian to Alejandro, 7 Nov 2022: please add references if necessary.
Total Absorption Gamma-ray Spectroscopy (TAGS) measurements can overcome these difficulties in order to provide the necessary mean beta and gamma decay energies for more comprehensive and satisfactory decay-heat calculations \cite{Greenwood1996,Greenwood1997,Algora2010}.

% By Vivian (07/12/2021): Based on Tadashi's feedback, beta-delayed neutron emission is comparable to fission product DH only up to 10s after shut-down for 239U-239Np. Need to see how this is compatible with below text. This issue needs to be clarified in the meeting.
While fission-product and actinide decay are the most significant contributors to decay heat, caution is required with respect to the first $\sim$ 200 s after reactor shutdown as other effects play immediate initial roles, such as a brief plant-driven period of continued thermal impact along with
%Short-term decay heat up to 200 s after reactor shutdown is primarily identified with 
declining fission multiplication driven by delayed-neutron precursors \cite{Hill2011}. 
%This impact of delayed-neutron emissions at early calculated cooling times of 10 and 100 s has been noted as a warning in the assessment of potential Pandemonium.  
% By Vivian (07/12/2021): Tadashi suggests removing alphas
Beyond 200 s after reactor shutdown, 
%$\alpha$-, 
%Vivian, 14 Nov 2022: modified slightly following AA comment.
%Alan, 14 Nov 2022: modified even more following further analysis and thought - quite simply, I used an incorrect word for the actinides ("dominate") at the wrong cooling times. The word is spot on for the fission products.
 \be- and \g-decay processes of the fission products play a dominant role leading up to approximately 10$^6$ s (12 days). While both $^{239}$U and $^{239}$Np make noteworthy contributions up to one hour and $<$ 10$^6$ s cooling times, respectively, other actinides such as $^{238}$Pu, $^{241}$Am and $^{242,244}$Cm take their place in this respect at much longer times.  
 % AA: There is also Tadashi´s work that inspired us from 1999, I wonder if we should include it. ALN: this work is included elsewhere.
 Following on from assessment studies discussed and undertaken in 2005-2007, 2009/2010 and late 2014 \cite{Yoshida2007,Nichols2009,Gupta2010,Dimitriou2015} and various highly relevant TAGS measurements completed as a consequence of such assessments, further efforts have now been made to re-assess the requirements for additional measurements that effectively embrace singles \g-ray spectroscopy, \g-\g~coincidence and TAGS.  One extended aim would be for TAGS to quantify the degree of Pandemonium in particular fission products through the existence of previously unknown and ill-defined nuclear levels populated in the \be$^{-}$ decay process. Such studies would also underline the requirements for more extensive singles \g-ray spectroscopy and \g-\g~coincidence techniques with appropriate types of detector by which to recognise and quantify the higher-energy gamma emissions in order to improve decay-heat assessments, as well as dosimetry and shielding calculations. 

 %Vivian to Alan, 21 Nov 2022: please check below added paragraph.
 %ALN, 21 Nov 2022: paragraph below heavily edited, 21 Nov 2022.
 Several exploratory studies have been carried out recently in order to improve specific \be$^-$ decay data as an aid in undertaking more accurate and precise antineutrino spectral calculations~\cite{Sonzogni2015,Sonzogni2016,Sonzogni2017,Sonzogni2018,Hayes2018,Estienne2019}. Recommendations have been made to perform TAGS measurements that differ in priority to current decay-heat requirements due to the fact that such antineutrino calculations are more relevant and sensitive to the nature of the shorter-lived fission products. While work is underway to extend this effort towards the determination of the antineutrino spectral requirements for \g-ray spectroscopy, \g-\g~coincidence and TAGS measurements, these particular assessments are beyond the scope of the current publication and will be the subject of a future article.

\section{Total Absorption Gamma-Ray Spectroscopy (TAGS/TAS)}
%As already outlined above, 
Nuclear structure and decay data are important facets and spectral aspects of nuclear physics. These properties and associated parameters aid extensively in the characterisation of the nucleus, and provide definitive descriptive features of all nuclei for direct adoption in a wide range of nuclear applications that include the safe control and normal operation of power reactor systems and the detection of any clandestine procedures, as well as extending our basic understanding of the more exotic 
features of nuclear physics. 
%Beta-strength functions 
% AA: Beta-strength is better used in this context to validate nuclear models
% we should decide if we prefer beta decay probabilities of beta feedings in 
% the text
The emission probabilities of beta-decay
are important in this respect from the point of view of quantifying the mean beta (most frequently the main contributor to the light-particle component of the decay process) and mean gamma decay heat (most frequently the main contributor to the electromagnetic component), along with the antineutrino spectral signatures that can potentially assist greatly in the detection of illegal procedures undertaken to produce weapons-grade materials.

A major problem in both decay-heat and antineutrino studies is determining both the \be$^{-}$ and \g~feeding with good accuracy and confidence, which has in the recent past relied most frequently on direct experimental studies by means of various forms and arrays of Ge detector. While these detector systems are entirely suitable for multiple coincidence studies to define and resolve nuclear levels and decay schemes, they have long been known to possess limited \g-ray detection efficiency that falls dramatically with increase in \g~energy particularly above 1.5 MeV, as noted in Section 1. Under these circumstances, an entirely different method of spectral study and analysis has been considered and developed, based on the attempted total absorption of all \g-ray emissions to determine full \be$^{-}$ feeding to all individual nuclear levels \cite{Duke1970,Johansen1973,Rubio2005,Rubio2017,Algora2018}. 
% AA: we should cite here the very first articles of total absorption. I will add them if you agree
Although complete 100$\%$ efficiency is not fully feasible, scintillator \g~detector systems with geometries as close to 4$\pi$ as possible possess the capabilities to determine \be$^{-}$ feeding up to an excitation energy of the daughter nucleus that could match the parent Q-value. While under development in the early 1970s, TAGS/TAS involved the adoption of two large cylindrical NaI detectors, before moving in the 1990s to one large NaI-well counter to host the radioactive source of interest along with Si detectors \cite{Greenwood1992,Rubio2005,Rubio2017}. Other larger NaI(Tl) crystals have subsequently been used with a high degree of success in conjunction with various other smaller NaI, Ge-planar and thin Si $\Delta$E positron detectors to determine specific forms of coincidence, for example, GSI TAS (GSI) \cite{Karny1997} and Lucrecia TAS (ISOLDE) \cite{Rubio2005}. 
These studies have been extended to significant arrays of Ge detectors in a compact geometry, such as Clustercube 
%EUROBALL
% AA: I will add some references here of the cluster cube results
%Alan, 5 Dec 2022 - above: identified GSI TAS (GSI) with requested Karny1997 reference, and Lucrecia TAS with only Rubio/Algora reference.
and Gammasphere to study a number of more complex beta spectra, providing opportunities to compare the results obtained by means of different techniques \cite{Hu2000,Algora2003}. 
% ,Rubio2017,Hu1998,Hu1999. Vivian, 10 Nov 2022: most important Refs are the two left in the text.
% AA: I will addd here as well some references, including 150Ho decay, and around 100Sn by the Polish group at MassSep (GSI). 

%leading on to the development of segmented detectors, such as the Lucrecia spectrometer at ISOLODE, and the introduction of multi-array BaF$_{2}$ and LaBr$_{3}$(Ce) crystals and NaI(Tl) detectors \cite{Rubio2017}. 

%ALN:
%Might be nice to have a combined photo and figurative drawing/representation of Lucrecia here?  ALN
% AA: yes, we can find one
%Alan, 5 Dec 2022 - below: Inserted Rasco2015b reference near end of paragraph below. Was already cited elsewhere, and now becomes ref. [28]
Recent measurements related to power reactor operations and antineutrino physics have been performed at the IGISOL facility in Jyv\"{y}skyl\"{a}, Finland. These studies have involved the use of the ion-guide technique in which many of the refractory fission products of interest have been successfully generated, and high isotopic purity has been achieved by means of the JYFL Penning trap operated as a high-resolution mass separator \cite{Algora2018}. Accurate and valid analyses of the spectral data have also been an important feature of this work, based on a sound starting knowledge of the level scheme up to a reasonable excitation energy. As outlined in Refs. \cite{Rubio2017,Algora2021,Rasco2015b}, these spectral analyses take into account non-linearity of the light output in the NaI(Tl) scintillator and pile-up in the electronic signals. 

%A segmented detector array has been designed and assembled as the Modular Total Absorption Spectrometer (MTAS) that consists of nineteen NaI(Tl) hexagonal modules arranged as a honeycomb-like structure. Radioactive sources are delivered by means of a moving tape system to the central module which also contains two Si-strip detectors for \be$^{-}$ coincidence studies. Initial measurements on a wide range of seventy fission products from $^{238}$U fast fission have focused on preliminary TAGS studies of $^{86,89}$Br, $^{89}$Kr, $^{137}$I, $^{139}$Xe, $^{137}$Cs, $^{142}$Ba and $^{142}$La ~\cite{Wolinska2014,Fijalkowska2014b,Karny2016}.
%ALN:Would be nice to have a combined photo and figurative drawing/representation of MTAS here.  
Exploiting the experience gained, various spectrometers have been commissioned to better absorb and quantify all gamma emissions. These studies have extended our understanding of the nucleus with respect to \be~decay and the weak interaction, and so aided in the development of related modelling calculations over the full chart of the nuclides. 
Improvements have also been made in the handling of the radioactive species to obtain clean beams for spectral studies by means of various forms of ancillary detector located close to the main counting area of the multi-array of segmented 4$\pi$ NaI(Tl) scintillators. A key aspect of this work is "segmentation" whereby additional information on the multiplicity of the gamma cascades can be used to test assumptions made during the analyses. Rocinante is an example of such a detector system, which is composed of twelve optically-isolated BaF$_{2}$ crystals assembled in a cylindrical geometry and operated in conjunction with an approximately 30$\%$ efficient Si detector or plastic detectors, together with the ion-guide source of the IGISOL mass separator and JYFLTRAP Penning trap at the University of Jyv\"{y}skyl\"{a} \cite{Algora2014,Valencia2014,Valencia2017}. 
A similar segmented arrangement of sixteen modular NaI(Tl) scintillators in a combined cuboid form has been developed as the DTAS (Decay Total Absorption Spectrometer) for DESPEC (DEcay SPECtroscopy)
%and is located  - Vivian, 14 Nov 2022: modified following AA remark
at FAIR (Facility of Antiproton and Ion Research, near Darmstadt, Germany) \cite{Tain2015,Guadilla2016,Guadilla2017a}. These detector systems
have been used to study various beta-decaying radionuclides of relevance to decay-heat assessments and the prediction of antineutrino spectra that arise from reactor operations, along with more basic nuclear structure research in facilities such as RIKEN and GSI(FAIR).  
%ALN:
%Would be nice to have combined two separate photos along with figurative drawings/representations of Rocinante and DTAS here.  ALN
%Any other TAGS facilities and equipment that should be mentioned here? – any new equipment/experimental ideas? What is VTAS (mentioned by Alejandro Algora at February 2018 meeting – does acronym just mean “Valencia TAS”)? Uses BaF2 crystals – further development of Rocinante, or totally different?  Also 93Rb is mentioned in the VTAS list within the Powerpoint presentation, but no visible data for this nuclide elsewhere (??)
%Alejandro: Dear Alan, Rocinante and VTAS is the same. I will look for photos. 93Rb is being analyzed by the Nantes group. We can check the status of this analysis. 

A segmented detector array has been designed, developed, and assembled as the Modular Total Absorption Spectrometer (MTAS) at the Holifield Radioactive Ion Facility of the Oak Ridge National Laboratory. MTAS consists of nineteen NaI(Tl) hexagonal modules arranged as a honeycomb-like structure. Radioactive sources are delivered by means of a moving-tape system to the central module which also contains two Si-strip detectors for \be$^{-}$ coincidence studies. Initial measurements of seventy fission products from $^{238}$U fast fission have focused on TAGS studies of $^{86,89}$Br, $^{89,90}$Kr, $^{88,89,90,90m,92}$Rb, $^{96}$Y, $^{98}$Nb, $^{137}$I, $^{137,139}$Xe, $^{142}$Cs, $^{142}$Ba and $^{142}$La ~\cite{Fijalkowska2014b,Wolinska2014,Karny2016,Rasco2016,Rasco2017a,Shuai2022,Rasco2022}.
%Alan, 5 Dec 2022 - below: {Laminack2022} reference lacks full list of authors, and has only been defined as "submitted to Phys. Rev. C (2022)". Requires more detail, etc.  Originates from Argonne National Lab?
%Vivian, 6 Dec 2022: have added a reference provided to me before midnight 5 Dec
%Alan, 7 Dec 2022 - have deleted 137Cs and added 88Kr, 88,92Rb, 96Y, 98Nb, 137Xe and 142Cs. All required eight refs. have been cited.
%Vivian 13 Dec 2022: references were modified according to CR. HRIF at ORNL was added.
Measurements have also been performed at Argonne National Laboratory with isotopically pure beams, see for example~\cite{Laminack2022}. The SuN spectrometer is another example of a segmented TAS detector that was originally designed for reaction studies \cite{Simon2013}, and has also been used for decay studies in fragmentation facilities \cite{Spyrou2014,Spyrou2016,Gombas2021,Dombos2021}. A new segmented central module for MTAS has been provided by the University of Warsaw~\cite{Karny2019}, and the array assembled at the Facility for Rare Isotope Beams (FRIB), Michigan State University, East Lansing, USA. This improved detector system has been located at the FRIB Decay Station initiator (FDSi), and is operational~\cite{Rykaczewski2022}.
%ALN, 7 Dec 2022: Vivian - fragmentation decay of 82Se is of no interest to us. This stable isotope does not feature in Tables 6 and 7 for the very good reason that its role in terms of fragmentation is nothing to do with fission-product decay-heat benchmarks. I have recently  suggested that we abbreviate this sentence somehow, for example:  "This improved detector system has been located at the FRIB Decay Station initiator (FDSi), and is operational~\cite{Rykaczewski2022}."  And this is what I have now done above.
%ALN - 28 July 2022: above single-sentence ANL and SUN statements are lacking in detail. Reader left thinking "so what?".
\section{Analytical procedures and assessments}\label{decaydata}
Extensive inventory and decay-heat calculations have been performed with the UKAEA FISPACT-II code by Fleming and Sublet \cite{Fleming2015,Fleming2015a,Fleming2015b}.  %These UKAEA reports are readily accessible in the public domain from the Culham (CCFE) website via: 
%introductory web page - http://www.ccfe.ac.uk/tech_reports.aspx 
%relevant listing - http://www.ccfe.ac.uk/tech_reports.aspx?date=2015 
%relevant report of main decay-heat contributors, June 2015 - http://www.ccfe.ac.uk/assets/Documents/CCFE-R(15)28_S1.pdf 
All relevant nuclear data addressed in these studies have involved the main fission-product contributors from thermal and fast fission identified with all the fissile radionuclides to be found in various types of irradiated fuel (Table~\ref{tab:1}).
%five particular fissile radionuclides: 
\begin{table}
\caption{Irradiated fuel inventories and decay-heat calculations \cite {Fleming2015,Fleming2015a,Fleming2015b}. \label{tab:1}}
\begin{tabular}{ll} \hline \hline
thermal-neutron pulse  & $^{235}$U, $^{238}$Pu, $^{239}$Pu, $^{240}$Pu, \\
(0.0253 eV)    & $^{241}$Pu, $^{242}$Pu, $^{241}$Am, $^{242m}$Am, \\
    & $^{243}$Am, $^{243}$Cm, $^{245}$Cm  \\ \hline
fast-neutron pulse  & $^{232}$Th, $^{233}$U, $^{238}$U, $^{237}$Np \\
(400 or 500 keV)  &         \\ \hline
cooling times (s) & 10, 100, 1000, 5011 and 10000  \\ \hline
\end{tabular}
\end{table}

Selection criteria for the current assessment of Pandemonium include the following: identification and detailed inspection of the nuclear structure of fission products with $\ge 2\%$ contribution to the total decay heat, along with both their light-particle and electromagnetic components; comparison of the Q-values recommended in AME2020 \cite{AME2020} with the highest known energies of daughter nuclear levels that possess either unknown or commensurate spin-parity with the ground state of the parent fission-product; and assessment of possible Pandemonium within the decay schemes as currently defined. Under such circumstances, such judgements are inevitably subjective in nature and are best described as indicative,
%entirely subjective
with emphasis placed on the possible existence of previously undetected high-energy \g-ray emissions. The primary aim has been to identify radionuclides that may possess the potential for Pandemonium so as to formulate recommendations concerning the need for further quantification of individual fission-product decay schemes by means of Total Absorption Gamma-ray Spectroscopy (TAGS/TAS) and/or Discrete Gamma-ray Spectroscopy (DGS). Furthermore, improved quantification of the related nuclear structure and antineutrino emissions will also be inevitable consequences of such dedicated studies.
These assessments may either support TAGS measurements for the first time, or reveal the need for further additional TAGS along with improved \g-ray spectroscopy and \g-\g~coincidence studies that will be more fundamentally necessary if TAGS reveals the existence of significant Pandemonium. %Subjective
% AA: Even though they are subjective, I think we have to be careful with this
% otherwise this will create a problem with referees. 
%ALN to AA: in what way is it problematic? When it comes to decay-scheme data (and other forms of numerical data) and reality, it sometimes has to be resorted to - leaving zeros in place within a database is not always a satisfactory/acceptable option as input to some forms of plant calculation.
Such judgements have been made on the basis of what we know about the nuclear structure of each of almost 120 fission products chosen for consideration from the FISPACT-II decay-heat calculations for the thermal fission of $^{235}$U, $^{238,239,240,241,242}$Pu, $^{241}$Am, $^{243,245}$Cm, and fast fission of $^{232}$Th, $^{233,238}$U and $^{237}$Np.

%%V: we will not include Table 6 in the paper.
%ALN to V: I absolutely do not agree at all. I end up busting a gut over many hours for nothing recognizable. 
%(I must also assume that you include Table 7 in your above exclusion statement) Table 6 is one of the major features of this paper, along with what is now going on measurement-wise, and the promised new calculations by Muriel and Yoshida-san.  As for any concern about the length of this paper - that should also be a function of the willingness of the journal to publish such a paper (and possibly the attitude of the referee(s)). We should submit in full, and then react if we have to.
A notable feature of the various sets of fission products is that they exhibit a fair degree of commonality. What we observe is that the more dominant decay-heat fission products for $^{232}$Th, $^{233,238}$U and $^{237}$Np fast fission and $^{241,242}$Pu and $^{245}$Cm thermal fission are those that do not appear in such a manner in the equivalent exercise for $^{235}$U, $^{238,239,240}$Pu, $^{241}$Am and $^{243}$Cm thermal fission as a consequence of differences in the individual fission yields and the subsequent impact of their decay chains. The potential for Pandemonium within the known \be$^{-}$ decay of a particular fission product are based on the documented nuclear properties, Q-value of the \be$^{-}$ decay, and the possible existence of higher-energy nuclear levels that could be populated by 
%ill-defined 
\be$^{-}$ emissions and depopulated by as yet unknown high-energy \g~rays, both of which would impact on their contribution to their decay-heat components.  A further issue in such assessments is the perceived correctness and accuracy of the absolute intensity of \be$^{-}$ decay feeding directly to the ground state of the daughter, which could possibly impact the complete decay scheme greatly through the normalization of the beta intensities (i.e., all \be$^{-}$ emissions, \g~rays, antineutrino spectra, and additional features of the decay process). 

TAGS/TAS measurements can improve our understanding of incomplete and therefore erroneous decay schemes as formulated mainly by \g-ray spectroscopy identified with Ge-based detectors. 
Comments concerning TAGS/TAS and other \g-decay studies are given in Tables~\ref{tab:app1} and \ref{tab:app2} of~\ref{app}. 
%ALN to Vivian, 10 Oct 2022: This Appendix (or Appendices) has not been properly formulated yet - the material has been collected at the end of this file, but has not been labelled and laid-out as an Appendix/Appendices.
A total of almost 120 fission products have been assessed as significant contributors to the decay heat of various fissile materials used to determine and benchmark the residual post-irradiation decay heat for the fast fission of $^{232}$Th, $^{233,238}$U and $^{237}$Np, and the thermal fission of $^{235}$U, $^{238,239,240,241,242}$Pu, $^{241}$Am and $^{243,245}$Cm. A similar statement can also be made when considering predictive spectral calculations of the resulting antineutrinos emitted by operational reactor systems. References assigned Nuclear Science References (NSR) keynumbers that appear in Tables~\ref{tab:app1} and \ref{tab:app2} are fully listed in Table~\ref{tab:app3} of~\ref{app}. Highly-specific recommendations have been made as to whether additional discrete \g-ray spectroscopy and/or TAGS/TAS measurements are merited, as comprehensively summarized in Table~\ref{tab:5}. 
%
% Vivian 10 Dec: revision of Table 2 by Alan following Rasco feedback
\afterpage{
\begin{table*}
\caption{Summary assessments/re-assessments of fission-product decay data for decay-heat calculations: requests for specific measurements by means of Discrete Gamma-ray Spectroscopy (DGS) or Total Absorption Gamma-ray Spectroscopy (TAGS) denoted by $‘$y' $\rightarrow$ 'yes' ($‘$?' $\rightarrow$ may need to be assessed further at a later date), with priorities judged in terms of 1 high, 2 intermediate, 3 low, and -- satisfactory status and therefore not assigned a priority.}\label{tab:5}
\begin{center}
\begin{minipage}{0.36\linewidth}
%\strut\vspace*{-\baselineskip}\newline
\begin{tabular}{l|l|l|l|} \hline \hline\noalign{\smallskip}
Radionuclide & DGS & TAGS        & Priority \\ \hline
%\noalign{\smallskip}
33-As-84     	&	 y         	&	 y                  	&	 2        \\
34-Se-83     	&	 y         	&	 –                  	&	 3        \\
34-Se-85     	&	 y         	&	 y                  	&	 2        \\
34-Se-86     	&	 ?         	&	 y                  	&	 2?        \\
34-Se-87     	&	 ?         	&	 y                  	&	 2?        \\
35-Br-84     	&	 y         	&	 y                  	&	 2        \\
35-Br-86     	&	 y         	&	 –                  	&	 2        \\
35-Br-87     	&	 y         	&	 –                  	&	 2        \\
35-Br-88     	&	 y         	&	 –                  	&	 2        \\
35-Br-89     	&	 y         	&	 y                  	&	 2        \\
36-Kr-87     	&	 –         	&	 –                  	&	 –        \\
36-Kr-88     	&	 –         	&	 –                  	&	 –        \\
36-Kr-89     	&	 y         	&	 –                  	&	 2        \\
36-Kr-90     	&	 y         	&	 –                  	&	 2        \\
36-Kr-91     	&	 y         	&	 y                  	&	 2        \\
37-Rb-88     	&	 –         	&	 –                  	&	 –        \\
37-Rb-89     	&	 –         	&	 –                  	&	 –        \\
37-Rb-90     	&	 y         	&	 –                  	&	 2        \\
37-Rb-90m    	&	 y         	&	 –                  	&	 2        \\
37-Rb-91     	&	 y         	&	 –                  	&	 2        \\
37-Rb-92     	&	 y         	&	 –                  	&	 2        \\
37-Rb-93     	&	 y         	&	 –                  	&	 2        \\
38-Sr-91     	&	 y         	&	 –                  	&	 3        \\
38-Sr-92     	&	 –         	&	 y                  	&	 2        \\
38-Sr-93     	&	 –         	&	 y                  	&	 2        \\
38-Sr-94     	&	 –         	&	 y                  &	 2 or 3        \\
38-Sr-95     	&	 y         	&	 y                  	&	 2        \\
39-Y-92      	&	 –         	&	 –                  	&	 –       \\
39-Y-93      	&	 –         	&	 –                  	&	 –        \\
39-Y-94      	&	 –         	&	 –                  	&	 –        \\
39-Y-95      	&	 –         	&	 –                  &	 –        \\
39-Y-96      	&	 y         	&	 –                  	&	 2        \\
39-Y-96m     	&	 y         	&	 –                  	&	 2        \\
39-Y-97      	&	 y         	&	 y                  	&	 2        \\
40-Zr-97     	&	 ?         	&	 –                  	&	 3?        \\
40-Zr-98     	&	 y         	&	 y                  	&	 3        \\
40-Zr-99     	&	 y         	&	 y                   	&	 1        \\
40-Zr-100    	&	 y         	&	 –                  	&	 2        \\
41-Nb-98     	&	 y         	&	 y               	&	 1        \\
41-Nb-99     	&	 y         	&	 y                 	&	 1        \\
41-Nb-100    	&	 y         	&	 –                 	&	 1        \\
41-Nb-100m   	&	 y         	&	 –                 	&	 1        \\
41-Nb-101    	&	 y         	&	 –                        	&	 1        \\
41-Nb-102    	&	 y         	&	 –                     	&	 1        \\
41-Nb-104m   	&	 y         	&	 y                    &	 3?        \\
41-Nb-105    	&	 y         	&	 y         	&	 2   \\
42-Mo-101    	&	 –         	&	 –                     	&	 –        \\
42-Mo-102    	&	 y         	&	 –                   	&	 2        \\
42-Mo-103    	&	 y         	&	 y                      	&	 2        \\
42-Mo-104    	&	 y         	&	 y                      	&	 2        \\
42-Mo-105    	&	 y         	&	 y                 	&	 2        \\
42-Mo-106    	&	 y         	&	 y                 	&	 3        \\
42-Mo-107    	&	 y         	&	 y                         	&	 2        \\ 
43-Tc-101    	&	 –         	&	 –                 	&	 –        \\
43-Tc-102    	&	 y         	&	 –                    	&	   2      \\
43-Tc-103    	&	 –         	&	 –                 	&	 –        \\
43-Tc-104    	&	 y         	&	 –                   	&	 1        \\
43-Tc-105    	&	 y         	&	 –              	&	 1        \\
43-Tc-106    	&	 y         	&	 y                      	&	 1        \\
43-Tc-107    	&	 y         	&	 y                    &	 2   \\
%\noalign{\smallskip}
 \hline
\end{tabular}
\end{minipage}
\begin{minipage}{0.36\linewidth}
%\addtocounter{table}{-1}
%\caption{ -- Continued from previous page} 
\begin{tabular}{l|l|l|l} \hline \hline\noalign{\smallskip}
Radionuclide & DGS & TAGS & Priority \\ 
\hline
%\noalign{\smallskip}
43-Tc-108    	&	 y              	&	 y                &	 2        \\
44-Ru-105       & –                     & –         & –        \\
44-Ru-107    	&	 –               	&	 –       	&	 –        \\
44-Ru-109    	&	 y               	&	 y      	&	 3        \\
45-Rh-107    	&	 y               	&	 –      	&	 3        \\
45-Rh-108    	&	 y               	&	 y        	&	 3        \\
45-Rh-109    	&	 y               	&	 y      	&	 3        \\
45-Rh-110    	&	 y               	&	 y      	&	 3        \\
45-Rh-110m   	&	 y               	&	 y          	&	 3        \\
45-Rh-111    	&	 y               	&	 y       	&	 3        \\
46-Pd-111    	&	 ?               	&	 ?      	&	 3?        \\
51-Sb-128m   	&	 y               	&	 y          	&	 3        \\
51-Sb-129    	&	 y               	&	 –          	&	 3        \\
51-Sb-130    	&	 y               	&	 –          	&	 2        \\
51-Sb-130m   	&	 y               	&	 y        	&	 1        \\
51-Sb-131    	&	 y               	&	 –          	&	 2        \\
51-Sb-132    	&	 y               	&	 y                         	&	 1        \\
51-Sb-133    	&	 y               	&	 y              &	 2        \\
52-Te-131   	&	 y               	&	 –              &	 3        \\
52-Te-133    	&	 y               	&	 –          &	 2 or 3       \\
52-Te-133m   	&	 y               	&	 –           	&	 2        \\
52-Te-134    	&	 –               	&	 –    	&	 –        \\
52-Te-135    	&	 y               	&	 –          	&	 3        \\
52-Te-136    	&	 y               	&	 y          	&	 2        \\
52-Te-137    	&	y                	&	 y         	&	 2        \\
53-I-132     	&	 y               	&	 –        	&	 2        \\
53-I-134     	&	 y               	&	 –    	&	 1        \\
53-I-135     	&	 y               	&	 –          	&	 2        \\
53-I-136     	&	 y           	    &	 y    	&	 2        \\
53-I-136m    	&	 y               	&	 y             	&	 2        \\
53-I-137     	&	 y               	&	 –                       	&	 2        \\
53-I-138     	&	 y               	&	 y             	&	 3        \\
54-Xe-137    	&	 –               	&	 –            	&	 –        \\
54-Xe-138    	&	 –               	&	 –             	&	 –        \\
54-Xe-139    	&	 y               	&	 –                      	&	 2        \\
54-Xe-140    	&	 y               	&	 y           	&	 2        \\
55-Cs-138    	&	 y               	&	 y     	&	 1        \\
55-Cs-139    	&	 y               	&	 y                     	&	 2        \\
55-Cs-140    	&	 y               	&	 y              	&	 2        \\
55-Cs-141    	&	 y               	&	 y             	&	 2        \\
56-Ba-139    	&	 –               	&	 y           	&	 2        \\
56-Ba-141    	&	 y               	&	 y           	&	 2        \\
56-Ba-142    	&	 –               	&	 –                    	&	 –        \\
56-Ba-143    	&	 y               	&	 y                       	&	 2        \\
56-Ba-144    	&	 y               	&	 y                     	&	 2        \\
56-Ba-145    	&	 ?               	&	 y                     	&	 3?        \\
57-La-141    	&	 –               	&	 –        	&	 –        \\
57-La-142    	&	 y               	&	 y     	&	 1        \\
57-La-143    	&	 y               	&	 y                  	&	 1        \\
57-La-144    	&	 y               	&	 y              	&	 2        \\
57-La-145    	&	 y               	&	 y               	&	 3        \\
57-La-146m   	&	 y               	&	 y                      	&	 3        \\
57-La-147    	&	 y               	&	 y                       	&	 2        \\
58-Ce-147    	&	 y               	&	 y                           	&	 3        \\
58-Ce-149    	&	 y               	&	 y                          	&	 3        \\
59-Pr-145    	&	 y               	&	 –                  	&	 3        \\
59-Pr-146    	&	 –               	&	 y                    	&	 2        \\
59-Pr-147    	&	 ?               	&	 –            	&	 3?        \\
60-Nd-149    	&	 –               	&	 –                     	&	 –        \\
 & & & \\
%  	&	  	&	  	&	 \\
%  	&	  	&	 	&	 \\ 
%\noalign{\smallskip}
\hline
\end{tabular}
\end{minipage}
\end{center}
\end{table*}
}

\section{Relevant TAGS/TAS measurements}\label{tags-data}
Various nuclear decay data derived from TAGS/TAS have been compared with the equivalent data to be found in recommended decay data libraries, and more specifically the Evaluated Nuclear Structure Data File (ENSDF) \cite{ENSDF}, 
% Vivian 10 Dec 2022: comparisons have been made with all evaluated data libraries, there is no reason to single out ENDF/B libraries especially since in the tables we compare wiht all the available libraries so i have added the JEFF and JENDL libraries here.
the Evaluated Nuclear Data File/B-VII and VIII \cite{Chadwick2006,Chadwick2011,Brown2018}, the Joint Evaluated Fission and Fusion Files 3.x~\cite{Kellett2009,Kellett2016,Plompen2020}, and the Japanese Evaluated Nuclear Data Files JENDL-Decay Data File-2011, 2015 and JENDL-5~\cite{Katakura2011,Katakura2015,Iwamoto2021}. ENSDF is primarily defined as an evaluation of experimental nuclear structure and decay data which constitutes a comprehensive database maintained on a mass-chain basis over an approximate eight- to ten-year re-evaluation cycle. The decay-data files within ENDF/B-VIII are incremental updates of the re-evaluated contents of ENSDF with various additions and a series of modest modifications as defined at the time of their assembly and issue for public access. An improved and updated re-release occurred in early 2018 as ENDF/B-VIII.0 \cite{Brown2018}. Similarly, the JEFF-3.3 decay data sub-library is the latest version of an assembly of decay data~\cite{Plompen2020} from ENSDF, the Decay Data Evaluation Project (DDEP)~\cite{ddep} and evaluations performed for the United Kingdom UKPADD-6.7 and UKHEDD-2.5 libraries ~\cite{Kellett2009,Perry2014}. The recently released JENDL-5 decay data sub-library includes evaluated data from ENSDF combined with theoretical calculations using the Skyrme-Hartree-Fock theory combined with the Quasi-particle Random-Phase Approximation (Skyrme HF+QRPA) approach of Minato~\cite{Minato2016}.

An impressive systematic study of TAGS spectra for almost 50 important fission products was undertaken in the 1990s by Greenwood and co-workers at the Idaho National Engineering Laboratory, USA \cite{Greenwood1996,Greenwood1997,Greenwood1992}. A tape system was used to collect and transport selected fission-isotope mass fractions from the spontaneous fission of a $^{252}$Cf-based ISOL system (Isotope Separation On-Line) for spectral analysis by means of a large single-ingot NaI(Tl) scintillation detector with an axial well that contained a Si detector to provide the means of obtaining \be-\g~ coincidence data along with \be$^{-}$ and \g~ singles spectra. Their pioneering work involved TAGS measurements ranging from rubidium (starting with $^{89}$Rb) to yttrium (ending with $^{95}$Y) and from caesium (starting with $^{138}$Cs) to samarium/europium (ending with $^{158}$Eu) that has been substantially reconfirmed in more recent years, as outlined below. A major gap in their work arose as a consequence of the inability at that time to volatilise effectively the fission products of elemental refractory zirconium, niobium, molybdenum, technetium, ruthenium and rhodium in order to achieve optimum/suitable source preparation. Some of the radionuclides identified with these particular elements are expected to be important contributors to resulting reactor inventories, decay heat and antineutrino emissions of commercial irradiated nuclear fuel.

During the course of earlier assessments of the nuclear data needs for decay-heat and antineutrino spectral calculations as encouraged by the OECD-NEA and IAEA from 2005 to 2014 \cite{Yoshida2007,Nichols2009,Dimitriou2015} and beyond, portions of the research programmes of a number of TAGS facilities were re-aligned towards the study of potential Pandemonium within the ill-defined decay schemes of specific chosen fission products (see for example, Refs. \cite{Algora2010,Jordan2013,Fijalkowska2017}). These more recent and on-going experimental studies are described below, along with earlier notable work that remains highly relevant to our existing objectives within both nuclear applications and basic nuclear physics research. 
%One 
First motivated by Yoshida~\textit{et al.}~\cite{Yoshida1999} and evaluation work undertaken as part of an OECD-NEA/IAEA initiative~\cite{Yoshida2007}, a series of dedicated experimental TAGS/TAS studies were undertaken at the University of Jyv\"{a}skyl\"{a}, Finland, by the TAGS team from the Instituto de Física Corpuscular, CSIC, Universidad de Valencia, Spain, and co-workers. Work began to assess and confirm known decay-heat benchmarks, and was rapidly extended to consider observed difficulties in reproducing and defining the antineutrino spectra emitted from commercial reactor systems. Experiments were performed at the IGISOL (Ion-Guide Isotope Separator On-Line) facility, University of Jyv\"{a}skyl\"{a}, in which 
%nuclear reaction products recoil out of a thin neutron-irradiated target
fission products produced by proton bombardment of 
$^{238}$U recoiled out of the target and were transported by helium gas flow to the early stages of a separator and Penning trap with a mass-resolving power of 10$^{5}$ (JYFLTRAP) located adjacent to a Total Absorption Gamma-ray Spectrometer. Adoption of the Penning trap as an 
%isobaric  %Changed by AA. In reality with the trap we separate isotopes
isotope 
separator resulted in high-purity sources that are of great benefit for this type of study. The first measurements were performed with a
TAGS spectrometer that consisted of two cylindrical NaI(Tl) crystals, of which the larger possessed a longitudinal hole along the axis in which a silicon detector could be placed for coincidence studies at the defined measuring point and source location. A large amount of \g~strength was observed for $^{105}$Mo and $^{104,105,106,107}$Tc that had not previously been detected in earlier high-resolution  
%Vivian, 14 Nov 2022 following AA remark: \g~singles
%OK, but Alan has also added "-ray" to give "\g-ray".
\g-ray measurements \cite{Algora2010}. Results are listed in Table~\ref{tab:meanenergies} for a number of related studies that focused strongly on $^{102,104,105}$Tc \cite{Algora2010,Jordan2013,Algora2009,Algora2011,Algora2011b,Algora2014b}. An important conclusion from this work was that five of the seven radionuclides studied were found to exhibit the Pandemonium effect, with the need to confirm, explore and improve quantification of their inadequate decay parameters in greater detail by means of Discrete Gamma-ray Spectroscopy (DGS).

% AA 8 december 2022. I recomend removing the gs to gs feedings from the tables. They are more relevant in the antineutrino calculations context, and better to keep them for the next publication. Apart from that these results have not been consequently discussed in the different articles. 
% AA 8 december 2022.  I also corrected that it is not MTAS, just TAGS in this table. 

\begin{sidewaystable*}\vspace{-13.5cm}
\small
\caption{Ground-state to ground-state \be$^{-}$ feedings, half-lives and average $\overline{E }_{\beta}$ and $\overline{E}_{\gamma}$ energies for $^{86,87,88}$Br, $^{91,94,95}$Rb, $^{100,100m,101,102,102m}$Nb, $^{105}$Mo, $^{102,104,105,106,107}$Tc, and $^{137}$I, determined from IGISOL-TAGS measurements published in the last decade~\cite{Algora2010,Valencia2017,Guadilla2016,Jordan2013,Rice2017,Guadilla2019a,Guadilla2019c}, and compared with equivalent data determined from ENSDF database of 2022~\cite{ENSDF}, ENDF/B-VIII.0 data library of 2018 (ENDF/B8)~\cite{Brown2018}, JEFF-3.3 data library of 2020~\cite{Plompen2020}, and JENDL-5 data library of 2021~\cite{Iwamoto2021}. Where no values are provided for the TAGS ground-state to ground-state \be$^-$ feedings, the ENSDF values available at the time were used in the analysis leading to a fair reproduction of the measured data.}\label{tab:meanenergies}
\begin{minipage}{0.8\textwidth}\hspace{-0.5cm}
\begin{tabular}{rr|rr|rrrrr|rrrrr}\hline \hline
Nuclide  &	$T_{1/2}$ (s)&	\multicolumn{2}{c|}{g.s.~$\rightarrow$~ g.s. ($\%$)} & \multicolumn{5}{c|}{$\overline{E}_{\beta}$ (keV)} & \multicolumn{5}{c}{$\overline{E}_{\gamma}$ (keV)} \\ 
	     &            &	TAGS      &	ENSDF       &	TAGS &	ENSDF & ENDF/B8  &JEFF-3.3 & JENDL-5    &	TAGS &	ENSDF & ENDF/B8 &JEFF-3.3 & JENDL-5\\ \hline
\T $^{86}$Br&	55.1(4)&20.2(50) &15(8)  &1687(60)&	1900(300)& 1944(350) & 1943(345)& 1687(30)&	3782(116)	& 3360(110) & 3300(160) &3300 &3782(108)\\
$^{87}$Br&	55.65(13)& 10.1(9) &12.0(19)  &	1170(+32-19) &	1660(80) & 1660(80)&1170(26) &1170(8) &	3938(+40-67) &	3100(40) & 3350(40) &3938(54) & 3938(35)\\
$^{88}$Br&	16.34(8) & 4.7(9)  &$<$11  &1706(+32-38)&	2240(240)& 1702(50)&1706(35) &1706(30) &	4609(+78-67)&	2920(50) & 3134(60) &4609(70) &4609(46)\\
$^{91}$Rb&	58.2(3) & 9.2(10)\textsuperscript{a}  &2(5)  &1389(44)&	1580(190)& 1370(40)& 1368(137)& 1389(13)& 	2669(95)&	2270(40) & 2708(76)& 2706(271)& 2669(44)\\
$^{92}$Rb&	4.48(3) & 87.5(25)	 &	95.2(7) &	&	3640(30) & 2870(700) &3640(30) & 3498(1)&   & 170(9) & 2150(190)&170 & 461(8) \\
$^{94}$Rb&  2.702(5)& $\sim$0 & 0  &2450(+32-30)&	2020(90)& 2020(90)&2450(30)	& 2450(5)& 4063(+62-66)&	1750(50) & 1895(49)&4063(64) &4063(50)\\
$^{95}$Rb&	0.3777(8)  & 0.03(+10-2)  &$\le$0.1  &2573(+18-8)& 2320(110)& 1380(70) &2824 &2573(8) &3110(+17-38) &2050(40)	& 2162(40)&2629 &3110(40)\\ 
$^{96}$Y  & 5.34(5)& 96.6(+3-21) & 95.5(5) &3193(+2-19) & 3181(20) &3184(17) &3181(20) & 2656(1)  &67(+12-2) & 80(44) &80(44)& 80& 1206(4)\\
$^{96m}$Y  & 9.6(2)& & &1721(+5-9) &1600(160) &1600(160) &1821 & 1600(170)  &4669(+21-12) &4308  &4308 & 4479&4308 \\
$^{100}$Nb  &	1.4(2)  &  40(6)\textsuperscript{b} & 50(7) &2414(154)& 2510(210)	&2540(213) &2540(213) & 2414(7)&959(318) &710(40)	& 708(40)  &708 &959(36)\\
$^{100m}$Nb &	2.99(11) &   &  &1706(13)& 1910(180) &2000(200)&2040 &1706(13)  &	2763(27)&1720(50) &	2210(70)  &2056 &2210(69)\\
$^{101}$Nb & 	7.1(3) &   &  &1797(133) &	1800(300) & 1830(300) & 1830(300) &1797(8)	& 450(280) &	244(22) & 244(22) &244(22) &445(21)\\
$^{102}$Nb &	4.3(4) &   &  &1948(27)&2280(170)&	2300(170)&2400 & 1948(6)& 2764(57)  &	2090(100)& 2090(100) &2400 &2764(97)\\
$^{102m}$Nb &	1.3(2)  & 44.3(28)\textsuperscript{b}  &  &2829(82) &-- &  2420&2276(169) & 2829(82)	& 1023(170) &--	 & 2420& 2094  & 1023(170)\\ %\hline \hline
$^{105}$Mo &	35.6(16)&	 &  &1049(44)&	1900(300)&	1049(44) &1049(44) &1049(3) & 2407(93)&	548(24) & 2407(93) & 2407(93)&2407(24)\\
$^{102}$Tc	&5.28(15)&  &  &1935(11)&	1945(16)& 1945(16) &1945(16) & 1935(1)&	106(23)&	81(4) & 81(5) &81(5) &106(5)\\
$^{104}$Tc	&1098(18) &  &  &931(10)&	1590(70)& 931(10) & 931(10)& 931(4)&	3229(24)&	1890(30) & 3229(24) &3229(24) &3229(31)\\
$^{105}$Tc	&456(6) &  &  &764(81)&	670(70) & 764(81) &764(81) &764(6) &	1825(174) &	86.6(23) & 1825(174) & 1825(174)&1825(19)\\
$^{106}$Tc	&35.6(6)&  &  &1457(30)&	1900(70)& 1457(30)&1457(30) &1457(4) &	3132(70)&	2190(50) & 3132(70) &3132(70) &3132(51)\\
$^{107}$Tc	&21.2(2) &  &  &1263(212)&	1890(240)& 1263(212) &1263(212) &1263(7) &	1822(450)&	511(11) & 1822(450) &1822(450) &1822(11)\\ 
$^{137}$I&	24.5(2)  & 45.8(13)\textsuperscript{b} & 45.2(5)  &1934(+35-56)&	22.8(7) &1920(30) &1861 &1934(1) & 1220(+121-74) &	4.0(3) & 1135(20) &1212 &1220\\ 
$^{140}$Cs& 65.7(3) &  36.0(15)\textsuperscript{b} & 35.9(17) &1903(+32-69) & 1950(70)& 1893(55)&1890(50) & 1910(2)& 1792(+149-68)& 1765(21) & 1796(21) & 1819(18) & 1864(25) \\
\hline 
\multicolumn{14}{l}{\textsuperscript{a}\footnotesize{Separation of the two \be$^-$ emissions that populate the ground and first excited states is difficult to achieve.}} \\
\multicolumn{14}{l}{\textsuperscript{b}\footnotesize{Determined by means of a revised Greenwood procedure~\cite{Guadilla2020} (general principals and original procedure described in Greenwood \textit{et al.}~\cite{Greenwood1992b}).}}  \\
\end{tabular}
\end{minipage}
\end{sidewaystable*}

%ALN to Vivian, 10 Oct 2022: Tables 3 and 5 - I have slightly elaborated two of the column headings a number of times to display as ENDF/B7,8 and JEFF3.3.
%ALN to Vivian, 18 Oct 2022: I have defined everything that I can within the main title of this table (Table 3), including JEFF-3.3 and JENDL-5 from the point of view of suitable references. Minato2021 could also be included in the title, if appropriate, along with [54] rather than instead of [54]. See explanation in the attachment to my covering email to you.
%ALN to Vivian, 18 Oct 2022: a few vertical lines would greatly help in the layout of this complete table: vertical line between half-life title/data and "TAGS"/data; vertical line between first "JENDL5"/data and second "TAGS"/data. This to be done within all five sub-tables of Table 3.
As shown in Refs.~\cite{Algora2010,Jordan2013}, the inclusion of TAGS average-energy values for $^{105}$Mo and $^{104,105,106,107}$Tc~(particularly for $^{104,105}$Tc) in decay-heat summation calculations for a thermal-fission pulse in $^{239}$Pu results in significant improvement and better agreement with compiled experimental decay-heat benchmark data over cooling times from 300 to 3000 s, supporting the importance of identifying potential Pandemonium as highlighted by Yoshida~\textit{et al.}~\cite{Yoshida1999}. Equivalent comparisons with a thermal-fission pulse in $^{235}$U exhibit much less impact on the summation calculations, that can be attributed to the reasonably large differences in their cumulative fission yields (these particular cumulative fission yields in $^{239}$Pu are a factor of 2.5 higher than for $^{235}$U). Another point of note is that the \be$^{-}$ decay of $^{102}$Tc involves a significant fraction of direct feeding to the daughter ground state,
%Vivian 12 Dec 2022: changed to describe
which is deemed to be difficult to describe by means of gross \be$^{-}$ theory. Efforts were also made to describe the Gamow-Teller \be$^{-}$ decay properties of both~$^{102,104}$Tc~\cite{Jordan2013} employing the complex excited VAMPIR code and the gross theory of \be$^{-}$ decay. 
% I commented here the following sentence and simplified the text
%, from which significant differences were identified with prolate and oblate configurations in the parent and daughter states~\cite{Guadilla2020}.

%ALN to Vivian, 18 Oct 2022: is/are there suitable reference(s) for what is stated immediately above concerning calculations identified with prolate and oblate configurations?
%Vivian to Alan, 4 Nov 2022: Alejandro should provide these references.
% A. Algora, 8 Dec 2022, I have modified slightly the text concerning the GT strength calculations and put the correct reference
%Vivian to A. Algora, 8 Dec 2022: we should be consistent with how we write gross theory throughout the text. is it Gross theory or gross theory? Also what do you mean by 'complex excited VAMPIR code'? is it a code thta calculates complex excitations? it's not clear to me. And since you mention the name of the code, we will need a referecnce for it.
%Alan, 9 Dec 2022, to Vivian: I have been expressing it as "gross \be$^{-}$ theory" - not starting each word with capital letters.
%Alan, 9 Dec 2022 - I have also vetted the changes made by AA at the end of the above paragraph - last sentence.

TAGS/TAS measurements by the Valencia-led team at the University of  Jyv\"{a}skyl\"{a} have also included the possibility of identifying Pandemonium within the decay schemes of beta-decaying isomers such as $^{100}$Nb, $^{100}$Nb$^{m}$ and $^{102}$Nb, $^{102}$Nb$^{m}$~\cite{Guadilla2019a,Guadilla2019b}. 
A natural uranium target was irradiated at the IGISOL facility with 25-MeV protons to induce fission prior to high-resolution mass separation within the JYFLTRAP double-Penning trap in order to isolate zirconium parents and niobium daughter nuclides for delivery to and study by means of the DTAS detector array (operated as eighteen modular NaI(Tl) scintillators in cuboid form).
Several techniques were used to separate the different contributions of the metastable and ground-state decays, and their results were consistently compared. 
Both $^{100}$Nb and $^{102}$Nb exhibited evidence for metastable states of similar half-lives to those of their ground states, whereby $^{100}$Zr (0$^{+}$) and $^{102}$Zr (0$^{+}$) populated the low-spin niobium levels of $^{100}$Nb$^{g}$ and $^{102}$Nb$^{m}$, respectively. Absolute \g-ray emission probabilities were determined on the basis of their depopulation of $^{100}$Mo and $^{102}$Mo nuclear levels that had been populated by the \be$^{-}$ decays of $^{100,100m}$Nb and $^{102,102m}$Nb, respectively, and these values were in good agreement with known equivalent data when the branching-ratio matrix employed in the analysis
% Vivian, 14 Nov 2022: implemented AA suggestion
%Alan: Deleted "that was"
was modified. Significant \be$^{-}$ intensity was found at high excitation energies for which there were no previously known data, and the resulting average \be~and \g~energies obtained in this work impacted significantly on calculations of both decay heat and antineutrino spectra (listed in Table~\ref{tab:meanenergies}).  Thus, the \g~component of decay heat for $^{235}$U and $^{239}$Pu exhibited an increase of $\sim 3\%$ at 10 s cooling time and a decrease of between 1$\%$ and 3$\%$ at shorter cooling times; reactor antineutrino spectral calculations revealed evidence of the significant impact of these measurements in the energy region of 5 to 7 MeV by up to 2$\%$ for $^{235}$U and 6$\%$ for $^{239}$Pu over what has previously been known as an area of excessive shape distortion~\cite{Guadilla2019b}. 

%ALN: Vivian - new references as indicted in above paragraph need sorting out, and also to be added to main ref list.  There is also a new table to be added in the text here abouts - Table ZZ from what I emailed to you a week ago.
%ALN: any good suggestions for Figures throughout this section by the originators of the various experimental measurements??
The Valencia-Nantes-Surrey-Jyväskylä team has explored \be$^{-}$ intensities to daughter states above the neutron separation energy, which has involved extensive determination of the \g-ray emissions from the \be$^{-}$ decay of $^{87,88}$Br and $^{94}$Rb by means of TAGS \cite{Valencia2014,Valencia2017,Tain2015b,Agramunt2016,Tain2017}. 
%\textcolor{myblue}{In these measurements} - ALN: not required for the description of these experiments follow straight on.
Rocinante is a twelve-fold segmented BaF$_{2}$ detector system that has been operated in conjunction with the IGISOL mass separator and JYFLTRAP Penning trap to achieve a total 
%\be$^{-}$  AA: the efficiency of 80% is for gamma cascades
$\gamma$
detection efficiency of greater than 80$\%$ for $\gamma$ cascades, while also possessing the advantage of reduced neutron sensitivity when compared with equivalent NaI(Tl) scintillator systems. Detailed contamination analyses of descendant \be$^{-}$ and \be$^{-}$n decay have been conducted, along with the derivation of overall uncertainty estimates. As shown in Table \ref{tab:meanenergies}, all three radionuclides exhibit a significant Pandemonium effect, and none more so than $^{94}$Rb with an overall average \g~energy that is a factor of 2.14 larger than the calculated value from the ENDF/B-VIII.0 database. However, the thermal fission yield of $^{94}$Rb is relatively small in $^{235}$U/$^{239}$Pu nuclear fuel, which mitigates against such a radionuclide playing a significant role in the generation of substantial decay heat. A further comparison shows that the integrated intensity of \g~decay above the neutron separation energy ($S_n$) 
in the case of $^{87}$Br is larger than $P_n$, and is eight times more intense than known previously from high-resolution \g~spectroscopy~\cite{Valencia2017}. TAGS data from all three of these fission products impact in various modest degrees on decay-heat summation calculations at short cooling times, particularly the photon component of $^{94}$Rb and $^{88}$Br to a lesser degree, while the calculated antineutrino spectrum exhibits a reduction of intensity by a maximum of 6$\%$ at 7.2 MeV for the thermal fission of $^{235}$U and $^{239}$Pu \cite{Valencia2017,Tain2015b}. As an important follow-up to the above, the new DTAS detector assembly of eighteen modular NaI(Tl) crystals has been used to study $^{93,95}$Rb, $^{103}$Mo, $^{103}$Tc and $^{137,138}$I, with the provision to install large ancillary detectors for improved energy resolution and enhanced detection efficiency \cite{Guadilla2016,Tain2017,Guadilla2017b}. 
As mentioned earlier, TAGS has been successfully used to study \g-ray emissions above $S_n$ in \be-delayed neutron emitters, and in doing so has provided accurate information to improve (n,\g) cross-section assessments far from~\be$^{-}$~stability. One surprising observation from this work has been the high ratio of the integrated~\g~intensity for emissions from the states above the neutron separation energy ($S_n$) to the total intensity for many of the radionuclides studied. Such behaviour points firmly towards the need to improve our understanding of photon-strength functions and neutron transmission coefficients \cite{Tain2017,Tain2017b}. These studies are complicated by contamination from \be-n decay branches, and therefore efforts were made to evaluate the systematic uncertainties with greater accuracy and confidence. As described above, the decay measurements of $^{87,88}$Br and $^{94}$Rb were part of an integral experimental effort to study these beta delayed-neutron emitters by means of a number of complementary experimental techniques.
%\begin{table}
%\caption{Comparison of mean \be~($\overline{E }_{\beta}$) and \g~($\overline{E}_{\gamma}$) energies for $^{87,88}$Br, $^{94}$Rb, $^{137}$I as derived from IGISOL-TAGS measurements with ENDF/B-VII database of 2016~\cite{Guadilla2016,Valencia2017}.}\label{tab:Valencia-meanenergies}
%\begin{tabular}{cccccc}\hline \hline
%\T Nuclide&	$T_{1/2}$ &	\multicolumn{2}{c}{$\overline{E }_{\beta}$}&\multicolumn{2}{c}{$\overline{E}_{\gamma}$} \\
%                     &    (s)           & \multicolumn{2}{c}{(keV)} & \multicolumn{2}{c}{(keV)} \\
%	& &	TAGS&	ENDF/B&	TAGS&	ENDF/B \\ \hline
%$^{87}$Br &	55.65 (12)&	1170$_{-19}^{+32}$&	1599 &	3938$_{-67}^{+40}$ &	3009 \\
%$^{88}$Br&	16.34 (8)&1706$_{-38}^{+32}$&	2491&	4609$_{-67}^{+78}$&	2892\\
%$^{94}$Rb	&2.702 (5)&	2450$_{-30}^{+32}$&	2019&	4063$_{-66}^{+62}$&	1729\\
%$^{137}$I&	24.5 (2)& 1934$_{-56}^{+35}$&	1964&	1220$_{-74}^{+121}$&	1075\\
%\hline 
%\end{tabular}
%\end{table}

Some of the experimental work was directed towards measurements of the \be-delayed-neutron emission probabilities for particular fission products of importance to decay-heat and antineutrino studies. As developed initially 
%at Valencia, 
in Spain, the BELEN-20 4$\pi$-neutron counter consists of twenty $^{3}$He proportional counters arranged as two rings of eight and twelve tubes within a large polyethylene neutron moderator
% Vivian, 14 Nov 2022: added a plastic detector following AA remark
%Alan: Why not "..... a plastic or 0.5-mm thick Si detector ....", or do they have to be both 0.5-mm thick?  If so, would be best written as "..... a 0.5-mm thick Si or plastic detector ....."
around a central hole that contained either a 0.5-mm thick Si detector or 3-mm thick plastic scintillator to count \be$^{-}$~emissions from sources of high-purity fission products implanted into a movable tape after passage through the IGISOL mass separator and JYFLTRAP at the University of Jyv\"{a}skyl\"{a}. Another noteworthy feature of this detector system is the incorporation of self-triggered digital data acquisition with reduced deadtime \cite{Agramunt2016}. Early studies were directed towards the delayed-neutron decay of $^{88}$Br, $^{94,95}$Rb and $^{137}$I \cite{Agramunt2016}, and have been further extended to $P_n$ measurements of $^{135}$Sb, $^{138,139,140}$I and $^{137,138}$Te decay following the development of the BELEN-48 array of forty-eight $^{3}$He proportional counters located in three rings of six, twelve and thirty tubes around the central hole \cite{Agramunt2017}. Measured P$_n$ data from these experiments are compared in Table \ref{tab:branchings} with recommended values evaluated during the course of an IAEA-coordinated research project~\cite{Liang}. All compiled and evaluated P$_n$ data contained within Ref.~\cite{Liang} are available in the form of an IAEA Reference Database of beta-delayed-neutron data (\url{http://www-nds.iaea.org/beta-delayed-neutron/database.html}).

%ALN: data for four others are not available in the two papers published.
\begin{table}
\caption{Experimentally determined branching fractions of \be-n decay modes \cite{Agramunt2016,Agramunt2017}, compared with recommended values from the evaluation of Liang~\textit{et al.}~\cite{Liang}.}\label{tab:branchings}
\begin{center}
\begin{tabular}{ccc}\hline 
\T Nuclide&	$P_n$&	Ref.~\cite{Liang}\\ \hline
$^{135}$Sb	&0.245(10)&0.200(29)	\\	
$^{137}$Te	&0.0260(30)&0.0291(16) \\	
$^{138}$Te	&0.0480(23)&0.0482(23)  \\   
$^{137}$I   &0.0776(14)&0.0763(14)  \\
$^{138}$I &0.0498(18)& 0.0530(21)\\
$^{139}$I &	0.0927(33)& 0.0974(33)\\
$^{140}$I &	0.0760(28)&  0.0788(43)\\
\hline 
\end{tabular}
\end{center}
\end{table}

TAGS/TAS measurements performed by the Valencia-Jyv\"{a}skyl\"{a}-Nantes team have expanded over the years in order to improve benchmark calculations of the resulting antineutrino spectra by undertaking individual experimental measurements of $^{86}$Br and $^{91}$Rb \cite{Rice2017,Algora2017}, $^{92}$Rb \cite{Zakari2014,Fallot2017,Zakari2015}, $^{100,100m,102,102m}$Nb and $^{140}$Cs \cite{Guadilla2017a}~(see Table \ref{tab:meanenergies}), $^{100}$Tc \cite{Guadilla2017} and  $^{142}$Cs \cite{Fallot2017}. 
The \be$^{-}$~decay of $^{91}$Rb is highly relevant in decay-heat studies because the decay of this radionuclide has been adopted as a calibration point in the average \g~energy measurements performed by Rudstam~\textit{et al.}~\cite{Rudstam1990B}, assuming that $^{91}$Rb does not suffer from the Pandemonium effect. The work of Rice~\textit{et al.}~\cite{Rice2017} shows that this is not the case, and that the average energies quoted by Rudstam~\textit{et al.}~should be re-normalised by an adjustment factor of 1.14. 

Rocinante with twelve segmented BaF$_2$ crystals and DTAS with eighteen 
modular NaI(Tl) scintillators in cuboid form have also been employed as \g~detector assemblies to improve our knowledge of the decay of particular fission-product nuclides in terms of the quantification of their total energies, average \be~ and \g~energies, and contributions to summed antineutrino spectra of operational interest. Both $^{95}$Rb and $^{137}$I are \be$^{-}$-delayed neutron emitters, and were prepared on the IGISOL facility coupled to the JYFLTRAP for isobaric separation prior to \g-ray spectral analysis by means of DTAS and HPGe, along with a plastic scintillator for the detection of \be$^{-}$ emissions for gated spectra~\cite{Guadilla2019c}. A significant amount of \be$^{-}$ intensity was observed to populate states above the neutron separation energy, which underwent subsequent \g-ray de-excitation with comparable emission probabilities to those of the delayed neutrons. While the TAGS data in Table~\ref{tab:meanenergies} for $^{95}$Rb show the average \g~energy to be significantly underestimated and the average \be~energy to be more modestly underestimated in ENDF/B calculations, the low cumulative fission yield of this particular fission product in $^{235}$U and $^{239}$Pu thermal fission will result in relatively low impact, and therefore has not been considered in Tables~\ref{tab:app1} and \ref{tab:app2}. Average energy data in ENDF/B calculations for $^{137}$I are in reasonable agreement with the average \be~ and \g~energies determined by TAGS (Table~\ref{tab:meanenergies}) such that the database already contains and reflects the \be$^{-}$ population of the higher-energy nuclear levels of $^{137}$Xe and delayed-neutron decay to $^{136}$Xe. Both the extracted ground-state feeding and average $\gamma$ energy validate the previous MTAS $^{137}$I values, representing another example of two distinct sets of TAGS/TAS measurements obtaining consistently similar results. 
% Vivian 18 Dec 2022: insrted latest addition 140Cs
The~\be$^{-}$ decay of $^{140}$Cs to $^{140}$Ba has been studied by means of the DTAS spectrometer at IGISOL IV as part of a measurement campaign undertaken in 2014. Since the half-lives of  $^{140}$Cs and daughter $^{140}$Ba are  63.7 s and 12.75 days, respectively, the only contamination in the measurements was summing pile-up. Results from a preliminary analysis have been presented by Guadilla  {\it et al.} \cite{Guadilla2017a}. The feeding distribution is in good agreement with previous results obtained by Greenwood {\it et al.} employing a different experimental set-up and analysis technique \cite{Greenwood1997}. These high-resolution data exhibit evidence of the Pandemonium effect. 
%%V: some discussion on the results from the two measurements of 137I?
%Also as part of this campaign the beta decay of 100Tc has been studied [?]. The main goal of this study was to assess if this decay suffers from pandemonium in the framework of double beta decay studies. Individual decays like this one, can be used to fix model parameters used in theoretical calculations of the double beta decay systems. The 100Tc decay is also of relevance for reactor applications. This decay is the product of a second order process which requires first fission and later neutron capture before the decay occurs. This phenomena have been explored as a possible explanation of the distortion of the antineutrino spectrum from reactors in the 4-6 MeV energy range [?].
%ALN to AA: you have discovered one of my "bete noires" - a hatred of the word "framework", as falsely applied 40/50 years ago in Brussels and now a permanent fixture in their vocabulary. What does it mean in Euro-speak? Certainly not what it means in English, which is: "Framework" from the Oxford English Dictionary (close to being bible in this country) is defined as "a system of rules, ideas, or beliefs (now that would be interesting and subjective!) used to plan or decide something" (i.e., in order to make decisions and judgements). I have changed things below in various parts of this entry, and hope what I have done does not distract from your meaning.
The \be$^{-}$ decay of $^{100}$Tc was also studied \cite{Guadilla2017} with the aim of assessing if the decay process suffers from Pandemonium, and as a consequence would impact on related double-beta decay studies. Furthermore, $^{100}$Tc \be$^{-}$ decay is of relevance for reactor applications as the product of a second-order process, which is based on fission followed by neutron-capture activation of the fission product before~\be$^{-}$~decay occurs. This phenomenon has been explored as a possible explanation for the distortion of the antineutrino spectrum from reactors in the 4-6 MeV energy range \cite{Huber2016}. 
%Continued TAGS/TAS measurements and more extensive data analyses are also being planned or are underway at Valencia-Jyv\"{a}skyl\"{a} that include $^{100}$Mo(\be\be-decay), $^{100}$Tc, further studies of $^{103}$Mo, and the mass region around $^{100}$Sn \cite{Algora2018b}. 
Continued TAGS/TAS measurements and more extensive data analyses are also being jointly planned or are underway that include the mass region around $^{100}$Sn by RIKEN-Valencia \cite{Algora2018b,Algora2016a} and the mass region around $^{208}$Pb at GSI(FAIR)-Valencia \cite{Tain2020}.
TAGS/TAS experimental studies commenced in early 2012 at the On-Line Test Facility (OLTF) of the Tandem accelerator of the Holifield Radioactive Ion Beam Facility (HRIBF) at the Oak Ridge National Laboratory (ORNL). The Modular Total Absorption Spectrometer (MTAS) has been designed, constructed and assembled, and work undertaken to measure and characterise the decay products from 40-MeV energy 50 nA proton-beam irradiations of a $^{238}$UC$_x$ target. An array of nineteen hexagonal-shaped NaI(Tl) detectors weighing 1000 kg are aligned in a %extremely 
honey-combed geometry covering 99$\%$ of the solid angle around the activity to be measured. Each NaI(Tl) crystal is 53 cm in length and just over 20 cm maximum diameter. Surrounded by bespoke shielding, including over 5000 kg of lead and borated high-density polyethylene neutron shielding, the \g~background has been reduced by a factor of 1000. Nuclei of interest are deposited on to a tape transport system monitored by HPGe detectors for isotope identification. The tape transports these nuclei to auxiliary \be$^{-}$ detector triggers, two 1-mm thick segmented Si crystals (each divided into seven 8.5-mm wide strips) surrounding the transport tape over 95\% of the solid angle, and resulting in greater than 90\% \be$^{-}$-trigger efficiency for most fission nuclides. The activity triggers the silicon detectors for \be-\g~coincidence studies between these detectors and the MTAS NaI(Tl) array. 
% Vivian 13 Dec 2022: chnaged above text and references according to CR
%Auxiliary detectors included two 1-mm thick segmented Si crystals (each divided into seven $\sim$ 8.5-mm wide strips) located around the tape that transports the activity to act as~\be$^{-}$~triggers for \be-\g~coincidence studies between these detectors and the NaI(Tl) array. 
Simulated response functions and spectral modelling combined with a series of preliminary experimental studies have assisted greatly in the development of the spectrometer towards nuclear applications and more basic nuclear physics research 
\cite{Rasco2015b,Fijalkowska2014b,Wolinska2014,Karny2016,Rasco2016,Fijalkowska2017,Fijalkowska2014a}.

Seventy-seven fission products were studied at various times over the course of 2012 to 2016 by means of MTAS. Relevant spectral analyses have been reported for $^{86}$Br, $^{89,90}$Kr, $^{89,90,90m,92}$Rb, $^{137}$I, $^{139}$Xe, and $^{142}$Cs \cite{Fijalkowska2014b,Rasco2016,Fijalkowska2017,Fijalkowska2014a,Fijalkowska2018}, and to a greater extent and emphasis for $^{89}$Br, $^{96}$Y, $^{137}$I, $^{137}$Xe, $^{142}$Ba and $^{142}$La \cite{Wolinska2014,Karny2016,Rasco2017a,Rasco2015a,Rasco2017b,Wolinska2017}. Table \ref{tab:meanenergies-ORNL} lists the various ground-state to ground-state and other selected \be$^{-}$ emission probabilities, as well as the average \be~and \g~energies for $^{86}$Br, $^{88,89,90}$Kr, $^{88,89,90,90m,92}$Rb, $^{96}$Y, $^{98}$Nb, $^{137}$I, $^{137,139}$Xe and $^{142}$Cs as derived from MTAS measurements~\cite{Rasco2016,Rasco2017a,Shuai2022,Rasco2022,Fijalkowska2017,Fijalkowska2018}, compared with equivalent data in the 
% Vivian 10 dec: have added the other two libraries that are inclodued in the table.
ENDF/B-VIII \cite{Brown2018}, JEFF-3.3~\cite{Plompen2020}, and JENDL-5~\cite{Iwamoto2021} libraries.

%ALN to Vivian, 10 Oct 2022: in its current form, Table 5 needs "re-fitting" to the page without changing anything other than the fitting layout (see extreme right-hand side of this page (data in final column have been cut-off)).
% AA 8 December 2022. I recomend removing the gs to gs feedings from the tables. They are more relevant in the antineutrino calculations context, and better to keep them for the next publication. Apart from that these results have not been consequently discussed in the different articles. The table is cited on page 10, the reference to gs to gs values should be removed there as well if these columns are removed.
%Alan, 9 December 2022: I have set down my argument(s) for maintaining the gs-gs column in an earlier email. If accepted this would result in the gs-gs column of Table 3 being effectively split in two (Valencia and ENSDF). Nuclides that were analysed through adoption of gs-gs ENSDF data would remain blank in the Valencia part of the column, while the ENSDF gs-gs data would be listed in the ENSDF part of the gs-gs column; and for gs-gs data based directly on your TAGS studies/measurements, gs-gs data for those radionuclides would go into the Valencia part of the gs-gs column. It also would be better than yet again wadding through the text to delete the mention of now sadly inapplicable gs-gs data within the text. Greenwood et al. did this to good effect in a different manner in Nucl. Instrum. Methods Phys. Res. A378, 312-320 (1996).  

\begin{sidewaystable*}\vspace{-15cm}
\small
\caption{ Ground-state to ground-state \be$^{-}$ feedings, half-lives and average $\overline{E }_{\beta}$ and $\overline{E}_{\gamma}$ energies for $^{86}$Br, $^{88,89,90}$Kr, $^{88,89,90,90m,92}$Rb, $^{98}$Nb, $^{139}$Xe and $^{142}$Cs, as derived from MTAS measurements \cite{Rasco2016,Rasco2017a,Shuai2022,Rasco2022,Fijalkowska2017,Fijalkowska2018}, and compared with equivalent data determined from ENSDF database of 2022~\cite{ENSDF}, ENDF/B-VIII.0 data library of 2018 (ENDF/B8)~\cite{Brown2018}, JEFF-3.3 data library of 2020~\cite{Plompen2020}, and JENDL-5 data library of 2021~\cite{Iwamoto2021}.}\label{tab:meanenergies-ORNL}
\begin{minipage}{0.85\textwidth}\hspace{-0.9cm}
\begin{tabular}{rr|rr|rrrrr|rrrrr}\hline \hline
Nuclide  &	$T_{1/2}$ (s)&	\multicolumn{2}{c|}{g.s.~$\rightarrow$~ g.s. ($\%$)} & \multicolumn{5}{c|}{$\overline{E}_{\beta}$ (keV)} & \multicolumn{5}{c}{$\overline{E}_{\gamma}$ (keV)} \\ 
	     &            &	MTAS      &	ENSDF       &	MTAS &	ENSDF & ENDF/B8  &JEFF-3.3 & JENDL-5    &	MTAS &	ENSDF & ENDF/B8 &JEFF-3.3 & JENDL-5\\ \hline
$^{86}$Br&	55.1(4)   &$\approx$ 20&	15(8)      &	1687(30)&	1900(300)& 1944(350) & 1943(345)& 1687(30)&	3782(116)	& 3360(110) & 3300(160) &3300 &3782(108)\\
$^{88}$Kr&10170(68)& 12.9(25) & 14(4)  &   &  370(50) & 370(50)    &  370(50)    &  370(50)   &    &   1950(40) & 1954(40)   &  1950(40)  &  1950(40)     \\
$^{89}$Kr&	189(2)    &	11(1) &	23(4)  & 1250(40) &	1470(100) &1470(100) &1470(100) &1250(6) &	2240(80) &	1844(23)  & 1930(20) &1836 & 2240(24)\\
$^{90}$Kr&	32.32(9) &	7(1) &	29(4) &	1130(20) &	1360(100) & 1360(100)&1325 &1130(5) & 	1690(20)&	1320(40) & 1320(40) &1277 &1690(40) \\
$^{88}$Rb&1066.4(11)& 78.2(20) & 76.51(11) &   &  2050(5)  & 2050(5)    & 2057(20)     & 2050(5)    &    &  677(3)  &  677(3)  & 667(20)   &  677(3)     \\
$^{89}$Rb&	918(6) &	ND\textsuperscript{a} &	18.8(13) &	930(30) &	970(50) & 970(50) & 928 &935(3) &	2260(70) &	2240(50) & 2240(50)&2231 & 2230(50) \\
$^{90}$Rb&	158(5)&    15(1)\textsuperscript{b} &	26(2)\textsuperscript{b} &	1920(50)&	2050(130) & 1900(110) & 2036& 1920(6)& 	2300(100) &	1980(40) & 2270(80) &2014 & 2230(40)\\ 
$^{90m}$Rb	& 258(4)&	5(1)\textsuperscript{b} &	15(4)\textsuperscript{b} &	1100(100)&	1400(110) & 1115(33) &1400(110) &1150 & 	4000(200) &	3240(60) & 3870(120) & 3240& 3240(60) \\
$^{92}$Rb&	4.48(3) &	91(3) &	95.2(7) &	3570(70)&	3640(30) & 2870(700) &3640(30) & 3498(1)&	385(8) & 170(9) & 2150(190)&170 & 461(8) \\
$^{96}$Y&	5.34(5) & 95.5(20) & 95.5(5) &3200	&3181(20)	 & 3184(20) & 3181(20)&2656(1)  &100	& 80(4) & 80 &80 &1206(4)\\
$^{98}$Nb&2.86(6) &  61(3)       &  57(6)       &       &1776(144)   & 1776(144)& 1778(144)&1630(4)&275(29)&321(14)&321(16)&321&855(13) \\
$^{137}$I&	24.5(2) & 49(1)	  &	 45.2(5) &  &1897(15)	&1920(30) &1861 &1934(1)&  1250&1070(18) & 1135(20) &1212 &1220.00(15)\\ 
$^{137}$Xe& 229.1(8)   &  67(2)      &   67(3)   &  510    &  1700(70)   &   1698(71) &  1703(71)      &  1703(1)     &190  &   189(14)    &     189(14)    &    189     &   189(14)     \\    
$^{139}$Xe&	39.68(14) &	2(1) & 15(10) &	1580(30) &	1760(230) & 1790(210) & 1700&1518(7)&	1340(30) &	965(14) & 1020(20) &1007 &1342(20) \\
$^{142}$Cs&	1.684(14) &	43(3) &	56(5) &	2480(90) &	2920(180) &	2460(100) & 2920(180)&2480(4) & 1720(50) &	676(25) & 1704(50) &676 &1720(25) \\ 
 &    &		$<$ 0.5\textsuperscript{c}&	7.2(12)\textsuperscript{c}	& & && & & &  & & &\\
\hline 	
\multicolumn{14}{l}{\textsuperscript{a}\footnotesize{ND - studied but not declared in Ref. \cite{Fijalkowska2017}.}} \\
\multicolumn{14}{l}{\textsuperscript{b}\footnotesize{$\beta^{-}$ emission probability from $^{90g,m}$Rb states to the first-excited 2$^{+}$ state of $^{90}$Sr at an energy of 831.68 keV.}}  \\
\multicolumn{14}{l}{\textsuperscript{c}\footnotesize{$\beta^{-}$  emission probability from $^{142}$Cs ground state to the first-excited 2$^{+}$ state of $^{142}$Ba at an energy of 359.60 keV.}}\\
\end{tabular}
\end{minipage}
\end{sidewaystable*}

%ALN to Vivian, 10 Oct 2022: Table 5 - I have slightly elaborated two of the column headings to display as ENDF/B7,8 and JEFF3.3.
%ALN to Vivian, 18 Oct 2022: I have defined everything that I can within the main title of this table (Table 3), including JEFF-3.3 and JENDL-5 from the point of view of suitable references. Minato2021 could also be included in the title, if appropriate, along with [54] (rather than just replacing [54]). See explanation in the attachment to my covering email to you.
%ALN to Vivian, 18 Oct 2022: a few vertical lines would greatly help in the layout of this table: vertical line between half-life title/data and first "MTAS"/data; vertical line between first "ENSDF"/data and second "MTAS"/data; vertical line between first "JENDL5"/data and third "MTAS"/data. 
% AA: The following structure of the text looks different from the previous ones. One might think of not itemizing this entries or itemize the earlier ones to highlight the most important results. 

\begin{itemize}

\item Studies of $^{86}$Br include the suggested introduction of 65 pseudo-levels starting at an excitation energy of 5 MeV, along with the observation of 
%487 additional \g~rays 
487 additional branches %AA: I wonder if branches is not a better word here.
that require incorporation into the proposed decay scheme \cite{Fijalkowska2014b}. Ground-state to ground-state \be$^{-}$ feeding was determined to be just over 20$\%$, compared with a previously assigned value of 15(8)$\%$. These results are also in good agreement with the studies of Rice $et~al.$ \cite{Rice2017}.

\item $^{89}$Kr studies indicate the need for substantial changes in the accepted decay scheme that would appear to be incomplete and erroneous. A shift of \be$^{-}$ feeding from lower-energy states of 1530-2400 keV to higher levels was observed, and ground-state to ground-state \be$^{-}$ feeding was determined to be 11(1)$\%$ compared with a previously accepted value of 23(4)$\%$~\cite{Fijalkowska2017}.

\item $^{90}$Kr also exhibits the need for substantial changes to be made to the accepted decay scheme. Ground-state to ground-state \be$^{-}$ feeding was determined to be 7(1)$\%$ compared with a previously assigned value of 29(4)$\%$~\cite{Fijalkowska2017}.
	
\item Reductions were observed in the \be$^{-}$ feeding by both $^{90}$Rb and $^{90}$Rb$^{m}$ to the first excited 2$^{+}$ state of $^{90}$Sr to be found at 831.68 keV (15(1)$\%$ for $^{90}$Rb, and 5(1)$\%$ for $^{90}$Rb$^{m}$, compared with previously adopted values of 26(2)$\%$ and 15(4)$\%$, respectively)~\cite{Fijalkowska2017}.
	
\item TAGS measurements of $^{92}$Rb by means of MTAS are consistent and in good agreement with the studies of Zakari-Issoufou  \textit{et al.}  \cite{Zakari2015}. Ground-state to ground-state \be$^{-}$ feeding was determined to be 91(3)$\%$ by Rasco \textit{et al.} \cite{Rasco2016}, compared with a similarly determined value of 87.5(25)$\%$ \cite{Zakari2015}.

\item Beta-decay studies of $^{98}$Nb by Rasco \textit{et al.} have quantified ground-state to ground-state feeding, along with beta feeding to the second $^{+}0$ state in $^{98}$Mo that de-excites via an E0 transition by exploiting the modularity of the MTAS detector and coincidences with the beta detector~\cite{Rasco2022}. An average gamma energy of 275(29) keV has also been deduced. \be$^{-}$ feedings to eight excited levels above the 2608 keV level in $^{98}$Mo are reported, along with an overall impact on antineutrino calculations.
%Alan to Alejandro - Table I of 2022Ra11 shows eight levels (not seven) above 2608 keV (?), so I have already changed this value to eight. 
% AA: you are rigth. They are eigth. I made a mistake with the last known level from high resolution. 

\item Direct MTAS studies of $^{137}$I \be$^{-}$~decay (T$_{1/2}$ of 24.5(2) s) show a dominant ground-state to ground-state \be$^{-}$~transition of 49(1)$\%$ \cite{Rasco2017a}, compared with a recommended value of 45.2(5)$\%$ in ENSDF, while the average $\overline{E}_{\gamma}$ increases by 19$\%$ from 1050 to 1250 keV. Measured \be$^{-}$~feeding at the neutron separation energy is also greater by a factor of two, and increases by a further factor of five or more above this separation energy when compared with the ENSDF database . Altogether, there are sufficient differences in MTAS measurements and the currently recommended decay scheme data for $^{137}$I to merit further well-defined \g~singles and \g-\g~coincidence studies. An additional noteworthy observation is that a P$_n$ branch of 0.079 $(\pm~0.004(sys))(\pm~0.002(fit))$ has been derived which is in good agreement with a value of 0.0776 $\pm$ 0.0014 obtained by Agramunt $et~al.$ based on the use of BELEN~\cite{Agramunt2016}.
%%V: shouldn't the I-137 results be compared with the Valencia group results from both campaigns?
%ALN to V: yes, common radionuclides should be reported to the reader as being in common.
 
\item $^{139}$Xe exhibits the need for substantial changes in the accepted decay scheme. \be$^{-}$~emission probabilities to nuclear levels above 2500 keV increased by over 15$\%$, along with the introduction of various new \be$^{-}$ feedings to proposed levels above 3600 keV. Ground-state to ground-state \be$^{-}$ feeding was determined to be 2(1)$\%$ compared with a previously accepted value of 15(10)$\%$.

\item Although the~\be$^{-}$ decay of $^{142}$Cs is insufficient to register as significant in decay-heat calculations, the decay chain from this radionuclide generates both $^{142}$Ba and $^{142}$La that have much greater impact. Under these circumstances, TAGS and DGS decay-scheme studies of the parent fission product are of some importance and relevance. $^{142}$Cs exhibits a need for substantial changes in the accepted decay scheme, whereby \be$^{-}$ emissions were observed to populate a number of new higher-energy nuclear levels. Ground-state to ground-state \be$^{-}$ feeding was determined to be 43(3)$\%$ compared with a previously adopted value of 56(5)$\%$, and ground-state  \be$^{-}$ feeding to the first excited 2$^{+}$, 359.60-keV nuclear level of $^{142}$Ba was reduced to $<$ 0.5$\%$ compared with a previously agreed value of 7.2(12)$\%$. While a major contributor to the high-energy component of $\overline{\nu}_e$ spectra, at energies below 1800 keV, $\overline{\nu}_e$ of $^{142}$Cs increases to 23(3)$\%$, thereby reducing $\overline{\nu}_e$ interaction with matter.

\item As studied by Wolińska-Cichocka~$et~al.$ by means of the MTAS \cite {Wolinska2014}, the decay scheme of $^{142}$La (T$_{1/2}$ of 5466(30) s) was found to be in good agreement with another TAGS study \cite{Greenwood1997}, but to differ somewhat from the recommended decay data within ENSDF. \be$^{-}$~feeding to the first excited state of daughter $^{142}$Ce (2$^{+}$, 641.282 keV) was determined to be a factor of two higher than the value of 1.4(4)$\%$ in ENSDF. Discrepancies were also observed in the \be-~population of levels over the energy region of 2-5 MeV, along with consideration of possible \be$^{-}$~feeding at even higher levels close to the energy limit. Further TAGS/TAS analyses of existing $^{142}$La spectra would seem to be merited as an important aid in resolving these disparities.

\end{itemize}
Definitive points of note from the above MTAS campaign are the requirements for in-depth \g~singles and both \be-\g~and \g-\g~coincidence measurements of $^{86}$Br, $^{89,90}$Kr, $^{90,90m,92}$Rb, $^{139}$Xe and $^{142}$Cs \be$^{-}$ decay, along with further TAGS and DGS studies of $^{98}$Nb, in order to strive for the determination of an appropriate and comprehensive set of decay schemes.   

MTAS studies of other radionuclides have included observations that are in good agreement with existing HPGe measurements of their \g-ray decay, and hence align with the recommended decay schemes to be found in the ENSDF database: 

%AA: Alan, I have the feeling that the discussion of the MTAS results is 
% somehow different, because there are compiled in a different way.
% I wonder if we should not do the same with the Valencia-Nantes results.
% Otherwise this will look odd. 
%ALN to AA: I agree; but which "layout" should we adopt to achieve such full consistency?  Easiest way (but not for you) would be for you to reformulate the discussion text for the Valencia-Nantes results. Would this also involve losing the associated Table 3 (which I think would be a bad idea)?
%AA to ALN: I will think what is the best way and try to implement it soon. Table 3 should be kept definitely. 
% AA to ALN: I have tried to add a few relevant comments embedded in the text in relation to our results. 

\begin{itemize}

\item Decay data identified with the \be-delayed neutron decay of $^{89}$Br have been obtained from MTAS spectral analyses involving the operation of the central NaI(Tl) crystal in coincidence with the ancillary Si detector, along with an energy gate of 6.6 MeV to the Q(\be$^{-}$)-value on the inner, middle and outer rings of the main NaI(Tl) detectors. Two \g~peaks were evident at approximately 750 and 1600 keV that correspond to \g~depopulation from the first two excited states of $^{88}$Kr with energies of 775.32 and 1577.43 keV. The response function of the MTAS detector system to mono-energetic neutrons of differing kinetic energies was also shown to agree well with various other forms of dedicated experimental study \cite{Karny2016}.

\item Ground-state to ground-state feedings of the \be$^{-}$  decay of $^{88}$Kr and $^{88}$Rb have been determined by Shuai \textit{et al.} \cite{Shuai2022}. Data precision was improved in the case of $^{88}$Kr, and the sensitivity to the shape of the first forbidden decay in the determination of the ground-state to ground-state \be$^{-}$ emission of $^{88}$Rb was also considered. Small differences in the feeding distributions were observed in both $^{88}$Rb and $^{88}$Kr decay with respect to earlier high-resolution studies. 
% Vivian 8 Dec 2022: add that these results confirm that these two nuclides are not pandemonium nuclei and that they are also of lesser importance for decay data calculations? 
% AA: Yes
%Alan 9 Dec 2022: yes, and that is what we have ALWAYS said in Table 2. FISPACT calculations identify both 88Kr and 88Rb as significant contributors to decay heat - and for this type of calculation we have judged them to be already well-quantified.  That is why we assigned priority (-) "unassigned", i.e., not 1, 2 or 3.

\item %ALN: 89Rb - very little is said about this nuclide in the refs, apart from the data tabulated in \cite{Fijalkowska2017}.  Could say: 
TAGS data for $^{89}$Rb are in good agreement with existing \g~singles spectroscopy; little to no evidence of the Pandemonium effect.

\item $^{96}$Y (T$_{1/2}$ of 5.34(5) s) was determined by MTAS to possess a high ground-state to ground-state \be$^{-}$ emission probability of 95.5(20)$\%$, in agreement with the recommended value of 95.5(5)$\%$ in ENSDF. Difficulties in detection arose as a consequence of the E0 transition from the 0$^{+}$, 1581.34-keV first excited state to the 0$^{+}$ ground state of $^{96}$Zr, which is effectively via conversion electrons that are too energetic to achieve an efficient response in the 1-mm thick silicon detectors \cite{Rasco2016,Rasco2017b}. The impact of the E0 transition in this beta decay has also been fully considered in the recent work of Guadilla~\textit{et al.}~\cite{Guadilla2022}. 
%AA: Just a comment on 96Y. We are about to submit a new article on that.
% We believe that the E0 treatment has to be done in a different way. I can 
% explain that if necessary. The statement of the 1-mm thick silicon is not 
% completely correct !!!.
%ALN to AA: you should mention what you are doing or have done with 96Y.
%AA to ALN: this will be a difficult issue, I can try to include a discussion in our part concerning 96Y. But I do not agree with the statement of the conversion electrons but it is written like that in the publication. 

\item $^{137}$Xe (T$_{1/2}$ of 229.1(8) s) was determined by MTAS to decay almost exclusively via \be$^{-}$~emission to both the first excited state (5/2$^{+}$, 455.491(3) keV) and ground state (7/2$^{+}$, 0.0 keV) of $^{137}$Cs with emission probabilities of 31(1)$\%$ and 67(2)$\%$, respectively, to give a total probability of 98(2)$\%$. All of these particular decay data are in agreement with the equivalent recommended values of 31(3)$\%$, 67(3)$\%$ and 98(4)$\%$ in ENSDF. Another decay parameter to be highlighted is the 
%\be$^{-}$~strength AA: Strength can be misleading here
\be$^{-}$~feeding 
ratio for the population of the 2850.04(9)- and 2849.11(13)-keV nuclear levels of daughter $^{137}$Cs with values of 8.0(9) measured by MTAS and 7.2(13) in ENSDF. 
%$$(\beta strength to 2850-keV level)/(\beta strength to 2849-keV level)=  8.0(9)  {\rm \,by\, MTAS\,  and\,} 7.2(13) {\rm \, in\, ENSDF} $$
This constitutes reasonably good agreement in a relatively complex decay scheme of predominantly low-intensity transitions, other than two higher-energy \be$^{-}$~emissions and their associated \g~ray \cite{Rasco2017a,Rasco2015a}.

\item $^{142}$Ba (T$_{1/2}$ of 636(12) s) decay scheme as determined by MTAS was found to be in good agreement with another TAGS study \cite{Greenwood1997} and existing HPGe \g-ray measurements. As would be expected under such circumstances, the MTAS data for the \be$^{-}$~decay of $^{142}$Ba are also in good agreement with the equivalent data in the ENSDF database~\cite{Wolinska2014,Wolinska2017}.

\end{itemize}

Over sixty fission products studied by means of MTAS remain to be fully analysed and evaluated, and this work continues \cite{Rykaczewski2018}. Further MTAS measurements and data analyses are also being planned that will involve the Californium Rare Isotope Breeder Upgrade (CARIBU) of the Argonne Tandem Linac Accelerator System (ATLAS facility) at the Argonne National Laboratory (ANL), USA. Along with existing lead and neutron shielding, a concrete wall separating the $^{252}$Cf spontaneous fission source of CARIBU from MTAS has reduced the background activity substantially over an energy range of 30 keV to 8 MeV, from 2.4 kHz at ORNL to 1.9 kHz at ANL.
%Alan 5 Dec 2022 - I have made the changes requested, but it seems over-detailed in content
%Alan, 5 Dec 20122 - previous wording: "..... from MTAS will substantially reduce background activity to achieve further improved levels of detection." 
The existing facilities at ANL include a gas cell and high resolution mass separator able to handle the isotopes of refractory elements that include $^{98,99,100,101,102}$Nb, $^{103,104,105}$Mo and $^{103,104,105,106,107}$Tc. New activities are envisaged along with some repeat measurements at lower background conditions with an MTAS of greater efficiency and modularity. 
%Alan, 5 Dec 2022 - changes from future to initial past-tense studies (no surprise these).
%Vivian, 6 Dec 2022: added reference
As noted in Section 2, MTAS in an improved form has also been installed as a research tool at the Facility of Rare Ion Beams (FRIB), Michigan State University, East Lansing, USA ~\cite{Karny2019,Rykaczewski2022}.
%Alan, 5 Dec 2022 - ORNL guy regularly states that MSU is in East Lansing.
%AA it is East Lansing, you are right
An extensive range of silicon PIN detectors, double-sided silicon-strip detector (DSSD), silicon surface barrier detector and a summing NaI(Tl) SuN detector have been used to monitor and determine the half-lives of a number of separated relevant fission fragments from the spallation reaction of a 120 MeV/u $^{124}$Sn beam on a $^{9}$Be target~\cite{Dombos2019}. TAGS/TAS studies have been performed by means of the SuN array of eight segments of NaI(Tl), with each segment containing three photomultiplier tubes. Both TAGS and the sum-of-segments spectra were used to identify the radionuclides of interest and determine their half-lives. Seven such nuclides were studied in this manner, of which $^{102}$Nb$^{m}$ with a half-life of 1.33(27) s and $^{104}$Nb$^{m}$ with a half-life of 0.97(10) s are of particular note. While not considered in this wide-ranging assessment because of their questionable existence in the earlier stages of our work, there is increased merit in re-considering their present status with respect to their roles in the calculation of decay heat and antineutrino spectra.

Properties of neutron-rich nuclei in the fission-product and light rare-earth regions have been explored experimentally at Argonne National Laboratory~\cite{Kondev2019,Hartley2018,Hartley2020,Orford2020,Orford2020-1}. 
Discrete \be-\g-\g~spectroscopy measurements were conducted by means of a newly-commissioned decay station at the Gammasphere facility, along with related studies employing a Canadian Penning trap and X-array spectrometer (five Ge clover detectors). 
Initial studies were motivated by nuclear structure research whereby the emphasis was placed on $^{160,162}$Eu, followed by DGS measurements of $^{98}$Y, $^{102,104}$Zr, $^{98}$Nb, $^{102}$Nb$^{g,m}$, $^{104}$Nb$^{g,m}$, $^{102,104}$Mo, $^{144,146}$Ba, $^{144}$La, $^{146}$La$^{g,m}$ and $^{146}$Ce. 
Neutron-rich $^{98}$Nb, $^{102}$Nb$^{g,m}$, $^{104}$Nb$^{g,m}$, $^{102,104}$Mo, $^{144}$Ba, $^{144}$La and 
$^{146}$La$^{g,m}$ are of particular interest with respect to decay-heat and antineutrino calculations, and these measurements will be further analyzed and assessed in order to improve our knowledge of their decay schemes.  

\section{Impact of TAGS measurements on decay-heat calculations }\label{decayheat}
%At the beginning of this section, we briefly review our first encounter with the Pandemonium problem in the realm of the reactor decay heat and the decades-long efforts to overcome it. The prototypes of the modern extensive FP decay-data libraries for summation calculations were completed at the very end of the 1970s almost simultaneously in Europe, Japan and the US.  These libraries were generated on the basis of the really measured decay data available in the form of decay schemes at that time.
%ALN to Vivian, 10 Oct 2022: so many unnatural abbreviations throughout the text of this section in particular so as to become irritating: in the main DD, FY, but there are a few others. You should note that I have left LP and EM alone (as they were).
%Vivian, 14 Nov 2022: LG suggests to replace "reactor decay-heat calculations" with "fuel decay-heat calculations" so I will also remove "As mentioned in Section 5"  
Nuclear fuel decay-heat calculations require individual quantitative inventory data for all the actinides, their heavy-element decay products, and nearly a thousand fission products produced within the fuel as a function of the cooling time after shutdown. After the changing inventories of all the fission-product nuclides and actinides have been calculated on the basis of their formation cross sections and subsequent decay, these data for a particular radionuclide can be multiplied by the relevant decay constant and the various average energies released per decay. These energies are summed over all the radionuclides and all possible decays to yield the resulting total decay heat at designated cooling times. The total average energy per decay $\overline{E}_{T}$ released by a fission product or actinide consists of contributions from the average energy released as the kinetic energy of light-particle transitions $\overline{E}_{LP}$ (most frequently as $\beta{^-}$ emissions), heavy particles $\overline{E}_{HP}$, and as electromagnetic radiation $\overline{E}_{EM}$:
\begin{eqnarray}
    \overline{E}_T = &\overline{E}_{LP} + \overline{E}_{EM} + \overline{E}_{HP},   \nonumber
\end{eqnarray}
where 
\begin{eqnarray}
        \overline{E}_{LP} & = & \overline{E}_{\beta^-} + \overline{E}_{\beta^+} + \overline{E}_{ce} + \overline{E}_{Auger},  \nonumber \\
    \overline{E}_{EM} & = & \overline{E}_\gamma + \overline{E}_{X-ray} + \overline{E}_{annih} + \overline{E}_{bremss}, \nonumber  \\
    \overline{E}_{HP} & = & \overline{E}_\alpha + \overline{E}_{n,p} + \overline{E}_{SF}~..., \nonumber
\end{eqnarray}
in which: \\
$\overline{E}_{\beta^{\pm}}$ is the average kinetic energy of $\beta^{\pm}$~particles, \\
$\overline{E}_{ce}$ is the average kinetic energy of the internal-conversion electrons, \\
$\overline{E}_{Auger}$ is the average kinetic energy of the Auger electrons, \\
$\overline{E}_\gamma$ is the average $\gamma$-ray energy, \\ 
$\overline{E}_{X-ray}$ is the average X-ray energy, \\
$\overline{E}_{annih}$ is the positron annihilation energy associated with $\beta^+$~decay, \\
$\overline{E}_{bremss}$ is the internal bremsstrahlung energy,  \\
$\overline{E}_\alpha$ is the average kinetic energy of the $\alpha$~particles,  \\
$\overline{E}_{n,p}$ is the average kinetic energy of neutron/proton emissions, \\
and $\overline{E}_{SF}$ is the total recoverable energy associated with spontaneous fission. \\
These average energies are normalized per decay of the fission-product nuclides and parent actinides (including spontaneous fission), and are determined or adopted directly from decay-data libraries containing best recommended data for the radioactive properties of the emitted light particles, heavy particles and electromagnetic radiation.

The first appropriate decay-data libraries were developed at the end of the 1970s as a consequence of separate efforts in Europe, Japan and the USA. These libraries were largely based on experimental discrete spectral data available at the time. However, they all failed to reproduce the integral-type sample-irradiation measurements of decay heat performed by Dickens~\textit{et al.}~\cite{Dickens1980} and Akiyama and An~\cite{Akiyama1982} that had become available as potential benchmarks to validate the above three libraries. Comparisons of the measured light-particle (LP) and electromagnetic (EM) decay heat with calculated values indicated that the LP (${\beta^-}$) component was overestimated, while the EM (${\gamma}$) component was significantly underestimated. 

Members of the Japanese Nuclear Data Committee (JNDC) implemented the gross theory of $\beta^-$~decay~\cite{Takahashi1969,Koyama1970,Takahashi1971} in calculations of the average $\overline{E}_{\beta}$ and $\overline{E}_{\gamma}$ energies per decay. After applying the gross theory to fission-product nuclei for which no experimental information was available, this approach was applied further to all fission products with high $Q_\beta$ values, leading to significant improvements in the ${\beta}$ and ${\gamma}$ components of the decay heat in reproducing the sample-irradiation experiments~\cite{Yoshida1981}. An NEANDC specialists meeting on fission yields and decay data held in 1983 reached the following conclusions regarding the status of the decay-heat calculations~\cite{Chrien1983}: (1) discrepancies between the calculated components of decay heat and their experimental measurements were defined as a manifestation of the Pandemonium effect in beta decay arising from incomplete high $Q_\beta$-value decay schemes used in the benchmark calculations, (2) adoption of gross $\beta^-$~theory was only free from the problem in an average or aggregate sense, and (3) targeted experiments such as TAGS were required to assist in solving the problem. Extensive TAGS measurements of an impressive range of fission products of interest were performed by Greenwood~\textit{et al.}~\cite{Greenwood1997} in the mid-1990s. 
% Vivian 9 Dec 2022: Alan made the following modifications based on Tadashi's text
The introduction of these resulting data into decay-heat summation calculations revealed the usefulness of this approach compared with the averaging inadequacies of the gross $\beta^-$~theory~\cite{Hagura2006}. More focused studies designed to address the decay heat of $^{239}$Pu~were completed and made available for analysis by means of further decay-heat calculations approximately thirteen years later~\cite{Algora2010}.

\begin{figure}[h!]
\centering
\includegraphics[width=1.0\columnwidth]{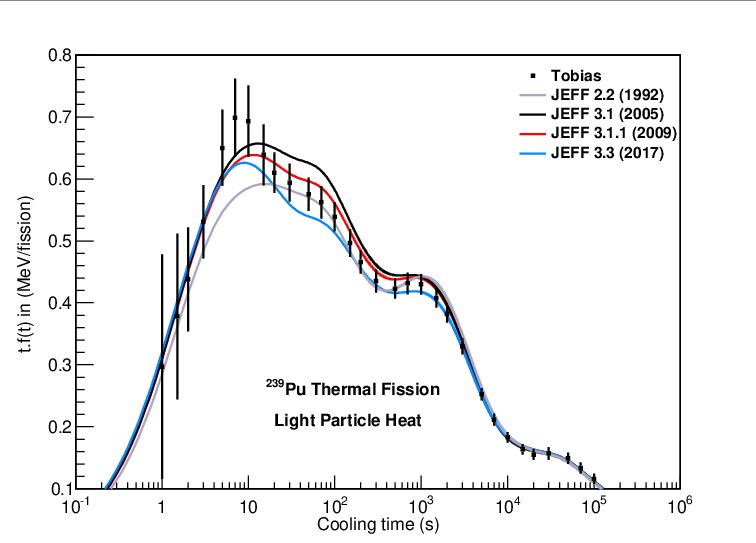}
\includegraphics[width=1.0\columnwidth]{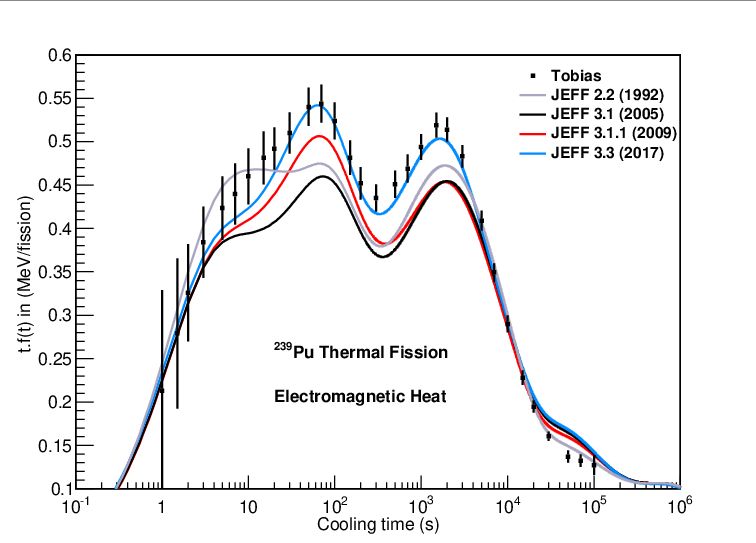}
\caption{Comparison of decay heat for $^{239}$Pu thermal fission obtained using different JEFF libraries that attempt to deal with the Pandemonium effect over the years listed. The evaluated data of Tobias~\cite{Tobias1980} were taken from the IAEA CoNDERC database~\cite{conderc}.}\label{fig:fig-history}
\end{figure}

Theoretical values obtained from gross $\beta^-$~theory were adopted for the mean decay energies of nuclides with $Q_\beta \geq$ 5 MeV in the JNDC fission-product decay data library~\cite{Tasaka1983}, which eventually became part of JENDL. A similar approach that involved fitting parameters in a model based on gross $\beta^-$~theory to replace $\beta^-$ and $\gamma$ experimental energies was tentatively applied to the ENDF/B-IV decay-data sub-library with some success, and has been used up to and including ENDF/B-VI~\cite{England1992}. Both JENDL and the ENDF/B decay-data libraries up to version VI are effectively "contaminated" by theoretical values, creating difficulties in disentangling and clearly demonstrating the importance of TAGS data. 
%The recent ENDF/B-VIII.0 decay data sublibrary is based mainly on ENSDF~\cite{ENSDF}, and includes several of the TAGS data that have been published to date:  Greenwood \textit{et. al.}~\cite{Greenwood1997}, Algora \textit{et al.}(2010)~\cite{Algora2010} and Fijakowska \textit{et al.}~\cite{Fijalkowska2017}. Theoretical mean energies~\cite{Moeller2016} have been adopted for certain FP nuclides for which no experimental information is available. 
On the other hand, average $\overline{E}_{\beta}$ and $\overline{E}_{\gamma}$ decay energies in the JEFF-3 decay-data files from the first release of JEFF-3.1 are based solely on experimentally-determined decay schemes. 
%ALN to Vivian, 10 Oct 2022: as an aside - I recall that TAGS data from Greenwood et al. were introduced with a fair amount of care in the mid-1990s as part of a number of things that I did involving UKPADD files (which were subsequently adopted in the JEFF-3 project), just before Mark Kellett started his JEFF-3 work at the NEA Data Bank under the external supervision of Olivier Bersillon.
When no reliable or complete experimental decay data are available, the empirical $Q_\beta /3$ rule has been used as an estimate of the partition of the total released energy ($Q_\beta)$ into  $\overline{E}_{LP}$, $\overline{E}_{EM}$ and the unrecoverable neutrino energy. The first TAGS data included in the JEFF-3.1.1 decay-data library~\cite{Kellett2009} were taken from Greenwood~\textit{et al.}~\cite{Greenwood1997}. A comparison of the decay-heat calculations obtained with different versions of the series of JEFF-3 libraries in their chronological order offers insights into the growing impact of TAGS. Partial decay-heat calculations based on different releases of the JEFF-3 library are compared in Fig.~\ref{fig:fig-history} with the Tobias evaluated decay-heat data for thermal pulse fission of $^{239}$Pu~\cite{Tobias1980}. JEFF-3.1~\cite{Kellett2009} does not include any gross $\beta^-$~theory or TAGS data, and overestimates the LP ($\beta$) component and underestimates the EM ($\gamma$) component. JEFF-3.1.1~\cite{Kellett2009} includes several of the Greenwood~\textit{et al.}~TAGS data~\cite{Greenwood1997} to give considerably improved agreement when compared with the experimental data for both the LP and EM components of decay heat. This is mainly attributed to the TAGS data being Pandemonium-free and more correctly representative of the relative contributions of $\beta^-$ and $\gamma$ emission energies per decay. Further improvement is obtained with JEFF-3.3~\cite{Plompen2020}, especially for the EM ($\gamma$) component between 30 and 4000 s cooling times after the fission burst, which now agrees with the evaluated decay-heat data. JEFF-3.3 includes the TAGS data for 
%Vivian to Alan, 4 Nov 2022: you mean Nb not Nd
Nb, Mo and Tc isotopes which were measured by Algora~\textit{et al.}~\cite{Algora2010} after they were identified as primary sources of the large Pandemonium problem affecting $^{239}$Pu decay heat. Figures appearing hereafter follow the same convention as Fig.~\ref{fig:fig-history} in which the vertical axis is decay heat (MeV/fission/s) $\times$ time after fission pulse (s) in order to achieve reasonably compact and comprehensive displays.

We have studied the impact of TAGS data on the calculated decay-heat components (LP/EM) for the different fission-based systems discussed in Section~\ref{decaydata} (see Table~\ref{tab:1}), focusing on the effect of the new TAGS data mentioned in Section~\ref{tags-data} (Tables~\ref{tab:meanenergies}, \ref{tab:meanenergies-ORNL}). Decay-heat calculations were performed on the basis of three available evaluated nuclear-data libraries (ENDF/B, JENDL and JEFF), with their associated decay-data and fission-yield sub-libraries. All decay-data sub-libraries by definition include the average decay energies $\overline{E}_{LP/EM/HP}$ followed by the individual components listed in the earlier equations above. However, in the latest versions of these sub-libraries, only the $\overline{E}_{LP/EM/HP}$ values are included in the data section, while the breakdown into individual components that include contributions from the atomic radiation processes ($\overline{E}_{ce/Auger/X-rays/...}$) are listed in the comments section. When TAGS data are incorporated in the ENDF/B, JENDL and JEFF libraries, the average $\overline{E}_{LP/EM}$ energies are replaced by the TAGS-derived average $\overline{E}_{\beta,\gamma}$ decay energies, omitting any possible contributions from the atomic radiation, internal bremsstrahlung, etc. 
%Vivian, 22 Nov 2022: slightly modified following comment
%Alan, 5 Dec 2022 - this long sentence got me lost!  Do you mean "....  i.e., $\approx$ 5% and 13% of the average TAGS-derived $\overline{E}_{\beta}$ for 101Nb and 105Tc, respectively."? Vivian - if so, you should make this change, and I will not get confused.
%Vivian, 6 Dec 2022: made the change
Contributions from the atomic radiation processes to the average energy emitted per decay are usually small and can normally be neglected. However, there are certain cases such as $^{101}$Nb and $^{105}$Tc, in which the sum of the conversion electron $\overline{E}_{ce}$ and Auger electron $\overline{E}_{Auger}$ mean energies as derived from the discrete $\gamma$-ray spectra in ENSDF~\cite{ENSDF} amounts to $\sim$ 100 keV, i.e., $\approx$ 5\% and 13\% of the average TAGS-derived $\overline{E}_{\beta}$ for $^{101}$Nb and $^{105}$Tc, respectively. Full consideration of the various atomic contributions requires sound knowledge of the decay scheme measured by TAGS, and re-calculation of these atomic data for inclusion in the corresponding average decay energies. 

%These contributions 
%are comparable with the differences between TAGS average energies and those derived from the discrete decay scheme, and 
%should be considered when evaluating the decay data for the decay-data sub-libraries.
%ALN to Vivian, 10 Oct 2022: when I read this, I was first inclined to think: "so what?"  Is there more to say about this observation?
Three sets of decay-heat calculations have been performed in combination with each one of the three fission-yield sub-libraries:
\begin{enumerate}
    \item decay data without measured Algora (2010) TAGS data (baseline); 
    \item decay data with measured Algora (2010) TAGS data~\cite{Algora2010} (+ TAGS 2010); 
    \item decay data with recent measured TAGS data published or communicated before the cut-off date of February 2022 (listed in Tables~(\ref{tab:meanenergies}, \ref{tab:meanenergies-ORNL}) (+ TAGS 2021).
\end{enumerate} 
Results have been generated for the thermal fission of $^{235}$U, $^{239}$Pu, $^{241}$Pu, and fast fission of $^{232}$Th, $^{233}$U, $^{237}$Np, $^{238}$U, for which experimental decay-heat measurements on neutron pulse irradiations exist.
%ALN to Vivian, 10 Oct 2022: These are some of the Figures yet to be fully incorporated with the text (Figs. 4, 5, etc.)  Surely these should be cited here as Figs. 4 to ? (to whatever?)
% Vivian to Alan, 4 Nov 2022: these figures should be cited in the subsections
Pulse irradiation decay-heat data for single-actinide targets were obtained from the IAEA CoNDERC database~\cite{conderc}. Further comparisons of the decay-heat calculations are discussed in Sub-section~\ref{comparison}, as obtained with the existing general purpose libraries ENDF/B-VIII.0, JEFF-3.3 and JENDL-5 for all sixteen fissioning systems mentioned in Section~\ref{decaydata}.

%Vivian to Alan, 7 Nov 2022: paragraph on short cooling times. Please check.
%Comparisons and conclusions drawn in the following subsections, with respect to all of the benchmark studies identified with the various single-actinide targets, are not expected to apply fully when considering decay-heat benchmark calculations for power reactors. Immediately after shutdown, such systems continue to operate for a short time at close to their high operational temperatures that arise originally from the $^{233,235}$U and $^{239}$Pu fission processes. Significant thermal-hydraulic and fluid-flow effects within the plant will dissipate this excess heat fairly rapidly before fission-product decay heat is seen to dominate, followed by actinide decay heat at a much later stage. Therefore, the fuel temperatures attained at short cooling times in the range from 0, 1, 10, 100 to 200 seconds will differ considerably from and exceed the type of peak observed at these times in pulse-irradiation measurements involving individual actinides.

\subsection{ENDF/B-VIII.0 library}\label{endfb}
ENDF/B-VIII.0 decay-data and fission-yield sub-libraries have been combined to calculate the LP and EM components of decay-heat for seven fissioning systems for which experimental pulse decay-heat data exist. Average $\overline{E}_\beta$ and $\overline{E}_\gamma$ decay energies determined mainly from the discrete decay schemes recommended in ENSDF~\cite{ENSDF} are included in the ENDF/B-VIII.0 decay-data sub-library. Conversion-electron, Auger-electron and X-ray contributions have also been derived from the decay schemes by means of dedicated codes with their input data obtained from atomic-data libraries~\cite{RADLIST}. The library also includes average $\overline{E}_{\beta}$ and $\overline{E}_{\gamma}$ energies from the TAGS measurements of Greenwood~\textit{et al.}~\cite{Greenwood1997} for the following fission products: $^{90,90m,91,93}$Rb, $^{93,95}$Sr, $^{95}$Y, $^{140,141}$Cs, $^{143,144,145}$Ba, $^{142,143,144,145}$La, $^{147}$Ce, $^{146,147}$Pr and $^{149}$Nd. Beta feedings taken from Tengblad~\textit{et al.}~\cite{Tengblad1987} were used to determine E$_{\beta}$ for $^{88,89}$Br and $^{138}$I. When no experimental decay scheme was available, average energies were obtained from the Finite-Range-Liquid-Drop {+} QRPA model of M\"{o}ller \textit{et al.}~\cite{Moeller2016}. From the most recent TAGS measurements listed in Tables~(\ref{tab:meanenergies}, \ref{tab:meanenergies-ORNL}), the data for $^{104,105,106,107}$Tc and $^{105}$Mo by Algora~\textit{et al.}~(2010)~\cite{Algora2010} and $^{142}$Cs by Rasco~\textit{et al.}~(2016)~\cite{Rasco2016} and Fijałkowska~\textit{et al.}~(2017)~\cite{Fijalkowska2017} have also been incorporated into the ENDF/B-VIII.0 decay-data sub-library.

\begin{figure*}[h!]
\centering
\includegraphics[width=0.9\columnwidth]{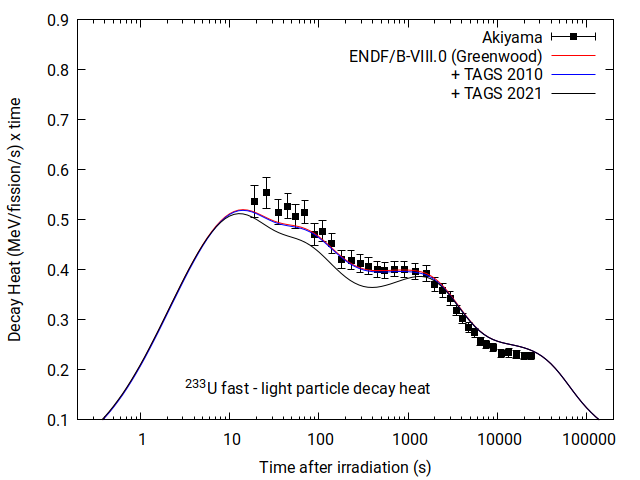}
\includegraphics[width=0.9\columnwidth]{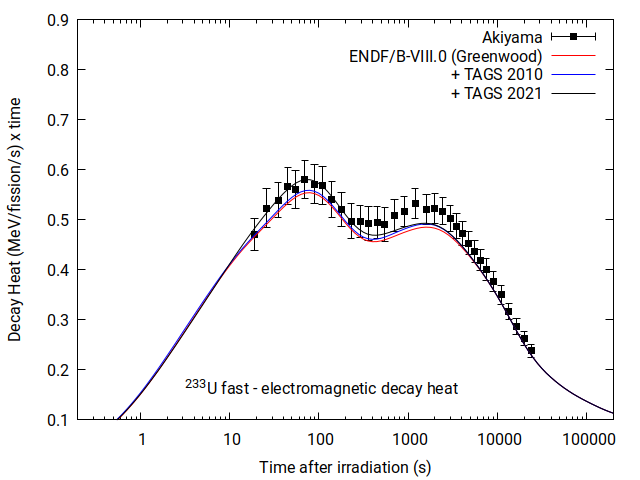}
\includegraphics[width=0.9\columnwidth]{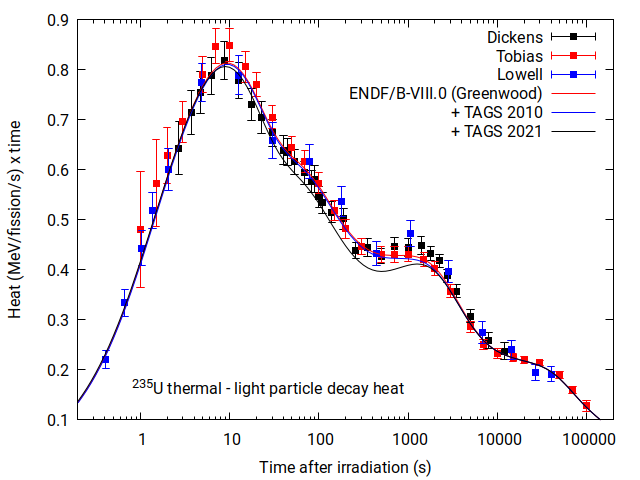}
\includegraphics[width=0.9\columnwidth]{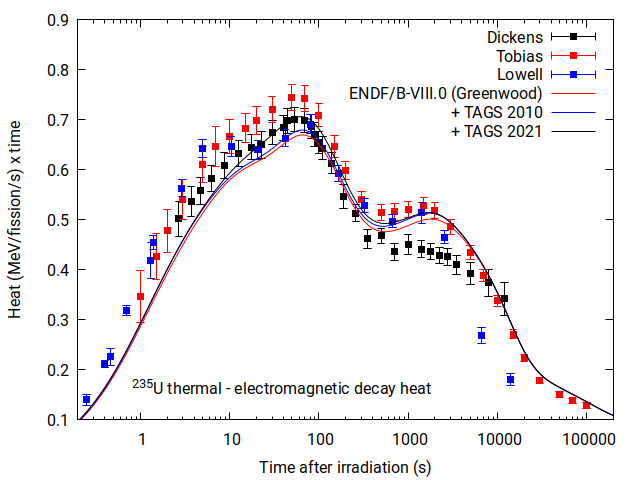}
\includegraphics[width=0.9\columnwidth]{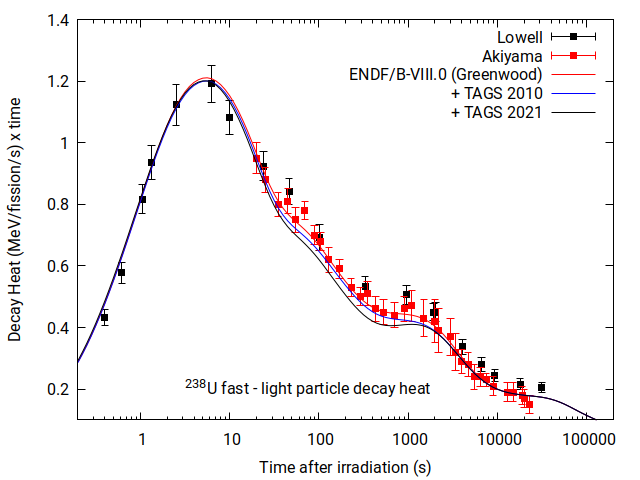}
\includegraphics[width=0.9\columnwidth]{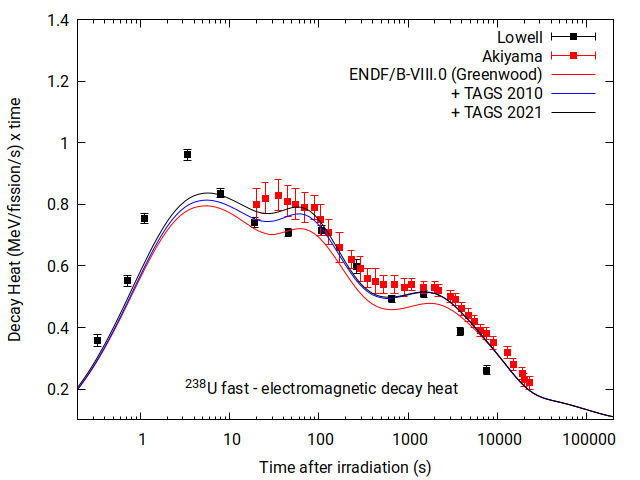}
\includegraphics[width=0.9\columnwidth]{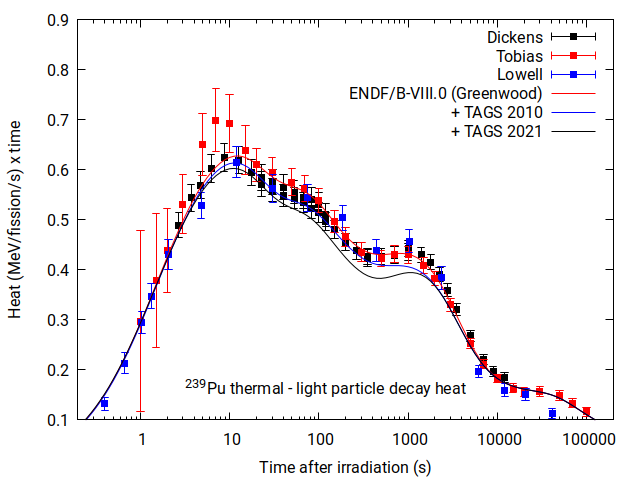}
\includegraphics[width=0.9\columnwidth]{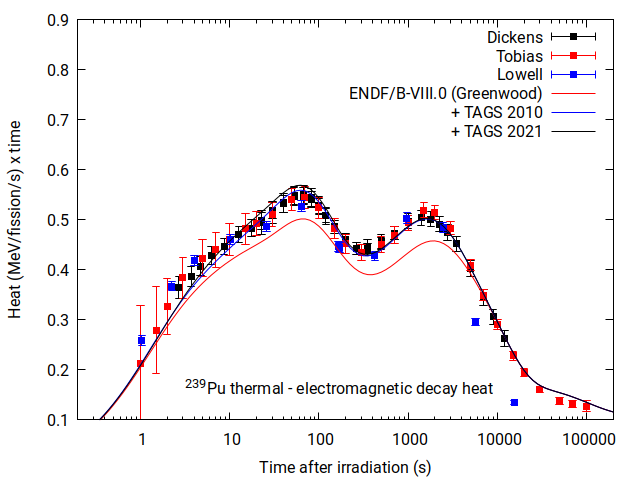}
\caption{Decay heat as a function of cooling time as obtained using the ENDF/B-VIII.0 fission-yield and decay-data sub-libraries~\cite{Brown2018}, with the addition of TAGS data from~\cite{Algora2010} (+ TAGS 2010) and Tables~(\ref{tab:meanenergies}, \ref{tab:meanenergies-ORNL}) (+ TAGS 2021). Experimental data have been taken from the CoNDERC database~\cite{conderc}.}\label{fig:fig-endfb-1}
\end{figure*}

\begin{figure*}[h!]
\centering
\includegraphics[width=0.9\columnwidth]{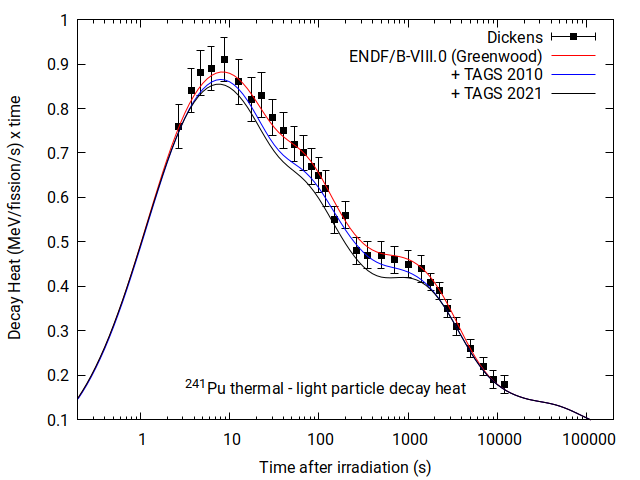}
\includegraphics[width=0.9\columnwidth]{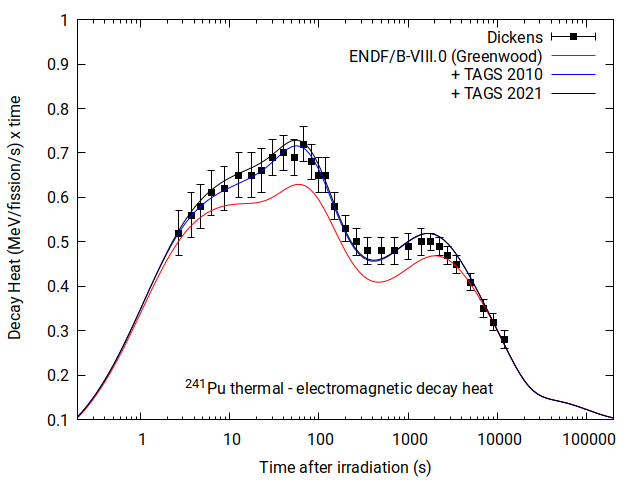}
\includegraphics[width=0.9\columnwidth]{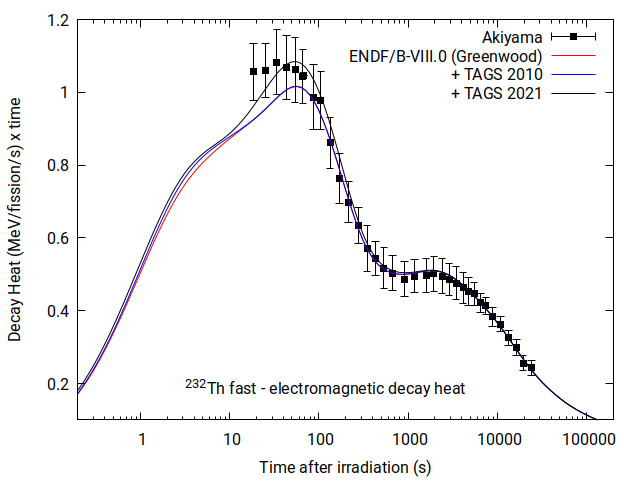}
\includegraphics[width=0.9\columnwidth]{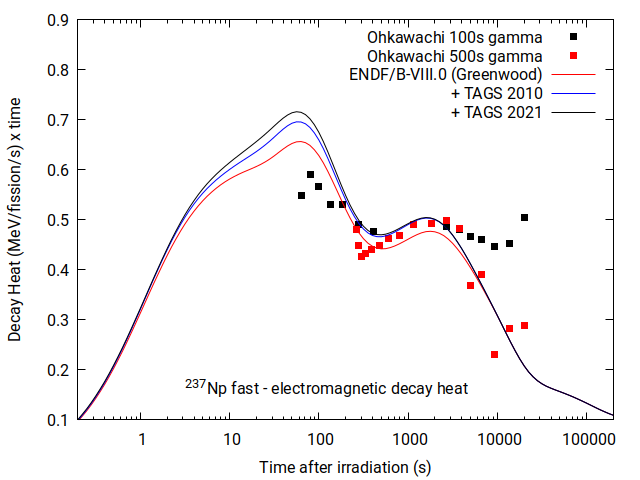}
\caption{Same as Fig.~\ref{fig:fig-endfb-1}. }\label{fig:fig-endfb-2}
\end{figure*}

The three-step process to assess the impact of the recent TAGS measurements described above was implemented as follows:
\begin{itemize}
    \item Step 1, baseline calculations: average $\overline{E}_{\beta}$ and $\overline{E}_{\gamma}$ decay energies of Algora~\textit{et al.}~(2010)~\cite{Algora2010} were replaced with average $\overline{E}_{LP}$ and $\overline{E}_{EM}$ energies derived from the discrete decay schemes in ENSDF~\cite{ENSDF}. Same was done with the average decay energies of Rasco~\textit{et al.}~(2016)~\cite{Rasco2016} and Fijałkowska~\textit{et al.}~(2017)~\cite{Fijalkowska2017} in the case of $^{142}$Cs. Therefore, baseline calculations with the ENDF/B-VIII.0 decay-data sub-library included only the Greenwood TAGS data~\cite{Greenwood1997} listed above (referred to as ENDF/B-VIII.0 (Greenwood)), data from Tengblad~\textit{et al.}~\cite{Tengblad1987} and model calculations~\cite{Moeller2016}. 
    \item Step 2: adopted ENDF/B-VIII.0 decay-data file as default, which includes five of the seven fission products measured by Algora~\textit{et al.}~(2010)~\cite{Algora2010}, and added TAGS average energies for $^{102}$Tc and $^{101}$Nb~\cite{Algora2010} - although not impacted by Pandemonium, with TAGS and ENSDF-derived average energies in agreement. TAGS average decay energies of $^{142}$Cs~\cite{Rasco2016,Fijalkowska2017} were replaced by ENSDF-derived values (+ TAGS 2010). 
    \item Step 3: all of the remaining TAGS data listed in Tables~\ref{tab:meanenergies}, \ref{tab:meanenergies-ORNL} (+ TAGS 2021) were added to the decay-data sub-library that had been modified in step 2. Hence, this sub-library included all of TAS/DTAS studies mentioned by Algora~\textit{et al.}~(2021)~\cite{Algora2021} and MTAS measurements on $^{89,90}$Kr, $^{89,90,90m}$Rb, $^{96}$Y, $^{137,139}$Xe and $^{142}$Cs from Rasco~\textit{et al.}~(2016)~\cite{Rasco2016} and Fijałkowska~\textit{et al.}~(2017)~\cite{Fijalkowska2017}. However, known TAGS data for $^{96m}$Y, $^{98}$Nb, and $^{140}$Cs, were not included as they were published or provided through private communication after the cut-off date (February 2022) for these wide-ranging inventory calculations.
%ALN to Vivian, 10 Oct 2022: state our cut-off date (not given above).
% Vivian to Alan, 4 Nov 2022: cut-off date was given in introductory sect.5. Have added it again here.
\end{itemize} 
Experimental pulse simulations and inventory calculations were performed with the FISPACT-II code~\cite{fispact} at incident neutron energies for which decay-heat and fission-yield data exist, and are available in the relevant ENDF/B-VIII.0 sub-libraries (i.e., at thermal (0.0235 eV) and fast (500 keV) neutron energies). The three sets of calculations are compared with experimental decay-heat data in Figs.~\ref{fig:fig-endfb-1}, \ref{fig:fig-endfb-2}. A consistent trend is emerging for all of the fissioning systems and energies whereby the inclusion of average $\overline{E}_{\beta}$ and $\overline{E}_{\gamma}$ energies derived from measured TAGS data leads to a decrease in the LP and an increase in the EM decay heat components, respectively. These gradual modifications in this manner to the relative contributions of the LP($\beta$) and EM($\gamma$) components of decay heat is direct confirmation of the fact that the TAGS-derived data are nominally free from the Pandemonium effect. 

%\clearpage
While the three sets of calculations reproduce the experimental decay-heat components fairly well, a particularly marked improvement is observed when including the TAGS data of Algora~\textit{et al.}~\cite{Algora2010} (+ TAGS 2010) in the decay-data sub-library. This much improved agreement had already been observed in the thermal decay-heat studies of $^{239}$Pu~\cite{Algora2010}, with the present results for the decay-heat components of $^{238}$U fast and $^{241}$Pu thermal also exhibiting better alignment with their equivalent experimental data. The decay-heat components of the heavier actinides are sensitive to the decay heat of five of the previously selected fission products: $^{105}$Mo, $^{104,105,106,107}$Tc~\cite{Yoshida2007}. These observations contrast sharply with the negligible impact of these TAGS data on decay-heat calculations for $^{232}$Th, $^{233}$U and $^{237}$Np fast-neutron fission. Consideration of $^{235}$U thermal fission is both important and noteworthy: the LP component of decay heat is not affected by the inclusion of the TAGS data and is in good agreement with the experimental data, while the EM component for all three available sets of decay-heat data exhibits discrepancies~\cite{Tobias1989,Dickens1980,Schier1997} . At cooling times ranging from 1 to 400 s, these data sets agree within the experimental uncertainties, while above 400 s there is disagreement. 
%ALN to Vivian, 10 Oct 2022: There are good technical reasons not to make too much of the first observable decay-heat peak from 20 to 200 seconds, particularly when considering that our interest is really all about potentially important applications concerned with power reactors.  This is because the existing decay-heat experiments that we are striving to identify as benchmarks are “unreal” at early cooling times as far as power-reactor systems are concerned.  Power reactors immediately after shutdown continue to operate for a very short time at very high operational temperatures arising from thermal-hydraulic and fluid-flow effects within the plant that need to dissipate before fission-product decay heat takes over (followed by actinide decay heat much later on).  There are a number of places in the text below when the cooling time ranges start with 1, 10, 100 seconds.  Their implied importance is misplaced.
Without the TAGS data of Algora~\textit{et al.}~\cite{Algora2010}, the calculated EM component agrees with the Dickens data below 400 s~\cite{Dickens1980}, and with the evaluated data of Tobias above 400 s~\cite{Tobias1989} - these observed differences between the Dickens data and Tobias evaluation remain unresolved to date.

Inclusion of all of the recent TAGS measurements (+ TAGS 2021) displayed in Tables~\ref{tab:meanenergies}, \ref{tab:meanenergies-ORNL} (with the exception of $^{96}$Y$^{m}$~and $^{98}$Nb) leads to a small underestimation of the LP decay heat at cooling times in the range 10 - 1000 s for $^{233,238}$U fast and $^{235}$U, $^{239,241}$Pu thermal fission. The EM components for $^{233,235,238}$U, $^{241}$Pu and $^{232}$Th fission exhibit improved agreement at shorter cooling times below 100 s, whereas the relatively large increase in the EM component of $^{239}$Pu approaches the upper experimental limits. These results indicate that the inclusion of the (+ TAGS 2021) data is insufficient such that additional TAGS measurements of the remaining high priority fission products are merited to arrive at a robust description of the experimental pulse decay-heat data. 

A re-assessment of the known decay-scheme data for further fission products should be undertaken to justify new TAGS measurements, along with an investigation of the experimental pulse-irradiation decay-heat data to determine the need for additional such studies. Apart from discrepancies in the experimental EM decay-heat data of $^{235}$U thermal fission that remain to be resolved, there are also issues with the experimental decay-heat data of $^{238}$U and $^{237}$Np. The EM decay-heat data of Akiyama and An~\cite{Akiyama1982} for $^{238}$U disagree with the data from Lowell measured by Schier and Couchell~\cite{Schier1997}, while the LP components are in agreement. Two sets of $^{237}$Np measurements were performed at the YAYOI reactor by Ohkawachi and Shono~\cite{Ohkawachi2001} in the form of 100 and 500 s irradiation bursts. These two separate sets of finite irradiation data were converted by the authors into instantaneous pulse data for adoption as benchmarks. The correction factor utilized in this conversion process was of the order of 2 at most, such that the experimental data possessed large uncertainties, especially at shorter cooling times (60 to 200 s) when a large scatter in the data was observed (Fig.~\ref{fig:fig-endfb-2}). Longer cooling times between 2500 and 20000 s were also affected by significant uncertainties that arose from contamination by $\gamma$~rays emitted from $^{238}$Np produced by neutron capture on $^{237}$Np.  Finally, there exists only one reliable set of data for $^{232}$Th, $^{233}$U measured at the fast-neutron reactor YAYOI by Akiyama and An~\cite{Akiyama1982}, while for $^{241}$Pu there is only one measurement at thermal-fission energy by Dickens~\textit{et al.}~\cite{Dickens1980}. The significant improvement in the decay-data sub-libraries that has been achieved in the past two to three decades means that we are now able to explore the impact of uncertainties of the input data on decay-heat calculations. A necessary pre-requisite is to possess accurate and precise decay-heat data, which implies a need for new measurements of the LP and EM components of the various fissioning systems discussed above. 
%Both the nature and handling of uncertainties in the decay data and fission-yield data also need to be considered. A full and respected treatment of these uncertainties, including correlation effects identified with the fission-yield data, would permit reliable and accurate assessments of the decay-data and fission-yield sub-libraries, and draw definitive conclusions on the impact of the TAGS data and the need and form of other additional measurements. Our present results indicate certain trends in such calculations, but without proper propagation of uncertainties they are not definitive. Ongoing efforts to develop suitable approaches for propagating the uncertainties in decay-heat calculations are expected to come to fruition in the near future, so that we should eventually be able to perform uncertainty quantification of such integral calculations with confidence.
%ALN to Vivian, 10 Oct 2022: I would assume that the paragraph above applies to JEFF-3.3 and JENDL-5 as well as ENDF/B-VIII. Therefore, why does it appear here in Sub-section 5.1?  There is also a large slice of conjecture (rather vague) within its content.  Is it justified, or even necessary?  If so, it might be best located as part and at the end of Sub-section 5.4?
%Vivian, 7 Nov 2022: moved to end of 5.4
\subsection{JEFF library}\label{jeff}
JEFF-3.1.1 decay-data and neutron-induced fission-yield sub-libraries were used to calculate the LP and EM components of decay heat for the seven fissioning systems irradiated under thermal or fast neutrons for which experimental pulse decay-heat data exist. Depletion calculations for the pulse experiments have been performed with the SERPENT2 code~\cite{leppanen2015}, while the JEFF-3.1.1 decay-data sub-library was chosen for reference calculations because all recent TAGS measurements presented in Tables~\ref{tab:meanenergies},~\ref{tab:meanenergies-ORNL} have not been included in this particular database.
%the Algora TAGS data from 2010~\cite{Algora2010}. 
Fission-yield data were also taken from the relevant JEFF-3.1.1 sub-library for thermal (0.0235 eV) and fast (400 keV) neutron energies. 

The JEFF-3.1.1 decay-data sub-library was released in November 2007~\cite{Kellett2009} after decay-heat studies
%benchmarking
of various decay-data libraries (including JEFF-3.1) by the OECD/NEA Working Party on International Evaluation Co-operation subgroup 25 (WPEC-25)~\cite{Yoshida2007}. 
%ALN to Vivian, 10 Oct 2022: I am asking myself whether it was really seen by WPEC-25 members as being "benchmarking"?  I always felt that we were just trying to determine whether there were any possibilities (involving "neglected/ill-defined" fission products) of improving our calculational fits by means of TAGS experimental measurements of these neglected fission products.
Members of WPEC-25 concluded that TAGS measurements to determine the average decay energies would improve the quality of the calculated $\beta$ and $\gamma$ decay-heat components. As a result, the mean energies of twenty-nine fission-product nuclides measured by Greenwood~\textit{et al.}~\cite{Greenwood1997} were initially included in JEFF-3.1.1. Furthermore, high-quality evaluations of fifty nuclides were taken from the United Kingdom libraries UKPADD-6.7 and UKHEDD-2.5~(\cite{Perry2014}, and Ref.~\cite{Kellett2009} for selection details), while the rest of the decay data were adopted from ENSDF~\cite{ENSDF} and the Decay Data Evaluation Project (DDEP)~\cite{ddep}.

\begin{figure*}[h!]
\centering
\includegraphics[width=0.9\columnwidth]{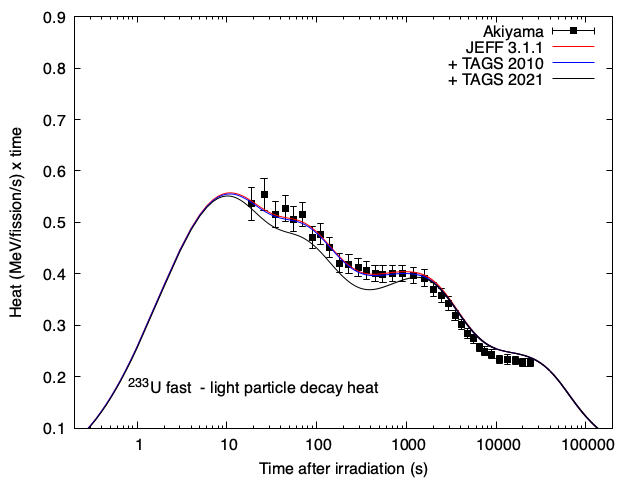}
\includegraphics[width=0.9\columnwidth]{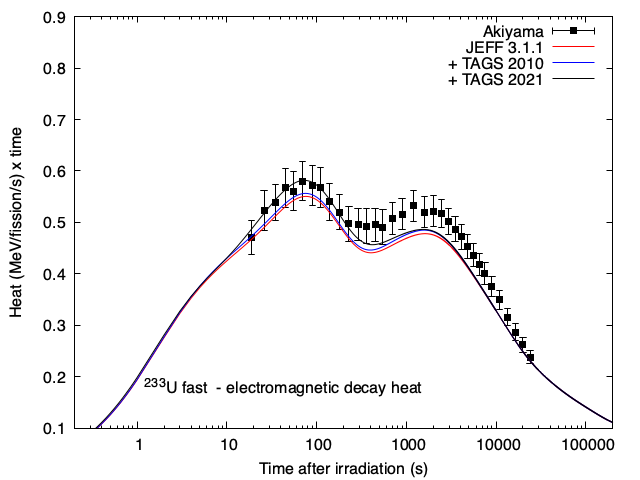}
\includegraphics[width=0.9\columnwidth]{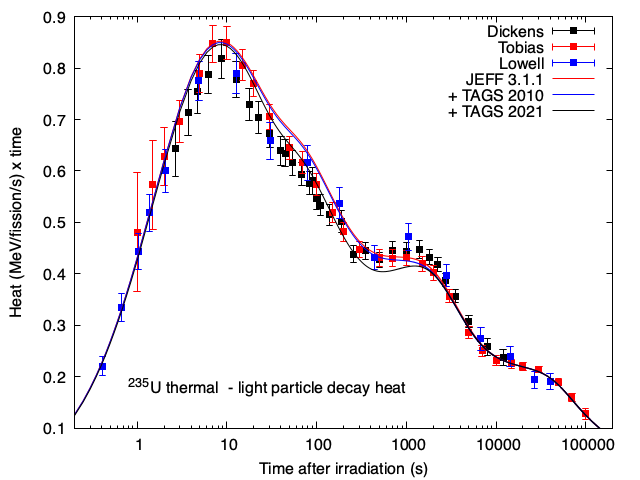}
\includegraphics[width=0.9\columnwidth]{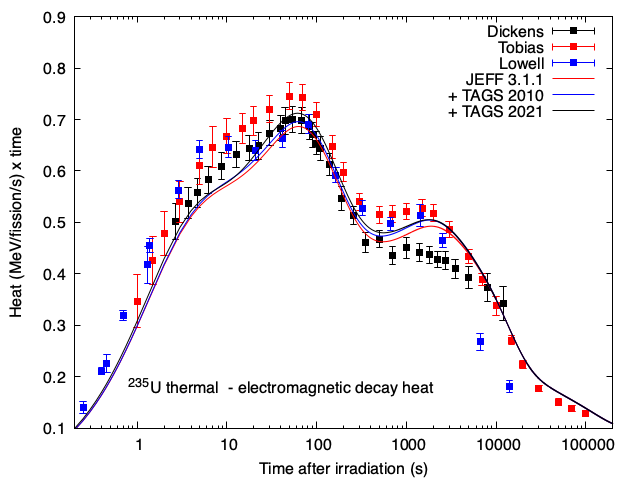}
\includegraphics[width=0.9\columnwidth]{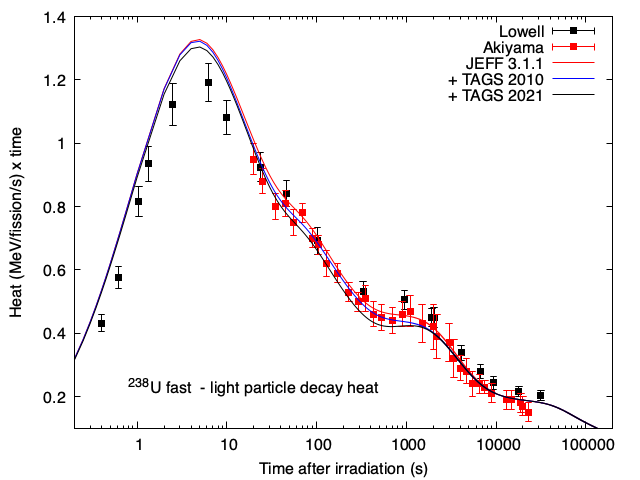}
\includegraphics[width=0.9\columnwidth]{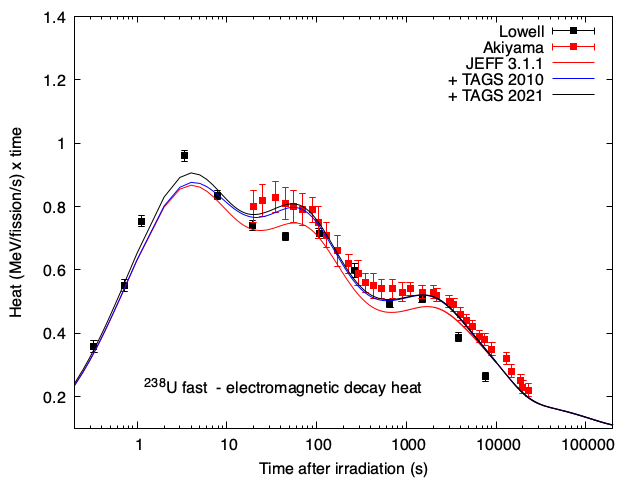}
\includegraphics[width=0.9\columnwidth]{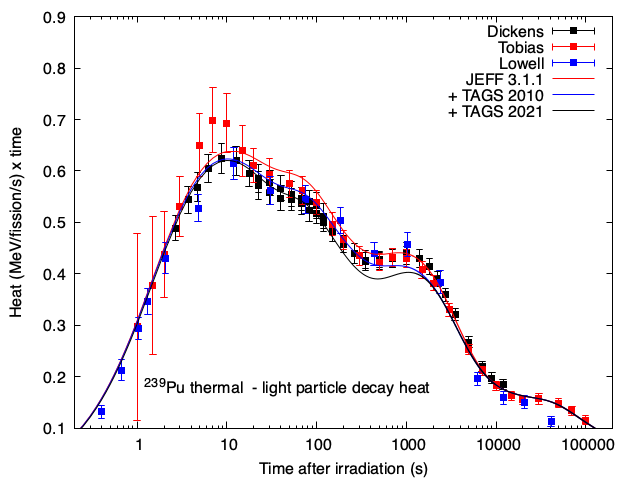}
\includegraphics[width=0.9\columnwidth]{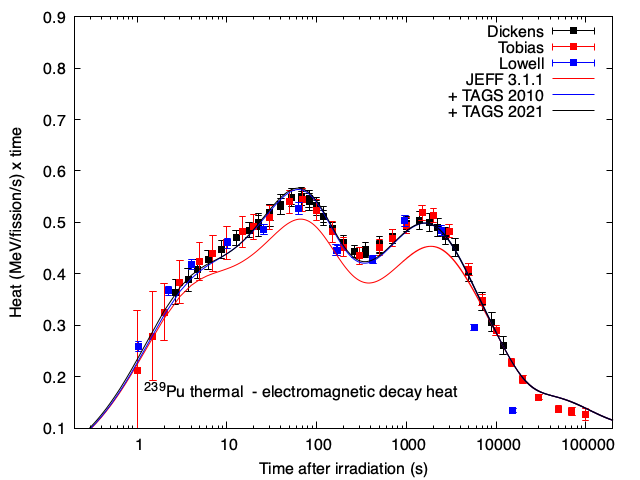}
\caption{Decay heat as a function of cooling time as obtained using JEFF-3.1.1 fission-yield and decay-data sub-libraries~\cite{Kellett2009}, with the addition of TAGS data from~\cite{Algora2010} (+ TAGS 2010) and Tables~(\ref{tab:meanenergies}, \ref{tab:meanenergies-ORNL}) (+ TAGS 2021). Experimental data have been taken from the CoNDERC database~\cite{conderc}. }\label{fig:fig-jeff-1}
\end{figure*}

\begin{figure*}[h!]
\centering
\includegraphics[width=0.9\columnwidth]{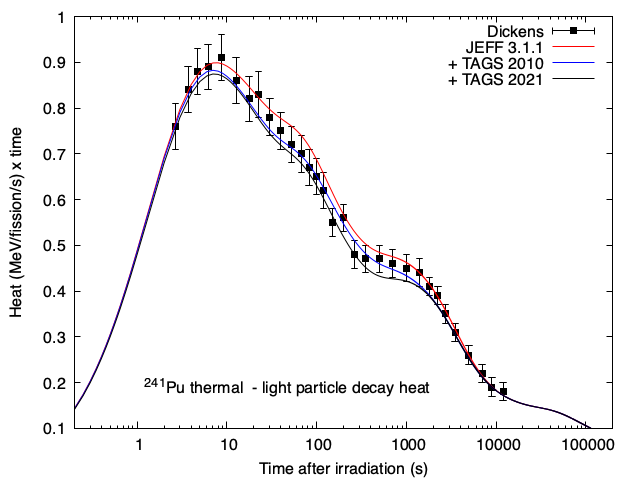}
\includegraphics[width=0.9\columnwidth]{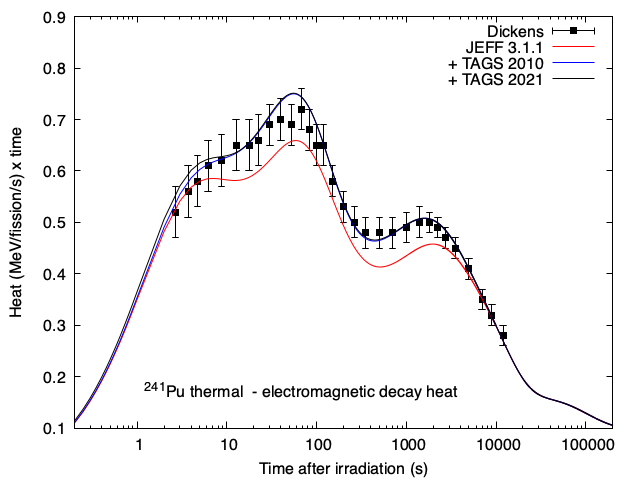}
\includegraphics[width=0.9\columnwidth]{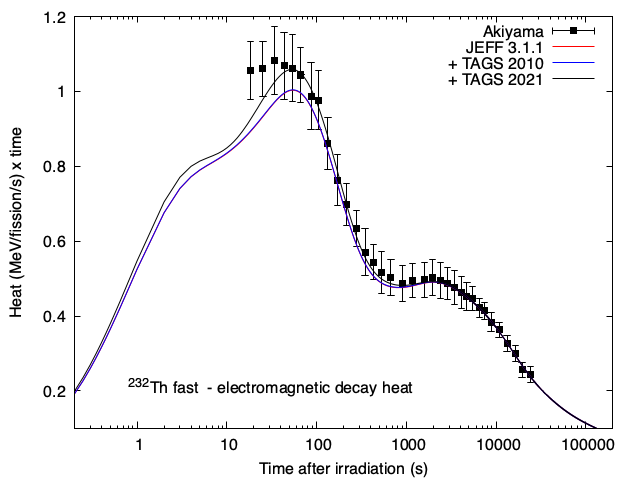}
\includegraphics[width=0.9\columnwidth]{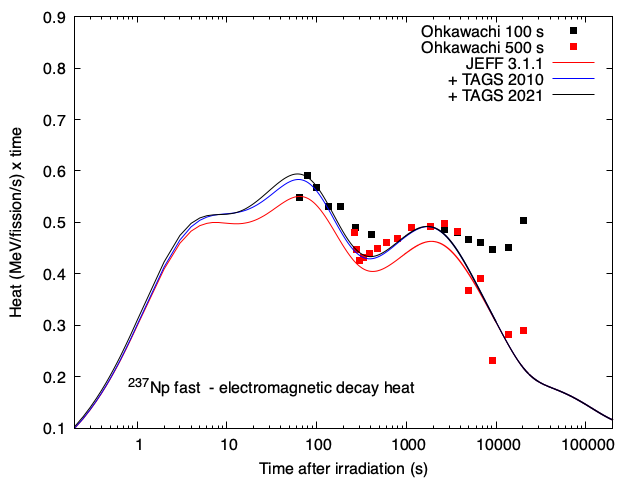}
\caption{Same as Fig.~\ref{fig:fig-jeff-1}. }\label{fig:fig-jeff-2}
\end{figure*}

Three sets of decay-heat calculations denoted earlier in the Section as (1), (2) and (3) were performed with the same JEFF-3.1.1 fission-yield sub-library each time.
The first set of calculations was based on the JEFF-3.1.1 decay data described above, and formed the baseline set of data. A second set of calculations (2) was associated with the replacement of the $\overline{E}_{LP}$ and $\overline{E}_{EM}$ energies of seven fission products $^{102,104,105,106,107}$Tc,$^{101}$Nb and $^{105}$Mo by the TAGS average $\overline{E}_{\beta}$ and $\overline{E}_{\gamma}$ energies measured by Algora~\textit{et al.}~(2010)~\cite{Algora2010} (referred to as {+} TAGS 2010 in Figs.~\ref{fig:fig-jeff-1}, \ref{fig:fig-jeff-2}).
The last set of calculations (3) included the addition of the remaining TAGS data listed in Table~3 of Algora~\textit{et al.}~(2021)~\cite{Algora2021}, and those for $^{89,90}$Kr, $^{89,90,90m}$Rb, $^{96}$Y, $^{137,139}$Xe and $^{142}$Cs from Rasco~\textit{et al.}~(2016)~\cite{Rasco2016} and Fijałkowska~\textit{et al.}~(2017)~\cite{Fijalkowska2017} to the JEFF-3.1.1 decay data library (referred to as + TAGS 2021 in Figs.~\ref{fig:fig-jeff-1}, \ref{fig:fig-jeff-2}). Corresponding average energies are displayed in Tables~\ref{tab:meanenergies}, \ref{tab:meanenergies-ORNL}. As mentioned in Sect.~\ref{endfb}, the TAGS mean energies of $^{96m}$Y, $^{98}$Nb and $^{140}$Cs were not considered as they were available after the cutoff date. 

%\clearpage
As shown in the Figures, results obtained for the LP and EM components of the decay heat for the seven fissioning systems exhibit the same features reported in the previous subsection devoted to the ENDF/B-VIII.0 decay-data and fission-yield sub-libraries. Inclusion of the average $\overline{E}_{\beta}$ and $\overline{E}_{\gamma}$ decay energies from the TAGS measurements (+ TAGS 2010) leads to a decrease in the LP and an increase in the EM decay-heat components, which is a direct consequence of the adoption of Pandemonium-free decay data. TAGS 2010 data improve the agreement with the pulse decay-heat measurements for $^{239,241}$Pu thermal and $^{238}$U fast fission, but has only a small impact on $^{232}$Th and $^{233}$U fast fission. While decay-heat calculations of $^{235}$U thermal fission are associated with an LP component that is not affected by the inclusion of the TAGS 2010 data, conclusions are more difficult to draw when considering the EM component because of discrepancies between the three sets of experimental pulse decay-heat data.

Insertion of all the recent TAGS measurements into the JEFF-3.1.1 decay-data sub-library (+ TAGS 2021) leads to a underestimation of the LP decay-heat component in the 10 to 1000 s cooling time for the following fissioning systems: $^{233,238}$U fast fission and $^{239,241}$Pu thermal fission. The LP component in $^{235}$U thermal fission exhibits a small improvement in the decay-heat calculations from 10 to 1000 s cooling time, along with a small yet continuous underestimation from 400 to 1000 s cooling time. EM decay-heat components for $^{232}$Th and $^{233,238}$U fast fission, and $^{235}$U and $^{241}$Pu thermal fission also improve for cooling times below 100 s. 
%ALN to Vivian, 10 Oct 2022: see above - more very short to short cooling times of little worth to power reactors.
The conclusions and recommendations presented in Subsection~\ref{endfb} are also valid for the impact of TAGS data into the JEFF-3.1.1 sub-library discussed in this subsection.
\subsection{JENDL library}\label{jendl}
As noted in the introduction to Section~\ref{decayheat}, JENDL fission-product decay-data files contain a mixture of experimental and theoretical average decay energies. JENDL/FPD-2000 includes experimental average energies from ENSDF~\cite{ENSDF} and theoretical values from gross \be$^{-}$ theory, all of which are combined to reproduce the experimental decay-heat data. An update from JENDL/FPD-2000 to JENDL/FPD-2011 includes the TAGS 2010 data of Algora~\textit{et al.}~(2010)~\cite{Algora2010}, while average $\overline{E}_\beta$ and $\overline{E}_\gamma$ energies for the other fission-product nuclides have been re-evaluated following a prescription used in all of the evaluations performed before and after FPD-2000 up to FPD-2011: adopted decay data should consistently reproduce, within the uncertainties of the experimental data and theoretical calculations, not only the available experimental decay-heat data, but also the aggregate $\gamma$-ray spectra of Ref.~\cite{Dickens1980,Akiyama1988} and individual $\beta$/$\gamma$ spectra of Ref.~\cite{Rudstam1990B}. Average decay energies were also carefully compared with the measured TAGS data of Greenwood~\textit{et al.}~\cite{Greenwood1997}.  

The impact of the Algora~\textit{et al.}~(2010)~\cite{Algora2010} as well as the most recent TAGS data listed in Tables~\ref{tab:meanenergies}, \ref{tab:meanenergies-ORNL} was assessed without any "contamination" from theoretical data. Two baseline calculations were undertaken separately with the JENDL FP decay-data file 2000(JENDL/FPD-2000) and JENDL FP decay-data file 2011 (JENDL/FPD-2011), and four sets of decay-heat calculations were also performed: one baseline calculation with JENDL/FPD-2000; second calculation in which Algora~\textit{et al.}~(2010) TAGS data were added to the JENDL/FPD-2000 baseline (referred to as + TAGS 2010); third calculation as a baseline with JENDL/FPD-2011 (which includes Algora~\textit{et al.}~(2010) TAGS data with minimum modification to keep consistency with the beta-ray spectrum data); and fourth calculation whereby the remaining TAGS data listed in Tables~\ref{tab:meanenergies}, \ref{tab:meanenergies-ORNL} were added, except for $^{96m}$Y, $^{98}$Nb and $^{140}$Cs (referred to as + TAGS 2021). One notable difference between the ENDF/B-VIII.0 and JEFF-3.1.1 used in the previous subsections and the above JENDL decay-data sub-libraries is that the latter does not include any of the Greenwood~\textit{et al.}~TAGS data~\cite{Greenwood1997}, whereas the other two sub-libraries incorporate them either completely (ENDF/B-VIII.0) or partially (JEFF-3.1.1). Under such circumstances, theoretical average decay energies have been calculated for these specific fission products that lead to good description of the experimental decay-heat data for their subsequent introduction into the JENDL decay-data sub-libraries. 
%\textcolor{blue}{The JENDL FP Decay Data File 2011 (JENDL/FPD-2011)~\cite{Katakura2011} was used as the baseline to explore the impact of TAGS average energies on DH calculations with JENDL libraries. Because this file, one of the Special Purpose Files annexed to JENDL, already included the Algora’s data (TAGS2010), we had to introduce a secondary baseline which does not contain TAGS2010.  For this purpose, we took the old JENDL FP Decay Data File 2000 (JENDL/FPD-2000)~\cite{Katakura2000} which was generated primarily based on ENSDF~\cite{ENSDF} and the improved version\cite{Tachibana1990} of gross theory of $\beta$-decay. During the evaluation process attention was paid to keep the consistency with the experimental  $\beta$\cite{Rudstam1990B}- and $\gamma$\cite{Akiyama1988}-ray spectra. There the theoretical energies still fully used for nuclides having $Q$-values larger than 5 MeV. These theory nuclides can easily be identified as their energies are assigned zero errors.  In updating from FPD-2000 to FPD-2011, they included the TAGS2010 (Algora, 2010) data and re-evaluated the $\beta$ and $\gamma$ energies of other nuclides paying a full and careful attention to the Greenwood's TAGS data.}
The JENDL Fission-product Yield Data File 2011 \cite{Katakura2011} was used in the baseline and subsequent calculations. Both the fission-product decay and fission-product yield files originate from the JNDC Nuclear Data Library of Fission Products, as mentioned earlier in Section~\ref{decayheat}~\cite{Tasaka1983}. 

\begin{figure*}[h!]
\centering
\includegraphics[width=0.9\columnwidth]{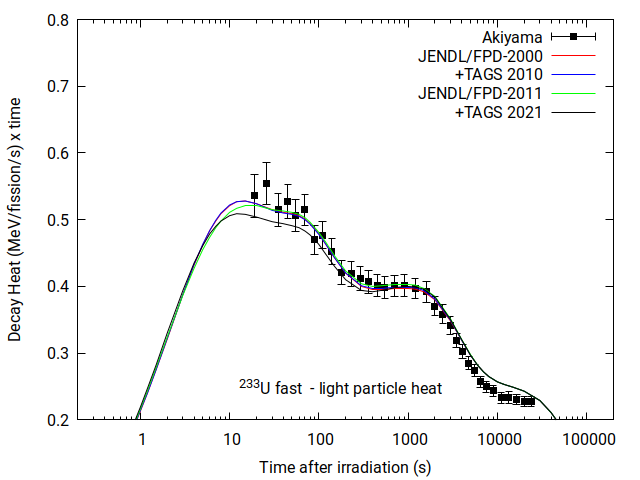}
\includegraphics[width=0.9\columnwidth]{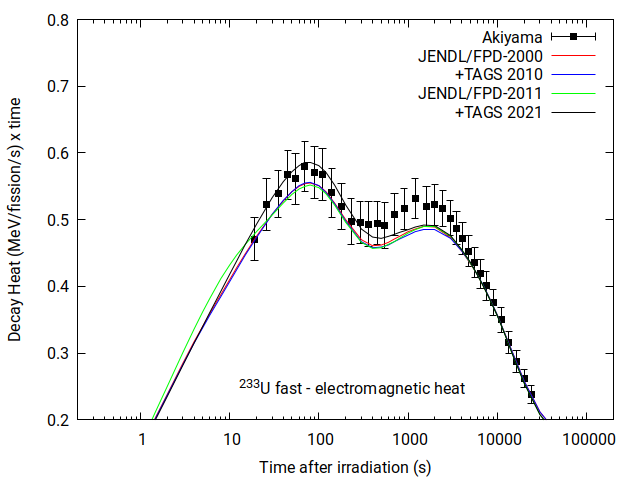}
\includegraphics[width=0.9\columnwidth]{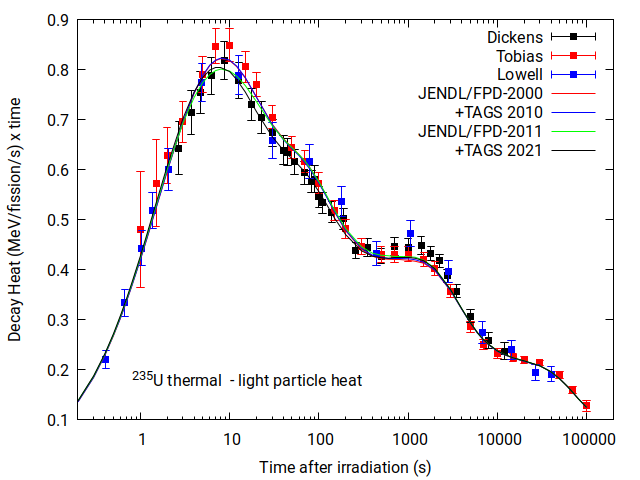}
\includegraphics[width=0.9\columnwidth]{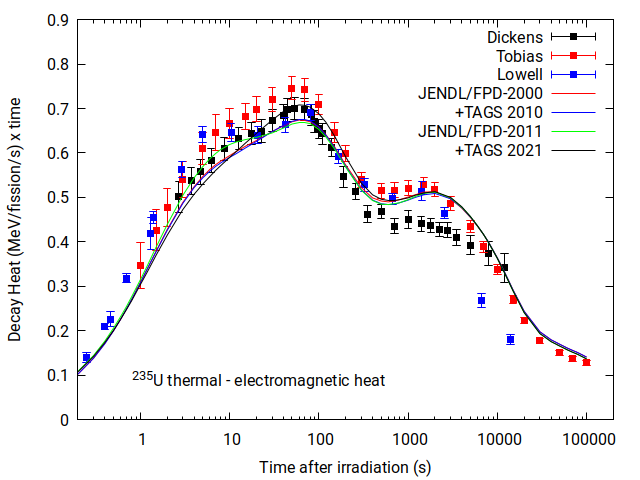}
\includegraphics[width=0.9\columnwidth]{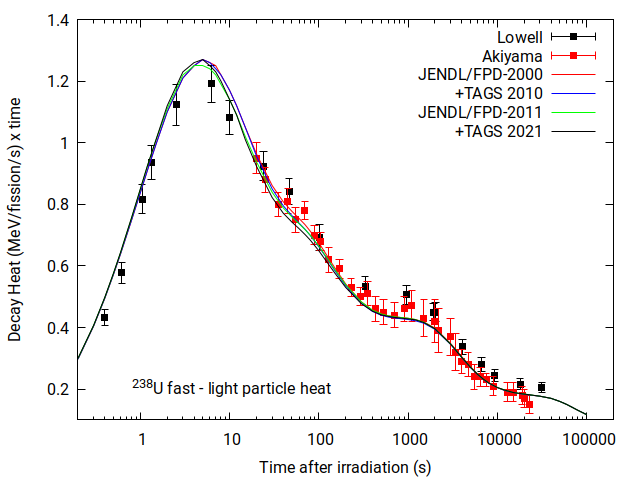}
\includegraphics[width=0.9\columnwidth]{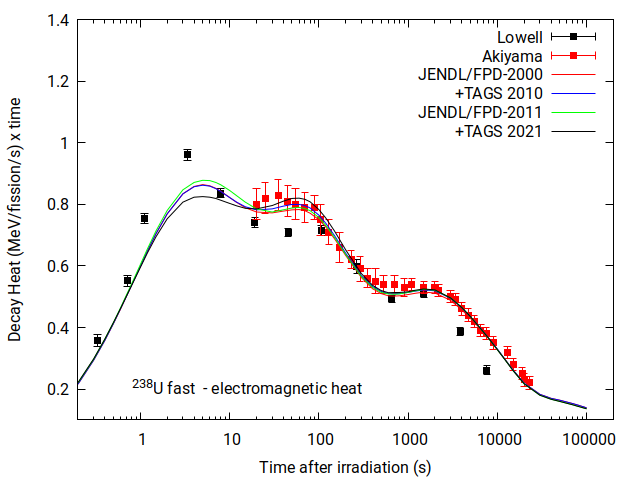}
\includegraphics[width=0.9\columnwidth]{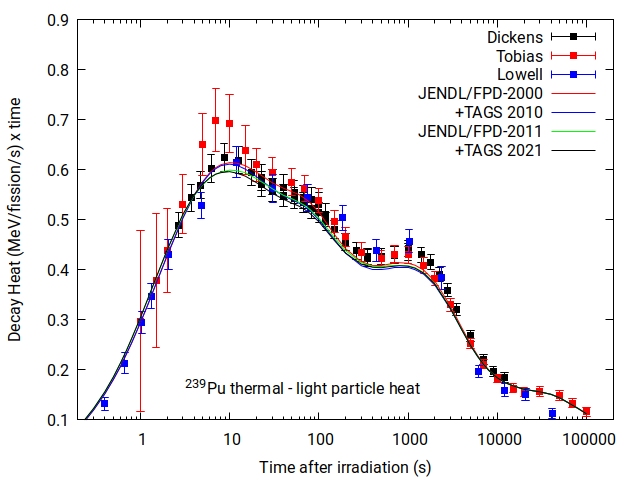}
\includegraphics[width=0.9\columnwidth]{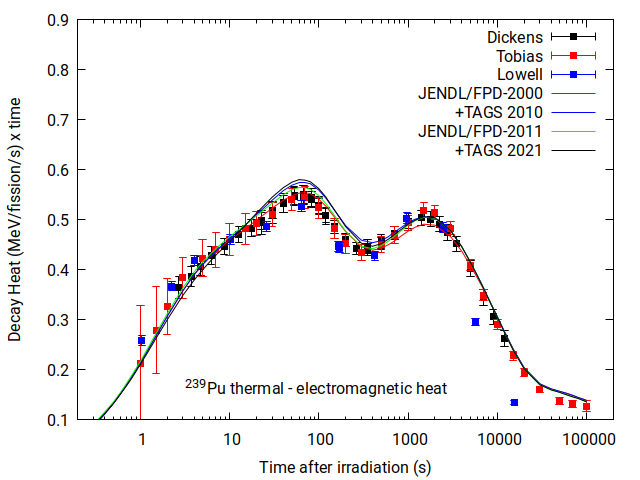}
\caption{Decay heat as a function of cooling time as obtained using JENDL fission-yield and decay-data sub-libraries~\cite{jendl4} with the addition of TAGS data from~\cite{Algora2010} (+ TAGS 2010) and~Tables~(\ref{tab:meanenergies}, \ref{tab:meanenergies-ORNL}) (+ TAGS 2021). Experimental data have been taken from the CoNDERC database~\cite{conderc}.}\label{fig:fig-jendl-1}
\end{figure*}

\begin{figure*}[h!]
\centering
\includegraphics[width=0.9\columnwidth]{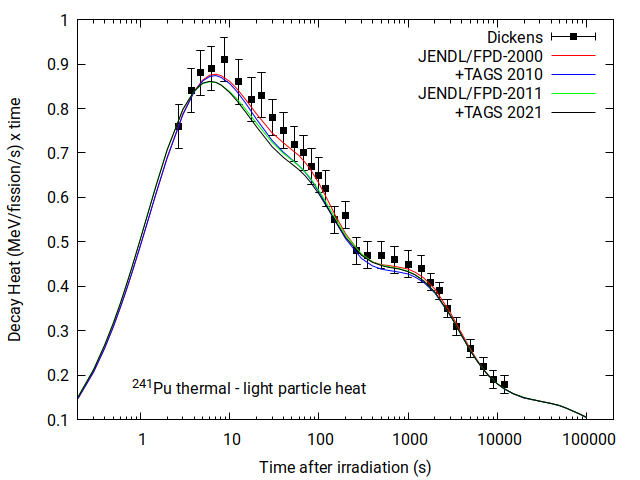}
\includegraphics[width=0.9\columnwidth]{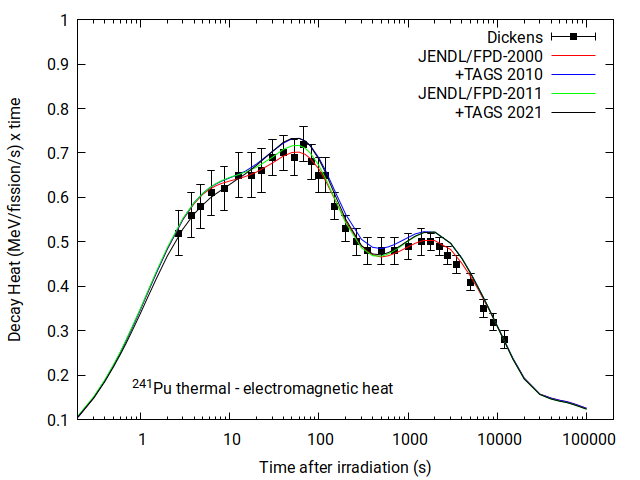}
\includegraphics[width=0.9\columnwidth]{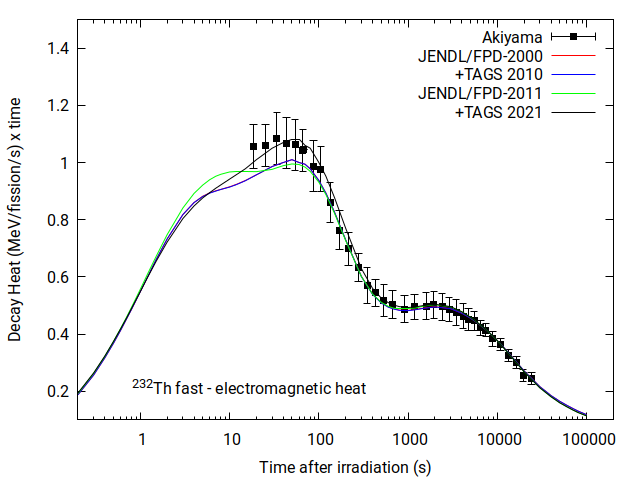}
\includegraphics[width=0.9\columnwidth]{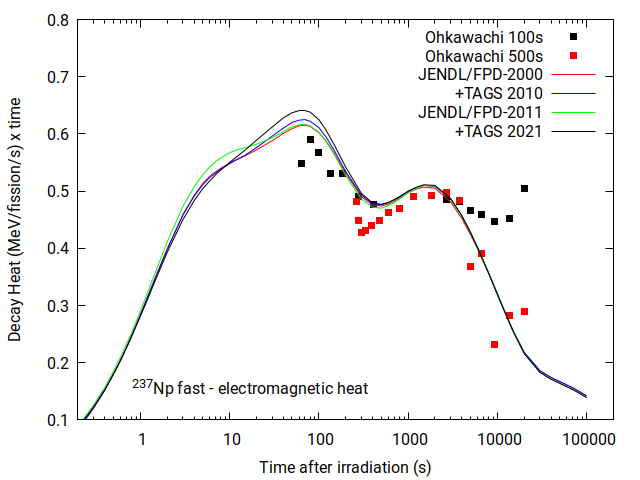}
\caption{Same as Fig.~\ref{fig:fig-jendl-1}. }\label{fig:fig-jendl-2}
\end{figure*}

All of decay-heat calculations undertaken in conjunction with the JENDL data files were performed by means of the OYAK98 code~\cite{Oyamatsu1999}, and the results have been compared with available experimental decay-heat data in Figs.~\ref{fig:fig-jendl-1}, \ref{fig:fig-jendl-2}. 

%\clearpage
Both of the baseline calculations for JENDL/FPD-2000 and JENDL/FPD-2011 lie predominantly within the experimental uncertainties, except for a few cases such as the EM component for $^{233}$U fast fission over cooling times of 500 to 3000 s and the LP component for $^{239}$Pu thermal fission at cooling times from 400 to 2000 s. These observations indicate that the gross \be$^{-}$ theory reproduces the average decay energies reasonably well, as applied and described in the introductory part of this section. Deviations from the baseline curves caused by the introduction of the average energies determined from TAGS measurements are relatively small. However, the introduction of TAGS data in the case of $^{239}$Pu thermal fission increases the EM component considerably beyond the experimental uncertainties at cooling times of 20 to 300 s. Possible reasons for this overestimation are discussed in Subsection~\ref{comparison}. Overall, all four sets of decay-heat calculations give a reasonable description of the experimental decay-heat data as shown in Figs.~\ref{fig:fig-jendl-1}, \ref{fig:fig-jendl-2}. However, this exception should not detract from the importance of including TAGS data in the decay-data sub-libraries - they are expected to represent the best recommended data for each individual nuclide, whereas gross \be$^{-}$ theory is limited to only average-energy properties. 
\subsection{Comparison of libraries}\label{comparison}
We have undertaken studies of the impact of recent TAGS measurements on pulse decay-heat calculations that were based upon comparisons of the available experimental decay-heat data with the results of three different modelling codes and differing input libraries of nuclear data. Assessments were made in order to judge the performance of the latest general-purpose evaluated libraries (ENDF/B-VIII.0~\cite{Brown2018}, JEFF-3.3~\cite{Plompen2020}, and JENDL-5~\cite{Iwamoto2021}), in calculations of pulse decay heat for the sixteen fissioning systems listed in Section~\ref{decaydata}. The relevant sub-libraries of these databases differ in the content of their recommended cross sections, decay data and fission-yield data. Our efforts have focused mainly on actinide and fission-product decay-heat calculations that are most strongly dependent on the adopted fission yields and decay data used as input to the inventory calculations. We have described in Section~\ref{endfb} the inclusion of measured TAGS data within the decay-data sub-library of ENDF/B-VIII.0; JEFF-3.3 includes TAGS data from Greenwood~\textit{et al.}~(1997)~\cite{Greenwood1997,Kellett2009}, Algora~\textit{et al.}~(2010)~\cite{Algora2010} (with the exception of $^{101}$Nb and $^{102}$Tc), and for $^{87}$Br, $^{88}$Br, $^{92}$Rb and $^{94}$Rb~\cite{Valencia2017,Zakari2015}, as defined in Subsection~\ref{jeff}; JENDL-5~\cite{Iwamoto2021} contains all of the TAGS measurements performed as part of 
% Vivian, 14 Nov 2022: modified following AA remark
the TAS/DTAS collaboration listed in the review article of Algora~\textit{et al.}~\cite{Algora2021}, and several nuclides from the MTAS collaboration published in Fijałkowska~\textit{et al.}~\cite{Fijalkowska2017} (see Tables~\ref{tab:meanenergies}, \ref{tab:meanenergies-ORNL}), as noted in Subsection~\ref{jendl}. The JENDL-5 decay-data sub-library is judged to be the most complete database for the inclusion of available TAGS data.

\begin{figure*}[h!]
\centering
\includegraphics[width=0.85\columnwidth]{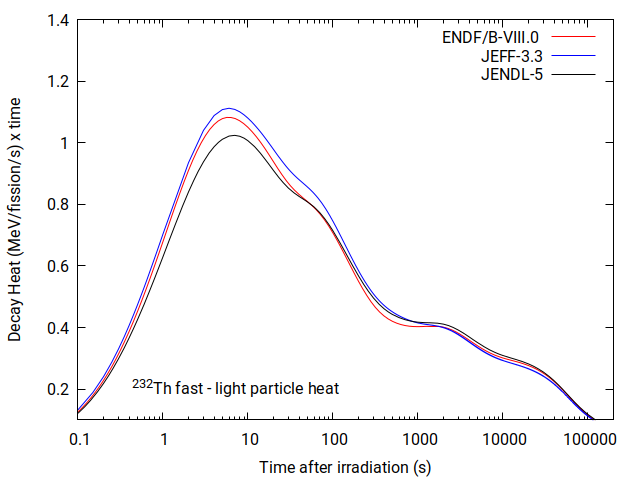}
\includegraphics[width=0.85\columnwidth]{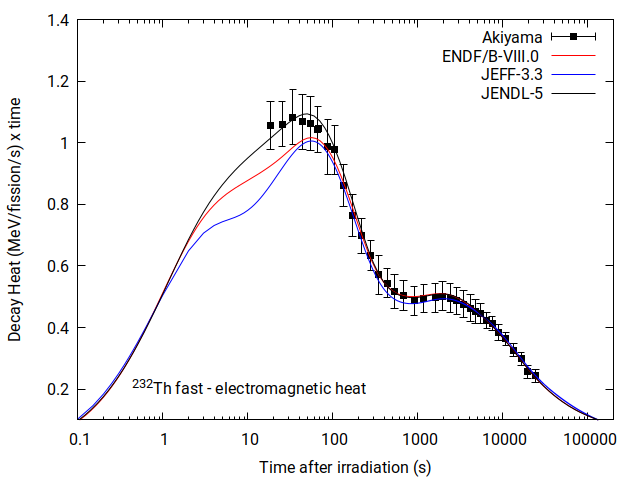}
\includegraphics[width=0.85\columnwidth]{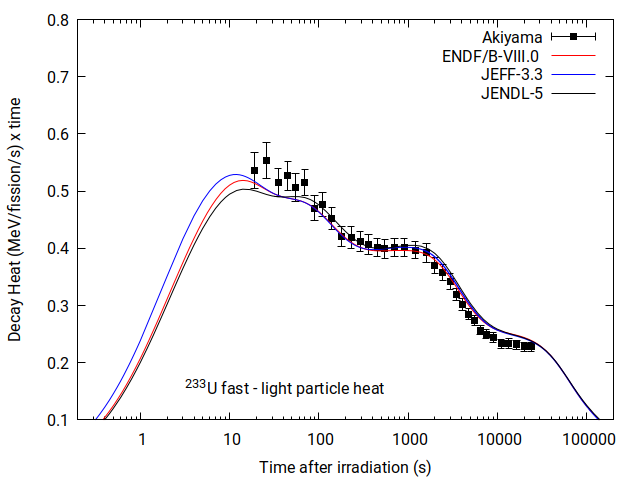}
\includegraphics[width=0.85\columnwidth]{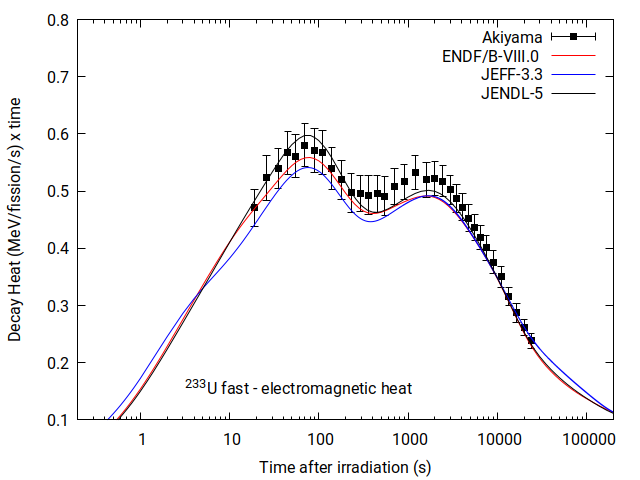}
\includegraphics[width=0.85\columnwidth]{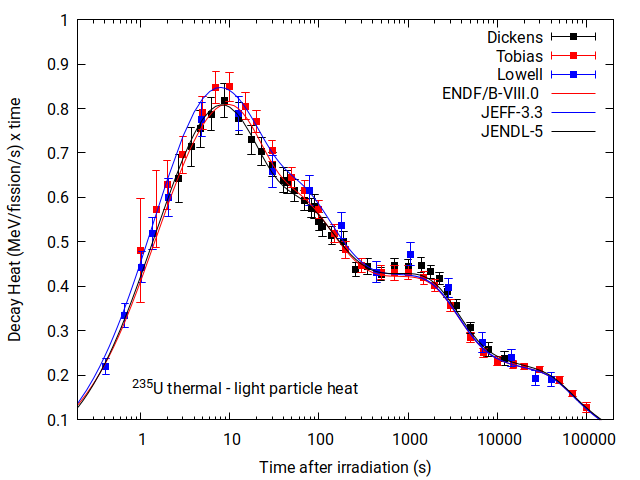}
\includegraphics[width=0.85\columnwidth]{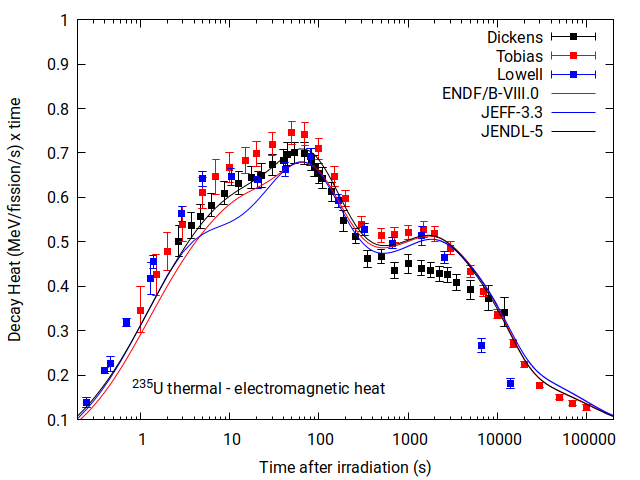}
\includegraphics[width=0.85\columnwidth]{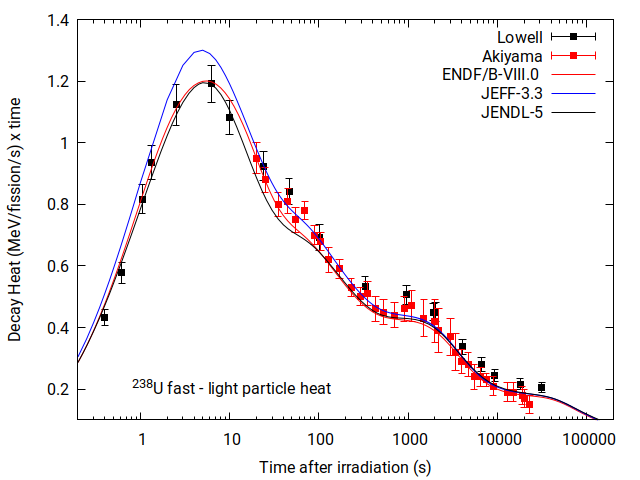}
\includegraphics[width=0.85\columnwidth]{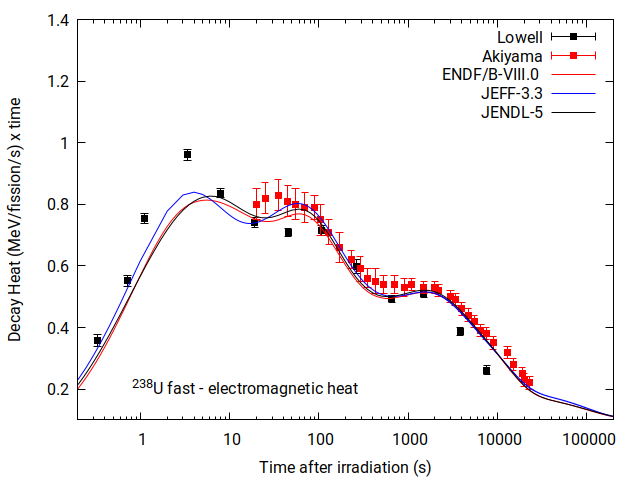}
\caption{Decay heat as a function of cooling time as obtained using ENDF/B-V111.0, JEFF-3.3 and JENDL-5 evaluated libraries. Experimental data have been taken from the CoNDERC database~\cite{conderc}. }\label{fig:fig-comp-1}
\end{figure*}

\begin{figure*}[h!]
\centering
\includegraphics[width=0.85\columnwidth]{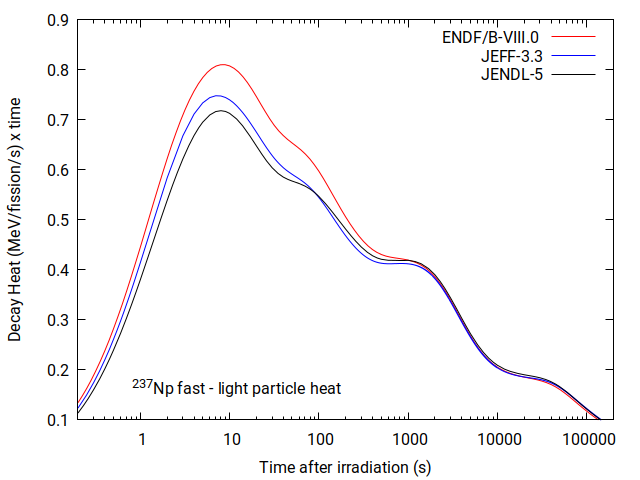}
\includegraphics[width=0.85\columnwidth]{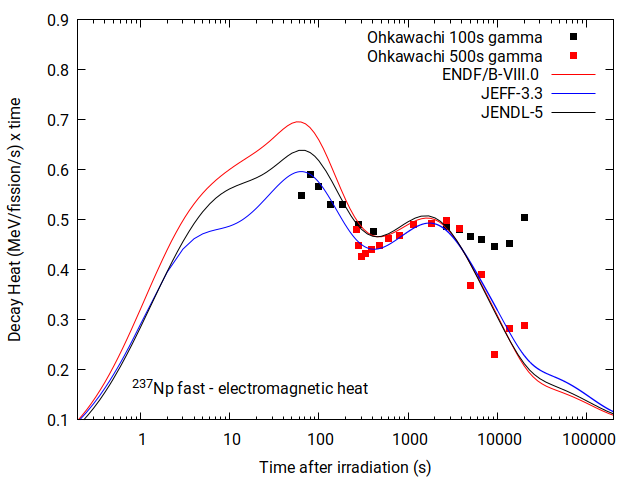}
\includegraphics[width=0.85\columnwidth]{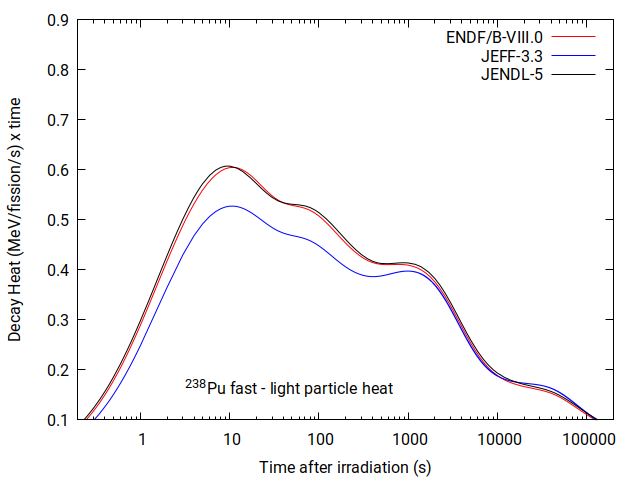}
\includegraphics[width=0.85\columnwidth]{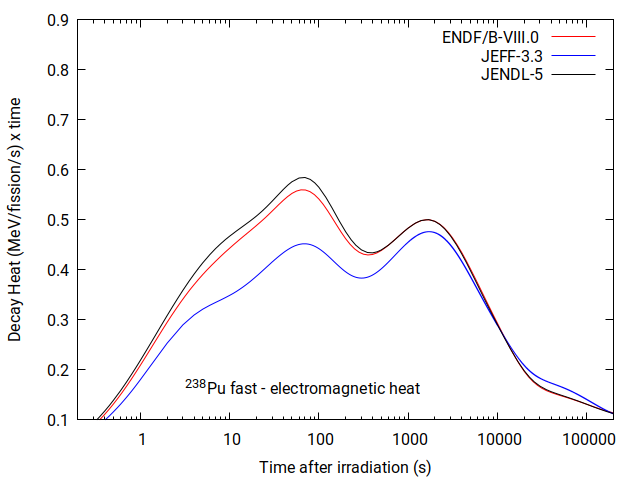}
\includegraphics[width=0.85\columnwidth]{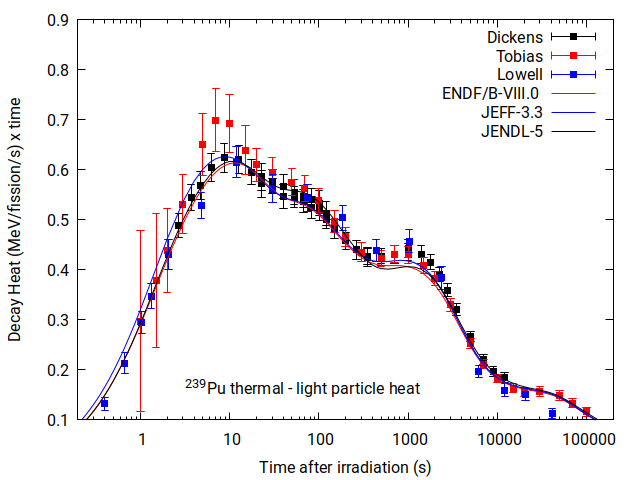}
\includegraphics[width=0.85\columnwidth]{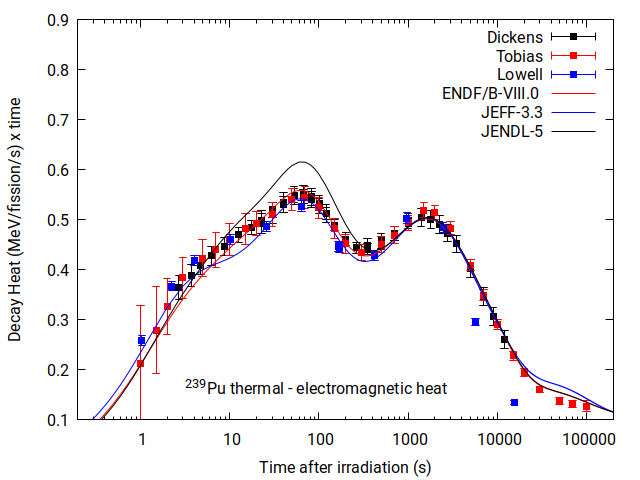}
\includegraphics[width=0.85\columnwidth]{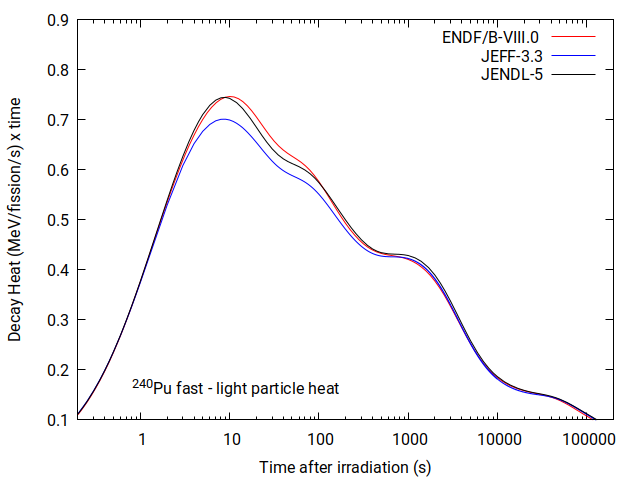}
\includegraphics[width=0.85\columnwidth]{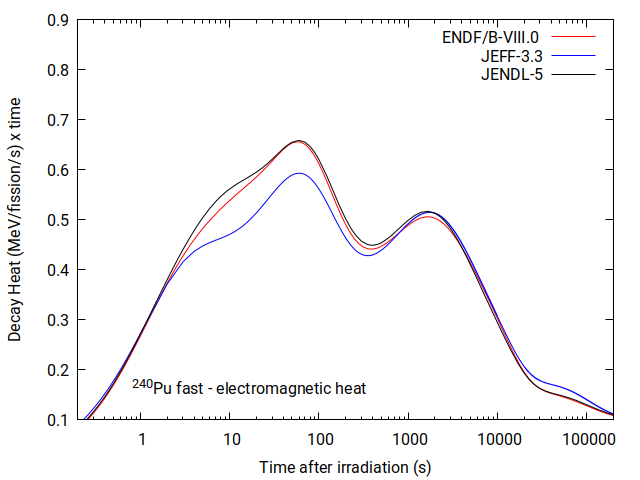}
\caption{Same as Fig.~\ref{fig:fig-comp-1}. }\label{fig:fig-comp-2}
\end{figure*}

\begin{figure*}[h!]
\centering
\includegraphics[width=0.85\columnwidth]{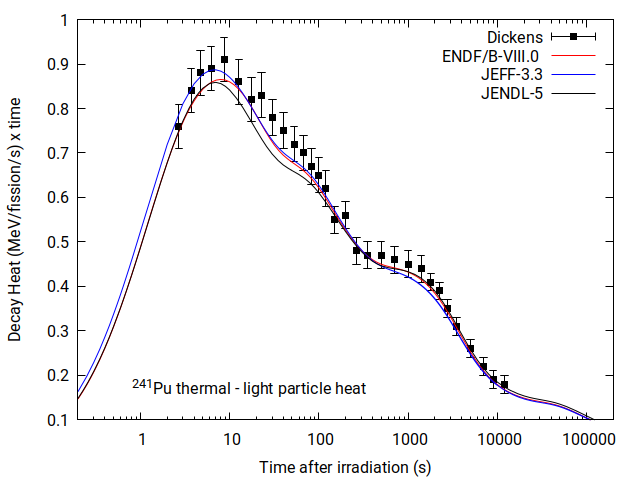}
\includegraphics[width=0.85\columnwidth]{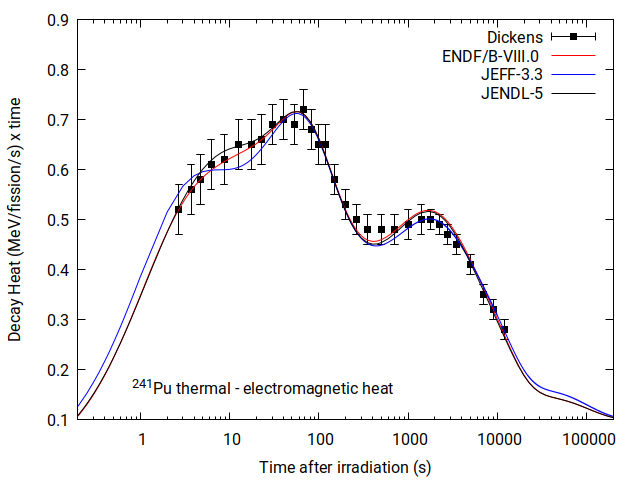}
\includegraphics[width=0.85\columnwidth]{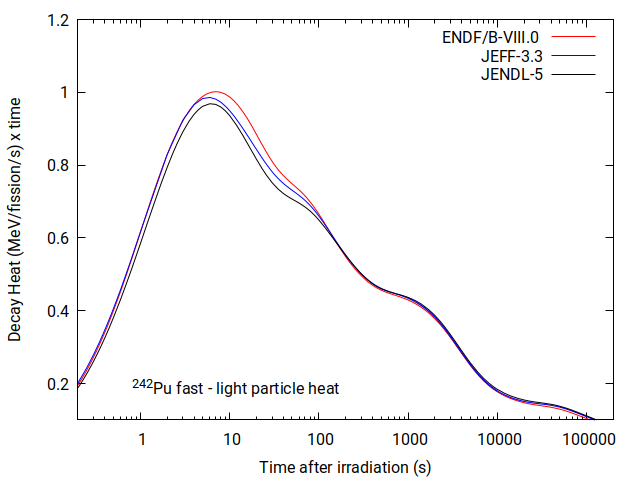}
\includegraphics[width=0.85\columnwidth]{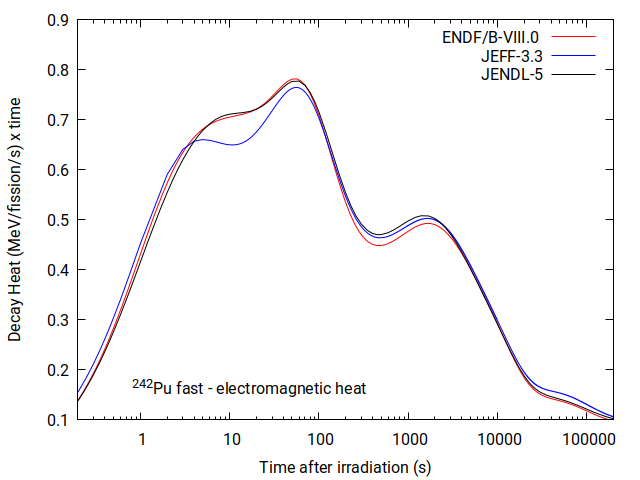}
\includegraphics[width=0.85\columnwidth]{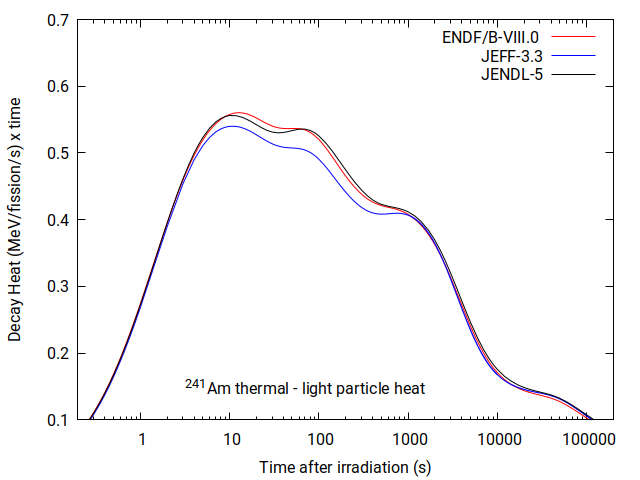}
\includegraphics[width=0.85\columnwidth]{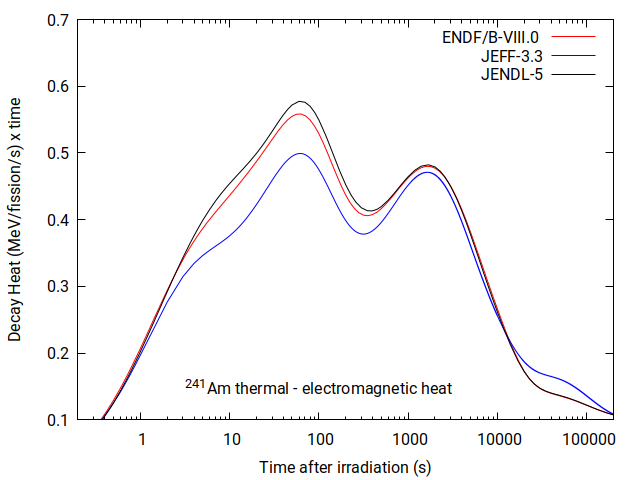}
\includegraphics[width=0.85\columnwidth]{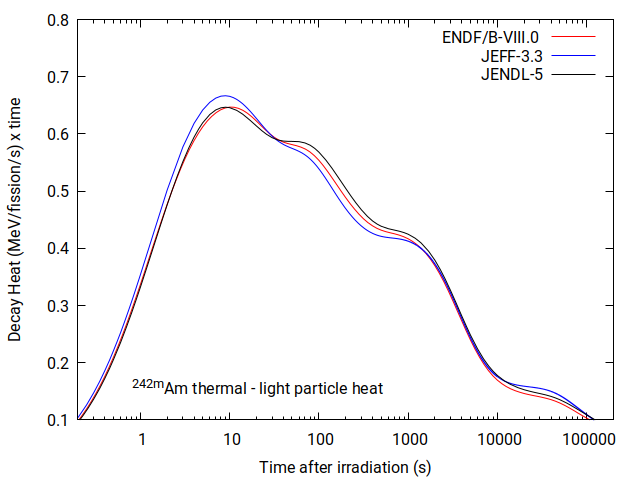}
\includegraphics[width=0.85\columnwidth]{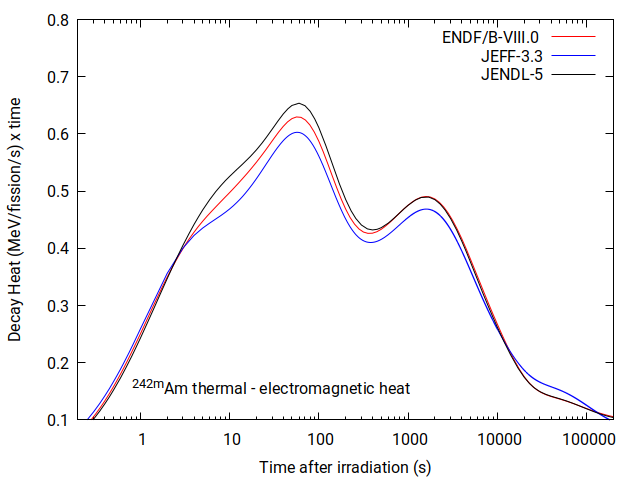}
\caption{Same as Fig.~\ref{fig:fig-comp-1}. }\label{fig:fig-comp-3}
\end{figure*}

\begin{figure*}[h!]
\centering
\includegraphics[width=0.9\columnwidth]{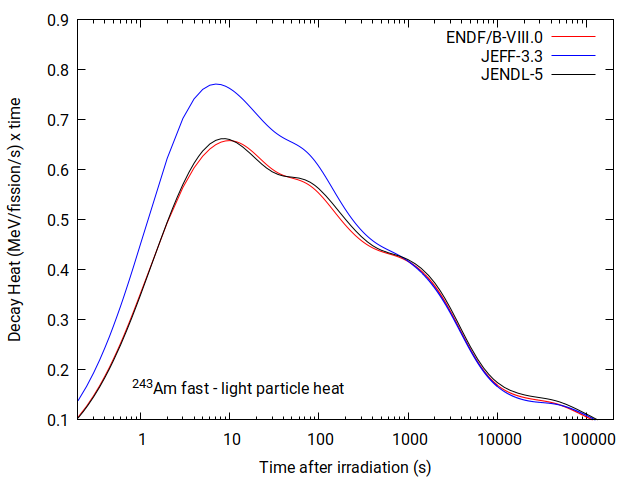}
\includegraphics[width=0.9\columnwidth]{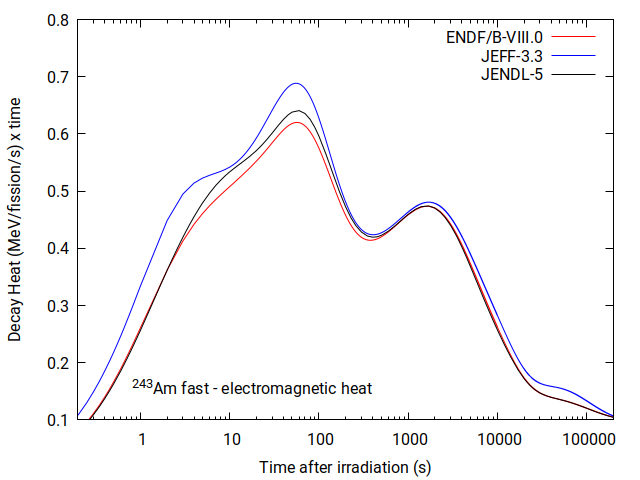}
\includegraphics[width=0.9\columnwidth]{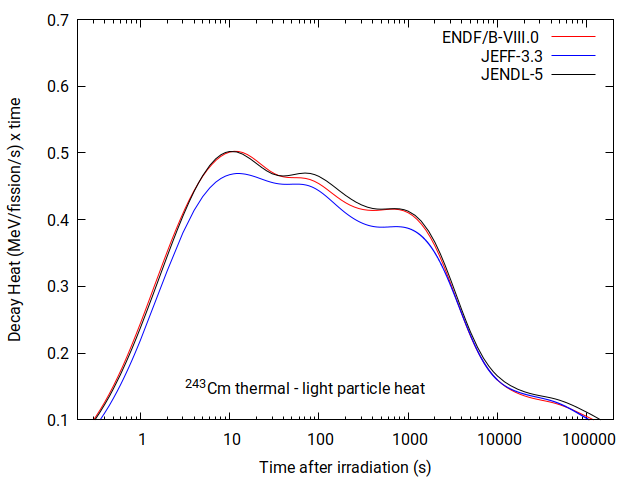}
\includegraphics[width=0.9\columnwidth]{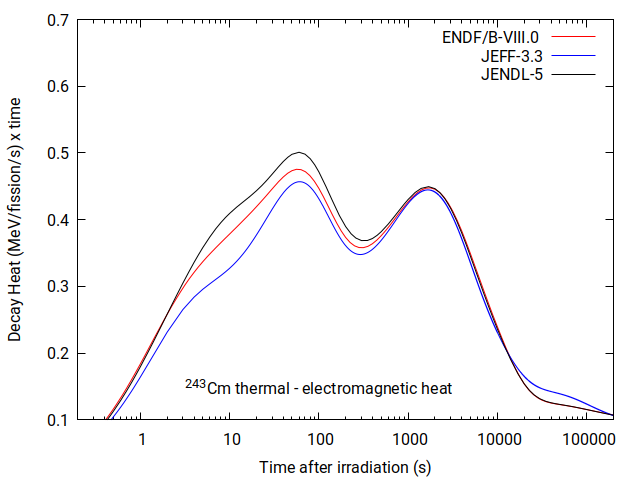}
\includegraphics[width=0.9\columnwidth]{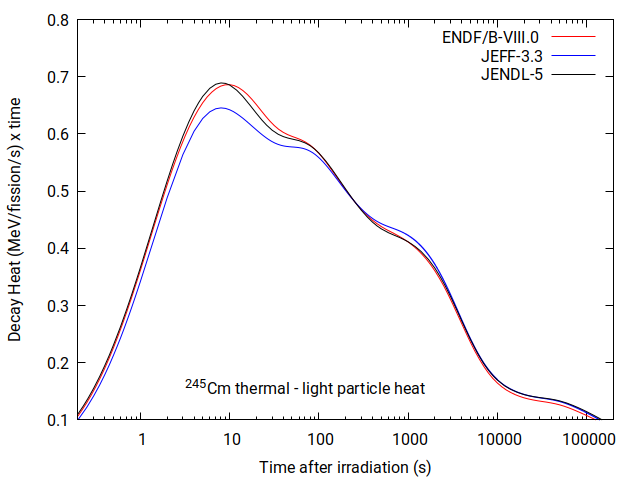}
\includegraphics[width=0.9\columnwidth]{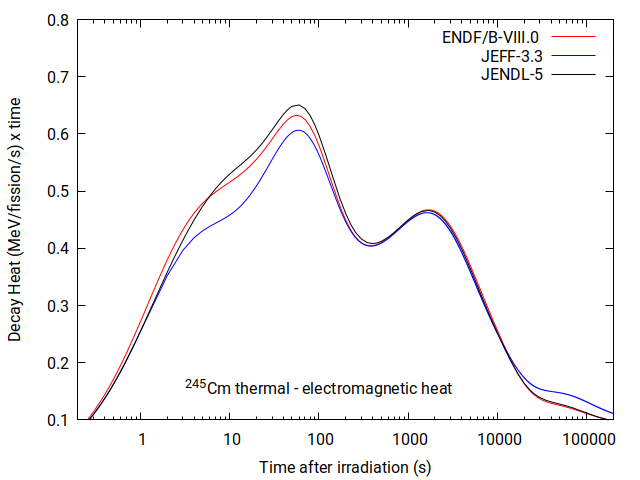}
\caption{Same as Fig.~\ref{fig:fig-comp-1}. }\label{fig:fig-comp-4}
\end{figure*}

Comparisons of the various calculated decay-heat studies with experimental pulse decay-heat data are shown in Figs.~\ref{fig:fig-comp-1}~to~\ref{fig:fig-comp-4}. The resulting decay heat curves are in reasonably good agreement at cooling times above 1000 s, while they can be rather discrepant at shorter cooling times below 1000 s. 
%ALN to Vivian, 10 Oct 2022: someone wrote: "At shorter cooling times where shorter-lived fission products are more important, one expects to see a stronger impact of the Pandemonium effect. Measured TAGS data are therefore more influential than at larger cooling times and the differences in the DH results could be attributed to the different TAGS data implemented in the different libraries."  As far as I am concerned (ALN), what is written above is NOT necessarily valid (does not always follow) - why should shorter-lived fission products possess larger Pandemonium effects than longer-lived fission products? Presence of Pandemonium is more likely to be dependent on the actual Q(beta) value and some of the detail within the evaluated upper-energy region of the daughter nuclear-level structure.
The most marked discrepancies with respect to experimental decay-heat data are observed in the following cases: $^{232}$Th fast fission, where the electromagnetic component from 2 to 100 s cooling times is underestimated by both ENDF/B-VIII.0 and JEFF-3.3; $^{238}$U fast-fission cooling times of 1 to 10 s whereby JEFF-3.3 overestimates the decay heat; $^{237}$Np fast fission, for which ENDF/B-VIII.0 and JENDL-5 overestimate the decay-heat data at around 100 s cooling time; $^{239}$Pu thermal fission, where JENDL-5 overestimates the peak at cooling times from 10 to 300 s. 
%ALN to Vivian, 10 Oct 2022: see above- more very short to short cooling times of little worth to power reactors.

\begin{figure}[H]
\centering
\includegraphics[width=1.0\columnwidth]{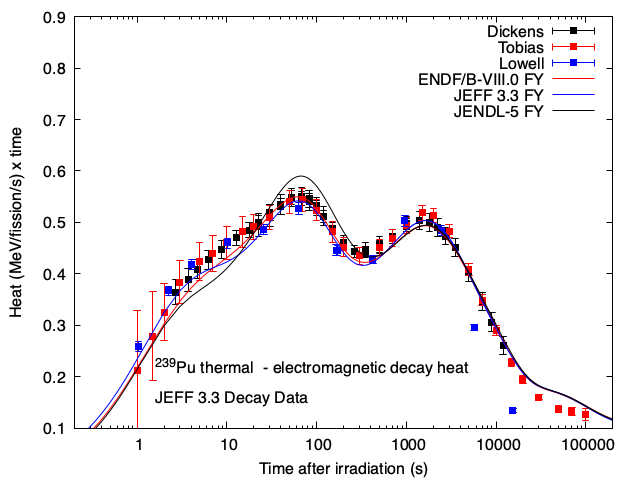}
\caption{Comparison of the EM component of the decay heat of $^{239}$Pu thermal fission obtained from calculations with the JEFF-3.3 decay-data sub-library and three fission-yield sub-libraries (ENDF/B-VIII.0, JEFF-3.3 and JENDL-5).}\label{fig:jeff-fy-1}
\end{figure}

\begin{figure}[H]
\centering
\includegraphics[width=1.0\columnwidth]{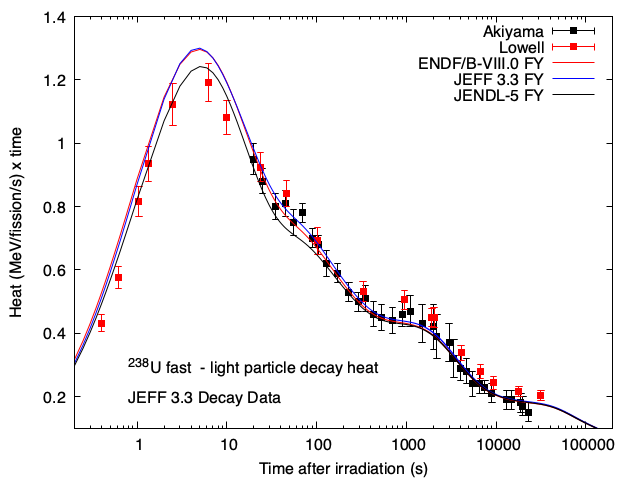}
\includegraphics[width=1.0\columnwidth]{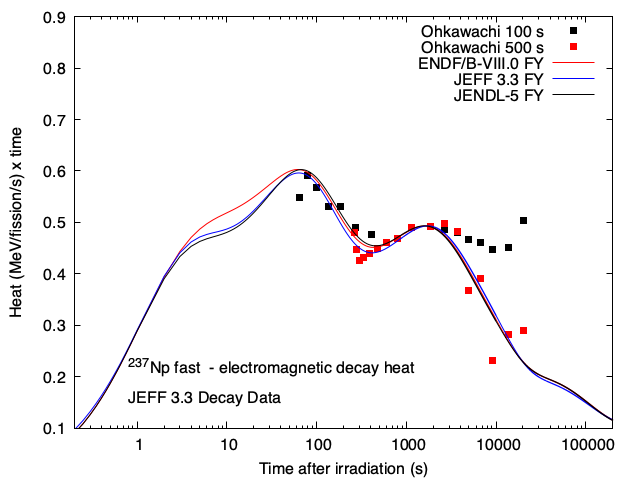}
\caption{Comparison of light-particle decay heat of $^{238}$U fast fission and electromagnetic decay heat of $^{237}$Np fast fission obtained from calculations with the JEFF-3.3 decay-data sub-library and three fission-yield sub-libraries (ENDF/B-VIII.0, JEFF-3.3 and JENDL-5).}\label{fig:jeff-fy}
\end{figure}

%\clearpage
Given the importance of $^{239}$Pu thermal fission for power-reactor applications, the results obtained with JENDL-5 were further investigated to identify the possible source of this overestimation. Decay-heat calculations will also be sensitive to the fission-yield libraries, and therefore the decay-heat calculations were repeated for $^{239}$Pu in terms of the same JEFF-3.3 decay-data sub-library combined with three different fission-yield sub-libraries: ENDF/B-VIII.0 fission yields, JEFF-3.3 fission yields, and JENDL-5 fission yields. The results are shown in Fig.~\ref{fig:jeff-fy-1}.  
%ALN to Vivian, 10 Oct 2022: How confident are you that these three fission-yield sub-libraries are significantly different?!!  Historically, I have been aware in the distant past of nuclear data sub-libraries that were sometimes claimed by their "authors" to be of national origin, but in fact were 80% to 90% taken from other nations' sub-libraries! China used to do this regularly with USA and European cross sections and fission yields, and Europe with USA/Japan cross sections%
%ALN to Vivian, 10 Oct 2022: You cite Fig. 2 here, followed later on in this paragraph by Fig. 1, then Fig. 3, and then Figs. 10 to 13.  I have not been following how you have been citing the Figs within this paper, but is this ordering OK throughout the whole paper?  Figs. should be numbered in the order with which you first cite them (Fig. 1, Fig. 2, Fig. 3, Fig. 4, etc.), as well as all the individual Figures being very closely located to where you first individually cite them in the actual paper (at present, the latter does not occur for Figs 4 to 13).%
%using the same JENDL-5 DD library and two different sets of FY data, the JENDL-5 FY and the JENDL/FPY2011 library, respectfully. The calculations were performed for Th, U and Pu isotopes: $^{232}$Th, $^{235,238}$U, $^{239,240,241,242}$Pu. A comparison of the results obtained for total decay heat is shown in Fig.~\ref{fig:jendl-fy}.
%\begin{figure}[h!]
%\centering
%\includegraphics[width=1.0\columnwidth]{Figures/JDL5DH_FY_ThU.pdf}
%\includegraphics[width=1.0\columnwidth]{Figures/JDL5DH_FY_Pu.pdf}
%\caption{Comparison of total (beta plus gamma) decay heat for Th, U and Pu isotopes obtained using JENDL-5 DD %and two sets of FY data libraries: JENDL-5 and JENDL/FPY2011.}\label{fig:jendl-fy}
%\end{figure}
As shown clearly in Fig.~\ref{fig:jeff-fy-1}, adoption of the ENDF/B-VIII.0 and JEFF-3.3 fission-yield sub-libraries give similar results that agree reasonably well at and around both peaks of the experimental decay-heat data, while the JENDL-5 fission-yield sub-library leads to a significant overestimation of the first peak of decay heat at cooling times between 30 and 200 s, similar to the results obtained with the JENDL-5 decay-data sub-library in Fig.~\ref{fig:fig-comp-2}. 
%ALN to Vivian, 10 Oct 2022: my regular point of note - is it OK to write this with respect to benchmarks that use pure single-actinide targets, but not when considering decay-heat calculations for power reactors ("invalid" nature of the first 200 secs)? - there are large impacts immediately after shutdown that affect gross heat output by the core that can be based on available heat-transfer and fluid-flow models, and thus power-plant operations ensure that necessary coolant flow is maintained for a long enough time to guarantee their safety.
%ALN to Vivian, 10 Oct 2022: another part of someone's above statement that puzzles me is "...... similar to the results obtained with the JENDL-5 decay-data library in Fig. 11" - does this imply anything at all about the adopted fission yields? But rather that some of the JENDL-5 decay-data sub-library files may be inferior to those of ENDF/B-VIII.0 and JEFF-3.3 decay-data sub-libraries!
%that the two different FY libraries have a strong impact at short cooling times, especially for $^{232}$Th fast fission, $^{238}$U fast fission, and $^{239}$Pu thermal fission. These are the cases where large discrepancies were also observed among the three evaluated libraries in Figs.~\ref{fig:fig-comp-1}-\ref{fig:fig-comp-4}. In the case of $^{239}$Pu, the largest differences between JENDL-5 and JENDL/FPY2011 arise at 20 -200 s cooling times which is the region where JENDL-5 overestimates the decay-heat data as can be seen in Fig.~\ref{fig:fig-comp-2}.
This overestimation would appear to be associated with the fission yields rather than the decay data, and is now being studied in detail with respect to the planned release of improved fission-yield data in the foreseeable future~\cite{IAEA0817}. A further investigation of the impact of the fission yields on decay-heat calculations was obtained with the JEFF-3.3 decay-data sub-library, and is shown in Fig.~\ref{fig:jeff-fy} for $^{238}$U and $^{237}$Np fast fission. These two cases were chosen because of the major differences observed between JEFF-3.3 and the other two libraries in Figs.~\ref{fig:fig-comp-1}~to~\ref{fig:fig-comp-4}. One can see from Fig.~\ref{fig:jeff-fy} that the effect of the three different fission-yield  sub-libraries differs for the two irradiated actinides. Adoption of JENDL-5 fission-yield sub-library leads to a significant improvement for $^{238}$U fast fission at cooling times around the first peak (i.e., 1 to 10 s cooling time); all three fission-yield sub-libraries give similar results for $^{237}$Np fast fission at cooling times greater than 80 s, while ENDF/B-VIII.0 deviates from JEFF-3.3 and JENDL-5 at shorter cooling times between 3 and 50 s.  
%Unfortunately, there are no available DH data at these shorter cooling times, while the existing data of Ohkawachi et al.~\cite{Ohkawachi2001} at longer cooling times are not that reliable as discussed in Subsection~\ref{endfb}. 
%ALN to Vivian, 10 Oct 2022: I have removed your last sentence (see immediately above). Stop pushing these short-lived cooling times below 200 s - decay heat at these times in a power reactor are overwhelmed by engineering-based heat transfer and fluid flow effects, decreasing as you go up to 200/300 seconds, at which decay heat has taken over domination. Any valid/worthwhile attempt to benchmark at short cooling times below 200 s would only apply to experiments in which separated actinide radioisotope targets had been satisfactorily prepared, irradiated and decay heat measured.

A detailed study of the fission-yield sub-libraries and their impact on decay-heat calculations is beyond the scope of this paper. Nevertheless, international effort coordinated by the IAEA is ongoing with the aim of updating and improving fission-yield data for the major actinides~\cite{IAEA0817}. We expect the output and conclusions of this effort will allow clear conclusions to be drawn on the recommended fission-yield data to be used in decay-heat calculations.

Both the nature and handling of uncertainties in the decay data and fission-yield data also need to be considered. A full and respected treatment of these uncertainties, including correlation effects identified with the fission-yield data, would permit reliable and accurate assessments of the decay-data and fission-yield sub-libraries, and draw definitive conclusions on the impact of the TAGS data and the need and form of other additional measurements. Our present results indicate certain trends in such calculations, but without proper propagation of uncertainties they are not definitive. Ongoing efforts to develop suitable approaches for propagating the uncertainties in decay-heat calculations are expected to come to fruition in the near future, so that we should eventually be able to perform uncertainty quantification of such integral calculations with confidence.
% Vivian, 21 Nov 2022: The articles below have been moved to the introduction.
%Nothing available from participants – while people were actioned, there have been no visible and constructive results yet, other than a series of short-ish papers identified solely with their own names (and exclusively about antineutrinos):}
%2015So01:  A.A. Sonzogni, T.D. Johnson, E.A. McCutchan, Nuclear Structure Insights into Reactor Antineutrino Spectra, Phys. Rev. C91, 011301, 1-5 (2015).
%2017So18:  A.A. Sonzogni, E.A. McCutchan, A.C. Hayes, Dissecting Reactor Antineutrino Flux Calculations, Phys. Rev. Lett. 119, 112501, 1-5 (2017).
%2018So13:  A.A. Sonzogni, M. Nino, E.A. McCutchan, Revealing Fine Structure in the Antineutrino Spectra from a Nuclear Reactor, Phys. Rev. C98, 014323, 1-7 (2018).
%2018Ha01:  A.C. Hayes, G. Jungman, E.A. McCutchan, A.A. Sonzogni, G.T. Garvey, X.B. Wang, Analysis of the Daya Bay Reactor Antineutrino Flux Changes with Fuel Burnup, Phys. Rev. Lett. 120, 022503, 1-5 (2018).
\section{Concluding remarks}
%\textcolor{red}{To be revised and updated once all other sections are complete. ALL}

Table~\ref{tab:5} constitutes a brief summary of the fission-product radionuclides considered in defining the requirements for TAGS studies to detect and quantify Pandemonium, as well as elaborate on the related needs for \g~singles and \g-\g~coincidence measurements to improve the definition of each individual decay scheme.  Much greater detail, various relevant information and fuller assessments are to be found within Tables~\ref{tab:app1} and \ref{tab:app2}:
\begin{itemize}
\item	relevant nuclear properties, 
\item	radionuclidic importance from the point of view of decay heat (based on each fissile actinide as a function of cooling time), and
\item	definitive recommendations as to whether to perform (or not perform) TAGS, \g~singles and/or \g-\g~coincidence studies, each defined in terms of priority from a highest of 1 decreasing to 2 and 3, or no assigned priority at all. 
\end{itemize}
Further extensive assessments need to accommodate consideration of the impact of delayed neutrons on decay heat up to $\sim$ 
% By Vivian (08/12/2021): 100 
10 seconds cooling time, along with an awareness of other unrelated requirements such as antineutrino spectral signatures and research studies within nuclear physics. 
%Vivian to Alan, 1 Nov. 2022: This was not my input...I merely changed the cooling time
%ALN to Vivian, 18 Oct 2022: I will repeat myself yet again - Stop pushing these short-lived cooling times below 200 s. Decay heat at these times in a power reactor are overwhelmed by engineering-based heat transfer and fluid flow effects, decreasing as you go up to 200/300 seconds, at which decay heat has taken over domination. Any valid/worthwhile attempt to benchmark at short cooling times below 200 s would only apply to experiments in which separated actinide radioisotope targets had been satisfactorily prepared, irradiated and decay heat measured.
% AA: Alan I think we should remove the "subjective" word to avoid problems 
% with referees, as well as the "fully consistency".
%ALN to AA: not sure whether your concern is justified - judgements always possess some "subjectivity".  So what?

All of the recommendations constitute subjective judgements by individual authors that when brought together in this manner will not necessarily be fully consistent with each other. Taking these subjective judgements at face value (Table~\ref{tab:5}), the following experimental measurements would seem to be most appropriate at the present time:
\begin{itemize}
\item priority 1 TAGS/TAS and DGS \g~singles and~\g-\g~(and \be-\g) coincidence measurements of the \be$^{-}$~decay of  $^{99}$Zr, $^{98,99}$Nb, $^{106}$Tc, $^{130m,132}$Sb, $^{138}$Cs and $^{142,143}$La;
\item priority 1 DGS measurements only of the \be$^{-}$~decay of $^{100,100m,101,102}$Nb, $^{104,105,106}$Tc and $^{134}$I;
\item priority 2 TAGS/TAS and DGS measurements of the \be$^{-}$~decay of $^{84}$As, $^{85}$Se, $^{84,89}$Br, $^{91}$Kr, $^{95}$Sr, $^{97}$Y, $^{105}$Nb, $^{103,104,105,107}$Mo, $^{107,108}$Tc, $^{133}$Sb, $^{136,137}$Te, $^{136,136m}$I, $^{140}$Xe, $^{139,140,141}$Cs, $^{141,143,144}$Ba and $^{144,147}$La;
\item priority 2 TAGS/TAS measurements only of the \be$^{-}$~decay of $^{92,93,94}$Sr, $^{139}$Ba and $^{146}$Pr;
\item priority 2 DGS measurements only of the \be$^{-}$~decay of $^{86,87,88}$Br, $^{89,90}$Kr, $^{90,90m,91,92,93}$Rb, $^{96,96m}$Y, $^{100}$Zr, $^{102}$Mo, $^{102}$Tc, $^{130,131}$Sb, $^{133,133m}$Te, $^{132,135,137}$I and $^{139}$Xe.
\end{itemize}
However, requirements and priorities can be expected to change as such work proceeds, so that consideration should also be given to some of the lesser fission products of Table~\ref{tab:5} at later ill-defined stages of such studies.
%ALN: if we get an antineutrino FP priority list that can be incorporated into Table 10, the above three FP lists will expand in length.  [[ALN: this is an old comment that can be dumped]]
One aim of this particular work has been to provide guidance in improving our knowledge of the energy distribution between the light-particle (LP) and electromagnetic (EM) components of decay heat. Along with the need to assess properly the impact of delayed-neutron emissions on the fuel inventory at early cooling times (LP decay heat vs EM decay heat at relatively short and intermediate cooling times (up to 10 to 20 seconds)), we must ensure that adopted fission-yield data are also fully fit for purpose in all of the important decay-heat calculations. As was shown in the previous section, some of the pronounced features in the calculated decay heat components (LP and EM) are sensitive to fission-product yield data rather than the decay data. Although the impact of the fission-yield sub-libraries (ENDF/B-VIII.0, JEFF-3.3 and JENDL-5) on decay-heat calculations was found to vary within the uncertainties of the experimental decay-heat data, larger discrepancies were observed in the thermal decay heat of $^{239}$Pu and fast decay heat of $^{238}$U which require further investigation. The impact of the energy dependence of the fission-yield data on decay heat also merits further exploration. 
% Vivian, added on 18 Nov 2022
%ALN: checked and modified, 18 Nov 2022
Envisaged improvements in some of the more important decay data and their uncertainties will inevitably necessitate better quality integral measurements of decay heat to give conclusive comparisons and more precise benchmarking. 

As for repeating this exercise at a future date, changes in the total decay heat, LP and EM components will be relatively modest to small because much of the relevant data should not subsequently alter much from year to year, or even over ten years, when considering half-lives, Q-values and decay radiation characteristics (i.e., key nuclear parameters in quantifying irradiated fuel inventories to determine total decay heat). This statement contrasts with the fission yields that are expected to improve significantly in the near future, supported fully by a concerted international effort coordinated by the IAEA to update the fission-yield database through consideration of all available experimental data measured at advanced facilities with better precision and higher resolution to be coupled with developments in modelling codes and more extensive validation exercises~\cite{IAEA0817}. Additional developments concerning the uncertainties in the recommended decay-data and fission-yield libraries should also include improved quantification of their correlation effects, along with propagation within inventory and summation calculations. Therefore, the current exercise should be repeated with improved fission-yield data files, as well as with codes that can accommodate full analyses and quantification of the overall uncertainties.
%Hopefully and eventually, this exercise will become a matter of data accuracy rather than continued concern about the validity of some of the required nuclear data and whether they are quite simply adequate and correct.

Assessments of the balance between performing further TAGS/TAS measurements or direct \g~singles and/or \be-\g, \g-\g~ coincidence spectral studies would appear best to depend upon the nature and degree of the inconsistencies and Pandemonium effects revealed by the systematic application of TAGS/TAS. Both types of measurement may be required, but there is always the possibility that only one approach is merited at a particular time in support or even contradiction of the other. Those undertaking calculations of the resulting decay heat or antineutrino emissions require reasonably comprehensive decay scheme data that are more reliably obtained directly, if possible. Under such circumstances, TAGS/TAS has proven to be highly effective at highlighting existing inadequacies within individual fission-product decay schemes, and in doing so strongly supports \g~singles and coincidence measurements by means of appropriate detector systems that may still be in the process of evolution (and are not necessarily Ge based). There are existing and will be future needs for TAGS/TAS studies that go hand-in-hand with \be$^{-}$ and \g-ray spectroscopy in order to improve various nuclear parameters in a complementary manner for their confident adoption in the evolution of nuclear physics research and a developing range of nuclear applications that includes power-reactor decay heat on shutdown.

\boldsymbol{$Data~uncertainties$}:
Data and their uncertainties are presented throughout the text and tables in the form 1234(x), where x is the uncertainty expressed in terms of the last digit or digits quoted with respect to the measured or evaluated number. This uncertainty is normally expressed at the 1$\sigma$ confidence level. Examples: 41.1(11) means 41.1$\pm$ 1.1, 6.783(8) means 6.783$\pm$0.008, 0.0820(17) means 0.0820$\pm$0.0017, 1.688(12)$\cdot 10^{-4}$ means (1.688$\pm$0.012)$\cdot 10^{-4}$, 26.3(+3-21) means 26.3$_{-2.1}^{+0.3}$, and 8.681(+18-25) means 8.681$_{-0.025}^{+0.018}$.

\begin{acknowledgements}

%\textcolor{red}{Contributions welcome from all authors.}

The content and preparation of this paper involved both the work and support of various individuals and their institutions. Our sincere thanks are extended to all colleagues who have contributed to this IAEA technical project since April 2005 when these studies were initially conceived and undertaken over a two-year period until mid-2007 as an OECD/NEA Working Party on International Evaluation Co-operation of the NEA Nuclear Science Committee. Furthermore, the IAEA is grateful to all participant laboratories for their assistance in this work since January 2009, and their support of individual staff to attend subsequent working meetings and perform related analytical and experimental activities. Futoshi Minato from the Japan Atomic Energy Agency is also gratefully acknowledged for providing decay-heat calculations for the sixteen fission systems studied herein, based on the recently released JENDL-5 libraries. 
%ALN to Vivian, 10 Oct 2022: what I have tried to do is acknowledge people in the first paragraph, and institutions and any related contracts in the later paragraph(s).

Work described in this paper would not have been possible without IAEA Member State contributions. Studies at ANL were supported by the US Department of Energy, Office of Science, Office of Nuclear Physics, under contract no. DE-AC02-06CH11357 and by the National Nuclear Security Administration, Office of Defense Nuclear Nonproliferation R$\&$D (NA-22).
Work at ORNL was partially supported by the Office of Nuclear Physics, US Department of Energy under Contract No. DE-AC05-00OR22725 and by the DOE Nuclear Data Program, within the FOA 18 1903 project. Thanks are also due to all collaborators who made these particular measurements possible - colleagues at HRIBF-ORNL (D.W. Stracener) and ANL (G. Savard, D. Santiago-Gonzales, J. Clark) for their continuous support and help, and in particular to PhD students and associates who worked in the analysis of the data and made this work possible including Toby T. King, Peng Shuai, Alexander Laminack (ORNL), A. Fijałkowska, M. Wolińska-Cichocka, M. Stepaniuk (Warsaw University), Robert Grzywacz,  Michael Cooper (UTK) and Thomas Ruland (LSU). Work at BNL was sponsored by the Office of Nuclear Physics, Office of Science of the US Department of Energy under Contract No. DE-AC02-98CH10886 as well as by the US Department of Energy, National Nuclear Security Administration, Office of Defense Nuclear Nonproliferation Research and Development (DNN R\&D).

Work undertaken by ALN was partially funded by the IAEA Nuclear Data Section under Special Service Agreement no. TAL-NAPC20170620-001 (work package 2). Other highly-related studies were supported by the CNRS challenge NEEDS and associated NACRE project, CHANDA FP7/EURATOM project (contract no. 605203) and SANDA project ref. 847552, the SAMPO program funded by the French Institute in Finland as well as via the CNRS/IN2P3 PICS TAGS programme between SUBATECH and IFIC, and Master projects Jyv\"{a}skyl\"{a}, OPALE and TAGS.
The measurements included in this work have also been supported by the Spanish Ministerio de Econom\'ia y Competitividad under Grants No. FPA2011-24553, No. AIC-A-2011-0696, No. FPA2014-52823-C2-1-P, No. FPA2015-65035-P, FPA2017-83946-C2-1-P, No. FPI/BES-2014-068222, Ministerio de Ciencia e Innovacion PID2019-104714GB-C21 grant and the program Severo Ochoa (SEV-2014-0398), by the Spanish Ministerio de Educaci\'on under the FPU12/01527 Grant, by the European Commission under the European Return Grant, MERG-CT-2004-506849, and by the Junta para la Ampliaci\'on de Estudios Programme (CSIC JAE-Doc contract) co-financed by FSE. Also support from the STFC(UK) council grant ST/P005314/1 is acknowledged. Thanks are due to all collaborators who participated in the measurements, colleagues at IGISOL and the University of Jyv\"askyl\"a for their continuous support and help, and in particular PhD students and colleagues who made this work possible and analysed the resulting data (D. Jordan, E. Valencia, S. Rice, V.M. Bui, A.A. Zakari-Issoufou, V. Guadilla, L. Le Meur, J. Briz-Monago and A. Porta).

\end{acknowledgements}

\clearpage

\appendix

\section{Assessment and reference tables}\label{app}

%\afterpage{

\begin{sidewaystable*}
\vspace*{-14cm}
\caption{Fission-product decay data: assessment of potential Pandemonium and need for further TAGS and DGS (discrete gamma-ray spectroscopy involving \g~singles and \g-\g~coincidence measurements) studies for $^{235}$U, $^{238,239,240}$Pu, $^{241}$Am and $^{243}$Cm thermal fission. Relevant references in the literature are defined in NSR keynumber format (e.g., 2017Fi06 (NSR), 2016ORNL (equivalent non-NSR)). 
Spin, parity (J$^{\pi}$) and half-life are from NUBASE2020~\cite{NUBASE2020}. Q-value for ground state-to-ground state decay (Q($\beta^{-}$)) and the neutron separation energy (S$_{n}$) are from AME2020~\cite{AME2020}. Q($\beta^{-}$) value for an excited isomer is reported as the sum of the ground state-to-ground state decay and the excitation energy of the isomer taken from NUBASE2020~\cite{NUBASE2020}. Beta-delayed neutron emission probabilities P$_n$(\%) are taken from the IAEA Reference Database for beta-delayed neutron emission~\cite{Liang,Dimitriou2021,bdnIAEA}.
(a)	Possible potential for Pandemonium effect as a consequence of large energy differential $\Delta$ (keV) = (Q($\beta^{-}$) – excitation energy of highest known relevant nuclear level); 
(b)	FISPACT-II calculations of fission-product inventories and their percentage decay-heat contributions as a function of the cooling time of irradiated actinides~\cite{Fleming2015,Fleming2015a,Fleming2015b}: fissile actinide, (post-irradiation cooling time, \%\g, \%\be, \%(total decay heat)), 
fission-product decay schemes assessed in terms of known \be$^{‒}$ decay and discrete nuclear levels of the daughter~\cite{ENSDF}; 
(c)	Request for specific measurements denoted by $‘$y' $\rightarrow$ 'yes' ($‘$?' $\rightarrow$ to be assessed further); 
(d)	Priorities defined as 1 $\rightarrow$ high, 2 $\rightarrow$ intermediate, 3 $\rightarrow$ low, ‒ $\rightarrow$ unassigned.}\label{tab:app1} 

 \begin{adjustbox}{width=0.9\textwidth}

\begin{tabular}{lclcccp{10cm}ccc}\hline \hline
\T Fission & J$^{\pi}$  & Half-life & Q($\beta^{-}$)  & $\Delta$\textsuperscript{a}   & S$_n$    & Assessment of potential Pandemonium\textsuperscript{b}    & DGS\textsuperscript{c}  & TAGS\textsuperscript{c} & Priority\textsuperscript{d} \\
\B   product & &   & (keV)     &(keV) &(keV) &   &   &    &   \\  \hline

\T 35-Br-86        & (1‒)    & 55.1(4) s   & 7633(3)   & 865(3)    & 9856.7(20) & $^{235}$U (100 s, 3.9\%\g, 2.7\%\be, 3.3\%); 
ENSDF (2015Ne01) reports 18 levels in $^{86}$Kr populated by \be$^{‒}$ decay, but also many unplaced \g-rays. 
TAGS measurements support possibility of Pandemonium (2014Fi09, 2016Ka65, 2017AlZY, 2017Fi06, 2017Ri08, 2018Fi03). 
\be$^{‒}$-decay feeding intensity to $^{86}$Kr ground state ((1‒) $\rightarrow$ 0+ transition) of 15(8)\% adopted in ENSDF (based on a value of $\approx$15\% derived from \be$^{‒}$-decay end-point studies), compared with a more precise value of 20.23(11)\% from TAGS (2017Ri08). 
DGS studies should be incorporated into ENSDF (2016Ur04), followed by consideration of further DGS measurements.   
& y & – & 2 \\ \hline

\T 35-Br-87        & 5/2‒  & 55.68(12) s & 6818(3)   & 1025(3)   & 5515.17(25) & $^{235}$U (100 s, 3.9\%\g, 2.2\%\be, 3.1\%); 
ENSDF (2015Jo11) reports 139 levels in $^{87}$Kr populated by \be$^{‒}$ decay. 
TAGS measurements support possibility of Pandemonium (2014VaZZ, 2015Ta17, 2016ORNL, 2017TaZZ, 2017Va04).  
\be$^{‒}$-decay feeding intensity to $^{87}$Kr ground state (5/2‒ $\rightarrow$ 5/2+ transition) of 12.0(19)\% adopted in ENSDF (based on \%I\g(1419.7\g) = 22.0(15)\%), compared with 10.1\% from TAGS (2017Va04). 
DGS studies merited. 
P$_n$($^{87}$Br) = 2.53(10)\%. 
& y & – & 2 \\
\hline

\end{tabular}

\end{adjustbox}
\end{sidewaystable*}

\begin{sidewaystable*}

\vspace*{-14cm}
\addtocounter{table}{-1}
\caption{ -- Continued from previous page} 
 \begin{adjustbox}{width=0.9\textwidth}
\begin{tabular}{llllllp{10cm}ccc}\hline \hline
\T Fission & J$^{\pi}$  & Half-life & Q($\beta^{-}$)  & $\Delta$\textsuperscript{a}   & S$_n$    & Assessment of potential Pandemonium\textsuperscript{b}    & DGS\textsuperscript{c}  & TAGS\textsuperscript{c} & Priority\textsuperscript{d} \\
\B   product & &   & (keV)     &(keV) &(keV) &  &    &     &   \\  \hline 

\T 36-Kr-90        & 0+      & 32.32(9) s   & 4406(7)    & 525(7)   & 5724(8)    & $^{235}$U (100 s, 2.4\%\g, 2.8\%\be, 2.6\%); 
ENSDF (2020Ba21) reports 32 levels in $^{90}$Rb populated by \be$^{‒}$ decay. 
TAGS measurements in support of existence of Pandemonium (2017Fi06).   
\be$^{‒}$-decay feeding intensity to $^{90}$Rb ground state (0+ $\rightarrow$ 0‒ transition) of 29(4)\% adopted in ENSDF, but factor of four smaller value of 7(1)\% determined from TAGS (2017Fi06).  
Comprehensive DGS studies merited.
& y  & – & 2 \\ \hline 

\T 36-Kr-91        & 5/2+  & 8.57(4) s   & 6771(8)   & 2073(8)   & 6451(10) & $^{235}$U (10 s, 3.6\%\g, 3.3\%\be, 3.4\%); 
ENSDF (2013Ba52) reports 61 levels in $^{91}$Rb populated by \be$^{‒}$ decay. 
TAGS measurements not reported, although potential for Pandemonium supported by large $\Delta$ energy differential.  
\be$^{‒}$-decay feeding intensity to $^{91}$Rb ground state (5/2+ $\rightarrow$ 3/2(‒) transition) of 9(4)\% adopted in ENSDF.  
Merits studies by means of TAGS and DGS.
& y  & y  & 2  \\
\hline
\end{tabular}
\end{adjustbox}
\end{sidewaystable*}

\begin{sidewaystable*}
\vspace*{-14cm}
\addtocounter{table}{-1}
\caption{ -- Continued from previous page} 
 \begin{adjustbox}{width=\textwidth}
\begin{tabular}{llllllp{10cm}ccc}\hline \hline
\T Fission & J$^{\pi}$  & Half-life & Q($\beta^{-}$)  & $\Delta$\textsuperscript{a}   & S$_n$    & Assessment of potential Pandemonium\textsuperscript{b}    & DGS\textsuperscript{c}  & TAGS\textsuperscript{c} & Priority\textsuperscript{d} \\
\B   product & &   & (keV)     &(keV) &(keV) &  &    &     &   \\  \hline

\T 37-Rb-92        & 0‒      & 4.48(3) s & 8095(6) & 732(6) & 7287(6) & $^{235}$U (10 s, 7.0\%\g, 6.6\%\be, 6.7\%), $^{238}$Pu (10 s, 4.0\%\g, 3.9\%\be, 3.9\%), $^{239}$Pu (10 s, 3.6\%\g, 3.4\%\be, 3.5\%), $^{240}$Pu (10 s, 3.7\%\g, 3.5\%\be, 3.6\%), $^{241}$Am (10 s, 2.6\%\g, 2.6\%\be, 2.6\%), $^{243}$Cm (10 s, 2.0\%\g, 2.0\%\be, 2.0\%); 
ENSDF (2012Ba51) reports 21 levels in $^{92}$Sr populated by \be$^{‒}$ decay. 
TAGS measurements by 2014ZaZZ, 2015Za10, 2016Ra25, 2017FaZY, 2017Fi06, 2017Ra06 support existence of Pandemonium.  
\be$^{‒}$-decay feeding intensity to $^{92}$Sr ground state (0‒ $\rightarrow$ 0+ transition) of 95.2(7)\% adopted in ENSDF, compared with 87.5(25)\% (2015Za10) and 91(3)\% (2016Ra25) from TAGS. 
Comprehensive DGS studies merited.  
P$_n$($^{92}$Rb) = 0.0109(7)\%.
& y & – & 2 \\
\hline

\end{tabular}
\end{adjustbox}
\end{sidewaystable*}

\begin{sidewaystable*}
\vspace*{-14cm}
\addtocounter{table}{-1}
\caption{ -- Continued from previous page} 
 \begin{adjustbox}{width=0.9\textwidth}
\begin{tabular}{llllllp{10cm}ccc}\hline \hline
\T Fission & J$^{\pi}$  & Half-life & Q($\beta^{-}$)  & $\Delta$\textsuperscript{a}   & S$_n$    & Assessment of potential Pandemonium\textsuperscript{b}    & DGS\textsuperscript{c}  & TAGS\textsuperscript{c} & Priority\textsuperscript{d} \\
\B   product & &   & (keV)     &(keV) &(keV) &  &    &     &   \\  \hline

\T 38-Sr-93        & 5/2+    & 7.43(3) min & 4141(12) & 246(12) & 7482(14)& $^{235}$U (100 s, 2.7\%\g, 2.0\%), (1000 s, 9.2\%\g, 3.9\%\be, 6.8\%), $^{238}$Pu (100 s, 2.2\%\g), (1000 s, 4.9\%\g, 2.2\%\be, 3.7\%), $^{239}$Pu (100 s, 2.0\%\g), (1000 s, 5.4\%\g, 2.4\%\be, 4.0\%), $^{240}$Pu (1000 s, 5.4\%\g, 2.4\%\be, 4.0\%), $^{241}$Am (1000 s, 4.0\%\g, 2.9\%), $^{243}$Cm (1000 s, 2.5\%\g): 
ENSDF (2011Ba40) reports 35 levels in $^{93}$Y populated via \be$^{‒}$ decay. 
TAGS measurements (1996Gr20/1997Gr09), although no uncertainties assigned to \be$^{‒}$-decay feeding intensities. Potential for Pandemonium judged to be low given relatively small $\Delta$ energy differential.
Ill-defined direct \be$^{‒}$ feeding to the $^{93}$Y ground state (5/2+ $\rightarrow$ 1/2‒ transition) of $\le$4.5\% adopted in ENSDF.
Combined \be$^{‒}$ decay feeding to ground and 758.7-keV isomeric state (half-life of 0.82(4) s; 5/2+ $\rightarrow$ 9/2+ transition) of 5.8(14)\% reported from TAGS (1996Gr20/1997Gr09). 
Further TAGS studies merited.
& – & y & 2 \\
\hline

\end{tabular}
\end{adjustbox}
\end{sidewaystable*}

\begin{sidewaystable*}
\vspace*{-14cm}
\addtocounter{table}{-1}
\caption{ -- Continued from previous page} 
 \begin{adjustbox}{width=0.9\textwidth}
\begin{tabular}{llllllp{10cm}ccc}\hline \hline
\T Fission & J$^{\pi}$  & Half-life & Q($\beta^{-}$)  & $\Delta$\textsuperscript{a}   & S$_n$    & Assessment of potential Pandemonium\textsuperscript{b}    & DGS\textsuperscript{c}  & TAGS\textsuperscript{c} & Priority\textsuperscript{d} \\
\B   product & &   & (keV)     &(keV) &(keV) &  &   &     &   \\ \hline

\T 39-Y-94         & 2‒      & 18.7(1) min & 4918(6) & 248(6)  & 8218.6(5) & $^{235}$U (1000 s, 3.5\%\g, 9.8\%\be, 6.4\%), (5011 s, 5.9\%\be, 3.3\%), $^{238}$Pu (1000 s, 2.6\%\g, 7.5\%\be, 4.8\%), (5011 s, 5.2\%\be, 2.8\%), $^{239}$Pu (1000 s, 2.4\%\g, 6.7\%\be, 4.4\%), (5011 s, 4.7\%\be, 2.5\%), $^{240}$Pu (1000 s, 2.2\%\g, 6.1\%\be, 3.9\%), (5011 s, 4.3\%\be, 2.3\%), $^{241}$Am (1000 s, 2.3\%\g, 6.1\%\be, 4.1\%), (5011 s, 4.4\%\be, 2.5\%), $^{243}$Cm (1000 s, 3.8\%\be, 2.5\%), (5011 s, 2.7\%\be); 
ENSDF (2006Ab37) reports 24 levels in $^{94}$Zr populated via \be$^{‒}$ decay.
TAGS studies, although no uncertainties assigned to \be$^{‒}$-decay feeding intensities (1996Gr20/1997Gr09). Potential for Pandemonium judged to be low given relatively small $\Delta$ energy differential.
\be$^{‒}$ feeding to $^{94}$Zr ground state (2‒ $\rightarrow$ 0+ transition) of 41(4)\% adopted in ENSDF, compared with 45.8(23)\% from TAGS (1996Gr20/1997Gr09).  
& – & – & –   \\
\hline

\end{tabular}
\end{adjustbox}
\end{sidewaystable*}

\begin{sidewaystable*}    
\vspace*{-14cm}
\addtocounter{table}{-1}
\caption{ -- Continued from previous page} 
 \begin{adjustbox}{width=0.9\textwidth}
\begin{tabular}{llllllp{10cm}ccc}\hline \hline
\T Fission & J$^{\pi}$  & Half-life & Q($\beta^{-}$)  & $\Delta$\textsuperscript{a}   & S$_n$    & Assessment of potential Pandemonium\textsuperscript{b}    & DGS\textsuperscript{c}  & TAGS\textsuperscript{c} & Priority\textsuperscript{d} \\
\B   product & &   & (keV)     &(keV) &(keV) &  &    &     &   \\  \hline

\T 40-Zr-97       & 1/2+    & 16.749(8) h & 2666(4) & 419(4) & 8071(4) & $^{238}$Pu (10000 s, 2.2\%\be), $^{239}$Pu (10000 s, 2.1\%\be), $^{240}$Pu (10000 s, 2.1\%\be), $^{241}$Am (10000 s, 2.1\%\be); 
ENSDF (2010Ni03) reports 15 levels in $^{97}$Nb populated via \be$^{‒}$ decay.
TAGS studies not reported, although low potential for Pandemonium given the relatively small $\Delta$ energy differential. 
Significant \be$^{‒}$ feeding to $^{97}$Nb isomer at 743.35-keV (half-life of 58.7(18) s; 1/2+ $\rightarrow$ 1/2‒ transition) of 87.8(4)\%) adopted in ENSDF.
\be$^{‒}$ feeding to $^{97}$Nb ground state (1/2+ $\rightarrow$ 9/2+ transition) of zero adopted in ENSDF. 
Weak $\gamma$-ray transitions unplaced in ENSDF level scheme. 
DGS studies by 2014Kr11 need to be incorporated into ENSDF, followed by an assessment of the need for further measurements.
& ? & – & 3?   \\
\hline

\end{tabular}
\end{adjustbox}
\end{sidewaystable*}

\begin{sidewaystable*}
\vspace*{-14cm}
\addtocounter{table}{-1}
\caption{ -- Continued from previous page} 
 \begin{adjustbox}{width=\textwidth}
\begin{tabular}{llllllp{10cm}ccc}\hline \hline
\T Fission & J$^{\pi}$  & Half-life & Q($\beta^{-}$)  & $\Delta$\textsuperscript{a}   & S$_n$    & Assessment of potential Pandemonium\textsuperscript{b}    & DGS\textsuperscript{c}  & TAGS\textsuperscript{c} & Priority\textsuperscript{d} \\
\B   product & &   & (keV)     &(keV) &(keV) &  &   &     &   \\  \hline 

\T 41-Nb-100       & 1+      & 1.5(2) s & 6402(8) & 3272(8)   & 8294.2(4)  & $^{235}$U (10 s, 3.0\%\g, 7.6\%\be, 5.7\%), $^{238}$Pu (10 s, 3.2\%\g, 8.4\%\be, 6.2\%), $^{239}$Pu (10 s, 3.6\%\g, 9.2\%\be, 6.9\%), $^{240}$Pu (10 s, 3.5\%\g, 8.9\%\be, 6.7\%), $^{241}$Am (10 s, 3.0\%\g, 7.9\%\be, 5.8\%), $^{243}$Cm (10 s, 3.2\%\g, 8.7\%\be, 6.4\%); 
ENSDF (2021Si08) reports 20 levels in $^{100}$Mo populated via \be$^{‒}$ decay.
2017GuZV/2019Gu03/2019Gu20 TAGS studies support Pandemonium. 
\be$^{‒}$ feeding to $^{100}$Mo ground state (1+ $\rightarrow$ 0+ transition) of 50(7)\% reported in ENSDF, compared with 46(+8-15)\% from TAGS (2019Gu20).
Comprehensive DGS studies merited, with unambiguous separation between ground state and isomer \be$^{‒}$ decays. 
& y & ‒ & 1   \\ \hline 

\T 41-Nb-100m      & (5+)    & 2.99(11) s & 6715(11)  & 3068(11)  & 8294.2(4) & $^{238}$Pu (10 s, 2.5\%\g, 2.0\%), $^{241}$Am (10 s, 2.5\%\g, 2.0\%), $^{243}$Cm (10 s, 3.5\%\g, 2.3\%\be, 2.8\%); 
ENSDF (2021Si08) reports 17 levels discrete levels in $^{100}$Mo populated via \be$^{‒}$ decay. 
$^{100m}$Nb(\be$^{‒}$) = 100\% in NUBASE2020.
TAGS studies that support Pandemonium reported in 2017GuZV/2019Gu03/2019Gu20. 
\be$^{‒}$ feeding to $^{100}$Mo ground state ((5+) $\rightarrow$ 0+ transition) of zero reported in ENSDF.  
Comprehensive DGS measurements merited, with unambiguous separation between ground-state and isomer \be$^{‒}$ decays. 
& y & ‒ & 1   \\
\hline

\end{tabular}
\end{adjustbox}
\end{sidewaystable*}

\begin{sidewaystable*}

\vspace*{-14cm}
\addtocounter{table}{-1}
\caption{ -- Continued from previous page} 
 \begin{adjustbox}{width=0.9\textwidth}
\begin{tabular}{llllllp{10cm}ccc}\hline \hline
\T Fission & J$^{\pi}$  & Half-life & Q($\beta^{-}$)  & $\Delta$\textsuperscript{a}   & S$_n$    & Assessment of potential Pandemonium\textsuperscript{b}    & DGS\textsuperscript{c}  & TAGS\textsuperscript{c} & Priority\textsuperscript{d} \\
\B   product & &   & (keV)     &(keV) &(keV) &  &   &     &   \\  \hline 

\T 42-Mo-101       & 1/2+    & 14.61(3) min & 2825(24) & 251(24) & 8395(24) & $^{235}$U (1000 s, 5.4\%\g, 2.4\%\be, 4.0\%), $^{238}$Pu (1000 s, 7.0\%\g, 3.4\%\be, 5.4\%), $^{239}$Pu (1000 s, 6.5\%\g, 3.0\%\be, 4.9\%), $^{240}$Pu (1000 s, 6.7\%\g, 3.1\%\be, 5.0\%), $^{241}$Am (1000 s, 7.4\%\g, 3.2\%\be, 5.5\%), $^{243}$Cm (1000 s, 7.5\%\g, 3.3\%\be, 5.5\%); 
ENSDF (2006BlENSDF) reports 44 levels in $^{101}$Tc populated via \be$^{‒}$ decay. 
TAGS studies not reported, with low potential for Pandemonium based on the relatively small $\Delta$ energy differential.
\be$^{‒}$ feeding to $^{101}$Tc ground state (1/2+ $\rightarrow$ 9/2+ transition) of zero adopted in ENSDF.  
& –  & –  & –   \\ \hline

\T 42-Mo-102       & 0+      & 11.3(2) min & 1012(12) & 652(12)   & 6300(26) & $^{238}$Pu (1000 s, 2.1\%\be), $^{240}$Pu (1000 s,   2.1\%\be), $^{241}$Am (1000 s, 2.1\%\be), $^{243}$Cm (1000 s, 2.3\%\be); 
ENSDF (2009De21) reports 5 levels in $^{102}$Tc populated via \be$^{‒}$ decay. 
TAGS studies not reported, with potential for Pandemonium difficult to assess other than likely to be minor. 
\be$^{‒}$ decay feeding intensities based on decay scheme and deduced absolute $\gamma$-ray emission probabilities (latter uncertainties greater than 13\%). 
\be$^{‒}$ feeding to $^{102}$Tc ground state (0+ $\rightarrow$ 1+ transition) of 94.1(6)\% adopted in ENSDF.
Comprehensive DGS measurements merited. 
& y  & – & 2   \\
\hline

\end{tabular}
\end{adjustbox}
\end{sidewaystable*}

\begin{sidewaystable*}
\vspace*{-14cm}
\addtocounter{table}{-1}
\caption{ -- Continued from previous page} 
 \begin{adjustbox}{width=0.9\textwidth}
\begin{tabular}{llllllp{10cm}ccc}\hline \hline
\T Fission & J$^{\pi}$  & Half-life & Q($\beta^{-}$)  & $\Delta$\textsuperscript{a}   & S$_n$    & Assessment of potential Pandemonium\textsuperscript{b}    & DGS\textsuperscript{c}  & TAGS\textsuperscript{c} & Priority\textsuperscript{d} \\
\B   product & &   & (keV)     &(keV) &(keV) &  &    &     &   \\  \hline

\T 43-Tc-102       & 1+      & 5.28(15) s   & 4534(9)   & 1625(9)   & 9219.64(5)  & $^{235}$U (1000 s, 7.4\%\be, 3.5\%), $^{238}$Pu (100 s, 2.6\%\be), (1000 s, 11.5\%\be, 5.3\%), $^{239}$Pu (100 s, 2.1\%\be), (1000 s, 10.7\%\be, 5.1\%), $^{240}$Pu (100 s, 2.2\%\be), (1000 s, 11.5\%\be, 5.4\%), $^{241}$Am (100 s, 2.5\%\be), (1000 s, 11.7\%\be, 5.7\%), $^{243}$Cm (100 s, 3.0\%\be), (1000 s, 13.1\%\be, 6.4\%); 
ENSDF (2009De21) reports 9 levels in $^{102}$Ru populated via \be$^{‒}$ decay. 
TAGS measurements exhibit no evidence of significant Pandemonium (2013Jo02/2010Al22).
\be$^{‒}$ feeding to $^{102}$Ru ground state (1+ $\rightarrow$ 0+ transition) of 92.9(6)\% adopted in ENSDF, compared with 92.9\% from TAGS (2013Jo02). 
Comprehensive DGS studies merited, including the need to determine absolute $\gamma$-ray emission probabilities with greater accuracy.
%{\bf not sure this nuclide is relevant to quoted 100 s and 1000 s cooling times - as daughter of Mo-102 (half-life of 678 s) explains relevance to 1000 s (ALN) - OK (FGK)}
& y  & –  & 2   \\
\hline

\end{tabular}
\end{adjustbox}
\end{sidewaystable*}

\begin{sidewaystable*}
\vspace*{-14cm}
\addtocounter{table}{-1}
\caption{ -- Continued from previous page} 
 \begin{adjustbox}{width=\textwidth}
\begin{tabular}{llllllp{10cm}ccc}\hline \hline
\T Fission & J$^{\pi}$  & Half-life & Q($\beta^{-}$)  & $\Delta$\textsuperscript{a}   & S$_n$    & Assessment of potential Pandemonium\textsuperscript{b}    & DGS\textsuperscript{c}  & TAGS\textsuperscript{c} & Priority\textsuperscript{d} \\
\B   product & &   & (keV)     &(keV) &(keV) &  &   &     &   \\  \hline

\T 43-Tc-107       & (3/2‒)  & 21.2(2) s  & 5113(12)  & 2433(12)  & 5611(10)  & $^{238}$Pu (10 s, 2.4\%\g), $^{239}$Pu (10 s, 2.4\%\g), $^{240}$Pu (10 s, 2.5\%\g), $^{241}$Am (10 s, 3.7\%\g, 2.6\%), $^{243}$Cm (10 s, 5.2\%\g, 2.7\%\be, 3.8\%), (100 s, 2.0\%\g); 
ENSDF (2008Bl12) reports 38 levels in $^{107}$Ru populated via \be$^{‒}$ decay.  
TAGS measurements have provided evidence of a fair degree of Pandemonium (2010Al22), as supported by large $\Delta$ energy differential.
\be$^{‒}$ feeding to $^{107}$Ru ground state ((3/2–) $\rightarrow$ (5/2)+ transition) of 22(8)\%) adopted with a large uncertainty in ENSDF.
Comprehensive DGS studies merited, along with improved quantification of direct \be$^{‒}$ feeding intensity to $^{107}$Ru ground state by means of TAGS. 
& y  & y  & 2  \\ \hline

\T 43-Tc-108       & (2)+    & 5.17(7) s  & 7739(12)  & 4641(12)  & 7870(12)  & $^{239}$Pu (10 s, 2.1\%\g), $^{240}$Pu (10 s, 2.8\%\g, 2.1\%), $^{241}$Am (10 s, 2.4\%\g), $^{243}$Cm (10 s, 2.3\%\g); 
ENSDF (2008BlENSDF) reports 21 levels in $^{108}$Ru populated via \be$^{‒}$ decay. 
TAGS studies not reported, although potential for Pandemonium supported by large $\Delta$ energy differential. 
\be$^{‒}$ feeding to $^{108}$Ru ground state ((2)+ $\rightarrow$ 0+ transition) of zero, and 242.23-keV level ((2)+ $\rightarrow$ 2+ transition) of 27.6(25)\% adopted in ENSDF.
Merits studies by means of TAGS and DGS.
& y  & y   & 2    \\
\hline

\end{tabular}
 \end{adjustbox}
\end{sidewaystable*}

\begin{sidewaystable*}

\vspace*{-14cm}
\addtocounter{table}{-1}
\caption{ -- Continued from previous page} 
 \begin{adjustbox}{width=0.9\textwidth}
\begin{tabular}{llllllp{10cm}ccc}\hline \hline
\T Fission & J$^{\pi}$  & Half-life & Q($\beta^{-}$)  & $\Delta$\textsuperscript{a}   & S$_n$    & Assessment of potential Pandemonium\textsuperscript{b}    & DGS\textsuperscript{c}  & TAGS\textsuperscript{c} & Priority\textsuperscript{d} \\
\B   product & &   & (keV)     &(keV) &(keV) &  &    &     &   \\  \hline

\T 45-Rh-108       & 1+      & 16.8(5) s   & 4493(14) & 2953(14)  & 9222.9(16) & $^{240}$Pu (1000 s, 2.3\%\be); $^{241}$Am (100 s,   2.3\%\be),(1000 s, 2.8\%\be), $^{243}$Cm (100 s, 2.7\%\be), (1000 s, 3.0\%\be); 
ENSDF (2008BlENSDF) reports 6 levels in $^{108}$Pd populated via \be$^{‒}$ decay.
TAGS studies not reported, although potential for Pandemonium supported by large $\Delta$ energy differential.
\be$^{‒}$ feeding to $^{108}$Pd ground state (1+ $\rightarrow$ 0+ transition) of 56(11)\% reported in ENSDF. 
Merits studies by means of TAGS and DGS. 
& y  & y    & 3    \\
\hline

\end{tabular}
\end{adjustbox}
\end{sidewaystable*}

\begin{sidewaystable*}

\vspace*{-14cm}
\addtocounter{table}{-1}
\caption{ -- Continued from previous page} 
 \begin{adjustbox}{width=0.9\textwidth}
\begin{tabular}{llllllp{10cm}ccc}\hline \hline
\T Fission & J$^{\pi}$  & Half-life & Q($\beta^{-}$)  & $\Delta$\textsuperscript{a}   & S$_n$    & Assessment of potential Pandemonium\textsuperscript{b}    & DGS\textsuperscript{c}  & TAGS\textsuperscript{c} & Priority\textsuperscript{d} \\
\B   product & &   & (keV)     &(keV) &(keV) &  &    &     &   \\  \hline

\T 51-Sb-130m      & 4+  & 6.3(2) min   & 5072(14)  & 1659(14)  & 8419.4(7) & $^{238}$Pu (1000 s, 2.6\%\g), $^{239}$Pu (1000 s, 2.2\%\g), $^{240}$Pu (1000 s, 2.1\%\g); 
ENSDF (2001Si26) reports 16 levels in $^{130}$Te populated via \be$^{‒}$ decay.
$^{130m}$Sb(\be$^{‒}$) = 100\% in NUBASE2020.
TAGS studies not reported, although potential for Pandemonium supported by large $\Delta$ energy differential.  
\be$^{‒}$ feeding to $^{130}$Te ground state (4+ $\rightarrow$ 0+ transition) of zero adopted in ENSDF.  
Merits studies by means of TAGS and DGS. 
& y & y     & 1   \\ \hline

\T 51-Sb-131       & 7/2+  & 23.03(4) min & 3229.6(21) & 567(2)    & 5929.38(6) & $^{235}$U (1000 s, 2.8\%\g, 2.1\%), (5011 s, 2.1\%\g), $^{238}$Pu (1000 s, 3.3\%\g, 2.5\%), (5011 s, 2.7\%\g, 2.3\%), $^{239}$Pu (1000 s, 3.0\%\g, 2.2\%), (5011 s, 2.4\%\g, 2.0\%), $^{240}$Pu (1000 s, 3.1\%\g, 2.3\%), (5011 s, 2.5\%\g, 2.1\%), $^{241}$Am (1000 s, 2.7\%\g, 2.0\%), (5011 s, 2.3\%\g), $^{243}$Cm (1000 s, 2.5\%\g), (5011 s, 2.0\%\g); 
ENSDF (2006Kh09) reports 25 levels in $^{131}$Te populated via \be$^{‒}$ decay. 
TAGS studies are not reported, with low potential for Pandemonium based on relatively small $\Delta$ energy differential. 
\be$^{‒}$ feeding to $^{131}$Te ground state (7/2+ $\rightarrow$ 3/2+ transition) of zero adopted in ENSDF.
Comprehensive DGS measurements recommended. 
& y  & ‒ & 2  \\
\hline

\end{tabular}
\end{adjustbox}
\end{sidewaystable*}

\begin{sidewaystable*}

\vspace*{-14cm}
\addtocounter{table}{-1}
\caption{ -- Continued from previous page} 
 \begin{adjustbox}{width=0.9\textwidth}
\begin{tabular}{llllllp{10cm}ccc}\hline \hline
\T Fission & J$^{\pi}$  & Half-life & Q($\beta^{-}$)  & $\Delta$\textsuperscript{a}   & S$_n$    & Assessment of potential Pandemonium\textsuperscript{b}    & DGS\textsuperscript{c}  & TAGS\textsuperscript{c} & Priority\textsuperscript{d} \\
\B   product & &   & (keV)     &(keV) &(keV) &  &    &     &   \\  \hline

\T 52-Te-134       & 0+      & 41.8(8) min  & 1510(5) & 404(5)    & 6257(8)  & $^{235}$U (1000 s, 2.5\%\g), (5011 s, 4.6\%\g, 3.5\%), (10000 s, 2.9\%\g, 2.2\%), $^{238}$Pu (5011 s, 2.7\%\g, 2.1\%), $^{239}$Pu (5011 s, 3.3\%\g, 2.6\%), (10000 s, 2.3\%\g),  $^{240}$Pu (5011 s, 3.2\%\g, 2.5\%), (10000 s, 2.3\%\g), $^{241}$Am (5011 s, 2.4\%\be); 
ENSDF (2004So32) reports 9 levels in $^{134}$I populated via \be$^{‒}$ decay.
TAGS studies not reported, with low potential for Pandemonium based on a relatively small $\Delta$ energy differential. 
\be$^{‒}$ feeding to $^{134}$I ground state (0+ $\rightarrow$ (4)+ transition) of zero adopted in ENSDF.
%Q$_{\beta^{-}}$=1502(22) keV (determined using RADLIST from the adopted in ENSDF decay scheme) in good agreement with Q(\beta$^{-}$=1510(5) keV from AME2020 (FGK). 
& ‒ & – & –  \\ \hline

\T 52-Te-135  & (7/2‒)  & 19.0(2) s    & 6050.4(27)   & 1278(3)   & 7807(5)  & $^{235}$U (10 s, 2.8\%\be), $^{239}$Pu (10 s, 2.1\%\be); 
ENSDF (2008Si03) reports 18 levels in $^{134}$I populated via \be$^{‒}$ decay. 
TAGS studies not reported, although potential for Pandemonium supported by reasonably large $\Delta$ energy differential.
\be$^{‒}$ feeding to $^{134}$I ground state ((7/2‒) $\rightarrow$ 7/2+ transition) of 62(3)\% adopted in ENSDF. 
Comprehensive DGS studies recommended.
%Q$_{\beta^{-}}$=5870(140) keV (determined using RADLIST from the adopted in ENSDF decay scheme) somewhat lower compared to  Q(\beta$^{-}$=6050.4(27) keV from AME2020 (FGK). 
& y  & –   & 3     \\
\hline

\end{tabular}
\end{adjustbox}
\end{sidewaystable*}

\begin{sidewaystable*}

\vspace*{-14cm}
\addtocounter{table}{-1}
\caption{ -- Continued from previous page} 
 \begin{adjustbox}{width=\textwidth}
\begin{tabular}{llllllp{10cm}ccc}\hline \hline
\T Fission & J$^{\pi}$  & Half-life & Q($\beta^{-}$)  & $\Delta$\textsuperscript{a}   & S$_n$    & Assessment of potential Pandemonium\textsuperscript{b}    & DGS\textsuperscript{c}  & TAGS\textsuperscript{c} & Priority\textsuperscript{d} \\
\B   product & &   & (keV)     &(keV) &(keV) &  &    &     &   \\  \hline

\T 53-I-136m       & (6‒)    & 46.6(11) s   & 7090(21)  & 999(21)   & 8087(4)  & $^{235}$U (100 s, 2.2\%\g, 3.2\%\be, 2.7\%), $^{238}$Pu (100 s, 2.5\%\g, 3.2\%\be, 2.9\%), $^{239}$Pu (100 s, 3.1\%\g, 4.0\%\be, 3.5\%), $^{240}$Pu (100 s, 3.2\%\g, 4.2\%\be, 3.7\%), $^{241}$Am (100 s, 3.7\%\g, 4.6\%\be, 4.2\%), $^{243}$Cm (100 s, 3.1\%\g, 3.7\%\be, 3.4\%); 
ENSDF (2018Mc07) reports 14 levels in $^{136}$Xe populated via \be$^{‒}$ decay. Large imbalance between Q$_{\beta}$ energy determined from proposed ENSDF decay scheme (7550(240) keV) and that recommended by AME2020 (7090(21) keV), indicating a problematic decay scheme.
$^{136m}$I(\be$^{‒}$) = 100\% in NUBASE2020.
TAGS studies not reported, with modest potential for Pandemonium based on $\Delta$ energy differential. %Consideration should be given to TAGS studies as a consequence of the uncertainty in the detail of the decay scheme.
\be$^{‒}$ feeding to $^{136}$Xe ground state ((6–) $\rightarrow$ 0+ transition) of zero adopted in ENSDF.
Merits comprehensive studies by means of TAGS and DGS to resolve significant issues associated with proposed decay scheme.
%( {\bf But is there really a LARGE imbalance between Q$_{\beta}$ energy in ENSDF (6884(14) + 201(26) = 7085(30) keV) and AME2020 (7090(21) keV), or is there a typing error in ENSDF that creates RADLIST Q$_{\beta}$ energy of 7550(240) keV? (ALN) In ENSDF, the excitation energy of the isomer is determined from the beta-decay end-point (BD-EP) measurement of 2007Fo02, while in NUBASE2020/AME2020 in addition to 2007Fo02, the high-precision Penning trap results from 2021Va02 was considered (FGK). } )
& y & y    & 2  \\
\hline

\end{tabular}
\end{adjustbox}
\end{sidewaystable*}

\begin{sidewaystable*}
\vspace*{-14cm}
\addtocounter{table}{-1}
\caption{ -- Continued from previous page} 
 \begin{adjustbox}{width=\textwidth}
\begin{tabular}{llllllp{10cm}ccc}\hline \hline
\T Fission & J$^{\pi}$  & Half-life & Q($\beta^{-}$)  & $\Delta$\textsuperscript{a}   & S$_n$    & Assessment of potential Pandemonium\textsuperscript{b}    & DGS\textsuperscript{c}  & TAGS\textsuperscript{c} & Priority\textsuperscript{d} \\
\B   product & &   & (keV)     &(keV) &(keV) &  &    &     &   \\  \hline

\T 54-Xe-139       & 3/2‒    & 39.68(14) s  & 5057(4)  & 829(4)  & 5885(10) & $^{235}$U (100 s, 2.4\%\g, 4.8\%\be, 3.5\%), $^{238}$Pu (100 s, 2.5\%\be), $^{239}$Pu (100 s, 2.0\%\g, 3.6\%\be, 2.8\%), $^{240}$Pu (100 s, 3.1\%\be, 2.4\%), $^{241}$Am (100 s, 2.8\%\be, 2.2\%); 
ENSDF (2016Jo10) reports 63 levels in $^{139}$Cs populated via \be$^{‒}$ decay.
TAGS studies reveal a reasonable degree of Pandemonium (2014Fi11/2017Fi06). 
\be$^{‒}$ feeding to $^{139}$Cs ground state (3/2‒ $\rightarrow$ 7/2+ transition) of 15(10)\% adopted in ENSDF, compared with 2(1)\% from TAGS (2014Fi11, 2017Fi06). 
Comprehensive DGS measurements recommended. 
& y  & –  & 2   \\ \hline

\T 54-Xe-140       & 0+      & 13.60(10) s  & 4063(9) & 1739(9)   & 4420(9) & $^{235}$U (10 s, 2.3\%\g); 
ENSDF (2018Ni16) reports 33 levels in $^{140}$Cs populated via \be$^{‒}$ decay.
TAGS studies not reported, but there is support for Pandemonium due to large $\Delta$ energy differential. 
\be$^{‒}$ feeding to $^{140}$Cs ground state (0+ $\rightarrow$ 1‒ transition) of zero reported in ENSDF: most unusual ‒ requires confirmation. 
Merits studies by means of TAGS and DGS.
& y & y   & 2  \\
\hline

\end{tabular}
\end{adjustbox}
\end{sidewaystable*}

\begin{sidewaystable*}
\vspace*{-14cm}
\addtocounter{table}{-1}
\caption{ -- Continued from previous page} 
 \begin{adjustbox}{width=\textwidth}
\begin{tabular}{llllllp{10cm}ccc}\hline \hline
\T Fission & J$^{\pi}$  & Half-life & Q($\beta^{-}$)  & $\Delta$\textsuperscript{a}   & S$_n$    & Assessment of potential Pandemonium\textsuperscript{b}    & DGS\textsuperscript{c}  & TAGS\textsuperscript{c} & Priority\textsuperscript{d} \\
\B   product & &   & (keV)     &(keV) &(keV) &  &    &     &   \\  \hline

\T 55-Cs-140       & 1‒      & 63.7(3) s   & 6218(10)  & 453(10)   & 6425(8)  & $^{235}$U (100 s, 7.0\%\g, 8.2\%\be, 7.5\%), $^{238}$Pu (100 s, 5.0\%\g, 5.2\%\be, 5.1\%), $^{239}$Pu (100 s, 6.1\%\g, 6.4\%\be, 6.3\%), $^{240}$Pu (100 s, 6.2\%\g, 6.6\%\be, 6.4\%), $^{241}$Am (100 s, 6.5\%\g, 6.6\%\be, 6.6\%), $^{243}$Cm (100 s, 5.2\%\g, 5.2\%\be, 5.2\%); 
ENSDF (2018Ni16) reports 55 levels in $^{140}$Ba populated via \be$^{‒}$ decay.
TAGS studies (1997Gr09), with no uncertainties assigned to \be$^{‒}$-decay feeding intensities. Recent TAGS measurements (2017GuZV) show that the feeding distribution is in good agreement with previous 1997Gr09 results, although definitive data yet to be published.   
\be$^{‒}$ feeding to $^{140}$Ba ground state (1‒ $\rightarrow$ 0+ transition) of 35.9(17)\% reported in ENSDF, as adopted from TAGS (1996Gr20/1997Gr09), agrees with value of 36.0(15)\% provided by 2017GuZV.  

Merits further studies by means of TAGS and DGS.
& y & y & 2  \\
\hline

\end{tabular}
\end{adjustbox}
\end{sidewaystable*}

\begin{sidewaystable*}

\vspace*{-14cm}
\addtocounter{table}{-1}
\caption{ -- Continued from previous page} 
 \begin{adjustbox}{width=\textwidth}
\begin{tabular}{llllllp{10cm}ccc}\hline \hline
\T Fission & J$^{\pi}$  & Half-life & Q($\beta^{-}$)  & $\Delta$\textsuperscript{a}   & S$_n$    & Assessment of potential Pandemonium\textsuperscript{b}    & DGS\textsuperscript{c}  & TAGS\textsuperscript{c} & Priority\textsuperscript{d} \\
\B   product & &   & (keV)     &(keV) &(keV) &  &   &     &   \\  \hline

\T 56-Ba-142       & 0+      & 10.6(2) min  & 2182(8) & 643(8)    & 5165(7)  & $^{235}$U (1000 s, 4.5\%\g, 2.0\%\be, 3.3\%), $^{238}$Pu (1000 s, 3.6\%\g, 2.7\%), $^{239}$Pu (1000 s, 3.7\%\g, 2.8\%), $^{240}$Pu (1000 s, 3.5\%\g, 2.6\%), $^{241}$Am (1000 s, 3.6\%\g, 2.6\%), $^{243}$Cm (1000 s, 4.2\%\g, 3.0\%); 
ENSDF (2011Jo05) reports 24 levels in $^{142}$La populated via \be$^{‒}$ decay.
TAGS studies (1997Gr09) with no uncertainties assigned to \be$^{‒}$-decay feeding intensities, as well as 2014Wo01/2017WoZZ without absolute values and uncertainties for \be$^{‒}$-decay feeding intensities.
Good agreement with DGS measurements for population/depopulation of all relevant nuclear levels up to 1593.26 keV, with no significant evidence of Pandemonium.
Combined \be$^{‒}$ feeding to $^{142}$La ground state (0+ $\rightarrow$ 2‒ transition) and 77.594-keV level (0+ $\rightarrow$ (2)‒ transition) of zero adopted in ENSDF, compared with 2.0(19)\% from TAGS (1997Gr09). 
\be$^{‒}$ feedings to 255.3-keV (0+ $\rightarrow$ 1‒ transition), 425.0-keV (0+ $\rightarrow$ 1(‒) transition) and 1204.3-keV (0+ $\rightarrow$ (1)+ transition) levels in $^{142}$La of 3.5(7)\%, 5.4(2)\% and 46(2)\%, respectively, adopted in ENSDF, while values of 4.6\%, 6.0\% and 46.0\% (1997Gr09) and 3.8\%, 5.8\% and 45.8\% (2017WoZZ) determined by TAGS.
& –  & –  & –  \\
\hline

\end{tabular}
\end{adjustbox}
\end{sidewaystable*}

\begin{sidewaystable*}

\vspace*{-14cm}
\addtocounter{table}{-1}
\caption{ -- Continued from previous page} 
 \begin{adjustbox}{width=\textwidth}
\begin{tabular}{llllllp{10cm}ccc}\hline \hline
\T Fission & J$^{\pi}$  & Half-life & Q($\beta^{-}$)  & $\Delta$\textsuperscript{a}   & S$_n$    & Assessment of potential Pandemonium\textsuperscript{b}    & DGS\textsuperscript{c}  & TAGS\textsuperscript{c} & Priority\textsuperscript{d} \\
\B   product & &   & (keV)     &(keV) &(keV) &  &    &     &   \\  \hline

\T 57-La-141       & (7/2+)  & 3.92(3) h   & 2501(4)  & 172(4)    & 5428.15(10) & $^{235}$U (5011 s, 3.8\%\be), (10000 s, 7.7\%\be, 3.2\%), $^{238}$Pu (5011 s, 4.3\%\be), (10000 s, 9.4\%\be, 3.7\%), $^{239}$Pu (5011 s, 4.0\%\be), (10000 s, 8.9\%\be, 3.5\%), $^{240}$Pu (5011 s, 3.6\%\be), (10000 s, 8.2\%\be, 3.2\%), $^{241}$Am (5011 s, 3.8\%\be), (10000 s, 8.9\%\be, 3.6\%), $^{243}$Cm (5011 s, 4.1\%\be), (10000 s, 9.7\%\be, 4.0\%); 
ENSDF (2014Ni18) reports 17 levels in $^{141}$Ce populated via \be$^{‒}$ decay. 
TAGS studies not reported, with low potential for Pandemonium based on a reasonably small $\Delta$ energy differential. 
\be$^{‒}$ feeding to $^{141}$Ce ground state ((7/2+) $\rightarrow$ 7/2‒ transition) of 98.12(16)\% adopted in ENSDF. 
& – & –   & – \\
\hline

\end{tabular}
\end{adjustbox}
\end{sidewaystable*}

\begin{sidewaystable*}

\vspace*{-14cm}
\addtocounter{table}{-1}
\caption{ -- Continued from previous page} 
 \begin{adjustbox}{width=\textwidth}
\begin{tabular}{llllllp{10cm}ccc}\hline \hline
\T Fission & J$^{\pi}$  & Half-life & Q($\beta^{-}$)  & $\Delta$\textsuperscript{a}   & S$_n$    & Assessment of potential Pandemonium\textsuperscript{b}    & DGS\textsuperscript{c}  & TAGS\textsuperscript{c} & Priority\textsuperscript{d} \\
\B   product & &   & (keV)     &(keV) &(keV) &  &    &     &   \\  \hline

\T 57-La-145       & (5/2+)  & 24.8(20) s  & 4230(40) & 530(40)   & 4710(30)  & $^{235}$U (10 s, 2.7\%\g), (100 s, 2.3\%\g), $^{238}$Pu (10 s, 2.6\%\g), $^{239}$Pu (10 s, 2.6\%\g), $^{240}$Pu (10 s, 2.4\%\g), $^{241}$Am (10 s, 2.8\%\g), $^{243}$Cm (10 s, 3.2\%\g);
ENSDF (2009Br01) reports 32 levels in $^{145}$Ce populated via \be$^{‒}$ decay. 
TAGS studies provide evidence of Pandemonium (1997Gr09), with no uncertainties assigned to \be$^{‒}$-decay feeding intensities. 
Combined \be$^{‒}$ feeding to $^{145}$Ce ground state ((5/2+) $\rightarrow$ (5/2‒) transition), 64.3-keV ((5/2+) $\rightarrow$ (‒) transition), 70.0-keV ((5/2+) $\rightarrow$ (7/2‒) transition) and 118.2-keV ((5/2+) $\rightarrow$ unknown J$^{\pi}$ transition) of zero adopted in ENSDF, compared with ‒2.2(31)\%  from TAGS (1996Gr20/1997Gr09).  
Merits studies by means of TAGS and DGS. 
%( {\bf 32 levels (not 33) (ALN) } )
& y   & y     & 3    \\
\hline

\end{tabular}
\end{adjustbox}
\end{sidewaystable*}
%}

%\afterpage{
\begin{sidewaystable*}
\vspace*{-14cm}

\caption{Fission-product decay data: assessment of potential Pandemonium and need for further TAGS and DGS (discrete gamma-ray spectroscopy involving \g~singles and \g-\g~coincidence measurements) studies for $^{232}$Th, $^{233,238}$U and $^{237}$Np fast fission, and $^{241,242}$Pu and $^{245}$Cm thermal fission. Relevant references in the literature are defined in NSR keynumber format (e.g., 2017Fi06 (NSR), 2016ORNL (equivalent non-NSR)). 
Spin, parity (J$^{\pi}$) and half-life are from NUBASE2020~\cite{NUBASE2020}. Q-value for ground state-to-ground state decay (Q($\beta^{-}$)) and the neutron separation energy (S$_{n}$) are from AME2020~\cite{AME2020}. Q($\beta^{-}$) value for an excited isomer is reported as the sum of the ground state-to-ground state decay and the excitation energy of the isomer taken from NUBASE2020~\cite{NUBASE2020}.  Beta-delayed neutron emission probabilities P$_n$(\%) are taken from the IAEA Reference Database for beta-delayed neutron emission~\cite{Liang,Dimitriou2021,bdnIAEA}.
(a)	Possible potential for Pandemonium effect as a consequence of large energy differential $\Delta$ (keV) = (Q($\beta^{-}$) – excitation energy of highest known relevant nuclear level); 
(b)	FISPACT-II calculations of fission-product inventories and their decay heat as a function of the cooling time of irradiated actinides~\cite{Fleming2015,Fleming2015a,Fleming2015b}: fissile actinide, (post-irradiation cooling time, \%\g, \%\be, \%(total decay heat)), 
fission-product decay schemes assessed in terms of known \be$^{‒}$ decay and discrete nuclear levels of the daughter~\cite{ENSDF};
(c)	Request for specific measurements denoted by $‘$y' $\rightarrow$ 'yes' ($‘$?' $\rightarrow$ to be assessed further); 
(d)	Priorities defined as 1 $\rightarrow$ high, 2 $\rightarrow$ intermediate, 3 $\rightarrow$ low, – $\rightarrow$ unassigned. }\label{tab:app2} 
 \begin{adjustbox}{width=\textwidth}
\begin{tabular}{llllllp{10cm}ccc}\hline \hline
\T Fission & J$^{\pi}$  & Half-life & Q($\beta^{-}$)  & $\Delta$\textsuperscript{a}   & S$_n$    & Assessment of potential Pandemonium\textsuperscript{b}    & DGS\textsuperscript{c}  & TAGS\textsuperscript{c} & Priority\textsuperscript{d} \\
\B   product & &   & (keV)     &(keV) &(keV) &  &   &     &   \\  \hline

\T 33-As-84        & (2‒)        & 3.2(6) s              & 10094(4)         & 3490(4)   & 8679(4)     & $^{232}$Th (10 s, 1.9\%\g, 1.5\%\be, 1.7\%), $^{233}$U (10 s, 0.5\%\g, \textless 0.4\%\be, 0.4\%); 
ENSDF (2009Ab25) reports 34 levels in $^{84}$Se populated via \be$^{‒}$ decay. 
TAGS studies not reported, although potential for Pandemonium supported strongly by large $\Delta$ energy differential.  
\be$^{‒}$ feeding to $^{84}$Se ground state ((2–) $\rightarrow$ 0+ transition) of $\le$5\% adopted in ENSDF.
Merits studies by means of TAGS and DGS.
P$_n$($^{84}$As) = 0.281(28)\%.
& y   & y     & 2 \\ \hline 

\T 34-Se-83        & 9/2+        & 22.25(4) min           & 3673(5)          & 535(5)    & 9586(4)     & $^{232}$Th (1000 s, 3.3\%\g, 0.7\%\be,   2.1\%),   $^{233}$U (1000 s, 1.5\%\g, 0.3\%\be, 1.0\%); 
ENSDF (2015Mc01) reports 25 levels in $^{83}$Br populated via \be$^{‒}$ decay. 
TAGS studies not reported, with low potential for Pandemonium based on a relatively small $\Delta$ energy differential. 
\be$^{‒}$ feeding to $^{83}$Br ground state (9/2+ $\rightarrow$ 3/2‒ transition) of zero adopted in ENSDF, with more than 80\% \be$^{‒}$ feeding within a narrow energy region from 2645 to 2950 keV. 
DGS studies recommended.
& y  & –  & 3 \\ \hline     

\T 34-Se-85        & (5/2)+      & 32.9(3) s              & 6162(4)          & 1651(4)   & 8864(26)    & $^{232}$Th (100 s, 1.8\%\g, 2.1\%\be,   1.9\%),   $^{233}$U (100   s, 1.2\%\g, 1.3\%\be, 1.2\%); 
ENSDF (2015SiENSDF) reports 24 levels in $^{85}$Br populated via \be$^{‒}$ decay.  
TAGS studies not reported, although potential for Pandemonium supported by large $\Delta$ energy differential.
\be$^{‒}$ feeding to $^{85}$Br ground state ((5/2)+ $\rightarrow$ 3/2‒ transition) of 31(6)\% adopted in ENSDF. 
Merits studies by means of TAGS and DGS.
& y   & y     & 2 \\  \hline    

\T 34-Se-86        & 0+          & 14.3(3) s              & 5129(4)          & 2464(4)   & 5128(4)     & $^{232}$Th (10 s, 4.4\%\g, 2.1\%\be, 3.2\%),   $^{233}$U (10 s, 1.7\%\g, 0.8\%\be,  1.2\%); 
ENSDF (2015Ne01) reports 10 levels in $^{85}$Br populated via \be$^{‒}$ decay.  
TAGS studies not reported, although potential for Pandemonium supported by large $\Delta$ energy differential.
\be$^{‒}$ feeding to $^{86}$Br ground state (0+ $\rightarrow$ (1‒) transition) of \textless20\% estimated in ENSDF, along with significant \be$^{‒}$ feedings of 69(7)\% and 21.6(23)\% assigned to 2446.3- and 2665.1-keV levels, respectively. 
DGS studies to be incorporated into ENSDF (2016Ur04), followed by assessment of need for further DGS measurements. 
Merits study by means of TAGS.
& ?   & y     & 2? \\  \hline      

\T 34-Se-87        & (3/2+)      & 5.50(6) s             & 7466(4)         & 3479(4)   & 6331(4)     & $^{232}$Th (10 s, 0.8\%\g, 3.0\%\be, 2.0\%), $^{233}$U (10 s, \textless{}0.4\%\g, 1.2\%\be, 0.8\%); 
ENSDF (2015Jo11) reports 9 levels in $^{87}$Br populated via \be$^{‒}$ decay. 
TAGS studies not reported, although potential for Pandemonium supported by large $\Delta$ energy differential.
\be$^{‒}$ feeding to $^{87}$Br ground state ((3/2)+ $\rightarrow$ (5/2‒) transition) of 32\% estimated in ENSDF, and decay scheme defined as incomplete. 
DGS studies to be incorporated into ENSDF (2019Wi11), followed by assessment of need for further DGS measurements. 
Merits study by means of TAGS.
P$_n$($^{87}$Se) = 0.60(12)\%. 
& ?   & y     & 2?   \\  \hline
\end{tabular}
 \end{adjustbox}
\end{sidewaystable*}

\begin{sidewaystable*}
\vspace*{-14cm}
\addtocounter{table}{-1}
\caption{ -- Continued from previous page} 
 \begin{adjustbox}{width=\textwidth}
\begin{tabular}{llllllp{9cm}ccc}\hline \hline
\T Fission & J$^{\pi}$  & Half-life & Q($\beta^{-}$)  & $\Delta$\textsuperscript{a}   & S$_n$    & Assessment of potential Pandemonium\textsuperscript{b}    & DGS\textsuperscript{c}  & TAGS\textsuperscript{c} & Priority\textsuperscript{d} \\
\B   product & &   & (keV)     &(keV) &(keV) &  &   &     &   \\  \hline

\T 35-Br-84        & 2‒          & 31.76(8) min           & 4656(26)         & 467(26)   & 10520.020(8) & $^{232}$Th (5011 s, 5.2\%\g, 5.0\%\be, 5.1\%), $^{233}$U (5011 s, 2.2\%\g, 2.0\%\be,   2.2\%); 
ENSDF (2009Ab25) reports 20 levels in $^{84}$Kr populated via \be$^{‒}$ decay. 
TAGS studies not reported, with low potential for Pandemonium based on a relatively small $\Delta$ energy differential.
\be$^{‒}$ feeding to $^{84}$Kr ground state (2‒ $\rightarrow$ 0+ transition) of 33(5)\% adopted in ENSDF.   
Merits studies by means of TAGS and DGS. 
& y  & y     & 2   \\  \hline    

\T 35-Br-89        & (3/2‒,5/2‒) & 4.357(22) s            & 8262(4)        & 3555(4)   & 4916(3)         & $^{232}$Th (10 s, 3.6\%\g, 3.3\%\be, 2.0\%), $^{233}$U (10 s, 1.7\%\g, 2.1\%\be, 1.9\%); 
ENSDF (2022SiENSDF) reports 45 levels in $^{89}$Kr populated via \be$^{‒}$ decay. Proposed decay scheme incomplete: large difference between calculated effective Q($\beta^{-}$) and that recommended in AME2020. 
TAGS studies indicate significant Pandemonium (2016Ka65), although no absolute values assigned to \be$^{‒}$-decay feeding intensities. 
Combined \be$^{‒}$ feeding to ground state ((3/2‒,5/2‒) $\rightarrow$ 3/2(+) transition) and 28.59-keV level ((3/2‒,5/2‒) $\rightarrow$ (5/2+) transition) in $^{89}$Kr of 32(3)\% adopted in ENSDF. 
Significant \be-n emitter P$_n$($^{89}$Br) = 13.7(6)\%. 
Merits studies by means of TAGS and DGS. 
& y & y     & 2  \\  \hline

\T 37-Rb-90m       & 3‒          & 258(4) s               & 6692(7)        & 864(7)    & 7813.2(15)    & $^{232}$Th (1000 s, 0.5\%\g,   \textless{}0.3\%\be, 0.3\%), $^{233}$U (1000 s, 2.7\%\g, 1.0\%\be, 1.9\%);  
ENSDF (2020Ba21) reports 44 levels in $^{90}$Sr populated via \be$^{‒}$ decay. 
$^{90m}$Rb(IT) = 2.5(4)\% in NUBASE2020.
TAGS measurements support Pandemonium (1997Gr09, 2017Fi06), although 1997Gr09 assigns no uncertainties to \be$^{‒}$-decay feeding intensities.
\be$^{‒}$ feeding to $^{90}$Sr ground state (3‒ $\rightarrow$ 0+ transition) of zero, with a value of 15(4)\% determined to the 831.67-keV level (3‒ $\rightarrow$ 2+ transition) as adopted in ENSDF, compared with 5(1)\% (2017Fi06) and 1.76\% (1997Gr09) from TAGS.
Merits DGS studies, with unambiguous separation between ground-state and isomer \be$^{‒}$ decays. 
& y  & –  & 2   \\  \hline

\T 40-Zr-98        & 0+          & 30.7(4) s              & 2243(10)       & 2243(10)  & 5993(7)       & $^{237}$Np (100 s, 1.1\%\g, 1.9\%\be, 1.5\%); 
ENSDF (2020Ch10) reports \be$^{‒}$ decay only to $^{98}$Nb ground state (0+ $\rightarrow$ 1+ transition) with feeding intensity of 100\%. 
TAGS studies not reported, although potential for Pandemonium supported by large $\Delta$ energy differential.
Merits studies by means of TAGS and DGS. 
& y  & y     & 3   \\  \hline

\T 41-Nb-105       & (5/2+) & 2.91(5) s  & 7415(10)      & 6766(10)  & 5058(13)      & $^{241}$Pu (1 s, 1.6\%\g, 1.2\%\be, 1.3\%), $^{242}$Pu (1 s, 2.7\%\g, 2.0\%\be, 2.3\%); 
ENSDF (2019La19) reports 14 levels in $^{105}$Mo populated via \be$^{‒}$ decay, with decay scheme ill-defined and \be$^{‒}$-decay feeding intensities unquantified.  
TAGS studies not reported, although potential for Pandemonium supported by large $\Delta$ energy differential.  
Relatively large \be$^{‒}$-decay feeding intensity to the $^{105}$Mo ground state ((5/2+) $\rightarrow$ (5/2-) transition) expected, whereas no value determined for adoption in ENSDF. 
Merits studies by means of TAGS and DGS. 
P$_n$($^{105}$Nb) = 1.7(9)\%.  
& y  & y    & 2   \\ \hline 

\end{tabular}
\end{adjustbox}
\end{sidewaystable*}

\begin{sidewaystable*}
\vspace*{-14cm}
\addtocounter{table}{-1}
\caption{ -- Continued from previous page} 
 \begin{adjustbox}{width=\textwidth}
\begin{tabular}{llllllp{10cm}ccc}\hline \hline
\T Fission & J$^{\pi}$  & Half-life & Q($\beta^{-}$)  & $\Delta$\textsuperscript{a}   & S$_n$    & Assessment of potential Pandemonium\textsuperscript{b}    & DGS\textsuperscript{c}  & TAGS\textsuperscript{c} & Priority\textsuperscript{d} \\
\B   product & &   & (keV)     &(keV) &(keV) &  &   &     &   \\  \hline

\T 42-Mo-106       & 0+          & 8.73(12) s             & 3648(15)       & 2911(15)  & 5560(40)      & $^{245}$Cm (10 s, 1.0\%\g, 2.0\%\be, 1.6\%); 
ENSDF (2008De09) reports 17 levels in $^{106}$Tc populated via \be$^{‒}$ decay. 
TAGS studies not reported, although potential for Pandemonium supported by large $\Delta$ energy differential. 
\be$^{‒}$-decay feeding intensity to $^{106}$Tc ground state (0+ $\rightarrow$ (1,2) transition) of \textless18\% adopted in ENSDF. 
Merits studies by means of TAGS and DGS.  
& y & y  & 3    \\  \hline  

\T 42-Mo-107       & (1/2+)      & 3.5(5) s               & 6205(13)       & 4830(13)  & 7045(15)    & $^{242}$Pu (1 s, 2.2\%\be, 1.5\%), (10 s, 2.2\%\be, 1.5\%); 
ENSDF (2008Bl12) reports 9 levels in $^{107}$Tc populated via \be$^{‒}$ decay, with decay scheme ill-defined and \be$^{‒}$-decay feeding intensities unquantified.
TAGS studies not reported, although potential for Pandemonium supported by large $\Delta$ energy differential.  
Relatively large \be$^{‒}$-decay feeding intensity to $^{107}$Tc ground state ((1/2+) $\rightarrow$ (3/2‒) transition) expected, whereas no value adopted in ENSDF. 
Merits studies by means of TAGS and DGS.  
& y    & y    & 2   \\  \hline    

\T 45-Rh-111       & (7/2+)      & 11(1) s                & 3682(7)        & 2475(7)   & 5726.3(4)     & $^{245}$Cm (10 s, 0.9\%\g, 2.7\%\be, 1.9\%); 
ENSDF (2009Bl02) reports 16 levels in $^{111}$Pd populated via \be$^{‒}$ decay, although decay scheme and \be$^{‒}$-decay feeding intensities are ill-defined. 
TAGS studies not reported, although potential for Pandemonium supported by large $\Delta$ energy differential. 
\be$^{‒}$-decay feeding intensity to $^{111}$Pd ground state ((7/2+) $\rightarrow$ 5/2+ transition) of zero, while a preliminary value of 75.5\% was determined for the 275.4-keV level ((7/2+) $\rightarrow$ 3/2+,5/2+ transition) as adopted in ENSDF. 
Merits studies by means of TAGS and DGS.  
& y & y   & 3   \\  \hline      

\T 46-Pd-111       & 5/2+        & 23.56(9) min            & 2229.6(16)     & 524.5(16) & 8829.5(19)    & $^{245}$Cm (10 s, \textless{}0.3\%\g, 2.0\%\be, 1.0\%); 
ENSDF (2009Bl02) reports 29 levels in $^{111}$Ag populated via \be$^{‒}$ decay.
TAGS studies not reported, with low potential for Pandemonium based on a relatively small $\Delta$ energy differential.
\be$^{‒}$-decay feeding intensity to $^{111}$Ag ground state (5/2+ $\rightarrow$ 1/2‒ transition) of zero, while a value of 95.3(4)\% has been determined for the 59.87-keV level (5/2+ $\rightarrow$ 7/2+ transition) as adopted in ENSDF. 
DGS studies to be incorporated into ENSDF (2015Kr07), followed by assessment of need for further TAGS and DGS measurements. 
& ? & ?  & 3?  \\ \hline

\T 51-Sb-132       & (4)+        & 2.79(7) min            & 5553(4)        & 1990(4)   & 8048(3)        & $^{238}$U (100 s, 2.8\%\g, 1.5\%\be, 2.2\%), $^{241}$Pu   (100 s, 2.2\%\g, 1.1\%\be, 1.7\%); 
ENSDF (2005Kh07) reports 15 levels in $^{132}$Te populated via \be$^{‒}$ decay, although \be$^{‒}$-decay feeding intensities unquantified. 
TAGS studies not reported, although potential for Pandemonium supported by large $\Delta$ energy differential.  
\be$^{‒}$ feeding to the $^{132}$Te ground state ((4)+ $\rightarrow$ 0+ transition) not determined, but expected to be zero.   
Merits studies by means of TAGS and DGS, with unambiguous separation between ground-state and isomer \be$^{‒}$ decays.  
& y   & y     & 1  \\  \hline

\T 52-Te-136       & 0+          & 17.63(9) s             & 5120(14)       & 1885(14)  & 3837(14)     & $^{238}$U (10 s, 2.7\%\g, 1.2\%\be, 1.8\%), $^{241}$Pu (10 s, 2.1\%\g, 1.4\%);  
ENSDF (2018Mc07) reports 10 levels in $^{136}$I populated via \be$^{‒}$ decay.
TAGS studies not reported, although potential for Pandemonium supported by reasonably large $\Delta$ energy differential.  
\be$^{‒}$ feeding to the $^{136}$I ground state (0+ $\rightarrow$ (1‒) transition) of \textless8.6\% adopted in ENSDF. 
Merits studies by means of TAGS and DGS. 
P$_n$($^{136}$Te) = 1.37(5)\%.
& y  & y    & 2   \\  \hline

\end{tabular}
\end{adjustbox}
\end{sidewaystable*}

\begin{sidewaystable*}
\vspace*{-14cm}
\addtocounter{table}{-1}
\caption{ -- Continued from previous page} 
 \begin{adjustbox}{width=\textwidth}
\begin{tabular}{llllllp{10cm}ccc}\hline \hline
\T Fission & J$^\pi$  & Half-life & Q($\beta^-$)  & $\Delta$\textsuperscript{a}   & S$_n$    & Assessment of potential Pandemonium\textsuperscript{b}    & DGS\textsuperscript{c}  & TAGS\textsuperscript{c} & Priority\textsuperscript{d} \\
\B   product & &   & (keV)     &(keV) &(keV) &  &   &     &   \\  \hline

\T 52-Te-137       & 3/2‒        & 2.49(5) s              & 7053(9)       & 5884(9)   & 4882(16)     & $^{232}$Th (1 s, 3.2\%\g, 1.4\%), $^{238}$U (1 s, 3.1\%\g, 1.4\%); 
ENSDF (2007Br23) reports 6 levels in $^{137}$I populated via \be$^{‒}$ decay, with decay scheme ill-defined and \be$^{‒}$-decay feeding intensities unquantified. 
TAGS studies not reported, although potential for Pandemonium supported by large $\Delta$ energy differential.  
\be$^{‒}$-decay feeding intensity to the $^{137}$I ground state (3/2‒ $\rightarrow$ (7/2+) transition) not quantified in ENSDF.
Merits studies by means of TAGS and DGS. 
P$_n$($^{137}$Te) = 2.99(16)\%.
& y  & y     & 2   \\  \hline 

%\T 57-La-146       & (5‒)          & 9.9 (1) s             & 6405 (15)      & 1883 (15)  & 6630 (40)    &        &      &      &     \\  \hline

\T 57-La-146m      & (1‒,2‒)       & 6.08(22) s           & 6547(15)      & 1857(15)  & 6630(40)    & $^{232}$Th (10 s, 1.9\%\g, 2.9\%\be, 2.4\%), $^{233}$U (10 s, 0.4\%\g, 0.7\%\be, 0.6\%); 
Defined as metastable state in NUBASE2020. ENSDF (2016Kh07) reports 60 levels in $^{146}$Ce populated via \be$^{‒}$ decay. Decay scheme ill-defined, with population/depopulation of 45 levels identified with both $^{146}$La and $^{146m}$La \be$^{‒}$ decay.
$^{146m}$La(\be$^{‒}$) = 100\% in NUBASE2020.
TAGS studies not reported, although potential for Pandemonium supported by large $\Delta$ energy differential.  
\be$^{‒}$-decay feeding intensity to $^{146}$Ce ground state ((1‒,2‒) $\rightarrow$ 0+ transition) of 17.2(8)\% adopted in ENSDF. 
Merits studies by means of TAGS and DGS, with unambiguous separation between ground-state and isomer \be$^{‒}$ decays.   
& y  & y    & 3   \\  \hline      

\T 57-La-147       & (5/2+)      & 4.026(20) s              & 5336(14)      & 4412(14)  & 4459(17)     & $^{232}$Th (10 s, \textless{}0.4\%\g, 2.0\%\be, 1.2\%), $^{233}$U (10 s, \textless{}0.4\%\g, 0.5\%\be, \textless{}0.4\%); 
ENSDF (2009Ni02) reports 28 levels in $^{147}$Ce populated via \be$^{‒}$ decay. 
TAGS studies not reported, although potential for Pandemonium supported by large $\Delta$ energy differential.  
\be$^{‒}$-decay feeding intensity to $^{147}$Ce ground state ((5/2+) $\rightarrow$ (5/2‒) transition) of 29(8)\% adopted in ENSDF.
Merits studies by means of TAGS and DGS.
P$_n$($^{147}$La) = 0.043(4)\%. 
& y  & y    & 2  \\  \hline       

\T 58-Ce-149       & 3/2‒        & 4.94(4) s              & 4369(14)      & 3476(14)  & 6575(18)    & $^{238}$U (10 s, 1.9\%\g, 1.0\%), $^{245}$Cm (10 s, 1.8\%\g, 1.0\%); 
ENSDF (2004Si16) reports 6 levels in $^{149}$Pr populated via \be$^{‒}$ decay, with decay scheme ill-defined and \be$^{‒}$-decay feeding intensities unquantified. 
TAGS studies not reported, although potential for Pandemonium supported by large $\Delta$ energy differential. 
\be$^{‒}$-decay feeding intensity to $^{149}$Pr ground state (3/2‒ $\rightarrow$ (5/2+) transition) unquantified in ENSDF. 
Merits studies by means of TAGS and DGS. 
& y   & y    & 3   \\  \hline     

\T 59-Pr-147       & 3/2+        & 13.39(4) min            & 2703(16)      & 280(16)   & 5292.20(9)  & $^{232}$Th (1000 s, 2.3\%\g, 2.0\%\be, 2.1\%), $^{233}$U (10 s, 1.3\%\g, 1.1\%\be,   1.2\%); 
ENSDF (2009Ni02) reports 28 levels in $^{147}$Nd populated via \be$^{‒}$ decay. 
TAGS studies provide limited evidence of Pandemonium (1997Gr09), although no uncertainties assigned to \be$^{‒}$-decay feeding intensities. 
Combined \be$^{‒}$-decay feeding intensity to $^{147}$Nd ground state (3/2+ $\rightarrow$ 5/2‒ transition), 49.90-keV level (3/2+ $\rightarrow$ 7/2‒ transition) and 127.87-keV level (3/2+ $\rightarrow$ 5/2‒ transition) of \textless 10\% adopted in ENSDF, compared with 3.2(21)\% from TAGS (1996Gr20/1997Gr09).
DGS studies to be incorporated into ENSDF (2015Ru09), followed by assessment of need for further DGS measurements.
%( {\bf Combined \be~feeding in ENSDF for levels 0.0-, 49.90- and 127.87-keV is (1(5) + 2.5(25)) = 3.5(60) = 4(6)\%, or 4+6 to give \textless 10\%.  So \textless 11\% was NOT adopted in ENSDF (see statement above). Adopting GTOL value for 127.87-keV level of $\le$8\%, as quoted in ENSDF, and adding 2.5\% would arguably give you $\le$11\%, but I prefer \textless 10\%  (ALN) } )
& ?    & –     & 3  \\  \hline  

\T 60-Nd-149       & 5/2–        & 1.728(1) h             & 1688.9(25)    & 120.3(25) & 7270(6)    & $^{245}$Cm (10000 s, 1.3\%\g, 2.3\%\be, 1.7\%); 
%while $^{235}$U (10000 s, 0.4\%\g, 0.8\%\be, 0.6\%), $^{238}$Pu (10000 s, 0.5\%\g, 0.9\%\be, 0.6\%), $^{240}$Pu (10000 s, 0.7\%\g, 1.3\%\be, 0.9\%), $^{241}$Am (10000 s, 0.7\%\g, 1.3\%\be, 1.0\%), $^{243}$Cm (10000 s, 0.8\%\g, 1.4\%\be, 1.0\%); 
ENSDF (2004Si16) reports 53 levels in $^{149}$Pm populated via \be$^{‒}$ decay.
TAGS studies in reasonable agreement with evaluated data in ENSDF (1997Gr09), although no uncertainties assigned to \be$^{‒}$-decay feeding intensities.
\be$^{‒}$-decay feeding intensity to $^{149}$Pm ground state (5/2‒ $\rightarrow$ 7/2+ transition) of \textless 7\% adopted in ENSDF, compared with 2.8(44)\% from TAGS (1996Gr20/1997Gr09).
& ‒  & ‒   & ‒   \\  \hline      

\end{tabular}
\end{adjustbox}
\end{sidewaystable*}
%}

%\afterpage{
\begin{table*}
\caption{List of references cited as NSR keynumbers in Tables~\ref{tab:app1} and \ref{tab:app2}.}\label{tab:app3}
\begin{tabular}{l|p{14cm}}\hline \hline
NSR  & Reference  \\ 
keynumber &       \\ \hline
& Mass chain evaluations (ENSDF):  \\
%1997Br34 & E. Browne, Nuclear data sheets for A = 90, Nucl. Data Sheets 82, 379-546 (1997); Erratum, Nucl. Data Sheets 83, 533 (1998).  \\
2001Si26 & Balraj Singh, Nuclear data sheets for A = 130, Nucl. Data Sheets 93, 33-242 (2001).  \\
2001So16 &A.A. Sonzogni, Nuclear data sheets for A = 144, Nucl. Data Sheets 93, 599-762 (2001).  \\
%2002So05 &A.A. Sonzogni, Nuclear data sheets for A = 136, Nucl. Data Sheets 95, 837-994 (2002).  \\
%2003So13 &A.A. Sonzogni, Nuclear data sheets for A = 138, Nucl. Data Sheets 98, 515-664 (2003).  \\
2004Si16 &Balraj Singh, Nuclear data sheets for A = 149, Nucl. Data Sheets 102, 1-291 (2004).  \\
2004So32 &A.A. Sonzogni, Nuclear data sheets for A = 134, Nucl. Data Sheets 103, 1-182 (2004).  \\
%2005De52 &D. De Frenne, E. Jacobs, Nuclear data sheets for A = 105, Nucl. Data Sheets 105, 775-958 (2005).  \\
2005Kh07 &Yu. Khazov, A.A. Rodionov, S. Sakharov and Balraj Singh, Nuclear data sheets for A = 132, Nucl. Data Sheets 104, 497-790 (2005).  \\
2006Ab37 &D. Abriola, A.A. Sonzogni, Nuclear data sheets for A = 94, Nucl. Data Sheets 107, 2423-2578 (2006).  \\
2006BlENSDF&  J. Blachot, Full evaluation for A = 101, ENSDF (July 2006); \\ 
& access via:  www.nndc.bnl.gov/ensdf/ensdf/ensdf.jsp \\
2006Kh09 &Yu. Khazov, I. Mitropolsky, A. Rodionov, Nuclear data sheets for A = 131, Nucl. Data Sheets 107, 2715-2930 (2006).  \\
2007Bl18 &J. Blachot, Nuclear data sheets for A = 104, Nucl. Data Sheets 108, 2035-2172 (2007).  \\ 
2007Br23 &E. Browne, J.K. Tuli, Nuclear data sheets for A = 137, Nucl. Data Sheets 108, 2173-2318 (2007).  \\
%2007Ni07 &N. Nica, Nuclear data sheets for A = 140, Nucl. Data Sheets 108, 1287-1470 (2007).  \\
2008Ab19 &D. Abriola, A.A. Sonzogni, Nuclear data sheets A = 96, Nucl. Data Sheets 109, 2501-2655 (2008).  \\
2008Bl12 &J. Blachot, Nuclear data sheets for A = 107, Nucl. Data Sheets 109, 1383-1526 (2008).  \\
2008BlENSDF&  J. Blachot, Full evaluation for A = 108, ENSDF (July 2008); \\
& access via:  www.nndc.bnl.gov/ensdf/ensdf/ensdf.jsp \\
2008De09 &D. De Frenne, A. Negret, Nuclear data sheets for A = 106, Nucl. Data Sheets 109, 943-1102 (2008).  \\
%2008Si01 &Balraj Singh, Nuclear data sheets for A = 100, Nucl. Data Sheets 109, 297-516 (2008).  \\
2008Si03 &Balraj Singh, A.A. Rodionov, Y.L. Khazov, Nuclear data sheets for A = 135, Nucl. Data Sheets 109, 517-698 (2008).  \\
2009Ni02 & N. Nica, Nuclear data sheets for A = 147, Nucl. Data Sheets 110, 749 (2009).  \\
2009Ab25 &D. Abriola, M. Bostan, S. Erturk, M. Fadil, M. Galan, S. Juutinen, T. Kibédi, F. Kondev, A. Luca, A. Negret, N. Nica, B. Pfeiffer, Balraj Singh, A. Sonzogni, J. Timar, J. Tuli, T. Venkova, K. Zuber, Nuclear data sheets for A = 84, Nucl. Data Sheets 110, 2815-2944 (2009).  \\
2009Bl02 &J. Blachot, Nuclear data sheets for A = 111, Nucl. Data Sheets 110, 1239-1407 (2009).  \\
2009Br01 & E. Browne, J.K. Tuli, Nuclear data sheets for A = 145, Nucl. Data Sheets 110, 507-680 (2009).  \\
2009De21 &D. De Frenne, Nuclear data sheets for A = 102, Nucl. Data Sheets 110, 1745-1915 (2009).  \\
2009De29 &D. De Frenne, Nuclear data sheets for A = 103, Nucl. Data Sheets 110, 2081-2256 (2009).  \\
2010Ba39 &S.K. Basu, G. Mukherjee, A.A. Sonzogni, Nuclear data sheets for A = 95, Nucl. Data Sheets 111, 2555-2737 (2010).  \\
2010Ni03 &N. Nica, Nuclear data sheets for A = 97, Nucl. Data Sheets 111, 525-716 (2010).  \\
2011Ba40 &C.M. Baglin, Nuclear data sheets for A = 93, Nucl. Data Sheets 112, 1163-1389 (2011).  \\
2011Jo05 &T.D. Johnson, D. Symochko, M. Fadil, J.K. Tuli, Nuclear data sheets for A = 142, Nucl. Data Sheets 112, 1949-2127 (2011).  \\
2011Kh02 &Yu. Khazov, A. Rodionov, F.G. Kondev, Nuclear data sheets for A = 133, Nucl. Data Sheets 112, 855-1113 (2011).  \\
2012Ba51 &C.M. Baglin, Nuclear data sheets for A = 92, Nucl. Data Sheets 113, 2187-2389 (2012).  \\
2012Br05 &E. Browne, J.K. Tuli, Nuclear data sheets for A = 143, Nucl. Data Sheets 113, 715-908 (2012).  \\
2012Gu09 &G. Gürdal, F.G. Kondev, Nuclear data sheets for A = 110, Nucl. Data Sheets 113, 1315-1561 (2012).  \\
2012NiENSDF&  N. Nica, Full evaluation for A = 147, ENSDF (June 2012); \\
& access via:  www.nndc.bnl.gov/ensdf/ensdf/ensdf.jsp \\
2013Ba52 &C.M. Baglin, Nuclear data sheets for A = 91, Nucl. Data Sheets 114, 1293-1495 (2013).  \\
2013Si03 &Balraj Singh, Nuclear data sheets for A = 89, Nucl. Data Sheets 114, 1-208 (2013).  \\
2014Mc01 &E.A. McCutchan, A.A. Sonzogni, Nuclear data sheets for A = 88, Nucl. Data Sheets 115, 135-304 (2014).  \\
2014Ni18 &N. Nica, Nuclear data sheets for A = 141, Nucl. Data Sheets 122, 1-204 (2014).  \\
2014Ti05 &J. Timar, Z. Elekes, Balraj Singh, Nuclear data sheets for A = 129, Nucl. Data Sheets 121, 143-394 (2014).  \\
2015El04 &Z. Elekes, J. Timar, Nuclear data sheets for A = 128, Nucl. Data Sheets 129, 191-436 (2015).  \\
2015Jo11 &T.D. Johnson, W.D. Kulp, Nuclear data sheets for A = 87, Nucl. Data Sheets 129, 1-190 (2015).  \\
2015Mc01 &E.A. McCutchan, Nuclear data sheets for A = 83, Nucl. Data Sheets 125, 201-394 (2015).  \\
2015Ne01 &A. Negret, Balraj Singh, Nuclear data sheets for A = 86, Nucl. Data Sheets 124, 1-156 (2015).  \\
2015SiENSDF&  Balraj Singh, Full evaluation for A = 85, ENSDF (May 2015); \\
& access via:  www.nndc.bnl.gov/ensdf/ensdf/ensdf.jsp \\
2016Jo10 &P.K. Joshi, Balraj Singh, Sukhjeet Singh, A.K. Jain, Nuclear data sheets for A = 139, Nucl. Data Sheets 138, 1-292 (2016).  \\
2016Kh07 &Yu. Khazov, A. Rodionov, G. Shulyak, Nuclear data sheets for A = 146, Nucl. Data Sheets 136, 163-452 (2016).  \\

 \hline
\end{tabular}
\end{table*}

\begin{table*}
\addtocounter{table}{-1}
\caption{ -- Continued from previous page} 
\begin{tabular}{l|p{14cm}}\hline \hline
NSR  & Reference  \\ 
keynumber &       \\ \hline
2016Ku22 &S. Kumar, J. Chen, F.G. Kondev, Nuclear data sheets for A = 109, Nucl. Data Sheets 137, 1-286 (2016).  \\
2017Br15 &E. Browne, J.K. Tuli, Nuclear data sheets for A = 99, Nucl. Data Sheets 145, 25-340 (2017).  \\
2017Ch47  & J. Chen, Nuclear data sheets for A = 138, Nucl. Data Sheets 146, 1-386 (2017).  \\
2018Mc07 & E.A. McCutchan, Nuclear data sheets for A = 136,  Nucl. Data Sheets 152, 331-667 (2018). \\
2018Ni16 & N. Nica, Nuclear data sheets for A = 140,  Nucl. Data Sheets 154, 1-403 (2018). \\
2019La19 & S. Lalkovski, J. Timar, Z. Elekes, Nuclear data sheets for A = 105, Nucl. Data Sheets 161-162, 1-353 (2019). \\
2020Ba21 & S.K. Basu, E.A. McCutchan, Nuclear data sheets for A = 90,  Nucl. Data Sheets 165, 1-329 (2020). \\
2020Ch10 & J. Chen, Balraj Singh, Nuclear data sheets for A = 98,  Nucl. Data Sheets 164, 1-477 (2020). \\
2021Si08 & Balraj Singh, J. Chen, Nuclear data sheets for A = 100,  Nucl. Data Sheets 172, 1-542 (2021). \\
2022SiENSDF& Balraj Singh, Nuclear data sheets for A = 89, ENSDF (January 2022); \\
& access via:  www.nndc.bnl.gov/ensdf/ensdf/ensdf.jsp \\
\\
& Other references cited:   \\
1953Am08 &D.P. Ames, M.E. Bunker, L.M. Langer, B.M. Sorensen, The disintegration of Sr$^{91}$ and Y$^{91m}$, Phys. Rev. 91, 68-74 (1953).  \\
1970Ca23 &P.G. Calway, H.D. Sharma, The excited states of $^{129}$Te populated in the \be$^{-}$ decay of $^{129}$Sb, Nucl. Phys. A156, 338-352 (1970). \\
1973Ha11 &J.K. Halbig, F.K. Wohn, W.L. Talbert Jr., J.J. Eitter, J.R. McConnell, The \be$^{-}$ decay of $^{91}$Sr, Nucl. Phys. A203, 532-538 (1973).  \\
1974Ke08 &A. Kerek, P. Carlé, S. Borg, Beta decay studies in the vicinity of $^{132}$Sn: Excited states in $^{130}$Sb, $^{130}$Te and $^{132}$Te, Nucl. Phys. A224, 367-395 (1974).  \\
1976He10 &W. Herzog, N. Trautmann, R. Denig, G. Herrmann, The decay of 31 sec $^{98}$Zr, 2.9 sec $^{98}$Nb and 51 min $^{98}$Nb, Z. Phys. A276, 393-402 (1976).  \\
1977Pf01 &B. Pfeiffer, J.P. Bocquet, J.A. Pinston, \textit{et al.}, Gamma spectroscopy of some short-lived fission products with the isotope separator Lohengrin, J. Phys., Paris 38, 9-16 (1977). \\
1977Sc21 &F. Schussler, J. Blachot, E. Monnand, \textit{et al.}, The decay of $^{136}$Te (20.9 s), Z. Phys. A283, 43-49 (1977). \\
1979Ke02 &U. Keyser, H. Berg, F. Münnich, \textit{et al.}, Experimental beta-decay energies of several neutron-rich nuclides with mass number 131 $\le$ A $\le$ 146, Z. Phys. A289, 407-413 (1979). \\
1980KeZQ &U. Keyser, H. Berg, F. Münnich, \textit{et al.}, Experimental beta-decay energies of neutron-rich isotopes of I, Xe, Ce and Pr, pp 485-492 in Proc. 6th Int. Conf. Atomic Masses and Fundamental Constants, East Lansing, 1979, Editors: J.A. Nolen Jr., W. Benenson, Plenum Press, New York (1980). \\
1984Sh03 &K. Shizuma, H. Ahrens, J.P. Bocquet, \textit{et al.}, Odd neutron nuclei near A = 100: Rotational bands in $^{103}$Mo and $^{105}$Mo populated in the \be$^{-}$ decays of $^{103}$Nb and $^{105}$Nb, Z. Phys. A315, 65-75 (1984).  \\
1985Sa15 &M. Samri, G.J. Costa, G. Klotz, \textit{et al.}, The \be~decay of tellurium 135 and 137, Z. Phys. A321, 255-263 (1985). \\
1986OhZZ &H. Ohm, U. Paffrath, G. Lhersonneau, K. Sistemich, The \be-decay of $^{107}$Mo to excited states in $^{107}$Tc, Julich report JUL-Spez-344, 30 (1986). \\
1989WaZJ &W.B. Walters, Gamma rays observed in the \be$^{-}$~decay of 4.4-h~$^{129}$Sb, Department of Chemistry, University of Maryland, USA, private communication to ENSDF/NNDC (1989).  \\
1995Li13 &M. Liang, H. Ohm, B. De Sutter-Pommé, K. Sistemich, Lifetime measurements and particle-rotor calculations for $^{105}$Mo$_{63}$, Z. Phys. A351, 13-18 (1995). \\
1996Gr20& R.C. Greenwood, M.H. Putnam, K.D. Watts, Ground-state \be-branching intensities of several fission-product isotopes measured using a total absorption \g-ray spectrometer, Nucl. Instrum. Methods Phys. Res. A378, 312-320 (1996).\\
1997Gr09& R.C. Greenwood, R.G. Helmer, M.H. Putnam, K.D. Watts, Measurement of \be-decay intensity distributions of several fission-product isotopes using a total absorption \g-ray spectrometer, Nucl. Instrum. Methods Phys. Res. A390, 95-154 (1997).\\ 
2004Hu08 &R.O. Hughes, N.V. Zamfir, R.F. Casten, \textit{et al.}, Test of calculations with single-particle density dependent pairing in $^{132}$Te, Phys. Rev. C69, 051303, 1-4 (2004) \\
2004HuZX &R.O. Hughes, N.V. Zamfir, D.C. Radford, \textit{et al.}, Gamma ray spectroscopy of $^{132}$Te through beta decay of $^{132}$Sb radioactive beam, Wright Nuclear Structure Laboratory, Yale University, USA, private communication to ENSDF/NNDC (2004). \\
2005Hu08 &R.O. Hughes, N.V. Zamfir, D.C. Radford, \textit{et al.}, \g-ray spectroscopy of $^{132}$Te through \be~decay of a $^{132}$Sb radioactive beam, Phys. Rev. C71, 044311, 1-15 (2005). \\
2010Al22& A. Algora, D. Jordan, J.L. Taín, \textit{et al.}, Reactor decay heat in $^{239}$Pu: Solving the \g~discrepancy in the 4–3000-s cooling period, Phys. Rev. Lett. 105, 202501, 1-4 (2010).\\
2013Jo02& D. Jordan, A. Algora, J.L. Taín, \textit{et al.}, Total absorption study of the \be~decay of $^{102, 104, 105}$Tc,  Phys. Rev. C87, 044318, 1-14 (2013).\\
2014Fi09& A. Fijałkowska, M. Karny, K.P. Rykaczewski, \textit{et al.}, First results from the modular total absorption spectrometer at the HRIBF - includes preliminary results for the \be~decay of $^{86}$Br, Acta  Phys. Pol. B45, 545-552 (2014).\\
\hline

\end{tabular}
\end{table*}

\begin{table*}
\addtocounter{table}{-1}
\caption{ -- Continued from previous page} 
\begin{tabular}{l|p{14cm}}\hline \hline
NSR  & Reference  \\ 
keynumber &       \\ \hline
2014Fi11& A. Fijałkowska, M. Karny, K.P. Rykaczewski, \textit{et al.}, First results from the modular total absorption spectrometer at the HRIBF - includes preliminary results for the \be~decay of $^{89}$Kr and $^{139}$Xe, Nucl. Data Sheets 120, 26-29 (2014).\\
2014Kr11 &K.S. Krane, Neutron capture by $^{94,96}$Zr and the decays of $^{97}$Zr and $^{97}$Nb, Appl. Radiat. Isot. 94, 60-66 (2014). \\
2014VaZZ& E. Valencia, A. Algora, J.L. Taín, \textit{et al.}, Gamma/neutron competition above the neutron separation energy in delayed neutron emitters, INPC 2013 – Int. Nucl. Physics Conf., 2-7 June 2013, IUPAP, Firenze, Italy, Editors: S. Lunardi, P.G. Bizzeti, C. Bucci, \textit{et al.}, EPJ Web Conf. 66, 02002, 1-4 (2014).\\
2014Wo01& M. Wolińska-Cichocka, K.P. Rykaczewski, A. Fijałkowska, \textit{et al.}, Modular total absorption spectrometer at the HRIBF (ORNL, Oak Ridge), Nucl. Data Sheets 120, 22-25 (2014).\\
2014ZaZZ& A.-A. Zakari-Issoufou, A. Porta, M. Fallot, A. Algora, J.L Taín, \textit{et al.}, Results of fission products \be~ decay properties measurement performed with a total absorption spectrometer, INPC 2013 – Int. Nucl. Physics Conf., 2-7 June 2013, IUPAP, Firenze, Italy, Editors: S. Lunardi, P.G. Bizzeti, C. Bucci, \textit{et al.}, EPJ Web Conf. 66, 10019, 1-4 (2014).\\
2015Kr07& K.S. Krane, The decays of $^{109,111}$Pd and $^{111}$Ag following neutron capture by Pd, Appl. Radiat. Isot. 105, 278-289 (2015).\\ 
2015Ru09& E. Ruchowska, H. Mach, M. Kowal, J. Skalski, W.A. Płóciennik, B. Fogelberg, Search for octupole correlations in $^{147}$Nd, Phys. Rev. C92, 034328, 1-12 (2015) \\
2015Ta17& J.L. Taín, E. Valencia, A. Algora, \textit{et al.}, Enhanced \g-ray emission from neutron unbound states populated in \be~decay, Phys. Rev. Lett. 115, 062502, 1-5 (2015). \\
2015Za10& A.-A. Zakari-Issoufou, M. Fallot, A. Porta, A. Algora, \textit{et al.}, Total absorption spectroscopy study of $^{92}$Rb decay: A major contributor to reactor antineutrino spectrum shape, Phys. Rev. Lett. 115, 102503, 1-6 (2015). \\
2016Ag03& J. Agramunt, J.L. Taín, \textit{et al.}, Characterization of a neutron-beta counting system with beta-delayed neutron emitters, Nucl. Instrum. Methods Phys. Res. A807, 69-78 (2016). \\
2016Gu22& V. Guadilla, A. Algora, J.L. Taín, \textit{et al.}, First experiment with the NUSTAR/FAIR Decay Total Absorption \g-ray Spectrometer (DTAS) at the IGISOL IV facility, Nucl. Instrum. Methods Phys. Res. B376, 334-337 (2016). \\
2016Ka65&  M. Karny, K.P. Rykaczewski, A. Fijałkowska, B.C. Rasco, M. Wolińska-Cichocka, R.K. Grzywacz, K.C. Goetz, D. Miller, E.F. Zganjar, Modular total absorption spectrometer, Nucl. Instrum. Methods Phys. Res. A836, 83-90 (2016). \\
2016ORNL& K.P. Rykaczewski, Physics Division, ORNL, Nuclear data on decay heat and reactor antineutrino spectra in fission fragments from total absorption spectroscopy, presentation at a Workshop for Nuclear Data Needs and Capabilities for Basic Science, 10-11 August 2016, University of Notre Dame, South Bend, Indiana, USA; contribution prepared for White Paper on Nuclear Data Needs and Capabilities for Basic Science, as commissioned by the US DOE Office of Science/Nuclear Physics to guide future nuclear data activities within the US nuclear physics research community.\\
2016Ra25& B.C. Rasco, M. Wolinska-Cichocka, A. Fijalkowska, K.P. Rykaczewski, \textit{et al.}, Decays of the three top contributors to the reactor $\overline{\nu}_e$ high-energy spectrum, $^{92}$Rb, $^{96gs}$Y, and $^{142}$Cs, studied with total absorption spectroscopy, Phys. Rev. Lett. 117, 092501, 1-6 (2016). \\
2016Ur04& W. Urban, K. Sieja, T. Materna, M. Czerwinski, T. Rzaca-Urban, A. Blanc, M. Jentschel, P. Mutti, U. Köster, T. Soldner, G. de France, G.S. Simpson, C.A. Ur, C. Bernards, C. Fransen, J. Jolie, J.-M. Regis, T. Thomas, N. Warr, Low-spin structure of $^{86}_{35}$Br$_{51}$ and $^{86}_{36}$Kr$_{50}$ nuclei: The role of the g$_{7/2}$  neutron orbital, Phys. Rev. C94, 044328, 1-19 (2016). \\
2017AgZZ& J. Agramunt, J.L. Taín, F. Albiol, A. Algora, \textit{et al.}, New accurate measurements of neutron emission probabilities for relevant fission products, EPJ Web Conf.  146, 01004, 1-4 (2017). \\
2017AlZY& A. Algora, S. Rice, V. Guadilla, J.L. Tain, \textit{et al.}, Total absorption studies of high priority decays for reactor applications: $^{86}$Br and $^{91}$Rb, EPJ Web Conf.  146, 10001, 1-4 (2017). \\
2017FaZY& M. Fallot, A. Porta, L. Le Meur, \textit{et al.}, Total absorption spectroscopy of fission fragments relevant for reactor antineutrino spectra, EPJ Web Conf.  146, 10002, 1-5 (2017). \\  
2017Fi06& A. Fijalkowska, M. Karny, K.P. Rykaczewski, \textit{et al.}, Impact of modular total absorption spectrometer measurements of \be~decay of fission products on the decay heat and reactor $\overline{\nu}_e$ flux calculation, Phys. Rev. Lett. 119, 052503, 1-6 (2017). \\
2017Gu07& V. Guadilla, A. Algora, J.L. Tain, \textit{et al.}, Study of the \be~decay of fission products with the DTAS detector, Acta  Phys. Pol. B48, 529-532 (2017). \\
2017GuZV& V. Guadilla, A. Algora, J.L. Taín, \textit{et al.}, TAGS measurements of $^{100}$Nb ground and isomeric states and $^{140}$Cs for neutrino physics with the new DTAS detector, EPJ Web Conf. 146, 10010, 1-4 (2017). \\
2017Ra06& B.C. Rasco, A. Fijalkowska, K.P. Rykaczewski, \textit{et al.}, \be~decays of $^{92}$Rb, $^{96gs}$Y, and $^{142}$Cs measured with the modular total absorption spectrometer and the influence of \g~multiplicity on total absorption spectrometry measurements, Acta  Phys. Pol. B48, 507-515 (2017). \\
2017Ra10& B.C. Rasco, K.P. Rykaczewski, A. Fijalkowska, \textit{et al.}, Complete \be-decay pattern for the high-priority decay-heat isotopes $^{137}$I and $^{137}$Xe determined using total absorption spectroscopy, Phys. Rev. C95, 054328, 1-9 (2017). \\
2017Ri08& S. Rice, A. Algora, J.L. Tain, \textit{et al.}, Total absorption spectroscopy study of the \be~decay of $^{86}$Br and $^{91}$Rb, Phys. Rev. C96, 014320, 1-10 (2017). \\
\hline
\end{tabular}
\end{table*}

\begin{table*}
\addtocounter{table}{-1}
\caption{ -- Continued from previous page} 
\begin{tabular}{l|p{14cm}}\hline \hline
NSR  & Reference  \\ 
keynumber &       \\ \hline
2017TaZS& J.L. Tain, V. Guadilla, E. Valencia, A. Algora, \textit{et al.}, r process (n,\g) rate constraints from the \g~emission of neutron unbound states in \be-decay, Proc. 14th Int. Symp. on Nuclei in the Cosmos (NIC2016), JPS Conf. Proc.  14, 010607, 1-5 (2017). \\
2017TaZZ& J.L. Tain, V. Guadilla, E. Valencia, A. Algora, \textit{et al.}, Strong \g-ray emission from neutron unbound states populated in \be-decay and impact on (n,\g) cross-section estimates, EPJ Web Conf.  146, 01002, 1-6 (2017). \\
2017Va04& E. Valencia, J.L. Tain, A. Algora, \textit{et al.}, Total absorption \g-ray spectroscopy of the \be-delayed neutron emitters $^{87}$Br, $^{88}$Br and $^{94}$Rb, Phys. Rev. C95, 024320, 1-18 (2017). \\
2017WoZZ& M. Wolińska-Cichocka, B.C. Rasco, K.P. Rykaczewski, \textit{et al.}, Beta-strength and antineutrino spectra from total absorption spectroscopy of a decay chain $^{142}$Cs $\rightarrow$ $^{142}$Ba $\rightarrow$ $^{142}$La, EPJ Web Conf. 146, 10005, 1-4 (2017). \\
2018Fi03&  A. Fijalkowska, Reactor $\overline{\nu}_e$ properties from \be-decay study of fission fragments by total absorption spectroscopy, Acta Phys. Pol. B49, 399-408 (2018). \\
2019Gu03& V. Guadilla, A. Algora, J.L. Tain, \textit{et al.}, Large impact of the decay of niobium isomers on the reactor $\overline{\nu}_e$ summation calculations, Phys. Rev. Lett. 122, 042502, 1-6 (2019). \\
2019Gu20& V. Guadilla, A. Algora, J.L. Tain, \textit{et al.}, Total absorption \g-ray spectroscopy of niobium isomers, Phys. Rev. C100, 024311, 1-15 (2019).  \\
2019Gu29& V. Guadilla, J.L. Tain, A. Algora, \textit{et al.}, Total absorption \g-ray spectroscopy of the \be-delayed neutron emitters $^{137}$I and $^{95}$Rb, Phys. Rev. C100, 044305, 1-17 (2019).  \\
2019Wi11& J. Wisniewski, W. Urban, M. Czerwiński, J. Kurpeta, A. Plochocki, M. Pomorski, T. Rzaca-Urban, K. Sieja, L. Canete, T. Eronen, S. Geldhof, A. Jokinen, A. Kankainen, I.D. Moore, D.A. Nesterenko, H. Penttilä, I. Pohjalainen, S. Rinta-Antila, A. de Roubin, M. Vilén, Excited states in $^{87}$Br populated in \be~decay of $^{87}$Se, Phys. Rev. C100, 054331, 1-13 (2019).  \\ 
2020Gu26& V. Guadilla, J.L. Taín, A. Algora, \textit{et al.}, Determination of \be-decay ground state feeding of nuclei of importance for reactor applications, Phys. Rev. C102, 064304, 1-12 (2020).  \\
2021Ca11& M.A. Cardona, D. Hojman, B. Roussière, I. Deloncle, N. Barré-Boscher, M.C. Mhamed, E. Cottereau, B.I. Dimitrov, G.Tz. Gavrilov, A. Gottardo, C. Lau, S. Roccia, S. Tusseau-Nenez, D. Verney, M.S. Yavahchova, $^{145}$Ba \be{-} decay: Excited states and half-lives in neutron-rich $^{145}$La, Phys. Rev. C103, 034308, 1-14 (2021). \\
2022Gu09& V. Guadilla, L. Le Meur, M. Fallot, \textit{et al.}, Total absorption \g-ray spectroscopy of the \be~decays of $^{96gs,m}$Y, Phys. Rev. C106, 014306, 1-14 (2022).  \\ 
2022Ra11& B.C. Rasco, K.P. Rykaczewski, A. Fijałkowska, \textit{et al.}, Deciphering $^{98}$Nb \be-decay with the Modular Total Absorption Spectrometer at ORNL, Phys. Rev. C105, 064301, 1-10 (2022). \\
2022Sh20& P. Shuai, B.C. Rasco, K.P. Rykaczewski, \textit{et al.}, Determination of \be-decay feeding patterns of $^{88}$Rb and $^{88}$Kr using the Modular Total Absorption Spectrometer at ORNL HRIBF, Phys. Rev. C105, 054312, 1-13 (2022). \\
\hline \hline
\end{tabular}
\end{table*}

%}

\clearpage

\bibliographystyle{spphys}
\bibliography{references}% Produces the bibliography 

\begin{thebibliography}{100}
\providecommand{\url}[1]{{#1}}
\providecommand{\urlprefix}{URL }
\expandafter\ifx\csname urlstyle\endcsname\relax
  \providecommand{\doi}[1]{DOI \discretionary{}{}{}#1}\else
  \providecommand{\doi}{DOI \discretionary{}{}{}\begingroup
  \urlstyle{rm}\Url}\fi

\bibitem{Tobias1980}
A.~Tobias, Prog. Nucl. Energy \textbf{5}, 1 (1980)

\bibitem{Nichols2002}
A.L. Nichols, {Nuclear data requirements for decay heat calculations}.
\newblock Tech. rep., Workshop on Nuclear Reaction Data and Nuclear Reactors:
  Physics, Design and Safety, ICTP Lectures Notes, Vol. 20, pp. 65-195,
  editors: M. Herman and N. Paver, 25 February - 28 March 2002, The Abdus Salam
  International Centre for Theoretical Physics (ICTP). ICTP Publ., Trieste,
  Italy IBSN 92-95003-30-6 (2005)

\bibitem{Tobias1989}
A.~Tobias, {Derivation of decay heat benchmarks for U-235 and Pu-239 by a least
  squares fit to measured data}.
\newblock Tech. rep., Central Electricity Generating Board report RD/B/6210/R89
  (1989)

\bibitem{Hardy1977}
J.C. Hardy, L.C. Carrez, B.~Jonson, P.G. Hansen, Phys. Lett. B \textbf{71}, 307
  (1977)

\bibitem{Hu2000}
Z.~Hu, L.~Batist, J.~Agramunt, A.~Algora, B.A. Brown, D.~Cano-Ott, R.~Collatz,
  A.~Gadea, M.~Gierlik, M.~G\'orska, H.~Grawe, M.~Hellstr\"om, Z.~Janas,
  M.~Karny, R.~Kirchner, F.~Moroz, A.~P\l{}ochocki, M.~Rejmund, E.~Roeckl,
  B.~Rubio, M.~Shibata, J.~Szerypo, J.L. Taín, V.~Wittmann, Phys. Rev. C
  \textbf{62}, 064315 (2000).
\newblock \doi{10.1103/PhysRevC.62.064315}.
\newblock \urlprefix\url{https://link.aps.org/doi/10.1103/PhysRevC.62.064315}

\bibitem{Algora2003}
A.~Algora, B.~Rubio, D.~Cano-Ott, J.L. Ta\'{\i}n, A.~Gadea, J.~Agramunt,
  M.~Gierlik, M.~Karny, Z.~Janas, A.~P\l{}ochocki, K.~Rykaczewski, J.~Szerypo,
  R.~Collatz, J.~Gerl, M.~G\'orska, H.~Grawe, M.~Hellstr\"om, Z.~Hu,
  R.~Kirchner, M.~Rejmund, E.~Roeckl, M.~Shibata, L.~Batist, J.~Blomqvist,
  Phys. Rev. C \textbf{68}, 034301 (2003).
\newblock \doi{10.1103/PhysRevC.68.034301}.
\newblock \urlprefix\url{https://link.aps.org/doi/10.1103/PhysRevC.68.034301}

\bibitem{Greenwood1996}
R.C. Greenwood, M.H. Putnam, K.D. Watts, Nucl. Instrum. Methods. Phys. Res. A
  \textbf{378}, 312 (1996)

\bibitem{Greenwood1997}
R.C. Greenwood, R.G. Helmer, M.H. Putman, K.D. Watts, Nucl. Instrum. Methods
  Phys. Res. A \textbf{390}, 95  (1997)

\bibitem{Algora2010}
A.~Algora, D.~Jordan, J.L. Taín, B.~Rubio, J.~Agramunt, A.B. Perez-Cerdán,
  F.~Molina, L.~Caballero, E.~Nácher, A.~Krasznahorkay, M.D. Hunyadi,
  J.~Gulyás, A.~Vitéz, M.~Csatlós, L.~Csige, J.~Äystö, H.~Penttilä, I.D.
  Moore, T.~Eronen, A.~Jokinen, A.~Nieminen, J.~Hakala, P.~Karvonen,
  A.~Kankainen, A.~Saastamoinen, J.~Rissanen, T.~Kessler, C.~Weber,
  J.~Ronkainen, S.~Rahaman, {V.-V. Elomaa}, S.~Rinta-Antila, U.~Hager,
  T.~Sonoda, K.~Burkard, W.~Hüller, L.~Batist, W.~Gelletly, A.L. Nichols,
  T.~Yoshida, A.A. Sonzogni, K.~Peräjärvi, Phys. Rev. Lett. \textbf{105},
  202501 (2010)

\bibitem{Hill2011}
R.N. Hill, {Role of decay heat in advanced fuel cycles}.
\newblock Tech. rep., in Report of Workshop on Decay Spectroscopy at CARIBU:
  Advanced Fuel Cycle Applications, Nuclear Structure and Astrophysics, 14-16
  April 2011, Argonne National Laboratory, USA, Editors: M.P. Carpenter, P.
  Chowdhury, J.A. Clark, F.G. Kondev, C.J. Lister, A.L. Nichols, D. Seweryniak,
  ANL report ANL/NDM-168, Nuclear Data and Measurements Series. Also available
  online: www.ne.anl.gov/capabilities/nd/reports/ANLNDM-168.pdf (2011)

\bibitem{Yoshida2007}
T.~Yoshida, A.L. Nichols, M.A. Kellett, O.~Bersillon, H.~Henriksson,
  R.~Jacqmin, B.~Roque, J.~Katakura, K.~Oyamatsu, T.~Tachibana, A.~Algora,
  B.~Rubio, J.L. Taín, C.J. Dean, W.~Gelletly, R.W. Mills, I.C. Gauld,
  P.~Möller, A.A. Sonzogni, {Assessment of fission product decay data for
  decay heat calculations}.
\newblock Tech. rep., A report by the Working Party on International Evaluation
  Co-operation of the NEA Nuclear Science Committee, Vol. 25, NEA/WPEC-25,
  OECD/NEA, Paris, ISBN 978-92-64-99034-0. Also available online:
  www.oecd-nea.org/science/wpec/volume25/volume25.pdf (2007)

\bibitem{Nichols2009}
A.L. Nichols, C.~Nordborg, {Summary report of consultants' meeting on Total
  Absorption Gamma-ray Spectroscopy (TAGS), current status of measurement
  programmes for decay heat calculations and other applications}.
\newblock Tech. rep., IAEA Headquarters, 27-28 January 2009, IAEA report
  INDC(NDS)-0551, IAEA, Vienna, Austria, February 2009.
\newblock Also available online:
  www-nds.iaea.org/publications/indc/indc-nds-0551.pdf

\bibitem{Gupta2010}
M.~Gupta, M.A. Kellett, A.L. Nichols, O.~Bersillon, {Assessment of fission
  product decay data requirements for Th/U fuel}.
\newblock Tech. rep., IAEA report INDC(NDS)-0577, IAEA, Vienna, Austria, May
  2010.
\newblock Also available online:
  www-nds.iaea.org/publications/indc/indc-nds-0577.pdf

\bibitem{Dimitriou2015}
P.~Dimitriou, A.L. Nichols, {Summary report of consultants' meeting on total
  absorption gamma-ray spectroscopy for decay heat calculations and other
  applications}.
\newblock Tech. rep., IAEA Headquarters, 15-17 December 2014, IAEA report
  INDC(NDS)-0676, IAEA, Vienna, Austria, February 2015.
\newblock Also available online:
  www-nds.iaea.org/publications/indc/indc-nds-0676.pdf

\bibitem{Sonzogni2015}
{A.A. Sonzogni, T.D. Johnson, E.A. McCutchan}, Phys. Rev. C \textbf{91}, 011301
  (2015)

\bibitem{Sonzogni2016}
{A.A. Sonzogni, E.A. McCutchan, T.D. Johnson, P. Dimitriou}, Phys. Rev. Lett
  \textbf{116}, 132502 (2016)

\bibitem{Sonzogni2017}
{A.A. Sonzogni, E.A. McCutchan, A.C. Hayes}, Phys. Rev. Lett. \textbf{119},
  112501 (2017)

\bibitem{Sonzogni2018}
{A.A. Sonzogni, M. Nino, E.A. McCutchan}, Phys. Rev. C \textbf{98}, 014323
  (2018)

\bibitem{Hayes2018}
{A.C. Hayes, G. Jungman, E.A. McCutchan, A.A. Sonzogni, G.T. Garvey, X.B.
  Wang}, Phys. Rev. Lett. \textbf{120}, 022503 (2018)

\bibitem{Estienne2019}
M.~Estienne, M.~Fallot, A.~Algora, {J. Briz-Monago}, {V.M. Bui}, S.~Cormon,
  W.~Gelletly, L.~Giot, V.~Guadilla, D.~Jordan, {L. Le Meur}, A.~Porta,
  S.~Rice, B.~Rubio, {J.L. Taín}, E.~Valencia, {A.-A. Zakari-Issoufou},
  Phys.Rev.Lett. \textbf{123}, 022502 (2019)

\bibitem{Duke1970}
C.L. Duke, P.G. Hansen, O.B. Nielsen, {G. Rudstam and ISOLDE collaboration
  (CERN)}, Nucl. Phys. A \textbf{151}, 609 (1970)

\bibitem{Johansen1973}
K.H. Johansen, K.B. Nielsen, G.~Rudstam, Nucl. Phys. A \textbf{203}, 481 (1973)

\bibitem{Rubio2005}
B.~Rubio, W.~Gelletly, E.~Nácher, A.~Algora, J.L. Taín, A.~Perez,
  L.~Caballero, J. Phys. G: Nucl. Part. Phys. \textbf{31}, 1477 (2005)

\bibitem{Rubio2017}
B.~Rubio, W.~Gelletly, A.~Algora, E.~Nácher, J.L. Taín, J. Phys. G: Nucl.
  Part. Phys. \textbf{44}, 084004 (2017)

\bibitem{Algora2018}
A.~Algora, B.~Rubio, J.L. Taín, Nucl. Phys. News \textbf{28}, 12 (2018)

\bibitem{Greenwood1992}
R.C. Greenwood, R.G. Helmer, M.A. Lee, M.H. Putnam, M.A. Oates, D.A.
  Struttmann, K.D. Watts, Nucl. Instrum. Methods Phys. Res. A \textbf{314}, 514
   (1992)

\bibitem{Karny1997}
M.~Karny, {J.M. Nitschke}, {L.F. Archambault}, K.~Burkard, D.~Cano-Ott,
  M.~Hellström, W.~Hüller, R.~Kirchner, S.~Lewandowski, E.~Roeckl, A.~Sulik,
  Int. Conf. Electromagnetic Isotope Separators and Techniques Related to their
  Applications, Nucl. Instrum. Methods Phys. Res. B \textbf{126}, 411 (1997).
\newblock \doi{https://doi.org/10.1016/S0168-583X(96)01007-5}.
\newblock
  \urlprefix\url{https://www.sciencedirect.com/science/article/pii/S0168583X96010075}

\bibitem{Algora2021}
A.~Algora, {J. L. Taín}, B.~Rubio, M.~Fallot, W.~Gelletly, Eur. Phys. J. A
  \textbf{57}, 85 (2021)

\bibitem{Rasco2015b}
B.C. Rasco, A.~Fijałkowska, M.~Karny, K.P. Rykaczewski, M.~Wolińska-Cichocka,
  K.C. Goetz, Nucl. Instrum. Methods Phys. Res. A \textbf{788}, 137 (2015)

\bibitem{Algora2014}
A.~Algora, E.~Valencia, J.L. Taín, M.D. Jordan, J.~Agramunt, B.~Rubio,
  E.~Estévez, F.~Molina, A.~Montaner, V.~Guadilla, M.~Fallot, A.~Porta, {A.-A.
  Zakari-Issoufou}, V.M. Bui, S.~Rice, {Zs. Podolyák}, P.H. Regan,
  W.~Gelletly, M.~Bowry, P.~Mason, G.F. Farrelly, J.~Rissanen, T.~Eronen,
  I.~Moore, H.~Penttilä, J.~Äystö, {V.-V. Elomaa}, J.~Hakala, A.~Jokinen,
  V.~Kolkinen, M.~Reponen, V.~Sonnenschein, D.~Cano-Ott, T.~Martínez,
  E.~Mendoza, A.R. Garcia, M.B. Gómez-Hornillos, V.~Gorlychev,
  R.~Caballero-Folch, F.G. Kondev, A.A. Sonzogni, Nucl. Data Sheets
  \textbf{120}, 12 (2014)

\bibitem{Valencia2014}
E.~Valencia, A.~Algora, J.L. Taín, S.~Rice, J.~Agramunt, {A.-A.
  Zakari-Issoufou}, J.~Äystö, M.~Bowry, V.M. Bui, R.~Caballero-Folch,
  D.~Cano-Ott, {V.-V. Elomaa}, T.~Eronen, E.~Estévez, G.F. Farrelly,
  M.~Fallot, A.~Garcia, W.~Gelletly, M.B. Gómez-Hornillos, V.~Gorlychev,
  J.~Hakala, A.~Jokinen, M.D. Jordan, A.~Kankainen, F.G. Kondev, T.~Martínez,
  E.~Mendoza, F.~Molina, I.~Moore, A.~Perez, {Zs. Podolyák}, H.~Penttilä,
  A.~Porta, P.H. Regan, J.~Rissanen, B.~Rubio, C.~Weber{ and IGISOL staff}, EPJ
  Web Conf. Int. Nucl. Phys. (INPC2013) \textbf{66}, 02002 (2014).
\newblock Editors: S. Lunardi, P.G. Bizzeti, C. Bucci, M. Chiari, A. Dainese,
  P. Di Nezza, R. Menegazza, A. Nannini, C. Signorini, J.J. Valiente-Dobon, 2-7
  June 2013, IUPAP, Firenze, Italy

\bibitem{Valencia2017}
E.~Valencia, J.L. Taín, A.~Algora, J.~Agramunt, E.~Estévez, M.D. Jordan,
  B.~Rubio, S.~Rice, P.~Regan, W.~Gelletly, {Zs. Podolyák}, M.~Bowry,
  P.~Mason, G.F. Farrelly, {A.-A. Zakari-Issoufou}, M.~Fallot, A.~Porta, V.M.
  Bui, J.~Rissanen, T.~Eronen, I.~Moore, H.~Penttilä, J.~Äystö, {V.-V.
  Elomaa}, J.~Hakala, A.~Jokinen, V.S. Kolhinen, M.~Reponen, V.~Sonnenschein,
  D.~Cano-Ott, A.R. Garcia, T.~Martínez, E.~Mendoza, R.~Caballero-Folch,
  B.~Gómez-Hornillos, V.~Gorlichev, F.G. Kondev, A.A. Sonzogni, L.~Batist,
  Phys. Rev. C \textbf{95}, 024320 (2017).
\newblock \doi{10.1103/PhysRevC.95.024320}.
\newblock \urlprefix\url{https://link.aps.org/doi/10.1103/PhysRevC.95.024320}

\bibitem{Tain2015}
{J.L. Taín}, A.~Algora, J.~Agramunt, V.~Guadilla, {M.D. Jordan}, {A.
  Montaner-Piza}, B.~Rubio, E.~Valencia, {D. Cano-Ott}, W.~Gelletly,
  T.~Martinez, E.~Mendoza, {Zs. Podolyak}, P.~Regan, J.~Simpson, {A.J. Smith},
  J.~Strachan, Nucl. Instrum. Methods Phys. Res. A \textbf{803}, 36 (2015).
\newblock \doi{https://doi.org/10.1016/j.nima.2015.09.009}.
\newblock
  \urlprefix\url{https://www.sciencedirect.com/science/article/pii/S016890021501058X}

\bibitem{Guadilla2016}
V.~Guadilla, A.~Algora, J.L. Taín, J.~Agramunt, J.~Äystö, J.A. Briz,
  D.~Cano-Ott, A.~Cucoanes, T.~Eronen, M.~Estienne, M.~Fallot, L.M. Fraile,
  E.~Ganioğlu, W.~Gelletly, D.~Gorelov, J.~Hakala, A.~Jokinen, D.~Jordan,
  A.~Kankainen, V.~Kolhinen, J.~Koponen, M.~Lebois, T.~Martínez,
  M.~Monserrate, {A. Montaner-Piza}, I.~Moore, E.~Nácher, S.~Orrigo,
  H.~Penttilä, {Zs. Podolyák}, I.~Pohjalainen, A.~Porta, P.~Regan,
  J.~Reinikainen, M.~Reponen, S.~Rinta-Antila, B.~Rubio, K.~Rytkönen,
  T.~Shiba, V.~Sonnenschein, {A. A. Sonzogni}, E.~Valencia, V.~Vedia, A.~Voss,
  J.N. Wilson, {A.-A. Zakari-Issoufou}, Nucl. Instrum. Methods Phys. Res. B
  \textbf{376}, 334 (2016)

\bibitem{Guadilla2017a}
V.~Guadilla, A.~Algora, J.L. Taín, J.~Agramunt, J.~Äystö, J.A. Briz,
  A.~Cucoanes, T.~Eronen, M.~Estienne, M.~Fallot, L.M. Fraile, E.~Ganioğlu,
  W.~Gelletly, D.~Gorelov, J.~Hakala, A.~Jokinen, D.~Jordan, A.~Kankainen,
  V.~Kolhinen, J.~Koponen, M.~Lebois, T.~Martínez, M.~Monserrate, {A.
  Montaner-Pizá}, I.~Moore, E.~Nácher, {S. E. A. Orrigo}, H.~Penttilä,
  I.~Pohjalainen, A.~Porta, J.~Reinikainen, M.~Reponen, {S. Rinta-Antila},
  B.~Rubio, K.~Rytkönen, T.~Shiba, V.~Sonnenschein, {A. A. Sonzogni},
  E.~Valencia, V.~Vedia, A.~Voss, J.N. Wilson, {A.-A. Zakari-Issoufou}, EPJ Web
  Conf. ND2016 Int. Conf. Nucl. Data for Science and Technology \textbf{146},
  10010 (2017).
\newblock Editors: A.J.M. Plompen, {F.-J. Hambsch}, P. Schillebeeckx, W.
  Mondelaers, J. Heyse, S. Kopecky, P. Siegler, S. Oberstedt, 11-16 September
  2016, Bruges, Belgium

\bibitem{Fijalkowska2014b}
A.~Fijałkowska, M.~Karny, {K. P. Rykaczewski}, M.~Wolińska-Cichocka,
  R.~Grzywacz, {C. J. Gross}, {J. W. Johnson}, {B. C. Rasco}, {E. F. Zganjar},
  {D. W. Stracener}, C.~Jost, {K. C. Goetz}, R.~Goans, E.~Spejewski,
  L.~Cartegni, M.~Madurga, K.~Miernik, D.~Miller, {S. W. Padgett}, {S. V.
  Paulauskas}, {M. Al-Shudifat}, {J. H. Hamilton}, {A. V. Ramayya}, Acta Phys.
  Pol. B \textbf{45}, 545 (2014).
\newblock Includes preliminary results for the \be~decay of $^{86}$Br.

\bibitem{Wolinska2014}
{M. Wolińska-Cichocka, K.P. Rykaczewski, A. Fijałkowska, M. Karny, R.K.
  Grzywacz, C.J. Gross, J.W. Johnson, B. C. Rasco, E.F. Zganjar}, Nucl. Data
  Sheets \textbf{120}, 22 (2014)

\bibitem{Karny2016}
M.~Karny, K.P. Rykaczewski, A.~Fija{\l}kowska, B.C. Rasco,
  M.~Woli\'{n}ska-Cichocka, R.K. Grzywacz, K.C. Goetz, D.~Miller, E.F. Zganjar,
  Nucl. Instrum. Methods Phys. Res. A \textbf{836}, 83  (2016).
\newblock \doi{http://dx.doi.org/10.1016/j.nima.2016.08.046}.
\newblock
  \urlprefix\url{http://www.sciencedirect.com/science/article/pii/S0168900216308646}

\bibitem{Rasco2016}
B.C. Rasco, M.~Wolińska-Cichocka, A.~Fijałkowska, K.P. Rykaczewski, M.~Karny,
  R.K. Grzywacz, K.C. Goetz, C.J. Gross, D.W. Stracener, E.F. Zganjar, J.C.
  Batchelder, J.C. Blackmon, N.T. Brewer, {Sh. Go}, B.~Heffron, T.~King, J.T.
  Matta, K.~Miernik, C.D. Nesaraja, S.V. Paulauskas, M.M. Rajabali, E.H. Wang,
  J.A. Winger, Y.~Xiao, C.J. Zachary, Phys. Rev. Lett. \textbf{117}, 092501
  (2016).
\newblock \doi{10.1103/PhysRevLett.117.092501}.
\newblock
  \urlprefix\url{http://link.aps.org/doi/10.1103/PhysRevLett.117.092501}

\bibitem{Rasco2017a}
B.C. Rasco, K.P. Rykaczewski, A.~Fijałkowska, M.~Karny, M.~Wolińska-Cichocka,
  R.K. Grzywacz, C.J. Gross, D.W. Stracener, E.F. Zganjar, J.C. Blackmon, N.T.
  Brewer, K.C. Goetz, J.W. Johnson, C.U. Jost, J.H. Hamilton, K.~Miernik,
  M.~Madurga, D.~Miller, S.~Padgett, S.V. Paulauskas, A.V. Ramayya, E.H.
  Spejewski, Phys. Rev. C \textbf{95}, 054328 (2017).
\newblock \doi{10.1103/PhysRevC.95.054328}.
\newblock \urlprefix\url{https://link.aps.org/doi/10.1103/PhysRevC.95.054328}

\bibitem{Shuai2022}
P.~Shuai, B.C. Rasco, K.P. Rykaczewski, A.~Fija\l{}kowska, M.~Karny,
  M.~Woli\ifmmode~\acute{n}\else\'{n}\fi{}ska{-}Cichocka, R.K. Grzywacz, C.J.
  Gross, D.W. Stracener, E.F. Zganjar, J.C. Batchelder, J.C. Blackmon, N.T.
  Brewer, S.~Go, M.~Cooper, K.C. Goetz, J.W. Johnson, C.U. Jost, T.T. King,
  J.T. Matta, J.H. Hamilton, A.~Laminack, K.~Miernik, M.~Madurga, D.~Miller,
  C.D. Nesaraja, S.~Padgett, S.V. Paulauskas, M.M. Rajabali, T.~Ruland,
  M.~Stepaniuk, E.H. Wang, J.A. Winger, Phys. Rev. C \textbf{105}, 054312
  (2022).
\newblock \doi{10.1103/PhysRevC.105.054312}.
\newblock \urlprefix\url{https://link.aps.org/doi/10.1103/PhysRevC.105.054312}

\bibitem{Rasco2022}
B.C. Rasco, K.P. Rykaczewski, A.~Fija\l{}kowska, M.~Karny,
  M.~Woli\ifmmode~\acute{n}\else\'{n}\fi{}ska{-}Cichocka, R.K. Grzywacz, D.W.
  Stracener, E.F. Zganjar, J.C. Batchelder, J.C. Blackmon, N.T. Brewer, M.P.
  Cooper, K.C. Goetz, J.W. Johnson, T.~King, A.~Laminack, J.T. Matta,
  K.~Miernik, M.~Madurga, D.~Miller, M.M. Rajabali, T.~Ruland, P.~Shuai,
  M.~Stepaniuk, J.~Winger, Phys. Rev. C \textbf{105}, 064301 (2022).
\newblock \doi{10.1103/PhysRevC.105.064301}.
\newblock \urlprefix\url{https://link.aps.org/doi/10.1103/PhysRevC.105.064301}

\bibitem{Laminack2022}
{A. Laminack, B.C. Rasco, K.P. Rykaczewski, J.C. Blackmon, N.T. Brewer, J.A.
  Clark, M.T. Cooper, A. Fija\l{}kowska, R.K. Grzywacz, J. Heideman, M. Karny,
  T.T. King, R.J. Lorek, S. Neupane, M.M. Rajabali, T. Ruland, G. Savard, P.
  Shuai, K. Siegl, M. Singh, M. Stepaniuk, D.W. Stracener, M. Woli\ifmmode
  \acute{n}\else\'{n}\fi{}ska{-}Cichocka, S. Zhu}, submitted to Phys. Rev. C
  (2022)

\bibitem{Simon2013}
A.~Simon, S.J. Quinn, A.~Spyrou, A.~Battaglia, I.~Beskin, A.~Best, B.~Bucher,
  M.~Couder, P.A. DeYoung, X.~Fang, J.~G{\"o}rres, A.~Kontos, Q.~Li, S.N.
  Liddick, A.~Long, S.~Lyons, K.~Padmanabhan, J.~Peace, A.~Roberts,
  D.~Robertson, K.~Smith, M.K. Smith, E.~Stech, B.~Stefanek, W.P. Tan, X.D.
  Tang, M.~Wiescher, Nucl. Instrum. Methods Phys. Res. A \textbf{703}, 16
  (2013).
\newblock \doi{http://dx.doi.org/10.1016/j.nima.2012.11.045}.
\newblock
  \urlprefix\url{http://www.sciencedirect.com/science/article/pii/S0168900212013824}

\bibitem{Spyrou2014}
A.~Spyrou, S.N. Liddick, A.C. Larsen, M.~Guttormsen, K.~Cooper, A.C. Dombos,
  D.J. Morrissey, F.~Naqvi, G.~Perdikakis, S.J. Quinn, T.~Renstr\o{}m, J.A.
  Rodriguez, A.~Simon, C.S. Sumithrarachchi, R.G.T. Zegers, Phys. Rev. Lett.
  \textbf{113}, 232502 (2014).
\newblock \doi{10.1103/PhysRevLett.113.232502}.
\newblock
  \urlprefix\url{https://link.aps.org/doi/10.1103/PhysRevLett.113.232502}

\bibitem{Spyrou2016}
A.~Spyrou, S.N. Liddick, F.~Naqvi, B.P. Crider, A.C. Dombos, D.L. Bleuel, B.A.
  Brown, A.~Couture, L.~Crespo~Campo, M.~Guttormsen, A.C. Larsen, R.~Lewis,
  P.~M\"oller, S.~Mosby, M.R. Mumpower, G.~Perdikakis, C.J. Prokop,
  T.~Renstr\o{}m, S.~Siem, S.J. Quinn, S.~Valenta, Phys. Rev. Lett.
  \textbf{117}, 142701 (2016).
\newblock \doi{10.1103/PhysRevLett.117.142701}.
\newblock
  \urlprefix\url{https://link.aps.org/doi/10.1103/PhysRevLett.117.142701}

\bibitem{Gombas2021}
J.~Gombas, P.A. DeYoung, A.~Spyrou, A.C. Dombos, A.~Algora, T.~Baumann,
  B.~Crider, J.~Engel, T.~Ginter, E.~Kwan, S.N. Liddick, S.~Lyons, F.~Naqvi,
  E.M. Ney, J.~Pereira, C.~Prokop, W.~Ong, S.~Quinn, D.P. Scriven, A.~Simon,
  C.~Sumithrarachchi, Phys. Rev. C \textbf{103}, 035803 (2021).
\newblock \doi{10.1103/PhysRevC.103.035803}.
\newblock \urlprefix\url{https://link.aps.org/doi/10.1103/PhysRevC.103.035803}

\bibitem{Dombos2021}
A.C. Dombos, A.~Spyrou, F.~Naqvi, S.J. Quinn, S.N. Liddick, A.~Algora,
  T.~Baumann, J.~Brett, B.P. Crider, P.A. DeYoung, T.~Ginter, J.~Gombas,
  S.~Lyons, T.~Marketin, P.~M\"oller, W.J. Ong, A.~Palmisano, J.~Pereira, C.J.
  Prokop, P.~Sarriguren, D.P. Scriven, A.~Simon, M.K. Smith, S.~Valenta, Phys.
  Rev. C \textbf{103}, 025810 (2021).
\newblock \doi{10.1103/PhysRevC.103.025810}.
\newblock \urlprefix\url{https://link.aps.org/doi/10.1103/PhysRevC.103.025810}

\bibitem{Karny2019}
{M. Karny, A. Fijałkowska, R.K. Grzywacz, B.C. Rasco, K.P. Rykaczewski, M.
  Stepaniuk}, Nucl. Instrum. Methods Phys. Res. B \textbf{463}, 390 (2020)

\bibitem{Rykaczewski2022}
{T.T. King, A. Laminack, B.C. Rasco, K.P. Rykaczewski, P. Shuai},  Invited
  paper at 55th Zakopane Conf. Nuclear Physics, Extremes of the Nuclear
  Landscape, 28 August - 4 September 2022, Zakopane, Poland

\bibitem{Fleming2015}
M.~Fleming, {J.-Ch. Sublet}, {Validation of FISPACT-II decay heat and inventory
  predictions for fission events}.
\newblock Tech. rep., UK Atomic Energy Authority, Culham Centre for Fusion
  Energy, Culham, UK (2015).
\newblock CCFE report CCFE-R(15)28. Also available online:
  scientific-publications.ukaea.uk/wp-content/uploads/CCFE-R1528.pdf

\bibitem{Fleming2015a}
M.~Fleming, {J.-Ch. Sublet}, {Decay data comparisons for decay heat and
  inventory simulations of fission events}.
\newblock Tech. rep., UK Atomic Energy Authority, Culham Centre for Fusion
  Energy, Culham, UK (2017).
\newblock CCFE report CCFE-R(15)28/S1, June 2015; all of the decay-heat data at
  cooling times of 5011 and 10000 s were corrected after finding a transfer
  error that occurred when these calculated datasets were assembled in
  tabulated form prior to publication. Also available online:
  https://scientific-publications.ukaea.uk/wp-content/uploads/CCFE-R1528\_S1.pdf

\bibitem{Fleming2015b}
M.~Fleming, {J.-Ch. Sublet}, {Fission yield comparisons for decay heat and
  inventory simulations of fission events}.
\newblock Tech. rep., UK Atomic Energy Authority, Culham Centre for Fusion
  Energy, Culham, UK (2015).
\newblock CCFE report CCFE-R(15)28/S2. Also available online:
  scientific-publications.ukaea.uk/wp-content/uploads/CCFE-R1528\_S2-1.pdf

\bibitem{AME2020}
M.~{Wang}, W.J. {Huang}, F.G. {Kondev}, G.~{Audi}, S.~{Naimi}, Chin. Phys. C
  \textbf{45}, 030003 (2021).
\newblock \doi{10.1088/1674-1137/abddaf}

\bibitem{ENSDF}
{Evaluated Nuclear Structure Data File (ENSDF)}.
\newblock \urlprefix\url{https://www.nndc.bnl.gov/ensdf/}.
\newblock Source: Nuclear Structure and Decay Data Evaluators Network

\bibitem{Chadwick2006}
M.B. Chadwick, P.~Obložinský, M.~Herman, N.M. Greene, R.D. McKnight, D.L.
  Smith, P.G. Young, R.E. MacFarlane, G.M. Hale, S.C. Frankle, A.C. Kahler,
  T.~Kawano, R.C. Little, D.G. Madland, P.~Möller, R.D. Mosteller, P.R. Page,
  P.~Talou, H.~Trellue, M.C. White, W.B. Wilson, R.~Arcilla, C.L. Dunford, S.F.
  Mughabghab, B.~Pritychenko, D.~Rochman, A.A. Sonzogni, C.R. Lubitz, T.H.
  Trumbull, J.P. Weinman, D.A. Brown, D.E. Cullen, D.P. Heinrichs, D.P. McNabb,
  H.~Derrien, M.E. Dunn, N.M. Larson, L.~Leal, A.D. Carlson, R.C. Block, J.B.
  Briggs, E.T. Cheng, H.C. Huria, M.L. Zerkle, K.S. Kozier, A.~Courcelle, V.G.
  Pronyaev, S.C. van~der Marck, Nucl. Data Sheets \textbf{107}, 2931  (2006)

\bibitem{Chadwick2011}
M.B. Chadwick, M.~Herman, P.~Obložinský, M.E. Dunn, Y.~Danon, A.C. Kahler,
  D.L. Smith, B.~Pritychenko, G.~Arbanas, R.~Arcilla, R.~Brewer, D.A. Brown,
  R.~Capote, A.D. Carlson, Y.S. Cho, H.~Derrien, K.~Guber, G.M. Hale,
  S.~Hoblit, S.~Holloway, T.D. Johnson, T.~Kawano, B.C. Kiedrowski, H.~Kim,
  S.~Kunieda, N.M. Larson, L.~Leal, J.P. Lestone, R.C. Little, E.A. McCutchan,
  R.E. MacFarlane, M.~MacInnes, C.M. Mattoon, R.D. McKnight, S.F. Mughabghab,
  G.P.A. Nobre, G.~Palmiotti, A.~Palumbo, M.T. Pigni, V.G. Pronyaev, R.O.
  Sayer, A.A. Sonzogni, N.C. Summers, P.~Talou, I.J. Thompson, A.~Trkov, R.L.
  Vogt, S.C. van~der Marck, A.~Wallner, M.C. White, D.~Wiarda, P.G. Young,
  Nucl. Data Sheets \textbf{112}, 2887  (2011)

\bibitem{Brown2018}
{D.A. Brown, M.B. Chadwick, R. Capote, A.C. Kahler, A. Trkov, M.W. Herman, A.A.
  Sonzogni, Y. Danon, A.D. Carlson, M. Dunn, D.L. Smith, G.M. Hale, G. Arbanas,
  R. Arcilla, C.R. Bates, B. Beck, B. Becker, F. Brown, R.J. Casperson, J.
  Conlin, D.E. Cullen, M.-A. Descalle, R. Firestone, T. Gaines, K.H. Guber,
  A.I. Hawari, J. Holmes, T.D. Johnson, T. Kawano, B.C. Kiedrowski, A.J.
  Koning, S. Kopecky, L. Leal, J.P. Lestone, C. Lubitz, J.I. Márquez Damián,
  C.M. Mattoon, E.A. McCutchan, S. Mughabghab, P. Navratil, D. Neudecker,
  G.P.A. Nobre, G. Noguere, M. Paris, M.T. Pigni, A.J. Plompen, B. Pritychenko,
  V.G. Pronyaev, D. Roubtsov, D. Rochman, P. Romano, P. Schillebeeckx, S.
  Simakov, M. Sin, I. Sirakov, B. Sleaford, V. Sobes, E.S. Soukhovitskii, I.
  Stetcu, P. Talou, I. Thompson, S. van der Marck, L. Welser-Sherrill, D.
  Wiarda, M. White, J.L. Wormald, R.Q. Wright, M. Zerkle, G. Žerovnik, Y.
  Zhu}, Nucl. Data Sheets \textbf{148}, 1 (2018)

\bibitem{Kellett2009}
M.A. Kellett, O.~Bersillon, R.W. Mills, The {JEFF}-3.1/-3.1.1 radioactive decay
  data and fission yields sub-libraries ({JEFF Report 20}).
\newblock Tech. rep., OECD/Nuclear Energy Agency, NEA No. 6287 (2009)

\bibitem{Kellett2016}
{M.A. Kellett}, O.~Bersillon, EPJ Web Conf. ND2016 Int. Conf. Nucl. Data for
  Science and Technology \textbf{146}, 02009 (2017).
\newblock Editors: A.J.M. Plompen, {F.-J. Hambsch}, P. Schillebeeckx, W.
  Mondelaers, J. Heyse, S. Kopecky, P. Siegler, S. Oberstedt, 11-16 September
  2016, Bruges, Belgium

\bibitem{Plompen2020}
{A.J.M. Plompen}, O.~Cabellos, {C. De Saint Jean}, M.~Fleming, A.~Algora,
  M.~Angelone, P.~Archier, E.~Bauge, O.~Bersillon, A.~Blokhin, F.~Cantargi,
  A.~Chebboubi, C.~Diez, H.~Duarte, E.~Dupont, J.~Dyrda, B.~Erasmus,
  L.~Fiorito, U.~Fischer, D.~Flammini, D.~Foligno, {M.R. Gilbert}, {J.R.
  Granada}, W.~Haeck, {F.-J. Hambsch}, P.~Helgesson, S.~Hilaire, I.~Hill,
  M.~Hursin, R.~Ichou, R.~Jacqmin, B.~Jansky, C.~Jouanne, {M.A. Kellett}, {D.H.
  Kim}, {H.I. Kim}, I.~Kodeli, {A.J. Koning}, {A.Yu. Konobeyev}, S.~Kopecky,
  B.~Kos, A.~Krása, {L.C. Leal}, N.~Leclaire, P.~Leconte, {Y.O. Lee}, H.~Leeb,
  O.~Litaize, M.~Majerle, {J.I. Márquez Damián}, {F. Michel-Sendis}, {R.W.
  Mills}, B.~Morillon, G.~Noguère, M.~Pecchia, S.~Pelloni, P.~Pereslavtsev,
  {R.J. Perry}, D.~Rochman, A.~Röhrmoser, P.~Romain, P.~Romojaro, D.~Roubtsov,
  P.~Sauvan, P.~Schillebeeckx, {K.H. Schmidt}, O.~Serot, S.~Simakov,
  I.~Sirakov, H.~Sjöstrand, A.~Stankovskiy, {J.-Ch. Sublet}, P.~Tamagno,
  A.~Trkov, {S. van der Marck}, {F. Álvarez-Velarde}, R.~Villari, {T.C. Ware},
  K.~Yokoyama, G.~Žerovnik, Eur. Phys. J. A \textbf{56}, 181 (2020).
\newblock \doi{https://doi.org/10.1140/epja/s10050-020-00141-9}

\bibitem{Katakura2011}
J.~Katakura, {JENDL FP Decay Data File 2011 and Fission Yields Data File 2011}.
\newblock Tech. rep., Japan Atomic Energy Agency, JAEA-Data/Code 2011-025
  (2012)

\bibitem{Katakura2015}
J.~Katakura, F.~Minato,   (2015).
\newblock {JAEA}-Data/Code 2015-030

\bibitem{Iwamoto2021}
O.~Iwamoto, N.~Iwamoto, S.~Kunieda, F.~Minato, S.~Nakayama, Y.~Abe,
  K.~Tsubakihara, S.~Okumura, C.~Ishizuka, T.~Yoshida, S.~Chiba, N.~Otsuka,
  {J.-Ch. Sublet}, H.~Iwamoto, K.~Yamamoto, Y.~Nagya, K.~Tada, C.~Konno,
  N.~Matsuda, K.~Yokoyama, H.~Taninaka, A.~Oizumi, S.~Okita, G.~Chiba, S.~Sato,
  M.~Ota, S.~Kwon, to be published J. Nucl. Sci. Technol.  (2022)

\bibitem{ddep}
{Decay Data Evaluation Project (DDEP)}.
\newblock \urlprefix\url{{http://www.lnhb.fr/ddep_wg/}}

\bibitem{Perry2014}
{R.J. Perry}, {C.J. Dean}, {A.L. Nichols}, Nucl. Data Sheets \textbf{120}, 261
  (2014)

\bibitem{Minato2016}
F.~Minato, O.~Iwamoto, EPJ Web of Conferences \textbf{122}, 10001 (2016)

\bibitem{Jordan2013}
D.~Jordan, A.~Algora, J.L. Taín, B.~Rubio, J.~Agramunt, A.B. Perez-Cerdán,
  F.~Molina, L.~Caballero, E.~Nácher, A.~Krasznahorkay, M.D. Hunyadi,
  J.~Gulyás, A.~Vitéz, M.~Csatlós, L.~Csige, J.~Äystö, H.~Penttilä, I.D.
  Moore, T.~Eronen, A.~Jokinen, A.~Nieminen, J.~Hakala, P.~Karvonen,
  A.~Kankainen, A.~Saastamoinen, J.~Rissanen, T.~Kessler, C.~Weber,
  J.~Ronkainen, S.~Rahaman, {V.-V. Elomaa}, S.~Rinta-Antila, U.~Hager,
  T.~Sonoda, K.~Burkard, W.~Hüller, L.~Batist, W.~Gelletly, A.L. Nichols,
  T.~Yoshida, A.A. Sonzogni, K.~Peräjärvi, A.~Petrovici, K.W. Schmid,
  A.~Faessler, Phys. Rev. C \textbf{87}, 044318 (2013)

\bibitem{Fijalkowska2017}
A.~Fijałkowska, M.~Karny, K.P. Rykaczewski, B.C. Rasco, R.~Grzywacz, C.J.
  Gross, M.~Wolińska-Cichocka, K.C. Goetz, D.W. Stracener, W.~Bielewski,
  R.~Goans, J.H. Hamilton, J.W. Johnson, C.~Jost, M.~Madurga, K.~Miernik,
  D.~Miller, S.W. Padgett, S.V. Paulauskas, A.V. Ramayya, E.F. Zganjar, Phys.
  Rev. Lett. \textbf{119}, 052503 (2017).
\newblock \doi{10.1103/PhysRevLett.119.052503}.
\newblock
  \urlprefix\url{https://link.aps.org/doi/10.1103/PhysRevLett.119.052503}

\bibitem{Yoshida1999}
T.~Yoshida, T.~Tachibana, F.~Storrer, K.~Oyamatsu, J.~Katakura, J. Nucl. Sci.
  Technol. \textbf{36}, 135 (1999)

\bibitem{Algora2009}
A.~Algora, D.~Jordan, E.~Estévez, J.L. Taín, B.~Rubio, J.~Agramunt, A.B.
  Perez-Cerdán, F.~Molina, L.~Caballero, E.~Nácher, J.~Bernabeu, A.~Gadea,
  A.~Krasznahorkay, M.D. Hunyadi, J.~Gulyás, A.~Vitéz, M.~Csatlós, L.~Csige,
  J.~Äystö, H.~Penttilä, I.D. Moore, T.~Eronen, A.~Jokinen, A.~Nieminen,
  J.~Hakala, P.~Karvonen, A.~Kankainen, A.~Saastamoinen, J.~Rissanen,
  T.~Kessler, C.~Weber, J.~Ronkainen, S.~Rahaman, {V.-V. Elomaa}, U.~Hager,
  S.~Rinta-Antila, T.~Sonoda, K.~Burkard, W.~Hüller, J.~Döring, M.~Gierlik,
  R.~Kirchner, I.~Mukha, C.~Plettner, E.~Roeckl, D.~Cano-Ott, L.~Batist, J.J.
  Valiente, W.~Gelletly, T.~Yoshida, A.L. Nichols, A.~Sonzogni, K.~Peräjärvi,
  Rev. Mex. Fis. \textbf{55}, 6 (2009)

\bibitem{Algora2011}
A.~Algora, D.~Jordan, J.L. Taín, B.~Rubio, J.~Agramunt, L.~Caballero,
  E.~Nácher, A.B. Perez-Cerdán, F.~Molina, A.~Krasznahorkay, M.D. Hunyadi,
  J.~Gulyás, A.~Vitéz, M.~Csatlós, L.~Csige, J.~Äystö, H.~Penttilä,
  S.~Rinta-Antila, I.~Moore, T.~Eronen, A.~Jokinen, A.~Nieminen, J.~Hakala,
  P.~Karvonen, A.~Kankainen, U.~Hager, T.~Sonoda, A.~Saastamoinen, J.~Rissanen,
  T.~Kessler, C.~Weber, J.~Ronkainen, S.~Rahaman, {V.-V. Elomaa}, K.~Burkard,
  W.~Hüller, L.~Batist, W.~Gelletly, T.~Yoshida, A.L. Nichols, A.A. Sonzogni,
  K.~Peräjärvi, Proc. Int. Conf. Nuclear Data for Science and Technology,
  26-30 April 2010, Jeju Island, Republic of Korea, J. Korean Phys. Soc.
  \textbf{59}, no. 2, 1479 (2011)

\bibitem{Algora2011b}
A.~Algora, D.~Jordan, J.L. Taín, B.~Rubio, J.~Agramunt, L.~Caballero,
  E.~Nácher, A.B. Perez-Cerdán, F.~Molina, E.~Estévez, A.~Krasznahorkay,
  M.D. Hunyadi, J.~Gulyás, A.~Vitéz, M.~Csatlós, L.~Csige, J.~Äystö,
  H.~Penttilä, S.~Rinta-Antila, I.~Moore, T.~Eronen, A.~Jokinen, A.~Nieminen,
  J.~Hakala, P.~Karvonen, A.~Kankainen, U.~Hager, T.~Sonoda, A.~Saastamoinen,
  J.~Rissanen, T.~Kessler, C.~Weber, J.~Ronkainen, S.~Rahaman, {V.-V. Elomaa},
  K.~Burkard, W.~Hüller, L.~Batist, W.~Gelletly, T.~Yoshida, A.L. Nichols,
  A.A. Sonzogni, K.~Peräjärvi, AIP Proc. 3rd Int. Conf. Frontiers in Nuclear
  Structure, Astrophysics and Reactions (FINUSTAR3) 1377, 157-163  (2011).
\newblock Editors: P. Demetriou, R. Julin, S. Harissopulos, 23-27 August 2010,
  Rhodes, Greece.

\bibitem{Algora2014b}
A.~Algora, D.~Jordan, J.L. Taín, B.~Rubio, J.~Agramunt, L.~Caballero,
  E.~Nácher, A.B. Perez-Cerdán, F.~Molina, E.~Estévez, E.~Valencia,
  A.~Krasznahorkay, M.D. Hunyadi, J.~Gulyás, A.~Vitéz, M.~Csatlós, L.~Csige,
  T.~Eronen, J.~Rissanen, A.~Saastamoinen, I.D. Moore, H.~Penttilä, V.S.
  Kolhinen, K.~Burkard, W.~Hüller, L.~Batist, W.~Gelletly, A.L. Nichols,
  T.~Yoshida, A.A. Sonzogni, K.~Peräjärvi, Hyperfine Interactions
  \textbf{223}, 245 (2014)

\bibitem{Rice2017}
S.~Rice, A.~Algora, J.L. Tain, E.~Valencia, J.~Agramunt, B.~Rubio, W.~Gelletly,
  P.H. Regan, {A.-A. Zakari-Issoufou}, M.~Fallot, A.~Porta, J.~Rissanen,
  T.~Eronen, J.~\"Ayst\"o, L.~Batist, M.~Bowry, V.M. Bui, R.~Caballero-Folch,
  D.~Cano-Ott, {V.-V. Elomaa}, E.~Estevez, G.F. Farrelly, A.R. Garcia,
  B.~Gomez-Hornillos, V.~Gorlychev, J.~Hakala, M.D. Jordan, A.~Jokinen, V.S.
  Kolhinen, F.G. Kondev, T.~Mart\'{\i}nez, P.~Mason, E.~Mendoza, I.~Moore,
  H.~Penttil\"a, Z.~Podoly\'ak, M.~Reponen, V.~Sonnenschein, A.A. Sonzogni,
  P.~Sarriguren, Phys. Rev. C \textbf{96}, 014320 (2017).
\newblock \doi{10.1103/PhysRevC.96.014320}.
\newblock \urlprefix\url{https://link.aps.org/doi/10.1103/PhysRevC.96.014320}

\bibitem{Guadilla2019a}
V.~Guadilla, A.~Algora, J.L. Taín, J.~Agramunt, J.~Äystö, J.A. Briz,
  A.~Cucoanes, T.~Eronen, M.~Estienne, M.~Fallot, L.M. Fraile, E.~Ganioğlu,
  W.~Gelletly, D.~Gorelov, J.~Hakala, A.~Jokinen, M.D. Jordan, A.~Kankainen,
  V.S. Kolhinen, J.~Koponen, M.~Lebois, T.~Martínez, M.~Monserrate,
  A.~Montaner-Pizá, I.~Moore, E.~Nácher, S.E.A. Orrigo, H.~Penttilä,
  I.~Pohjalainen, A.~Porta, J.~Reinikainen, M.~Reponen, S.~Rinta-Antila,
  B.~Rubio, K.~Rytkönen, T.~Shiba, V.~Sonnenschein, A.A. Sonzogni,
  E.~Valencia, V.~Vedia, A.~Voss, J.N. Wilson, {A.-A. Zakari-Issoufou}, Phys.
  Rev. C \textbf{100}, 024311 (2019)

\bibitem{Guadilla2019c}
V.~Guadilla, J.L. Taín, A.~Algora, J.~Agramunt, D.~Jordan, M.~Monserrate,
  A.~Montaner-Pizá, E.~Nácher, {S.E.A. Orrigo}, B.~Rubio, E.~Valencia,
  M.~Estienne, M.~Fallot, {L. Le Mur}, {J.A. Briz}, A.~Cucoanes, A.~Porta,
  T.~Shiba, {A.-A. Zakari-Issoufou}, {A.A. Sonzogni}, J.~Äystö, T.~Eronen,
  D.~Gorelov, J.~Hakala, A.~Jokinen, A.~Kankainen, {V.S. Kolhinen}, J.~Koponen,
  {I.D. Moore}, H.~Penttilä, I.~Pohjalainen, J.~Reinikainen, M.~Reponen,
  S.~Rinta-Antila, K.~Rytkönen, V.~Sonnenschein, A.~Voss, {L.M. Fraile},
  V.~Vedia, E.~Ganioğlu, W.~Gelletly, M.~Lebois, {J.N. Wilson}, T.~Martínez,
  Phys. Rev. C \textbf{100}, 044305 (2019)

\bibitem{Guadilla2020}
V.~Guadilla, J.L. Taín, A.~Algora, J.~Agramunt, D.~Jordan, M.~Monserrate,
  A.~Montaner-Piz\'a, S.E.A. Orrigo, B.~Rubio, E.~Valencia, J.A. Briz,
  A.~Cucoanes, M.~Estienne, M.~Fallot, L.~Le~Meur, A.~Porta, T.~Shiba, A.A.
  Zakari-Issoufou, J.~\"Ayst\"o, T.~Eronen, D.~Gorelov, J.~Hakala, A.~Jokinen,
  A.~Kankainen, V.~Kolhinen, J.~Koponen, I.~Moore, H.~Penttil\"a,
  I.~Pohjalainen, J.~Reinikainen, M.~Reponen, S.~Rinta-Antila, K.~Rytk\"onen,
  V.~Sonnenschein, A.~Voss, L.M. Fraile, V.~Vedia,
  E.~Ganio\ifmmode~\breve{g}\else \u{g}\fi{}lu, W.~Gelletly, M.~Lebois, J.N.
  Wilson, T.~Martinez, E.~N\'acher, A.A. Sonzogni, Phys. Rev. C \textbf{102},
  064304 (2020)

\bibitem{Greenwood1992b}
{R.C. Greenwood, D.A. Struttmann, K.D. Watts}, Nucl. Instrum. Methods Phys.
  Res. A \textbf{317}, 175 (1992).
\newblock \doi{https://doi.org/10.1016/0168-9002(92)90607-6}.
\newblock
  \urlprefix\url{https://www.sciencedirect.com/science/article/pii/0168900292906076}

\bibitem{Guadilla2019b}
V.~Guadilla, A.~Algora, J.L. Taín, M.~Estienne, M.~Fallot, {A.A. Sonzogni},
  J.~Agramunt, J.~Äystö, {J.A. Briz}, A.~Cucoanes, T.~Eronen, {L.M. Fraile},
  E.~Ganioğlu, W.~Gelletly, D.~Gorelov, J.~Hakala, A.~Jokinen, {M.D. Jordan},
  A.~Kankainen, {V.S. Kolhinen}, J.~Koponen, M.~Lebois, {L. Le Meur},
  T.~Martínez, M.~Monserrate, A.~Montaner-Pizá, I.~Moore, E.~Nácher, {S.E.A.
  Orrigo}, H.~Penttilä, I.~Pohjalainen, A.~Porta, J.~Reinikainen, M.~Reponen,
  S.~Rinta-Antila, B.~Rubio, K.~Rytkönen, T.~Shiba, V.~Sonnenschein,
  E.~Valencia, V.~Vedia, A.~Voss, {J.N. Wilson}, {A.-A. Zakari-Issoufou}, Phys.
  Rev. Lett. \textbf{122}, 042502 (2019)

\bibitem{Tain2015b}
J.L. Taín, E.~Valencia, A.~Algora, J.~Agramunt, B.~Rubio, S.~Rice,
  W.~Gelletly, P.~Regan, {A.-A. Zakari-Issoufou}, M.~Fallot, A.~Porta,
  J.~Rissanen, T.~Eronen, J.~Äystö, L.~Batist, M.~Bowry, V.M. Bui,
  R.~Caballero-Folch, D.~Cano-Ott, {V.-V. Elomaa}, E.~Estévez, G.F. Farrelly,
  A.R. Garcia, B.~Gómez-Hornillos, V.~Gorlychev, J.~Hakala, M.D. Jordan,
  A.~Jokinen, V.S. Kolhinen, F.G. Kondev, T.~Martínez, E.~Mendoza, I.~Moore,
  H.~Penttilä, {Zs. Podolyák}, M.~Reponen, V.~Sonnenschein, A.A. Sonzogni,
  Phys. Rev. Lett. \textbf{115}, 062502 (2015)

\bibitem{Agramunt2016}
J.~Agramunt, J.L. Taín, {M. B. Gómez-Hornillos}, A.R. Garcia, F.~Albiol,
  A.~Algora, R.~Caballero-Folch, F.~Calviño, D.~Cano-Ott, G.~Cortés,
  C.~Domingo-Pardo, T.~Eronen, W.~Gelletly, D.~Gorelov, V.~Gorlychev,
  H.~Hakala, A.~Jokinen, M.D. Jordan, A.~Kankainen, V.~Kolhinen, L.~Kucuk,
  T.~Martínez, P.J.R. Mason, I.~Moore, H.~Penttilä, {Zs. Podolyák},
  C.~Pretel, M.~Reponen, A.~Riego, J.~Rissanen, B.~Rubio, A.~Saastamoinen, {A.
  Tarifeño-Saldivia}, E.~Valencia, Nucl. Instrum. Methods. Phys. Res. A
  \textbf{807}, 69 (2016)

\bibitem{Tain2017}
J.L. Taín, V.~Guadilla, E.~Valencia, A.~Algora, {A.-A. Zakari-Issoufou},
  S.~Rice, {L. Le Meur}, J.~Agramunt, J.~Äystö, L.~Batist, M.~Bowry, J.A.
  Briz, V.M. Bui, R.~Caballero-Folch, D.~Cano-Ott, A.~Cucoanes, {V.-V. Elomaa},
  T.~Eronen, E.~Estévez, M.~Estienne, M.~Fallot, G.F. Farrelly, L.M. Fraile,
  E.~Ganioğlu, A.R. Garcia, W.~Gelletly, B.~Gómez-Hornillos, D.~Gorelov,
  V.~Gorlychev, J.~Hakala, A.~Jokinen, M.D. Jordan, A.~Kankainen, V.S.
  Kolhinen, F.G. Kondev, J.~Koponen, M.~Lebois, T.~Martínez, P.~Mason,
  E.~Mendoza, M.~Monserrate, A.~Montaner-Pizá, I.~Moore, E.~Nácher, S.E.A.
  Orrigo, H.~Penttilä, {Zs. Podolyák}, I.~Pohjalainen, A.~Porta, P.H. Regan,
  J.~Reinikainen, M.~Reponen, S.~Rinta-Antila, J.~Rissanen, B.~Rubio,
  K.~Rytkönen, T.~Shiba, V.~Sonnenschein, A.A. Sonzogni, V.~Vedia, A.~Voss,
  J.N. Wilson, EPJ Web Conf. ND2016 Int. Conf. Nucl. Data for Science and
  Technology \textbf{146}, 01002 (2017).
\newblock Editors: A.J.M. Plompen, {F.-J. Hambsch}, P. Schillebeeckx, W.
  Mondelaers, J. Heyse, S. Kopecky, P. Siegler, S. Oberstedt, 11-16 September
  2016, Bruges, Belgium

\bibitem{Guadilla2017b}
V.~Guadilla, A.~Algora, J.L. Taín, J.~Agramunt, J.~Äystö, J.A. Briz,
  A.~Cucoanes, T.~Eronen, M.~Estienne, M.~Fallot, L.M. Fraile, E.~Ganioğlu,
  W.~Gelletly, D.~Gorelov, J.~Hakala, A.~Jokinen, M.D. Jordan, A.~Kankainen,
  V.S. Kolhinen, J.~Koponen, M.~Lebois, T.~Martínez, M.~Monserrate,
  A.~Montaner-Pizá, I.~Moore, E.~Nácher, S.E.A. Orrigo, H.~Penttilä,
  I.~Pohjalainen, A.~Porta, J.~Reinikainen, M.~Reponen, S.~Rinta-Antila,
  B.~Rubio, K.~Rytkönen, T.~Shiba, V.~Sonnenschein, A.A. Sonzogni,
  E.~Valencia, V.~Vedia, A.~Voss, J.N. Wilson, {A.-A. Zakari-Issoufou}, Acta
  Phys. Pol. B \textbf{48}, 529 (2017)

\bibitem{Tain2017b}
J.L. Taín, V.~Guadilla, E.~Valencia, A.~Algora, {A.-A. Zakari-Issoufou},
  S.~Rice, {L. Le Meur}, J.~Agramunt, J.~Äystö, L.~Batist, M.~Bowry, J.A.
  Briz, V.M. Bui, R.~Caballero-Folch, D.~Cano-Ott, A.~Cucoanes, {V.-V. Elomaa},
  T.~Eronen, E.~Estévez, M.~Estienne, M.~Fallot, G.~Farrelly, L.~Fraile,
  E.~Ganioğlu, A.R. Garcia, W.~Gelletly, B.~Gómez-Hornillos, D.~Gorelov,
  V.~Gorlychev, J.~Hakala, A.~Jokinen, M.D. Jordan, A.~Kankainen, V.S.
  Kolhinen, F.G. Kondev, J.~Koponen, M.~Lebois, T.~Martínez, P.~Mason,
  E.~Mendoza, M.~Monserrate, A.~Montaner-Pizá, I.~Moore, E.~Nácher, S.E.A.
  Orrigo, H.~Penttilä, {Zs. Podolyák}, I.~Pohjalainen, A.~Porta, P.H. Regan,
  J.~Reinikainen, M.~Reponen, S.~Rinta-Antila, J.~Rissanen, B.~Rubio,
  K.~Rytkönen, T.~Shiba, V.~Sonnenschein, A.A. Sonzogni, V.~Vedia, A.~Voss,
  J.N. Wilson, JPS Conf. Proc. 14th Int. Symp. Nuclei in the Cosmos (NIC2016)
  \textbf{14}, 010607 (2017)

\bibitem{Agramunt2017}
J.~Agramunt, J.L. Taín, F.~Albiol, A.~Algora, R.~Caballero-Folch, F.~Calviño,
  G.~Cortés, I.~Dillmann, T.~Eronen, A.R. Garcia, E.~Ganioğlu, W.~Gelletly,
  D.~Gorelov, V.~Guadilla, H.~Hakala, A.~Jokinen, A.~Kankainen, A.~Montaner,
  M.~Marta, E.~Mendoza, I.~Moore, C.~Nobs, {S.E.A. Orrigo}, H.~Penttilä,
  M.~Reponen, S.~Rinta-Antila, A.~Riego, B.~Rubio, A.~Saastamoinen,
  P.~Salvador-Castiñeira, A.~Tarifeño-Saldivia, A.~Tolosa, E.~Valencia, EPJ
  Web Conf. ND2016 Int. Conf. Nucl. Data for Science and Technology
  \textbf{146}, {01004} (2017).
\newblock \doi{10.1051/epjconf/201714601004}.
\newblock \urlprefix\url{https://doi.org/10.1051/epjconf/201714601004}.
\newblock Editors: A.J.M. Plompen, {F.-J. Hambsch}, P. Schillebeeckx, W.
  Mondelaers, J. Heyse, S. Kopecky, P. Siegler, S. Oberstedt, 11-16 September
  2016, Bruges, Belgium

\bibitem{Liang}
{J. Liang, Balraj Singh, E.A. McCutchan, I. Dillmann, M. Birch, A.A. Sonzogni,
  X. Huang, M. Kang, J. Wang, G. Mukherjee, K. Banerjee, D. Abriola, A. Algora,
  A.A. Chen, T.D. Johnson, K. Miernik}, Nucl. Data Sheets \textbf{168}, 1
  (2020).
\newblock \doi{https://doi.org/10.1016/j.nds.2020.09.001}

\bibitem{Algora2017}
A.~Algora, S.~Rice, V.~Guadilla, J.L. Taín, E.~Valencia, {A.-A.
  Zakari-Issoufou}, J.~Agramunt, J.~Äystö, L.~Batist, J.A. Briz, M.~Bowry,
  V.M. Bui, R.~Caballero-Folch, D.~Cano-Ott, A.~Cucoanes, T.~Eronen, {V.-V.
  Elomaa}, E.~Estévez, M.~Estienne, M.~Fallot, G.F. Farrelly, L.M. Fraile,
  M.~Fleming, E.~Ganioğlu, A.R. Garcia, W.~Gelletly, B.~Gómez-Hornillos,
  D.~Gorelov, V.~Gorlychev, J.~Hakala, A.~Jokinen, D.~Jordan, A.~Kankainen,
  V.S. Kolhinen, F.G. Kondev, J.~Koponen, M.~Lebois, T.~Martínez, P.~Mason,
  E.~Mendoza, M.~Monserrate, {A. Montaner-Piza}, I.~Moore, E.~Nácher, S.E.A.
  Orrigo, H.~Penttilä, {Zs. Podolyák}, I.~Pohjalainen, A.~Porta, P.H. Regan,
  J.~Reinikainen, M.~Reponen, S.~Rinta-Antila, J.~Rissanen, B.~Rubio,
  K.~Rytkönen, T.~Shiba, V.~Sonnenschein, A.A. Sonzogni, {J.-Ch. Sublet},
  V.~Vedia, A.~Voss, J.N. Wilson, EPJ Web Conf. ND2016 Int. Conf. Nucl. Data
  for Science and Technology \textbf{146}, 10001 (2017).
\newblock Editors: A.J.M. Plompen, {F.-J. Hambsch}, P. Schillebeeckx, W.
  Mondelaers, J. Heyse, S. Kopecky, P. Siegler, S. Oberstedt, 11-16 September
  2016, Bruges, Belgium

\bibitem{Zakari2014}
{A.-A. Zakari-Issoufou}, A.~Porta, M.~Fallot, A.~Algora, J.L. Taín,
  E.~Valencia, S.~Rice, J.~Agramunt, J.~Äystö, M.~Bowry, V.M. Bui,
  R.~Caballero-Folch, D.~Cano-Ott, V.~Eloma, E.~Estévez, G.F. Farrelly,
  A.~Garcia, W.~Gelletly, M.B. Gómez-Hornillos, V.~Gorlychev, J.~Hakala,
  A.~Jokinen, M.D. Jordan, A.~Kankainen, F.G. Kondev, T.~Martínez, E.~Mendoza,
  F.~Molina, I.~Moore, A.~Perez, {Zs. Podolyák}, H.~Penttilä, P.H. Regan,
  J.~Rissanen, B.~Rubio, {C. Weber and IGISOL staff}, EPJ Web Conf. Int. Nucl.
  Phys. (INPC2013) \textbf{66}, 10019 (2014).
\newblock {Editors: S. Lunardi, P.G. Bizzeti, C. Bucci, M. Chiari, A. Dainese,
  P. Di Nezza, R. Menegazza, A. Nannini, C. Signorini, J.J. Valiente-Dobon, 2-7
  June 2013, IUPAP, Firenze, Italy}

\bibitem{Fallot2017}
M.~Fallot, A.~Porta, {L. Le Meur}, J.A. Briz, {A.-A. Zakari-Issoufou},
  V.~Guadilla, A.~Algora, J.L. Taín, E.~Valencia, S.~Rice, V.M. Bui,
  S.~Cormon, M.~Estienne, J.~Agramunt, J.~Äystö, L.~Batist, M.~Bowry,
  R.~Caballero-Folch, D.~Cano-Ott, A.~Cucoanes, {V.-V. Elomaa}, T.~Eronen,
  E.~Estévez, G.F. Farrelly, L.M. Fraile, M.~Fleming, E.~Ganioğlu, A.R.
  Garcia, W.~Gelletly, M.B. Gómez-Hornillos, D.~Gorelov, V.~Gorlychev,
  J.~Hakala, A.~Jokinen, M.D. Jordan, A.~Kankainen, P.~Karvonen, V.S. Kolhinen,
  F.G. Kondev, J.~Koponen, M.~Lebois, T.~Martínez, P.~Mason, E.~Mendoza,
  F.~Molina, M.~Monserrate, A.~Montaner-Pizá, I.~Moore, E.~Nácher, S.E.A.
  Orrigo, H.~Penttilä, A.~Perez, {Zs. Podolyák}, I.~Pohjalainen, P.H. Regan,
  J.~Reinikainen, M.~Reponen, S.~Rinta-Antila, J.~Rissanen, B.~Rubio, T.~Shiba,
  V.~Sonnenschein, A.A. Sonzogni, {J.-Ch. Sublet}, V.~Vedia, A.~Voss, C.~Weber,
  J.N. Wilson, EPJ Web Conf. ND2016 Int. Conf. Nucl. Data for Science and
  Technology \textbf{146}, 10002 (2017).
\newblock Editors: A.J.M. Plompen, {F.-J. Hambsch}, P. Schillebeeckx, W.
  Mondelaers, J. Heyse, S. Kopecky, P. Siegler, S. Oberstedt, 11-16 September
  2016, Bruges, Belgium

\bibitem{Zakari2015}
{A.-A. Zakari-Issoufou}, M.~Fallot, A.~Porta, A.~Algora, J.L. Taín,
  E.~Valencia, S.~Rice, V.M. Bui, S.~Cormon, M.~Estienne, J.~Agramunt,
  J.~Äystö, M.~Bowry, J.A. Briz, R.~Caballero-Folch, D.~Cano-Ott,
  A.~Cucoanes, {V.-V. Elomaa}, T.~Eronen, E.~Estévez, G.F. Farrelly, A.R.
  Garcia, W.~Gelletly, M.B. Gómez-Hornillos, V.~Gorlychev, J.~Hakala,
  A.~Jokinen, M.D. Jordan, A.~Kankainen, P.~Karvonen, V.S. Kolhinen, F.G.
  Kondev, T.~Martínez, E.~Mendoza, F.~Molina, I.~Moore, A.B. Perez-Cerdán,
  {Zs. Podolyák}, H.~Penttilä, P.H. Regan, M.~Reponen, J.~Rissanen, B.~Rubio,
  T.~Shiba, A.A. Sonzogni, C.~Weber, Phys. Rev. Lett. \textbf{115}, 102503
  (2015).
\newblock \doi{10.1103/PhysRevLett.115.102503}.
\newblock
  \urlprefix\url{http://link.aps.org/doi/10.1103/PhysRevLett.115.102503}

\bibitem{Guadilla2017}
V.~Guadilla, A.~Algora, J.L. Taín, J.~Agramunt, D.~Jordan, A.~Montaner-Piz\'a,
  S.E.A. Orrigo, B.~Rubio, E.~Valencia, J.~Suhonen, O.~Civitarese,
  J.~\"Ayst\"o, J.A. Briz, A.~Cucoanes, T.~Eronen, M.~Estienne, M.~Fallot, L.M.
  Fraile, E.~Ganio\ifmmode~\breve{g}\else \u{g}\fi{}lu, W.~Gelletly,
  D.~Gorelov, J.~Hakala, A.~Jokinen, A.~Kankainen, V.~Kolhinen, J.~Koponen,
  M.~Lebois, T.~Martinez, M.~Monserrate, I.~Moore, E.~N\'acher, H.~Penttil\"a,
  I.~Pohjalainen, A.~Porta, J.~Reinikainen, M.~Reponen, S.~Rinta-Antila,
  K.~Rytk\"onen, T.~Shiba, V.~Sonnenschein, A.A. Sonzogni, V.~Vedia, A.~Voss,
  J.N. Wilson, A.A. Zakari-Issoufou, Phys. Rev. C \textbf{96}, 014319 (2017).
\newblock \doi{10.1103/PhysRevC.96.014319}.
\newblock \urlprefix\url{https://link.aps.org/doi/10.1103/PhysRevC.96.014319}

\bibitem{Rudstam1990B}
G.~Rudstam, P.~Johansson, O.~Tengblad, P.~Aagaard, J.~Eriksen, At. Data Nucl.
  Data Tables \textbf{45}, 239 (1990)

\bibitem{Huber2016}
P.~Huber, P.~Jaffke, Phys. Rev. Lett. \textbf{116}, 122503 (2016).
\newblock \doi{10.1103/PhysRevLett.116.122503}.
\newblock
  \urlprefix\url{https://link.aps.org/doi/10.1103/PhysRevLett.116.122503}

\bibitem{Algora2018b}
A.~Algora, J.L. Taín,   (2018).
\newblock IFIC, Universidad de Valencia, Spain; presentations at IAEA
  consultants' meeting

\bibitem{Algora2016a}
A.~Algora, B.~Rubio,   (2016).
\newblock RIKEN Experiment NP1612-RIBF147: Studies of the beta decay of
  $^{100}$Sn and its neighbours with a Total Absorption Spectrometer (TAS).
  Spokespersons: A. Algora, B. Rubio

\bibitem{Tain2020}
J.L. Taín, A.I. Morales, E.~Nácher,   (2020).
\newblock GSI Experiment S505: Investigation of the beta-strength crossing
  N=126 and the formation of the 3rd r-process abundance peak. Spokespersons:
  J.L. Taín, A.I. Morales and E. Nácher

\bibitem{Fijalkowska2014a}
A.~Fijałkowska, M.~Karny, K.P. Rykaczewski, M.~Wolińska-Cichocka,
  R.~Grzywacz, C.J. Gross, J.W. Johnson, B.C. Rasco, E.F. Zganjar, D.W.
  Stracener, C.~Jost, K.C. Goetz, R.~Goans, E.~Spejewski, L.~Cartegni,
  M.~Madurga, K.~Miernik, D.~Miller, S.W. Padgett, S.V. Paulauskas,
  M.~Al-Shudifat, J.H. Hamilton, A.V. Ramayya, Nucl. Data Sheets \textbf{120},
  26 (2014).
\newblock Includes preliminary results for the \be~decay of $^{89}$Kr and
  $^{139}$Xe.

\bibitem{Fijalkowska2018}
A.~Fijałkowska, Acta Phys. Pol. B \textbf{49}, 399 (2018)

\bibitem{Rasco2015a}
B.C. Rasco, A.~Fijałkowska, M.~Karny, K.P. Rykaczewski, M.~Wolińska-Cichocka,
  K.C. Goetz, R.K. Grzywacz, C.J. Gross, K.~Miernik, S.V. Paulauskas, JPS Conf.
  Proc. Advances in Radioactive Isotope Science (ARIS2014), 6, 030018  (2015)

\bibitem{Rasco2017b}
B.C. Rasco, A.~Fijałkowska, K.P. Rykaczewski, M.~Wolińska-Cichocka, M.~Karny,
  R.K. Grzywacz, K.C. Goetz, C.J. Gross, D.W. Stracener, E.F. Zganjar, J.C.
  Batchelder, J.C. Blackmon, N.T. Brewer, T.~King, K.~Miernik, S.V. Paulauskas,
  M.M. Rajabali, J.A. Winger, Acta Phys. Pol. B \textbf{48}, 507 (2017)

\bibitem{Wolinska2017}
M.~Wolińska-Cichocka, B.C. Rasco, K.P. Rykaczewski, N.T. Brewer, D.~Stracener,
  R.~Grzywacz, C.J. Gross, A.~Fijałkowska, K.C. Goetz, M.~Karny, T.~King, {Sh.
  Go}, E.A. McCutchan, C.~Nesaraja, A.A. Sonzogni, E.~Wang, J.A. Winger,
  Y.~Xiao, C.J. Zachary, E.F. Zganjar, EPJ Web Conf. ND2016 Int. Conf. Nucl.
  Data for Science and Technology \textbf{146}, 10005 (2017).
\newblock Editors: A.J.M. Plompen, {F.-J. Hambsch}, P. Schillebeeckx, W.
  Mondelaers, J. Heyse, S. Kopecky, P. Siegler, S. Oberstedt, 11-16 September
  2016, Bruges, Belgium

\bibitem{Guadilla2022}
V.~Guadilla, L.~Le~Meur, M.~Fallot, J.A. Briz, M.~Estienne, L.~Giot, A.~Porta,
  A.~Cucoanes, T.~Shiba, A.A. Zakari-Issoufou, A.~Algora, J.L. Taín,
  J.~Agramunt, D.~Jordan, M.~Monserrate, A.~Montaner-Piz\'a, E.~N\'acher,
  S.E.A. Orrigo, B.~Rubio, E.~Valencia, J.~\"Ayst\"o, T.~Eronen, D.~Gorelov,
  J.~Hakala, A.~Jokinen, A.~Kankainen, V.~Kolhinen, J.~Koponen, I.~Moore,
  H.~Penttil\"a, I.~Pohjalainen, J.~Reinikainen, M.~Reponen, S.~Rinta-Antila,
  K.~Rytk\"onen, V.~Sonnenschein, A.~Voss, L.M. Fraile, V.~Vedia,
  E.~Ganio\ifmmode~\breve{g}\else \u{g}\fi{}lu, W.~Gelletly, M.~Lebois, J.N.
  Wilson, T.~Martinez, A.A. Sonzogni, Phys. Rev. C \textbf{106}, 014306 (2022).
\newblock \doi{10.1103/PhysRevC.106.014306}.
\newblock \urlprefix\url{https://link.aps.org/doi/10.1103/PhysRevC.106.014306}

\bibitem{Rykaczewski2018}
K.P. Rykaczewski,   (2018).
\newblock Division of Physics, Oak Ridge National Laboratory, USA; presentation
  at IAEA consultants' meeting

\bibitem{Dombos2019}
A.C. Dombos, A.~Spyrou, F.~Naqvi, S.J. Quinn, S.N. Liddick, A.~Algora,
  T.~Baumann, J.~Brett, B.P. Crider, P.A. DeYoung, T.~Ginter, J.~Gombas,
  E.~Kwan, S.~Lyons, W.J. Ong, A.~Palmisano, J.~Pereira, C.J. Prokop, D.P.
  Scriven, A.~Simon, M.K. Smith, C.S. Sumithrarachchi, Phys. Rev. C
  \textbf{99}, 015802 (2019).
\newblock \doi{10.1103/PhysRevC.99.015802}.
\newblock \urlprefix\url{https://link.aps.org/doi/10.1103/PhysRevC.99.015802}

\bibitem{Kondev2019}
F.G. Kondev, D.J. Hartley, R.~Orford, J.A. Clark, G.~Savard, K.~Auranen, A.D.
  Ayangeakaa, S.~Bottoni, M.P. Carpenter, P.~Copp, K.~Hicks, C.R. Hoffman,
  R.V.F. Janssens, B.P. Kay, T.~Lauritsen, T.~Li, S.T. Marley, G.E. Morgan,
  G.~Mukherjee, S.~Nandi, W.~Reviol, J.~Sethi, D.~Seweryniak, S.~Stolze, J.~Wu,
  R.~Yadav, S.~Zhu, EPJ Web Conf. IV Int. Conf. Nuclear Structure and Dynamics
  (NSD2019) \textbf{223}, 01028 (2019).
\newblock Editors: G. de Angelis, L. Corradi, 13-17 May 2019, Venice, Italy

\bibitem{Hartley2018}
D.J. Hartley, F.G. Kondev, R.~Orford, J.A. Clark, G.~Savard, A.D. Ayangeakaa,
  S.~Bottoni, F.~Buchinger, M.~Burkey, M.P. Carpenter, P.~Copp, D.~Gorelov,
  K.~Hicks, C.R. Hoffman, C.~Hu, R.V.F. Janssens, J.~Klimes, T.~Lauritsen,
  J.~Sethi, D.~Seweryniak, K.~Sharma, S.~Zhu, Y.~Zhu, Phys. Rev. Lett.
  \textbf{120}, 182502 (2018)

\bibitem{Hartley2020}
D.J. Hartley, F.G. Kondev, J.A. Clark, G.~Savard, A.D. Ayangeakaa, S.~Bottoni,
  M.P. Carpenter, P.~Copp, K.~Hicks, C.R. Hoffman, R.V.F. Janssens,
  T.~Lauritsen, R.~Orford, J.~Sethi, S.~Zhu, Phys. Rev. C \textbf{101}, 044301
  (2020)

\bibitem{Orford2020}
R.~Orford, F.G. Kondev, G.~Savard, J.A. Clark, W.~Porter, D.~Ray, F.~Buchinger,
  M.~Burkey, D.~Gorelov, D.J. Hartley, J.~Klimes, K.~Sharma, A.~Valverde,
  X.L.Yan, Phys. Rev. C \textbf{102}, 011303 (2020)

\bibitem{Orford2020-1}
R.~Orford, J.A. Clark, G.~Savard, A.~Aprahamian, F.~Buchinger, M.~Burkey,
  D.~Gorelov, J.~Klimes, G.~Morgan, A.~Nystrom, W.~Porter, K.~Sharma, D.~Ray,
  K.~Sharman, Nucl. Instrum. Methods Phys. Res. B \textbf{463}, 491 (2020)

\bibitem{Dickens1980}
{J.K. Dickens, T.A. Love, J.W. McConnell, R.W. Peelle}, Nucl. Sci. Eng.
  \textbf{74}, 106 (1980)

\bibitem{Akiyama1982}
M.~Akiyama, S.~An, Proc. Int. Conf. Nuclear Data for Science and Technology,
  6-10 September 1982, Antwerp, Belgium, pp. 237--244 (1982)

\bibitem{Takahashi1969}
K.~Takahashi, M.~Yamada, Prog. Theor. Phys. \textbf{41}, 1470 (1969)

\bibitem{Koyama1970}
{S.-I. Koyama}, K.~Takahashi, M.~Yamada, Prog. Theor. Phys. \textbf{44}, 663
  (1970)

\bibitem{Takahashi1971}
K.~Takahashi, Prog. Theor. Phys. \textbf{45}, 1466 (1971)

\bibitem{Yoshida1981}
T.~Yoshida, R.~Nakasima, J. Nucl. Sci. Technol. \textbf{18}, 393 (1981)

\bibitem{Chrien1983}
R.E. Chrien, {T.W. Burrows (editors)}, Tech. rep., NEANDC specialists' meeting
  on yields and decay data of fission product nuclides, Brookhaven National
  Laboratory, Upton, NY (USA); 24-27 October 1983; BNL-51778.
\newblock Available from NTIS as DE85000967

\bibitem{Hagura2006}
N.~Hagura, T.~Yoshida, T.~Tachibana, J. Nucl. Sci. Technol. \textbf{43}, 497
  (2006)

\bibitem{Tasaka1983}
K.~Tasaka, H.~Ihara, M.~Akiyama, T.~Yoshida, Z.~Matumoto, R.~Nakasima, {JNDC
  nuclear data library of fission products}.
\newblock Tech. rep., Japan Atomic Energy Research Institute, JAERI 1287 (1983)

\bibitem{England1992}
{T.R. England}, {W.B. Wilson}, J.~Katakura, {F.M. Mann}, {R.E. Schenter}, {C.W.
  Reich}, {Decay data evaluation for ENDF/B-VI}.
\newblock Tech. Rep. LA-UR-92-3785 (1992).
\newblock \urlprefix\url{https://www.osti.gov/biblio/6965838}

\bibitem{conderc}
{CoNDERC: Compilation of Nuclear Data Experiments for Radiation
  Characterisation} (2019).
\newblock \urlprefix\url{http://www-nds.iaea.org/conderc/}

\bibitem{RADLIST}
{T.W. Burrows}, Tech. rep., NNDC, BNL, Brookhaven, USA: Medical Internal
  Radiation Dose (MIRD) as calculated by RadList program.
\newblock \urlprefix\url{https://www.nndc.bnl.gov/radlist}.
\newblock Online version/calculations

\bibitem{Tengblad1987}
O.~Tengblad, {K.-H. Beimer}, G.~Nyman, Nucl. Instrum. Methods Phys. Res. A
  \textbf{258}, 230  (1987).
\newblock \doi{https://doi.org/10.1016/0168-9002(87)90061-1}

\bibitem{Moeller2016}
P.~M{\"o}ller, A.J. Sierk, T.~Ichikawa, H.~Sagawa, At. Data Nucl. Data Tables
  \textbf{109-110}, 1 (2016)

\bibitem{fispact}
G.~Bailey, D.~Foster, P.~Kanth, M.~Gilbert, {The FISPACT-II user manual}.
\newblock Tech. Rep. UKAEA-CCFE-RE(21)02, UK Atomic Energy Authority, Culham,
  UK (2021)

\bibitem{Schier1997}
W.A. Schier, G.P. Couchell, {\be$^{-}$ and gamma decay heat measurements
  between 0.1 s - 50,000 s for neutron fission of $^{235}$U, $^{238}$U, and
  $^{239}$Pu: final report}.
\newblock Tech. Rep. DOE/ER/40723-4, University of Massachusetts Lowell (1997)

\bibitem{Ohkawachi2001}
Y.~Ohkawachi, A.~Shono, in \emph{Proc. 2000 Symposium on Nuclear Data.} (2001),
  Japan Atomic Energy Research Institute report JAERI-Conf 2001-006, IAEA
  report INDC(JPN)-188/U, pp. 121--125

\bibitem{leppanen2015}
J.~Leppänen, M.~Pusa, T.~Viitanen, V.~Valtavirta, T.~Kaltiaisenaho, Ann. Nucl.
  Energy \textbf{82}, 142 (2015).
\newblock \doi{10.1016/j.anucene.2014.08.024}.
\newblock \urlprefix\url{https://doi.org/10.1016/j.anucene.2014.08.024}

\bibitem{Akiyama1988}
M.~Akiyama, J.~Katakura, {Measured data of delayed gamma-ray spectra from
  fissions of $^{232}$Th, $^{233}$U, $^{235}$U, $^{238}$U and $^{239}$Pu by
  fast neutrons: tabular data}.
\newblock Tech. rep., Japan Atomic Energy Research Institute, JAERI-M 88-252
  (1988)

\bibitem{Oyamatsu1999}
K.~Oyamatsu, in \emph{Proc. 1998 Symposium on Nuclear Data} (1999), Japan
  Atomic Energy Research Institute report JAERI-Conf 99-002, IAEA report
  INDC(JPN)-182/U, pp. 234--239

\bibitem{jendl4}
{K. Shibata, O. Iwamoto, T. Nakagawa, N. Iwamoto, A. Ichihara, S. Kunieda, S.
  Chiba, K. Furutaka, N. Otuka, T. Ohsawa, T. Murata, H. Matsunobu, A. Zukeran,
  S. Kamada, J. Katakura}, J. Nucl. Sci. Technol. \textbf{48}, 1 (2011)

\bibitem{IAEA0817}
B.~Pritychenko, S.~Oberstedt, O.~Cabellos, R.~Vogt, {R. Capote Noy},
  S.~Okumura, T.~Kawano, {Summary Report of 1st Research Coordination Meeting
  of CRP on Updating Fission Yield Data for Applications}.
\newblock Tech. Rep. virtual meeting, 31 August - 4 September 2020, IAEA report
  INDC(NDS)-0817, IAEA, Vienna, Austria, April 2021.
\newblock
  \urlprefix\url{https://www-nds.iaea.org/publications/indc/indc-nds-0817.pdf}

\bibitem{NUBASE2020}
F.G. {Kondev}, M.~{Wang}, W.J. {Huang}, S.~{Naimi}, G.~{Audi}, Chin. Phys. C
  \textbf{45}, 030001 (2021).
\newblock \doi{10.1088/1674-1137/abddae}

\bibitem{Dimitriou2021}
P.~Dimitriou, I.~Dillmann, {Balraj Singh}, V.~Piksaikin, K.P. Rykaczewski, J.L.
  Taín, A.~Algora, K.~Banerjee, I.N. Borzov, D.~Cano-Ott, S.~Chiba, M.~Fallot,
  D.~Foligno, R.~Grzywacz, X.~Huang, T.~Marketin, F.~Minato, G.~Mukherjee, B.C.
  Rasco, A.~Sonzogni, M.~Verpelli, A.~Egorov, M.~Estienne, L.~Giot,
  D.~Gremyachkin, M.~Madurga, E.A. McCutchan, E.~Mendoza, K.V. Mitrofanov,
  M.~Narbonne, P.~Romojaro, A.~Sanchez-Caballero, N.D. Scielzo, Nucl. Data
  Sheets \textbf{173}, 144 (2021)

\bibitem{bdnIAEA}
{Reference database for beta-delayed neutron emission},
  \url{https://www-nds.iaea.org/beta-delayed-neutron/database.html}

\end{thebibliography}

%\begin{thebibliography}
%\end{thebibliography}

\end{document}